\documentclass[11pt,preprint]{aastex}

\newcommand{\nhi}{\mbox{$N({\rm H\,I})$}}

\newcommand{\lya}{\mbox{${\rm Ly}\alpha$}}
\newcommand{\apg}{\gtrsim}

\newcommand{\fesc}{\mbox{$f_{\rm esc}$}}

\usepackage{graphicx,amsmath,amssymb,epsfig,psfig,arydshln}
\usepackage[T1]{fontenc}
\usepackage{multirow}

\shorttitle{Optical Afterglow Spectroscopy}
\shortauthors{Fynbo et al.}

\begin{document}


\title{Low-Resolution Spectroscopy of Gamma-ray Burst Optical Afterglows: 
Biases in the \emph{Swift} Sample and Characterization of the 
Absorbers\thanks{Based on
observations collected at the European Organisation for Astronomical Research
in the Southern Hemisphere, Chile, under programs 275.D-5022, 075.D-0270,
077.D-0661, 077.D-0805, 078.D-0416, 079.D-0429, 080.D-0526, 081.A-0135,
281.D-5002, and 081.A-0856. Also based on observations made with the Nordic
Optical Telescope, operated on the island of La Palma jointly by Denmark,
Finland, Iceland, Norway, and Sweden, in the Spanish Observatorio del Roque de
los Muchachos of the Instituto de Astrofisica de Canarias. Some of the data
obtained herein were obtained at the W.M. keck Observatory, which is operated
as a scientific partnership among the California Institute of Technology,
the University of California and the National Aeronautics and Space
Adminisration. The Observatory was made possible by the generous financial
support of the W.M. Keck foundation.}}

\author{
J. P. U. Fynbo\altaffilmark{2,3},
P. Jakobsson\altaffilmark{3},
J. X. Prochaska\altaffilmark{4},
D. Malesani\altaffilmark{2},
C. Ledoux\altaffilmark{5},
A. de Ugarte Postigo\altaffilmark{5},  
M. Nardini\altaffilmark{6},
P. M. Vreeswijk\altaffilmark{2,5},
K. Wiersema\altaffilmark{7},
J. Hjorth\altaffilmark{2},
J. Sollerman\altaffilmark{2,8},
H.-W. Chen\altaffilmark{9},
C. C. Th\"one\altaffilmark{2,10}, 
G. Bj\"ornsson\altaffilmark{3},
J. S. Bloom\altaffilmark{11},
A. Castro-Tirado\altaffilmark{12},
L. Christensen\altaffilmark{13},
A. De Cia\altaffilmark{3},
A. S. Fruchter\altaffilmark{14},
J. U. Gorosabel\altaffilmark{12},
J. F. Graham\altaffilmark{14},
A. O. Jaunsen\altaffilmark{2},
B. L. Jensen\altaffilmark{2},
D. A. Kann\altaffilmark{15},
C. Kouveliotou\altaffilmark{16},
A. Levan\altaffilmark{17},
J. Maund\altaffilmark{2},
N. Masetti\altaffilmark{18},
B. Milvang-Jensen\altaffilmark{2},
E. Palazzi\altaffilmark{18},
D. A. Perley\altaffilmark{11},
E. Pian\altaffilmark{19},
E. Rol\altaffilmark{7},
P. Schady\altaffilmark{20},
R. Starling\altaffilmark{7},
N. Tanvir\altaffilmark{7},
D. J. Watson\altaffilmark{2},
D. Xu\altaffilmark{2},
T. Augusteijn\altaffilmark{21},
F. Grundahl\altaffilmark{22},
J. Telting\altaffilmark{21},
P.-O. Quirion\altaffilmark{22}
}

\altaffiltext{2}{Dark cosmology centre, Niels Bohr Institute, University of
Copenhagen, Juliane Maries Vej 30, DK-2100 Copenhagen O}
\altaffiltext{3}{Centre for Astrophysics and Cosmology, Science Institute,
University of Iceland, Dunhagi 5, IS-107 Reykjav\'ik, Iceland}
\altaffiltext{4}{Department of Astronomy and Astrophysics,
UCO/Lick Observatory;
University of California, 1156 High Street,
Santa Cruz, CA 95064}
\altaffiltext{5}{European Southern Observatory, Alonso de C\'ordova 3107,
Vitacura, Casilla 19001, Santiago 19, Chile}
\altaffiltext{6}{SISSA - Via Beirut 2/4, I-34014 Trieste, Italy}
\altaffiltext{7}{Department of Physics and Astronomy, University of Leicester, University Road, Leicester, LE1 7RH, UK}
\altaffiltext{8}{Department of Astronomy, The Oskar Klein Centre, Stockholm
University, 106 91 Stockholm, Sweden}
\altaffiltext{9}{Department of Astronomy \& Astrophysics and Kavli Institute for Cosmological Physics, University of Chicago, Chicago, IL, 60637, USA}
\altaffiltext{10}{INAF - Osservatorio Astronomico di Brera, Via Bianchi 46
I-23806 Merate, Italy}
\altaffiltext{11}{Department of Astronomy, University of California, Berkeley, CA 94720-3411, USA}
\altaffiltext{12}{IAA-CSIC, PO Box 03004, E-18080 Granada, Spain}
\altaffiltext{13}{European Southern Observatory, Karl-Schwarzschildstrasse 2, 
D-85748 Garching, Germany}
\altaffiltext{14}{Space Telescope Science Institute, Department of Physics and Astronomy, Johns Hopkins University, 3700 San Martin Drive, Baltimore, MD 21218, USA}
\altaffiltext{15}{Th\"uringer Landessternwarte Tautenburg, Sternwarte 5, 
D-07778 Tautenburg, Germany}
\altaffiltext{16}{NASA Marshall Space Flight Center, Huntsville, Alabama 35805, USA}
\altaffiltext{17}{Department of Physics, University of Warwick, Coventry CV4 7AL, UK}
\altaffiltext{18}{INAF - Istituto di Astrofisica Spaziale e Fisica Cosmica di Bologna, via Gobetti 101, 40129 Bologna,Italy}
\altaffiltext{19}{INAF-Trieste Astronomical Observatory, 34143 Trieste, Italy}
\altaffiltext{20}{The UCL Mullard Space Science Laboratory, Holmbury St Mary, Dorking, Surrey RH5 6NT, UK}
\altaffiltext{21}{Nordic Optical Telescope Apartado 474, 38700 Santa Cruz de La
Palma, Santa Cruz de Tenerife, Spain}
\altaffiltext{21}{Institute of Physics and Astronomy, University of Aarhus,
Ny Munkegade, DK-8000 Aarhus C}

\begin{abstract}
We present a sample of 77 optical afterglows (OAs) of {\it Swift} detected 
gamma-ray bursts (GRB) for which spectroscopic follow-up observations have 
been secured. Our first objective is to measure the redshifts of the bursts.  
For the majority (90\%) of the afterglows the redshifts have been determined 
from the spectra. We provide line-lists and equivalent widths (EWs) for all 
detected lines redward of Ly$\alpha$ covered by the spectra. In addition to 
the GRB absorption systems these lists include line strengths for a total of 
33 intervening absorption systems. We discuss to what extent the current 
sample of {\it Swift} bursts with OA spectroscopy is a biased subsample of 
all {\it Swift} detected GRBs. For that purpose we define an X-ray selected 
statistical sample of {\it Swift} bursts with optimal conditions for 
ground-based follow up from the period March 2005 to September 2008; 146 
bursts fulfill our sample criteria. 
We derive the redshift distribution for the statistical (X-ray selected) sample
and conclude that less than 18\% of {\it Swift} bursts can be at $z>7$.  We
compare the high energy properties (e.g. $\gamma$-ray (15--350 keV) fluence and
duration, X-ray flux and excess absorption) for three sub-samples of bursts in
the statistical sample: {\it i)} bursts with redshifts measured from OA
spectroscopy, {\it ii)} bursts with detected optical and/or near-IR afterglow,
but no afterglow-based redshift, and {\it iii)} bursts with no detection of the
OA. The bursts in group {\it i)} have slightly higher $\gamma$-ray fluences and
higher X-ray fluxes and significantly less excess X-ray absorption than bursts
in the other two groups. In addition, the fractions of dark bursts, defined as
bursts with an optical to X-ray slope $\beta_\mathrm{OX}<0.5$, is 14\% in group
{\it i)}, 38\% in group {\it ii)} and $>39$\% in group {\it iii)}.  For the
full sample the dark burst fraction is constrained to be in the range
25\%--42\%. From this we conclude that the sample of GRBs with OA spectroscopy
is not representative for all {\it Swift} bursts, most likely due to a bias
against the most dusty sight-lines. This should be taken into account when
determining, e.g., the redshift or metallicity distribution of GRBs and when
using GRBs as a probe of star formation. Finally, we characterize GRB
absorption systems as a class and compare them to QSO absorption systems, in
particular the damped Ly$\alpha$ absorption (DLA) systems. On average GRB
absorbers are characterized by significantly stronger EWs for \ion{H}{1} as
well as for both low and high ionization metal lines than what is seen in
intervening QSO absorbers.  However, the distribution of line strengths is very
broad and several GRB absorbers have lines with EWs well within the range
spanned by QSO-DLAs.  Based on the 33 $z>2$ bursts in the sample we place a
95\% confidence upper limit of $7.5$\% on the mean escape fraction of ionizing
photons from star-forming galaxies.

\end{abstract}

\keywords{dust, extinction --- galaxies: high-redshift --- gamma rays: bursts}

\section{Introduction}

The exploration of gamma-ray bursts (GRBs) has been among the most fascinating
one in the last decade of astrophysical research. After the breakthrough in
1997, when X-ray and optical afterglows were discovered \citep{C1997,PGG1997},
progress has been rapid. The connection between long-duration GRBs and
star-forming galaxies has been empirically well established. There is an
exclusive coincidence of long-duration GRBs with actively star-forming galaxies
\citep[e.g.,][]{hogg99, bloom02,fruchter06}, the majority of which show
elevated specific star-formation rates \citep{chg04}. In several cases
long-duration GRBs have been directly associated with SNe
\citep[e.g.,][]{hsm+03,smg+03,campana,WoosleyandBloom}. Only a decade after the
first afterglow detection GRBs have allowed to probe the Universe to redshifts
from almost 0 to larger than 8, representing larger look-back times than
accessible with any other class of astrophysical objects 
\citep{jakobsson06a,tanvir:090423,salvaterra:090423}.

An important objective is to use GRBs as cosmological probes to study star
formation primarily in the distant Universe. GRBs may be ideal probes as each
burst pinpoints the location of a single massive star.  Hence, GRBs may allow a
census of where massive stars are formed throughout the observable Universe
\citep[e.g.,][]{wijers98}. Through optical spectroscopy of the afterglows we
can measure metallicities, molecular content and kinematical properties of the
sight-lines to bursts within their hosts.  After the bursts have faded away,
imaging allows a detailed study of the host galaxies in emission, providing a census of the
classes of galaxies contributing to the global star-formation density as a
function of redshift. There are three main issues to consider in this context:
observational bias, intrinsic bias, and the influence the GRBs and their
afterglows may have on their environments.  Concerning intrinsic bias it is
believed that massive stars require a low metallicity to retain enough angular
momentum to form a GRB \citep[e.g.,][]{hirschi,WH06} and the GRBs hence are
biased tracers of star formation \citep[e.g.,][]{guetta07,kistler08}.
Nevertheless, to establish if the evidence supports a low metallicity bias it is
necessary to first establish the importance and possible implications of
observational bias, in particular as such a bias, if it exists, unavoidably
will impact the measured distribution of metallicities for GRBs \citep[see
also][]{berger07}. Finally, concerning the issue of the impact of the GRBs and
their afterglows on the gas along the line-of-sight it is now well established
that UV pumping affects intervening gas in the hosts
\citep{mirka06,vreeswijk07,delia09,delia09b,cedric:050730}. It is also expected on
theoretical grounds that GRB afterglows may destroy dust in the near
environment of the burst possibly out to as far as 100 pc from the GRB
\citep{fruchter01}.

To address the issue of observational bias it is necessary to build a sample
of GRBs for which the incompleteness is well understood and can be quantified.
In this paper we present a spectroscopic sample of long-duration GRBs 
discovered by the {\it Swift}
mission \citep{gehrels} for which follow-up optical spectroscopy has been
secured. The purpose of the paper is twofold: {\it i)} to discuss the GRB
absorbers as a class and compare this class of absorbers with QSO absorbers,
and {\it ii)} to discuss to which extent our current sample of GRBs with
measured redshifts from optical afterglow (OA) spectroscopy is a biased sub-sample
of all GRBs. GRB absorbers have been compared with other classes of absorbers
in several papers in the literature (e.g., Jensen et al.\ 2001; Savaglio et
al.\ 2003, Jakobsson et al.\ 2006b, Fynbo et al.\ 2006b, Savaglio 2006; Prochaska et al.\ 2007a, 2008b;
Fynbo et al.\ 2008a). However, most of these studies have been based on
samples containing only few and/or predominantly optically bright afterglows.
The issue of bias in the samples of GRBs with measured redshift has been
discussed in the literature before (e.g., Bloom 2003; Fiore et al.\ 2007; Coward
2009). Fiore et al.\ discussed the importance of selection effects in the
properties of the detection of the GRBs themselves, e.g.\ the sensitivity of
the triggering detector as a function of energy. Coward argued that there is a
learning curve at work in the following sense: over the years since the launch
of {\it Swift} the mean time taken to acquire spectroscopic redshifts for a GRB
afterglow has evolved to shorter times. He also finds a correlation
between the mean time before a spectroscopic redshift is secured and the
measured redshift suggesting that low redshift bursts were preferentially
missed in the first years of {\it Swift} operation.
We have been following the same follow-up
strategy since the launch of {\it Swift}. Furthermore, for the present study we
only include {\it Swift} bursts detected after March 2005 where {\it Swift}
operations were well established. Hence, for the present study there should be
no significant issue with a learning curve. Here we will rather try to infer to which
extent the sample of bursts with follow-up optical spectroscopy could be biased
against, e.g., dusty sight-lines.

The paper is organized in the following way: In Sect.~\ref{sampledef} we define
our sample and provide a list of the GRBs for which we present spectra in this
paper. Sect.~\ref{obs} describes our observations and data reduction.  
In Sect.~\ref{redshifts} we present the
redshift distribution of the statistical sample. 
In sect.~\ref{bias} we discuss biases in the sample of bursts with optical
afterglow spectroscopy.
In Sect.~\ref{grbsasclass} we discuss GRB absorbers as a class and compare them
to QSO absorbers. 
In Sect.~\ref{intervening} we briefly discuss the intervening absorbers in the
sample. 
Finally in Sect.~\ref{conclusion} we offer our conclusions.  
In Appendices~\ref{notes} and \ref{tablesandspectra} we provide line-lists and 
plots of the spectra for the 77 optical GRB afterglows in our sample and 
provide notes on individual objects.

Throughout this paper we assume a flat cosmology with $\Omega_{\Lambda}=0.70$,
$\Omega_m = 0.30$, and a Hubble constant of $H_0 = 70$ km s$^{-1}$ Mpc$^{-1}$.

\section{The Sample}
\label{sampledef}

The aim of the sample selection is to construct a sample of long GRBs that is
selected independent of the optical properties of the afterglows and at the same
time has as high completeness in high quality optical follow-up as possible.
We find that the optimal way to build such a sample is by including all {\it
Swift} GRBs fulfilling the following criteria \citep{jakobsson06a}: 

\begin{enumerate} 
\item{A {\it Swift} detected GRB with observed duration $T_\mathrm{90} > 2$ s}

We wish to build a sample of long-duration GRBs known to be associated
with massive stellar death. We could choose to make a stronger cut,
e.g.\ $T_\mathrm{90} > 5$~s, in order to avoid contamination from the long tail
of the short-duration bursts. However, the number of bursts with $2 <  T_{90} <
5$ s is so small that this would not make a significant difference for the 
statistical properties of the sample. Therefore we choose to keep the 
standard operational definition of long GRBs \citep{chryssa93}.

\item{XRT afterglow position distributed within 12 hr}

With this criterion we secure that a precise afterglow position is available 
quickly which is crucial for efficient ground based follow-up. This criterion
also excludes bursts close to the Moon.

\item{Small foreground Galactic extinction: $A_\mathrm{V}<0.5$ mag}

With this criterion we remove from the sample bursts with high Galactic
extinction. These bursts are typically also located in very crowded fields.
Removing these bursts from the sample does not introduce any 
bias on the intrinsic properties of the bursts.

\item{Favorable declination: $-70^{\circ} < \delta < 70^{\circ}$}

For bursts close to the poles the probability to secure ground based follow-up
is smaller and we therefore apply this declination cut to the sample.

\item{Sun-to-field distance larger than 55$^{\circ}$}

Bursts that are too close to the Sun cannot be observed from the ground
for very long. With this criterion we will have at least one hour during night
time to secure a spectrum within 24 hr after the burst.

\end{enumerate}

About 50\% of all {\it Swift} GRBs do not fulfill these criteria, primarily for
two reasons: first {\it Swift} has to point close to the Sun a significant
fraction of the time, and second the fraction of the sky with Galactic
A$_\mathrm{V} >
0.5$ mag is about 34\%. For bursts fulfilling the above criteria, we have
attempted to detect optical and near-infrared afterglows and to measure their
redshifts. Also, these bursts will have a high probability of being well
observed by other follow-up teams. 

\section{Observations and Data Reduction}
\label{obs}

In this paper we include bursts in the time interval from March 2005 to
September 2008. In this period, 146 bursts fulfilled our sample criteria (see
Table~A1). These we will refer to as the {\it statistical sample} in the 
following. For 69 of these we present spectroscopic observations in
this paper. 
Most of the spectra have been obtained through our
target-of-opportunity programs, but we also include a few spectra which we have
obtained from the ESO and Gemini archives.

The spectra were obtained using one of the following instruments: 
{\it i)} the Nordic Optical Telescope (NOT)
equipped with the Andalusian Faint Object Spectrograph and Camera (AlFOSC),
{\it ii)} the ESO-VLT equipped with either one of the two FOcal Reducer and low
dispersion Spectrographs (FORS1 and FORS2), or in rare cases the Ultraviolet and
Visual Echelle Spectrograph (UVES), 
{\it iii)} one of the Gemini telescopes equipped with one of the 
Gemini Multi-Object Spectrographs (GMOS-N and GMOS-S),
{\it iv)} the 3m Shane Telescope at Lick Observatory equipped with the
dual-channel Kast spectrometer,
{\it v)} the Keck telescope equipped with the Low Resolution Imaging 
Spectrograph (LRIS, Oke et al.\ 1995).
In Table~\ref{instruments} we provide further details
of each of the instrumental setups applied in this work.

The longslit spectra were reduced using standard methods for bias subtraction,
flat-fielding and wavelength calibration.  Most of the spectra have been flux
calibrated using observations of spectrophotometric standard stars observed
with the same setup as the afterglow spectra. For a few of the spectra no
standard star spectra were secured and here we instead provide normalized
spectra. Afterglows observed with UVES will be discussed in detail elsewhere
\citep[see, e.g.,][]{fox08,cedric:050730},
and for these bursts we here only show the spectra and provide equivalent
width (EW) measurements of the \ion{Si}{2},1526 and/or \ion{C}{4} lines.

\begin{deluxetable}{@{}llrrr@{}}
\tablecaption{Instrumental setups used for the spectroscopic data presented
in this work. Resolutions are given for a 1 arcsec slit even though we some
times used a slightly wider slit.
\label{instruments}}
\tablewidth{0pt}
\tablehead{
\colhead{Telescope} & \colhead{Instrument} & \colhead{Grating} & \colhead{Resolution} & \colhead{Range [\AA]}
}
\startdata
VLT                 & FORS1/2              & 300V            & 440   & 3400--9500 \\
VLT                 & FORS1/2              & 1400V           & 2100  & 4600--5900 \\
VLT                 & FORS1/2              & 1200R           & 2140  & 5800--7300 \\
VLT                 & FORS1/2              & 600I            & 1500  & 6600--9400 \\
VLT                 & UVES                 &                 & 45000 & \\
Gemini-S/N          & GMOS-S/N             & R831            & 2200  & 5300--9000 \\
                    &                      & R400            & 965   & 5300--9000 \\
                    &                      & R150            & 315   & 4500--10000 \\
                    &                      & B600            & 844   & 3400--6000 \\
Keck 1              & LRIS                 & R300            & 540   & 3900--8000 \\
                    &                      & R400            & 1100  & 5800--9200 \\
                    & HIRES                &                 & 45000 & \\
Shane-3m            & Kast                 & 600/7500        & 1000  & 3800--10000 \\
                    &                      & 800/3460        & 1500  & 3800--7310 \\
NOT                 & AlFOSC               & G4              & 355   & 3600--9000 \\

\enddata

\end{deluxetable}

The spectra have been obtained under very diverse observing conditions (see
Table~\ref{sample}). Given the transient nature of GRBs the afterglows often
have to be observed at high airmass, with poor seeing, through clouds and/or
with a large Moon phase. In a few cases we observed the afterglows in twilight. 

The full list of spectroscopic targets is given in Table~\ref{sample}. In the
table we also include 8 bursts that do not fulfill the sample criteria for
which we also have secured spectra. In the following we refer to the bursts in
Table~\ref{sample} as the {\it spectroscopic sample}.

\begin{deluxetable}{@{}lllcccclc@{}}
\tablecaption{The spectroscopic sample. We here list the burst names and details
of the spectroscopic observations. The column $\Delta t$ shows the time after trigger
when the spectroscopic observation was started. Mag$_\mathrm{acq}$ gives the approximate
magnitude (typically in the $R$-band) of the afterglow in the acquisition image.
\label{sample}}
\tablewidth{0pt}
\tablehead{
\colhead{GRB} &  \colhead{Instrument} & \colhead{Exptime} & \colhead{Airmass} & \colhead{Seeing} & \colhead{$\Delta t$} & \colhead{Mag$_\mathrm{acq}$} & \colhead{Redshift} & \colhead{Ref}\\
              &                       &  \colhead{(ks)}   &                   & \colhead{(arcsec)} & \colhead{(hr)}       &  & &  
}
\startdata
050319  & AlFOSC  &  2.4   &    1.1  & 1.3  &  34.5   &   21.0  & 3.2425 & (1) \\
050401  & FORS2   & 11.6   & 1.1--1.7& 0.7  &  14.7   &   23.3  & 2.8983 & (2) \\
050408  & GMOS-N  &  3.6   &         &      &         &   21.0  & 1.2356 & (3) \\
050730  & FORS2   &  1.8   &    1.2  & 1.5  &   4.1   &   17.8  & 3.9693 & (4) \\
050801  & LRIS    &  1.8   &    1.9  & --   &   5.7   &   20.7  & 1.38   & (5) \\
050802  & AlFOSC  &  4.8   &    1.2  & 0.7  &  11.4   &   20.5  & 1.7102 & (6) \\
050820A & UVES    & 12.1   &    2.1  & 1.0  &   0.5   &   16.0  & 2.6147 & (7) \\
050824  & FORS2   &  3.0   &    1.8  & 0.7  &   9.5   &   20.6  & 0.8278 & (8) \\
050908  & FORS1   &  3.6   &    1.1  & 0.6  &   1.6   &   20.5  & 3.3467 & (9) \\
050922C & AlFOSC  &  2.4   &    0.9  & 1.3  &   1.0   &   16.5  & 2.1995 & (1) \\
060115  & FORS1   &  3.6   & 1.3--1.6& 0.7  &   8.9   &   22.0  & 3.5328 & (10) \\
060124  & LRIS    &  1.0   &    1.6  & 1.0  &  16.1   &   19.5  & 2.3000 & (11) \\
060206  & AlFOSC  &  2.4   &    1.0  & 1.2  &   0.3   &   17.5  & 4.0559 & (12) \\
060210  & GMOS-N  &  3.0   &    1.1  & --  &   1.2   &   20.6  & 3.9133& (13) \\
060502A & GMOS-N  &  3.6   &    1.6  & --  &   5.2   &   21.2  & 1.5026 & (14) \\
060512  & FORS1   &  3.6   &    2.5  & 1.6  &   3.0   &   19.9  & 2.1   & (15) \\
060526  & FORS1   &  9.9   & 1.1--1.4& 1.3  &   8.8   &   19.5  & 3.2213& (1) \\
060604  & AlFOSC  &  1.2   &    1.7  & 1.0  &  10.0   &   21.5  & $\lesssim3$ & (16) \\
060607A & UVES    &  12.0  & 1.9--1.0& 1.0  &   0.1   &   14.7  & 3.0749 & (17) \\
060614  & FORS2   &  1.8   &    1.2  & 0.7  &  21.1   &   19.8  & 0.1257 & (18) \\
060707  & FORS2   &  5.4   &    1.0  & 1.1  &  34.4   &   22.4  & 3.4240 & (1) \\
060708  & FORS2   &  3.6   &    1.2  & 0.6  &  43.0   &   22.9  & 1.92   & (19) \\
060714  & FORS1   &  5.4   &    1.1  & 0.7  &   8.5   &   20.4  & 2.7108 & (1) \\
060719  & FORS2   &  2.4   &    1.1  & 2.2  &  50.0   &   24.5  & $\lesssim4.6$ & (5) \\
060729  & FORS2   &  5.4   & 2.0--2.6& 1.5  &  13.2   &   17.5  & 0.5428 & (20) \\
060807  & FORS1   &  7.2   &    1.8  & 0.8  &   9.5   &   22.9  & $\lesssim3.4$ & (21) \\
060908  & GMOS-N  &  1.8   &    1.2  & 1.6  &   2.0   &   19.8  & 1.8836 &  (22) \\
060927  & FORS1   &  5.4   &    1.2  & 1.5  &  12.5   &   24.0  & 5.4636 & (23) \\
061007  & FORS1   &  5.4   & 1.2--1.3& 0.9  &  17.4   &   21.5  & 1.2622 & (24) \\
061021  & FORS1   &  1.8   &    1.9  & 0.8  &  16.5   &   20.5  & 0.3463 & (25) \\
061110A & FORS1   &  5.4   & 1.4--1.8& 0.8  &  15.0   &   22.0  & 0.7578 & (26) \\
061110B & FORS1   &  3.6   & 1.3--1.5& 0.7  &   2.5   &   22.5  & 3.4344 & (27) \\
061121  & LRIS    &  1.2   &    1.2  & --  &   0.2   &   17.8  & 1.3145 & (28)  \\
070110  & FORS2   &  5.4   & 1.5--1.9& 1.0  &  17.6   &   20.8  & 2.3521 & (29) \\
070129  & FORS2   &  1.8   &    2.2  & 1.0  &   2.2   &   21.3  & $\lesssim3.4$ & (1) \\
070306  & FORS2   &  5.4   & 1.2--1.3& 1.0  &  34.0   &   23.1  & 1.4965 & (30) \\
070318  & FORS1   &  1.8   &    1.6  & 0.7  &  16.7   &   20.2  & 0.8397 & (31) \\
070419A & GMOS-N  &  2.4   & 1.2--1.3& --  &   0.8   &   20.4  & 0.9705 & (5) \\
070506  & FORS1   &  2.7   & 1.6--1.8& 1.1  &   4.0   &   21.0  & 2.3090 & (32) \\
070611  & FORS2   &  3.6   & 1.1--1.2& 1.0  &   7.7   &   21.0  & 2.0394 & (33) \\
070721B & FORS2   &  5.4   & 1.2--1.5& 1.2  &  21.6   &   24.3  & 3.6298 & (34) \\
070802  & FORS2   &  5.4   &    1.2  & 0.5  &   1.9   &   21.9  & 2.4541 & (35) \\
071020  & FORS2   &  0.6   &    2.0  & 1.0  &   2.0   &   20.4  & 2.1462 & (36) \\
071025  & HIRES   &  1.8   &    1.35 & --   &   --    &   --    & 5.2    & (1)  \\
071031  & FORS2   &  1.8   &    1.2  & 1.0  &   1.2   &   18.9  & 2.6918 & (37) \\
071112C & FORS1   &  3.6   &    1.7  & 1.2  &   9.0   &   21.9  & 0.8227 & (38) \\
071117  & FORS1   &  5.4   &    1.4  & 0.9  &   9.0   &   23.0  & 1.3308 & (39) \\
080210  & FORS2   &  1.2   & 1.8--2.4& 1.4  &   0.7   &   18.8  & 2.6419 & (40) \\
080310  & Kast    &  1.8   &    --   & --   &   --    &   17.0  & 2.4274 & (41) \\
080319B & FORS2   &  3.6   & 2.1--2.3& 1.0  &  26.0   &   20.5  & 0.9382 & (42) \\
080319C & GMOS-N  &  3.6   &    1.2  & 1.7  &   2.4   &   21.1  & 1.9492 & (43) \\
080330  & AlFOSC  &  1.8   &    1.8  & 1.5  &   0.8   &   17.6  & 1.5119 & (44) \\
080413B & FORS1   &  0.6   &    1.5  & 0.5  &   0.8   &   19.3  & 1.1014 & (45) \\
080520  & FORS2   &  5.4   &    1.2  & 0.7  &   7.3   &   23.0  & 1.5457 & (46) \\
080523  & FORS2   &  5.4   & 1.8--2.0& 1.0  &  10.9   &   23.   & $\lesssim3.0$ & (47) \\
080603B & AlFOSC  &  2.0   &    1.3  & 1.0  &   2.0   &   17.5  & 2.6892 & (48) \\
080604  & GMOS-N  &  1.8   &    1.3  & 1.0  &   1.6   &   21.9  & 1.4171 & (49) \\
080605  & FORS2   &  0.6   & 1.4--1.9& 0.9  &   1.7   &   20.4  & 1.6403 & (50) \\
080607  & LRIS    &  5.1   &    1.0  & 0.8  &   0.2   &   18.9  & 3.0368 &  (51) \\
080707  & FORS1   &  2.4   & 2.1--2.5& 2.2  &   1.1   &   19.6  & 1.2322 & (52) \\
080710  & GMOS-S  &  2.4   &    1.6  & --  &   2.2   &   --  & 0.8454 & (53) \\
080721  & FORS1   &  1.2   &    1.5  & 1.7  &  10.2   &   20.0  & 2.5914 & (54) \\
080804  & UVES    &  4.7   & 2.1--2.5& 0.9  &   0.8   &   19.0  & 2.2045 & (55) \\
080805  & FORS2   &  1.2   &    1.7  & 1.0  &   1.0   &   21.5  & 1.5042 & (56)  \\
080810  & AlFOSC  &  2.4   &    1.8  & 1.0  &  10.6   &   19.1  & 3.3604 & (57) \\
080905B & FORS2   &  0.6   &    1.3  & 1.7  &   8.3   &   20.2  & 2.3739 & (58) \\
080913  & FORS2   &  1.8   &    1.0  & 1.1  &   2.0   &   24.2  &  6.7  & (59) \\
080916A & FORS2   &  3.6   &    1.2  & 0.9  &  17.1   &   22.3  & 0.6887 & (60) \\
080928  & FORS2   &  1.8   &    1.6  & 1.1  &  15.5   &   20.4 & 1.6919 & (61) \\
060904B\tablenotemark{1} & FORS1  &  3.6   & 1.1--1.3&  0.7  &   5.1  &   19.9  & 0.7029 & (62) \\
060906\tablenotemark{1} & FORS1   &  4.8   &   2.0   &  0.9  &   1.0  &   20.0  & 3.6856 &  (1) \\
060926\tablenotemark{1} & FORS1   &  4.8   & 1.7--2.4&  1.0  &   7.7  &   23.0  & 3.2086 & (1) \\
070125\tablenotemark{1} & FORS2   &  1.8   &   1.8   &  0.8  &   21.0 &   18.8  & 1.5471 & (5) \\
070411\tablenotemark{1} & FORS2   &  3.0   & 1.4--1.7&  1.0  &   5.0  &   20.8  & 2.9538 & (63) \\
070508\tablenotemark{1} & FORS1   &  5.4   &   1.7   &  1.3  &   3.8  &   22.0  & $\lesssim3.0$ & (64) \\
080411\tablenotemark{1} & FORS1   &  0.6   & 2.3--2.6&  1.1  &   2.4  &   17.5  & 1.0301 & (65) \\
080413A\tablenotemark{1}& UVES    &  2.7   & 1.2--1.6&  0.7  &   3.7  &   19.0  & 2.4330 & (66) \\
\enddata
\tablenotetext{1}{Not part of the statistical sample}
\tablerefs{
(1) Jakobsson et al.\ (2006b); (2) Watson et al.\ (2006); (3) \citet{foley:050408};
(4) D'Elia et al.\ (2005); (5) This work;  (6) Fynbo et al.\ (2005); 
(7) Ledoux et al.\ (2005); (8) Sollerman et al.\ (2007); (9) Fugazza et al.\ (2005); 
(10) Piranomonte et al.\ (2006a); (11) \citep{X:060124}; (12) Fynbo et al.\ (2006b); 
(13) \citep{cucchiara:060210}; (14) \citep{cucchiara:060502A}; (15) \citep{starling:060512a};
(16) Castro-Tirado et al.\ (2006);
(17) \citet{cedric:060607A}; 
(18) Della Valle et al.\ (2006); (19) \citet{palli:060708,oates09}; 
(20) \citet{ct:060729}; 
(21) \citet{piranomonte:060926}; 
(22) \citep{rol:060908}; (23) \citep{alma:060927}; 
(24) \citet{palli:061007}; (25) \citet{ct:061021}; 
(26) \citet{ct:061110A,johan:061110A}; (27) \citet{johan:061110B}; 
(28) \citep{bloom:061121}; (29) \citet{andreas:070110}; (30) \citet{andreas:070306}; 
(31) \citet{andreas:070318}; (32) \citet{ct:070506}; (33) \citet{ct:070611};
(34) \citet{daniele:070721B}; (35) \citet{ardis:070802}; 
(36) \citet{palli:071020}; (37) \citet{cedric:071031}; 
(38) \citet{palli:071112C}; (39) \citet{palli:071117};
(40) \citet{palli:080210}; (41) \citep{X:080310};
(42) \citet{vreeswijk:080319B}; (43) \citep{wiersema:080319C}; 
(44) \citet{guidorzi};
(45) \citet{vreeswijk:080413B}; (46) \citet{palli:080520};
(47) \citet{fynbo:080523}; (48) \citet{fynbo:080603B}; 
(49) \citep{wiersema:080604};
(50) \citet{palli:080605}; (51) \citep{X:080607};
(52) \citet{fynbo:080707}; (53) \citep{perley:080710};
(54) \citet{starling:080721};
(55) \citet{thoene:080804}; (56) \citet{palli:080805};
(57) \citet{antonio:080810}; (58) \citet{vreeswijk:080905B};
(59) \citet{fynbo:080913,greiner:080913}; (60) \citet{fynbo:080916A};
(61) \citet{vreeswijk:080928}; (62) \citet{fugazza:060904}; (63) \citet{palli:070411};  
(64) \citet{palli:070508}; 
(65) \citet{thoene:080411}; (66) \citet{thoene:080413A}; 
}
\end{deluxetable}


\section{Results}
\label{results}

In Appendix~\ref{notes} we provide notes on each burst in our spectroscopic 
sample. In Tables~\ref{050319}--\ref{080928} in appendix B we provide 
linelists for all lines
detected redward of Ly$\alpha$ both from the GRB absorbers and for intervening
absorption systems. We provide measurements for all lines for which we can
measure the EWs with a signal-to-noise ratio higher than 2.
The EWs are measured in normalized spectra using an aperture
given by 2 times the resolution full-width-at-half-maximum (for unresolved
lines) or from where the profile reaches 1 on each side of the profile (for
resolved or blended lines). The error bar includes the statistical noise and
the error from the normalization. For the Ly$\alpha$ lines we do not provide
EWs. Instead we in Table~\ref{HItab} provide the \ion{H}{1} column densities
derived from Voigt-profile fits to the Ly$\alpha$ lines. In cases where several
lines are blended we provide the EWs for the full blend. In cases where three
or more lines are blended we have drawn a dashed line around the blend in the
tables to ease the reading.
In Fig.~\ref{fig050319} in the appendix we show 1- and 2-dimensional spectra of
each of the 77 bursts listed in Table~\ref{sample}.


\subsection{Redshift Distribution}
\label{redshifts}

Our first objective is to measure the true redshift distribution of {\it Swift}
GRBs and in particular to constrain the fraction of {\it Swift} bursts that
could be at very high redshifts (here $z>6$--7).
The error on the redshifts are typically a few permille.

As mentioned in Sect.~\ref{sample}, 146 {\it Swift} bursts fulfill our
selection criteria for the statistical sample in Sect.~\ref{sample}.
Table~\ref{statsample} in the appendix includes all 146 bursts and their
redshift constraints.  The optical and/or near-IR afterglow is detected for 108
of the bursts. This completeness of 74\% is much higher than for pre-{\it
Swift} samples where only about 30\% of the bursts have detected
optical/near-IR afterglows \citep[e.g.,][]{fynbo:2001,lazzati:2002}.  For the
full sample of all {\it Swift} bursts observed in the same period from March
2005 to September 2008, 198 out of 371 (53\%) have detected optical/near-IR
afterglows and this difference between the statistical sample and the full {\it
Swift} sample illustrates the motivation for our sample criteria.

The redshift is determined from the OA for 72 bursts in the statistical sample
(5 of these are photometric redshifts based on the detection of the Lyman-break
or Ly$\alpha$ break in the spectral energy distribution of the afterglow). For
an additional 12 the redshift is determined from the likely host galaxy. For 25
bursts an upper limit can be placed on the redshift through detection of the OA
and hence establishing an upper limit to the position of the Lyman-limit or
Ly$\alpha$ breaks. For the remaining 37 we have no constraints on the redshift
from the OA or host galaxy (four of these are detected in the J, H and/or
K-band, but these detection do not provide constraining, i.e.\ $z<10$, redshift
limits). For these bursts we follow \citet{grupe:2007} and assign a redshift
upper limit to bursts with excess (over Galactic) X-ray absorbing column
density above an equivalent \ion{H}{1} column density of $2\times10^{21}$
cm$^{-2}$ (also taking uncertainty into account). Grupe et al.\ used an upper
limit of $z=2$, but we will be slightly more conservative and assign an upper
limit of $z=3.5$ to these bursts (corresponding to an upper limit on the
intrinsic absorbing column of about $10^{23}$ cm$^{-2}$ for Solar metallicity).
10 bursts fulfill this criterion at 90\% confidence. We will return to the
issue of X-ray absorption in Sect.~\ref{bias} below. In Fig.~\ref{zdist} we
show the resulting redshift distribution for the full sample of 146 bursts.
Both the median and the mean of the measured redshifts is 2.2. The fraction of
$z>6$ bursts is constrained to be in the range 1--23\% (2--34 out of 146)
and the fraction of $z>7$ bursts are less than about 18\% (27 out of 146).
Based on detections of likely host galaxies of dark bursts\footnote{Note that
this work uses a somewhat different definition of dark bursts than the one we
use in Sect.~\ref{darkburstssec},}, \citet{perley09} constrain the fraction of
$z>7$ bursts further to $<7$\% at 90\% confidence. A similar conclusion is
reached from a study of a complete sample of GRB host galaxies (Hjorth et al.\
in preparation).

It has been argued that the redshift measurements of {\it Swift} GRBs show
evidence for a ``learning curve" in the sense that the mean time before a
spectrum is secured is decreasing and that the mean redshift is also decreasing
as a function of time since the launch of {\it Swift} \citep{coward}. In
Fig.~\ref{magdt} we plot the magnitude of the afterglow at acquisition against
the time of the burst and color code the points by redshift. The mean (median)
times after burst at the start of integration are 9.7 (9.5) hr, 9.6 (8.5) hr,
10.4 (7.7) hr, and 5.3 (2.0) hr for 2005, 2006, 2007 and 2008 respectively. The
corresponding median redshifts are 2.9, 2.7, 2.0, and 2.0. Hence, there does
seem to be a tendency for the bursts to be spectroscopically observed earlier
and for the median redshift to decrease.  However, the effect is very small and
the scatter is large. If we simply split the sample in two at the median time
after trigger of 5 hr then we find mean redshifts and standard deviations for
triggers before and after 5 hr are $<z> = 1.9$, $\sigma(z)= 1.2$ and
$<z> = 2.5$, $\sigma(z) = 1.2$, respectively. 

\begin{figure}
\psfig{figure=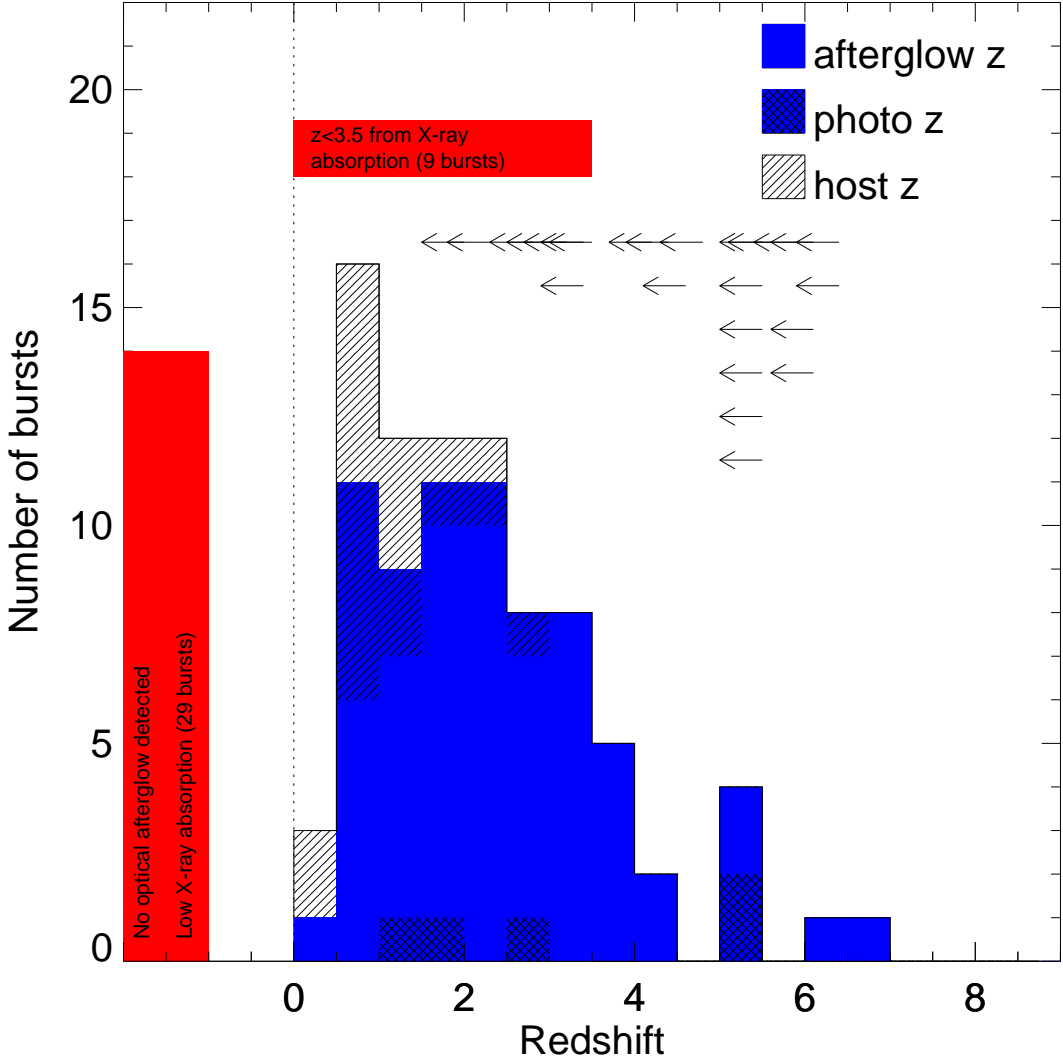,width=15cm}
\caption[]{Redshift distribution of 146 {\it Swift} GRBs localized with the
X-ray telescope and with low foreground extinction $A_\mathrm{V}<0.5$.  We
indicate with different colors and shadings bursts depending on whether the
redshift is based on afterglow spectroscopy, host emission lines (or both), or
a photometric redshift based on the OA broadband colors.  Bursts, for which
only an upper limit on the redshift could be established from photometry of the
OA, are indicated by arrows. The red histogram at the left indicates the 28
bursts with no OA detection, weak or absent X-ray absorption and no
redshift measurement from the underlying host galaxy. For these
bursts no redshift constraint could be inferred. The red block at the top
indicates the 10 bursts for which the OA was not detected, but an upper limit 
of $z = 3.5$ could be placed based on excess X-ray absorption.
}
\label{zdist}
\end{figure} 

\begin{figure}
\psfig{figure=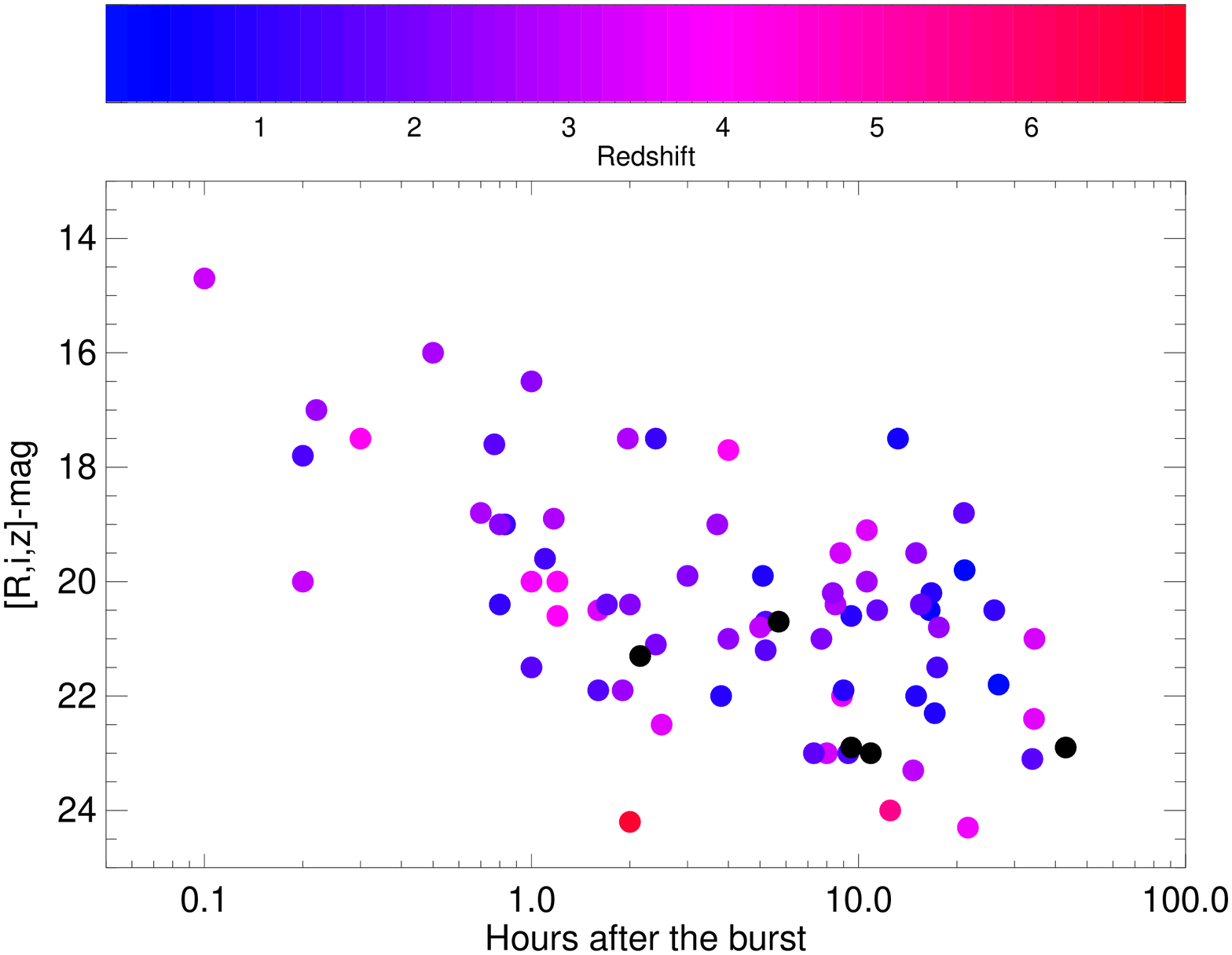,width=15cm}
\caption[]{
The $R$, $i$ or $z$-band magnitude of the OAs in the acquisition
image for the spectroscopy as a function of the time when the spectroscopic
observations was started. The color bar in the top indicates the color code
for the measured redshifts. Black points represents spectra for which we where
not able to determine the redshift.
}
\label{magdt}
\end{figure} 


\subsection{How Biased is the Optical Afterglow Spectroscopy Sample?}
\label{bias}

Whereas there is evidence that the QSO-DLAs drawn from samples of optically
selected QSOs are fairly representative for the full underlying population
\citep{ellison01,akerman05,ellison08} this may well not be the case for the
sample of GRB absorbers with OA spectroscopy. It was found soon after the
discovery of OAs that some bursts are very faint in the optical
\citep[e.g.,][]{Groot98}. There is growing evidence that such bursts are
mainly associated with dusty sight-lines
\citep[e.g.,][]{levan06,darach:050401,berger07,andreas:070306,tanvir:08,ardis:070802,X:080607,perley09,Zheng09}.
Due to its high fraction of optical follow-up (and X-ray selection) our
statistical sample is ideal to address this issue quantitatively.

In the following, we divide our
statistical sample into three groups: {\it i)} bursts with redshifts measured
from optical spectroscopy of the afterglow, {\it ii)} bursts with a detected
OA, but no afterglow-based redshift measurement, and
{\it iii)} bursts with no detection of the OA. 

\subsubsection{Distribution of the Sun Angle}
\begin{figure}
\psfig{figure=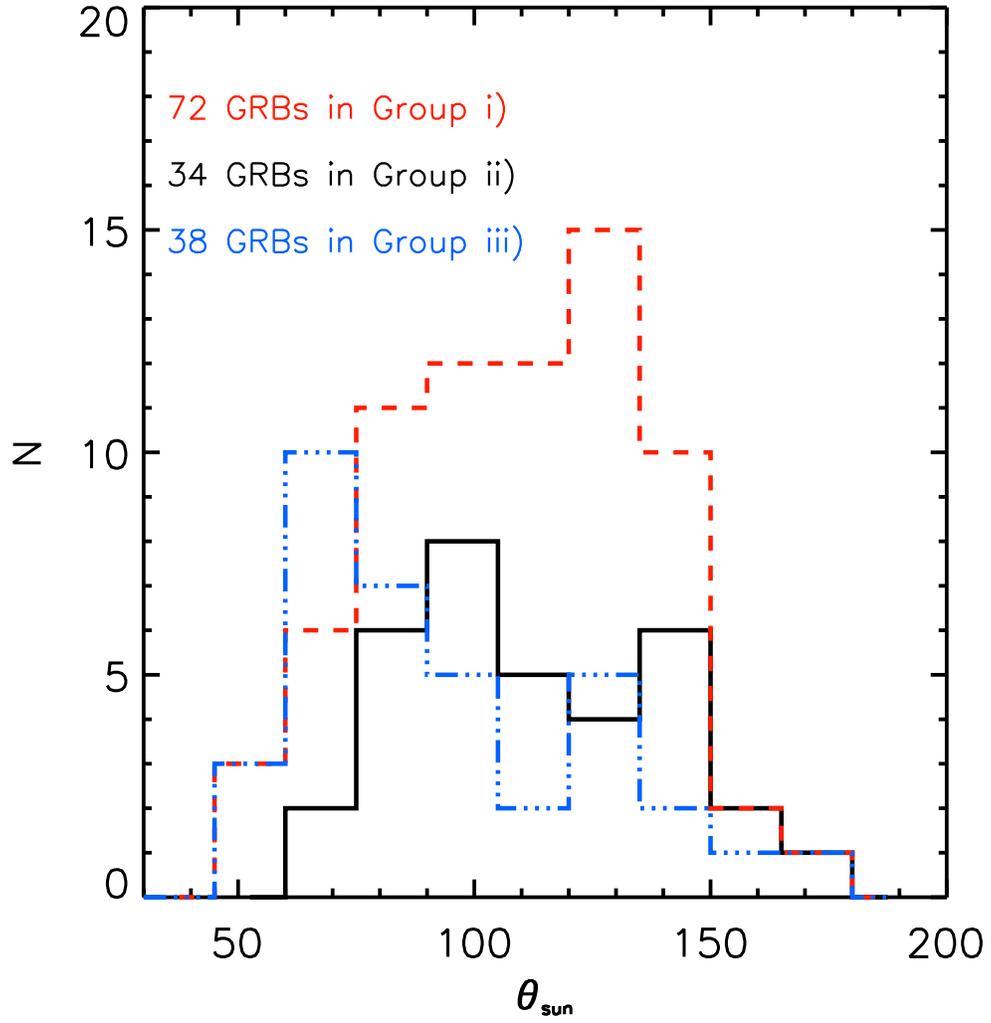,width=15cm}
\caption[]{Sun angle distribution for members of the statistical sample.}
\label{sunangle}
\end{figure}

As described in Sect.~\ref{sample} we have defined our sample to only contain
bursts with good conditions for optical/near-IR follow-up conditions.
Nevertheless, it is clear that all the bursts in the statistical sample did not
have equally good conditions for being observed from ground.  In
Fig.~\ref{sunangle} we compare the Sun angle distribution for the three
samples. The Sun angle is a good measure for how long a burst can be observed
during night time from the ground. It is clear that bursts with no OA detections
tends to be closer to the Sun and hence are more difficult to observe.  This is
one of the contributing reasons why these bursts do not have detected OAs.

\subsubsection{Distributions of High-Energy Properties}

\begin{figure}
\psfig{figure=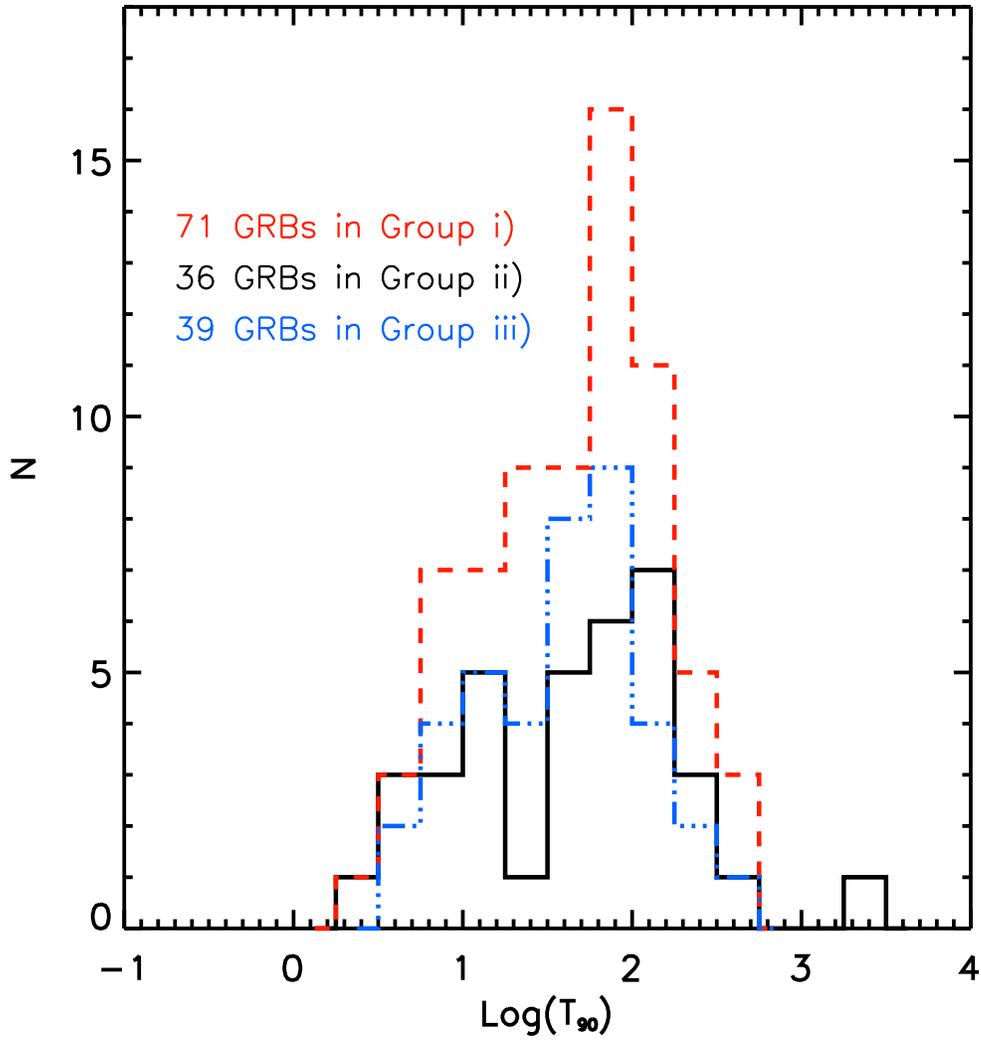,width=15cm}
\caption[]{BAT T$_{90}$ distribution for members of the statistical sample.}
\label{H21}
\end{figure}

\begin{figure}
\psfig{figure=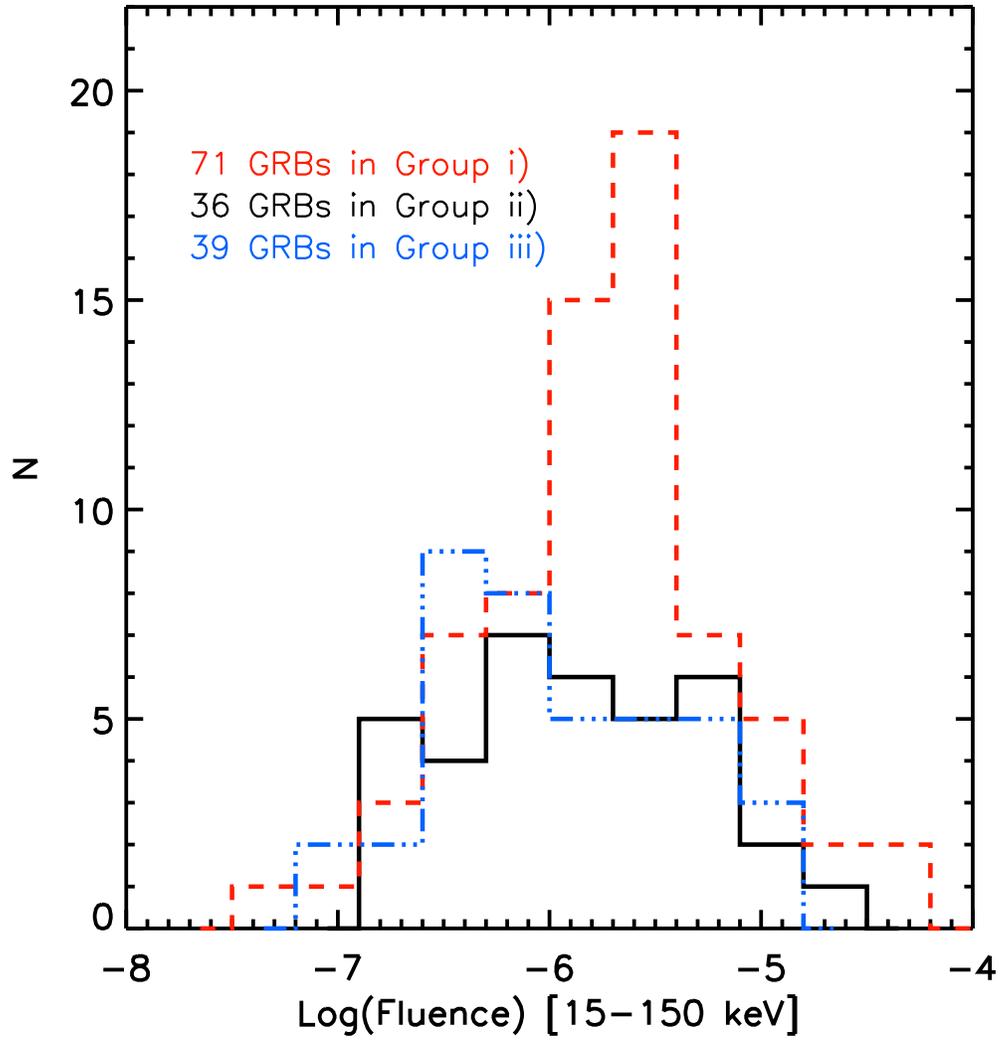,width=15cm}
\caption[]{BAT Fluence distribution for members of the statistical sample.}
\label{Fluence}
\end{figure}

\begin{figure}
\psfig{figure=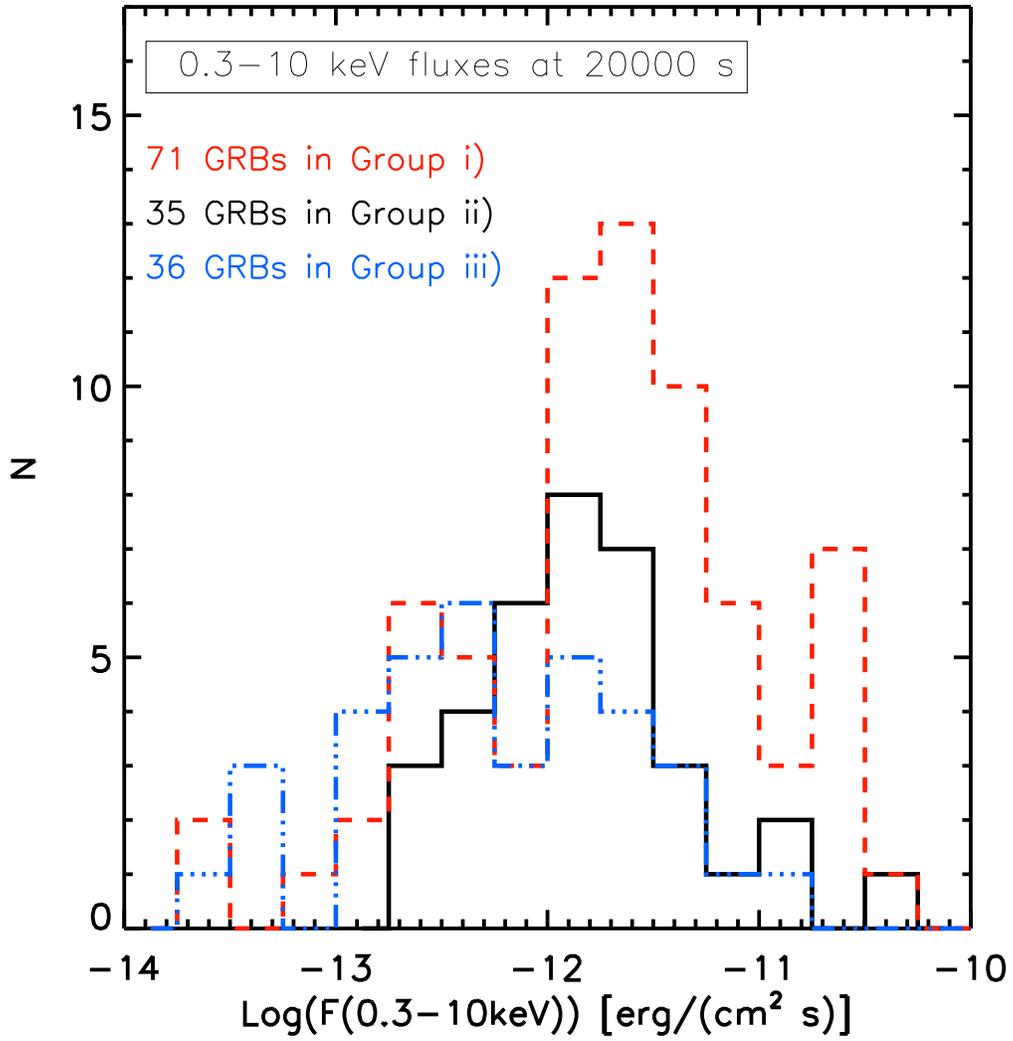,width=15cm}
\caption[]{XRT flux distribution at 20000 s post burst for members of the statistical sample.}
\label{XRT20000}
\end{figure}

\begin{figure}
\psfig{figure=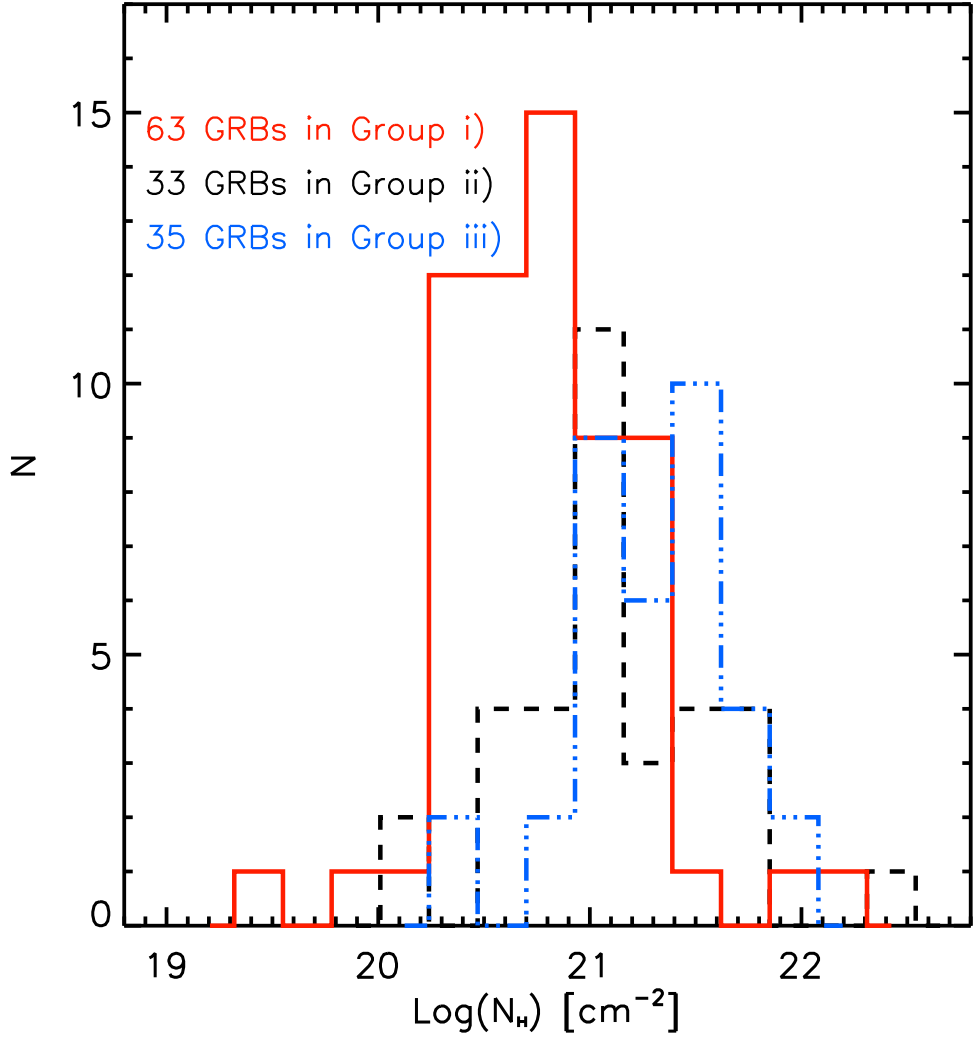,width=15cm}
\caption[]{XRT excess absorption distribution (assuming $z=0$) for members of the statistical sample.}
\label{XRTabs}
\end{figure}

In addition to the observational differences we are interested in revealing 
astrophysical differences between the bursts in the statistical sample.
Essentially all long GRBs have an X-ray afterglow. In particular, in our
statistical sample all bursts have available X-ray spectra from {\it Swift}
\citep[e.g.,][]{evans}.
It is therefore interesting to compare the X-ray properties such as excess
absorption and flux at fixed observed times for the the sample of GRBs with
follow-up optical spectroscopy with those that do not. 
It has
been found that the fluence of the prompt emission correlates with the flux of
the X-ray afterglow at fixed rest-frame times \citep{nysewander}. We therefore
compare both gamma-ray and X-ray properties of the three sub-samples. The results are
shown in Fig.~\ref{H21}--\ref{XRTabs} (note that the number of bursts in each group 
is slightly different in the three plots as some of the measurements were unavailable for
a few of the bursts). In Table~\ref{KStest} we provide the
KS-test based probabilities that these three sub-samples are drawn from the
same underlying distributions. It is clear that bursts in groups {\it ii)} and
{\it iii)} have fainter afterglows and in particular more excess absorption
than bursts in group {\it i)}. The most striking difference is the X-ray
excess for which it is firmly excluded that groups {\it i)} and {\it iii)}
are drawn from the same underlying distribution. A similar conclusion was
reached by \citet{schady07} based on a smaller sample.


\subsubsection{Dark Bursts}
\label{darkburstssec}
\begin{figure}
\psfig{figure=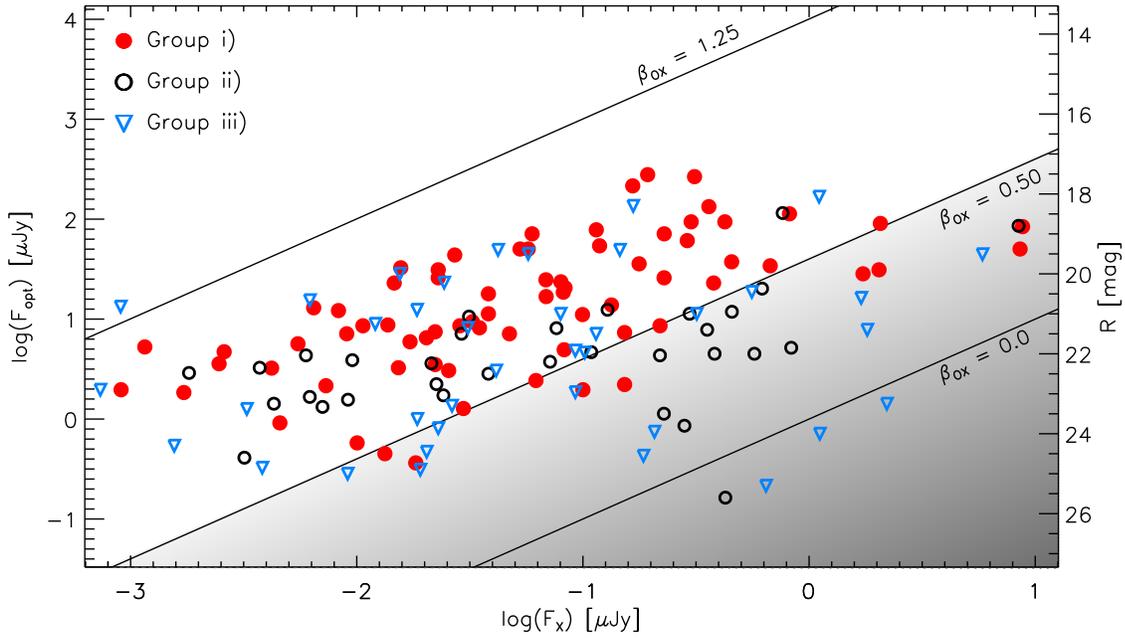,width=15cm}
\caption[]{
The dark burst diagram (F$_\mathrm{opt}$ vs. F$_\mathrm{X}$), first presented
in \citet{palli:dark}), for the statistical sample. GRB with 
$\beta_\mathrm{OX} < 0.5$ are defined as dark bursts. The $\beta_\mathrm{OX}$
values were calculated in an almost identical way as in \citet{palli:dark}. 
The only difference is that here we did not use 11\,hr as a reference time.
Rather, when possible, we selected measurements obtained a few hours after
a burst to avoid the early stage of the X-ray canonical behavior. For
low-$z$ bursts we also avoided late-time measurements to prevent any host
contamination in the OA flux.
}
\label{darkbursts}
\end{figure}

We follow the definition of dark bursts advocated by \citet{palli:dark},
whereby dark bursts are defined as bursts with an optical to X-ray spectral
slope $\beta_\mathrm{OX}<0.5$.
The fact that group {\it i)} is not representative for all bursts in the
statistical sample is confirmed by the fraction of dark bursts in the three
sub-samples (see Fig.~\ref{darkbursts}). For 4 bursts in the sample it was
not possible to calculate $\beta_\mathrm{OX}$ due to insufficient data (3 from
group {\it ii)} and 1 from group {\it iii)}). In group {\it i)} the fraction
of dark bursts is 14\% (10/72), in group {\it ii)} it is 38\% (12/32) and in
group {\it iii)} the fraction has a lower limit of 39\% (15/38). In the full
statistical sample the fraction of dark bursts is constrained to be in the
range 25--42\%.  Note that the $\beta_\mathrm{OX}<0.5$ definition of dark bursts
is conservative in the sense that a burst with an intrinsic
$\beta_\mathrm{OX}=1.25$ can suffer from $\sim 6$ mag extinction in
the observed $R$-band and still have an observed $\beta_\mathrm{OX}>0.5$.

\begin{figure}
\psfig{figure=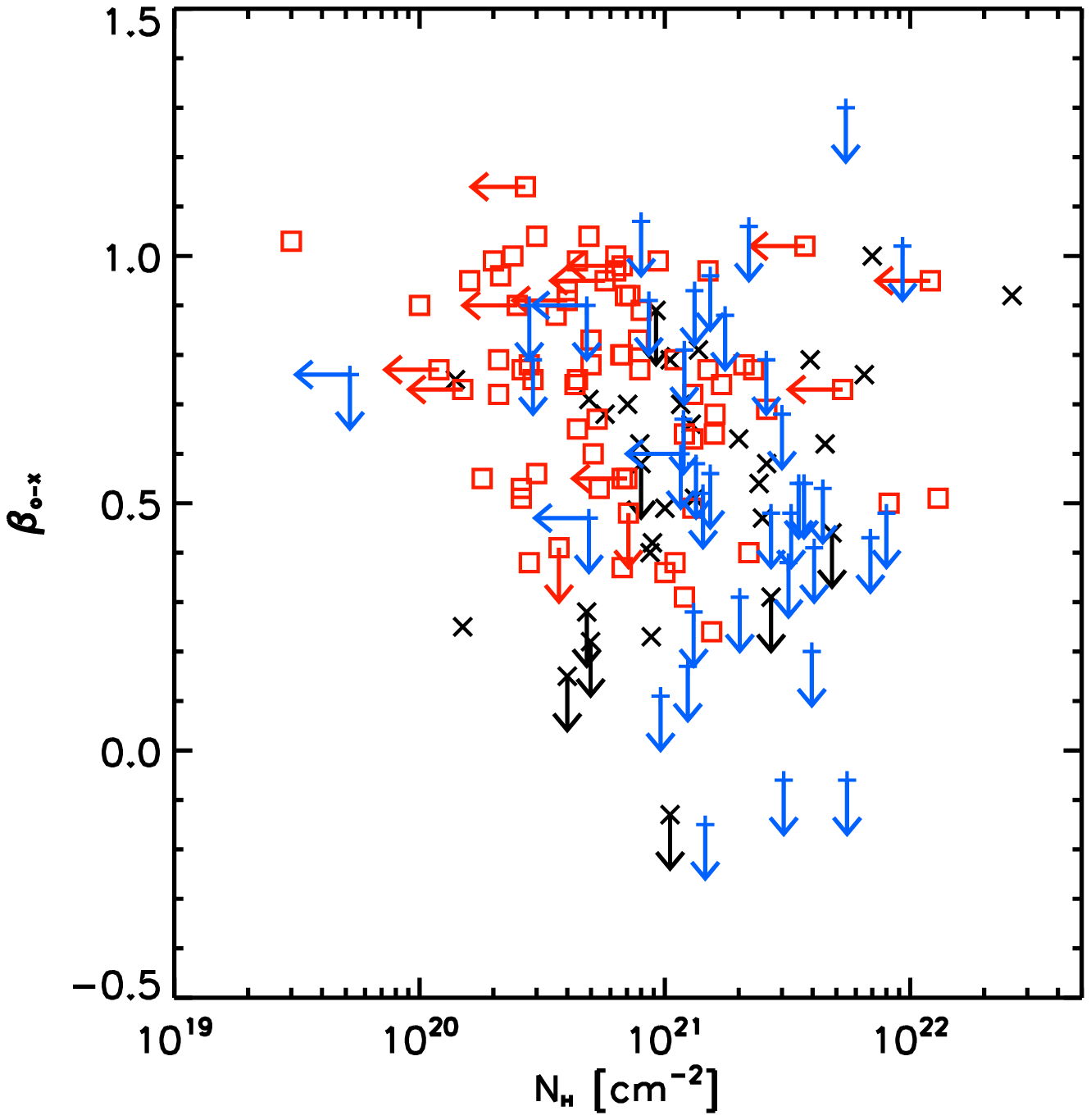,width=15cm}
\caption[]{
$\beta_\mathrm{OX}$ plotted against X-ray excess absorption (assuming $z=0$). 
Dark bursts 
(defined by $\beta_\mathrm{OX}<0.5$) tend to have large X-ray excess absorption
consistent with the interpretation that these bursts are obscured by dust.
}
\label{box}
\end{figure}

A contributing factor to the increasing ''darkness'' from groups {\it i)} to 
{\it iii)} could be a different fraction of bursts with cooling breaks 
between the
optical and X-ray bands \citep{pedersen:1025}. However, the fact that GRBs in
groups {\it ii)} and {\it iii}) have more excess absorption than bursts in group
{\it i)} shows that dust extinction of the optical light most likely is the
dominating factor for the higher dark burst fraction in groups {\it ii)} and
{\it iii)}. This is made clearer from Fig.~\ref{box} where we we plot
$\beta_\mathrm{OX}$ against excess absorption for the full statistical sample.
Bursts with $\beta_\mathrm{OX}<0.5$ also have higher excess absorption than
bursts with $\beta_\mathrm{OX}>0.5$.
The nine dark bursts in group {\it i)} are GRBs 050401, 050904, 060210,
070802, 080319C, 080605, 080607, 080805, and 080913. Two of these are dark due
to high redshifts ($z>6$) and the rest probably due to a combination of high
column densities and high metallicities and hence high dust column densities.

%

\begin{deluxetable}{@{}lcc@{}}
\tablecaption{KS test probabilities that the bursts in group {\it ii)} (OA
detected, but no OA based redshift) and group {\it iii)} (no OA detection) are
drawn from the same distribution as group {\it i)} (bursts with OA
based redshift measurement).
\label{KStest}}
\tablewidth{0pt}
\tablehead{
Property/group & {\it ii)} & {\it iii)}  
}

\startdata

0.3--10 keV flux 20000 s & 6.1$\times$10$^{-2}$ & 4.8$\times$10$^{-4}$ \\
X-ray excess abs & 2.1$\times$10$^{-3}$ & 7.0$\times$10$^{-7}$ \\
15-350 keV fluence & 0.21 & 3.5$\times$10$^{-2}$ \\
15-350 peak flux & 0.20 & 0.29 \\
T$_{90}$ & 0.76 & 0.71 \\

\enddata


\end{deluxetable}


\subsection{GRB Absorption Systems Compared to QSO Absorption Systems}
\label{grbsasclass}

QSO absorption systems have been studied since the discovery of QSOs and for
these systems a classification is well established \citep{Weymann81}.
Weymann et al.\ divided the QSO absorbers into four classes: {A)} associated
broad absorption line systems, {B)} associated narrow line systems, {C)}
intervening narrow metal line systems, and {D)} intervening Ly$\alpha$ forest
systems. Classes C and D are further divided into subclasses such as \ion{Mg}{2}
absorbers, \ion{C}{4} absorbers, DLAs, sub-DLAs, etc. Clearly, these classes
are not disjunct, i.e., DLAs are also \ion{Mg}{2} and \ion{C}{4} absorbers, but
the opposite is generally not true. The physical selection mechanism behind GRB
absorption systems is fundamentally different from those of intervening QSO
absorbers. Hence, the fact that GRB absorbers in many ways appear similar to
mainly DLA absorbers provides interesting new information not only of the
properties of the GRB host galaxies, but also about the physical origin of
intervening QSO absorbers. Specifically, the fact that QSO-DLAs and GRB
absorbers look so similar strongly suggests that DLAs also originate from the
interstellar medium of high-redshift galaxies. 

Interestingly, GRB absorbers 
have overlapping
properties not only with class C of Weymann et al., but also with their class
B. There are even similarities between the afterglow spectrum of GRB\,021004 
and associated QSO absorbers from class A, namely both very high blueshifts and
evidence for line-locking 
\citep[][but see also Chen et al.\ 2007b]{savaglio:021004, moller:021004}.

Using our large sample we will here try to characterize the GRB absorbers 
as a class in terms of their absorption line strengths. The low resolution 
of most of our data prevents us from comparing properties such
as metallicities or kinematical structure. 

\subsubsection{Comparison of GRB- and QSO-DLA Absorption Systems}

\begin{figure}
\psfig{figure=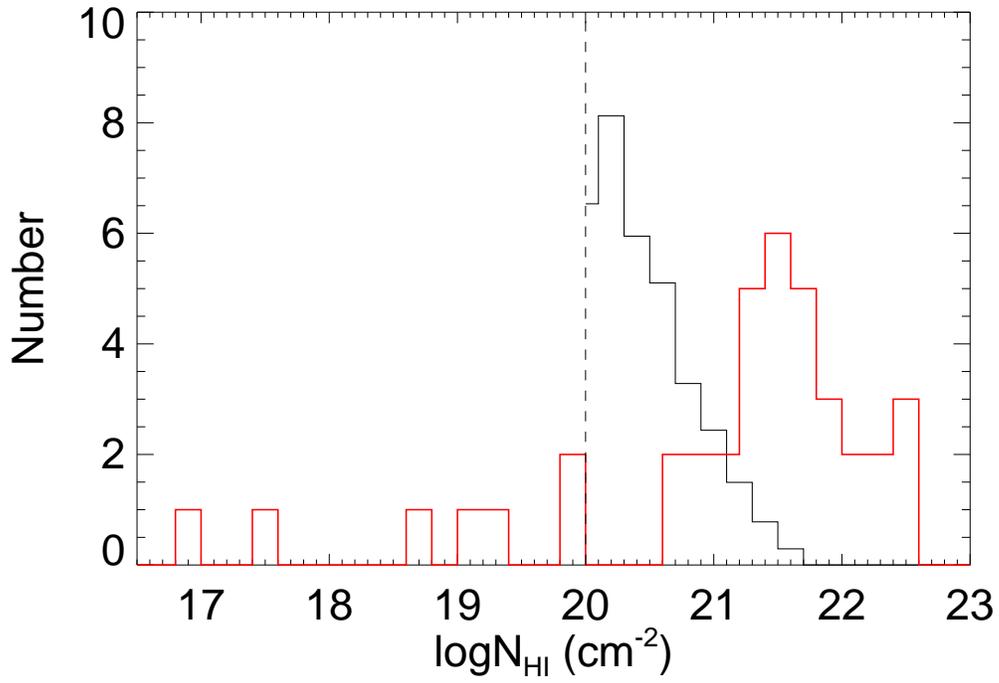,width=15cm}
\caption[]{Distribution of \ion{H}{1} column densities. Black: QSO-DLAs 
from the sample of Noterdaeme et al.\ (2009). Red: GRB absorbers from
the spectroscopic sample. The numbers of QSO-DLAs have been renormalized
to the same number of objects as the GRB absorbers.}
\label{HI}
\end{figure}

The issue of the distribution of \ion{H}{1} column densities in
{\it Swift} detected long-duration GRBs has already been addressed in
\citet{palli:NH}. Compared to that study we here provide 14 additional
measurements. In addition we have improved measurements for some
of the bursts in Table 3 of \citet{palli:NH}, e.g., GRB\,060607A and
GRB\,060124. The column densities for all 33 Ly$\alpha$ lines 
detected in our spectroscopic sample are listed in Table~\ref{HItab}
and plotted in Fig.~\ref{HI}. The \ion{H}{1} distribution covers
five orders of magnitude from 10$^{17}$ to $\lesssim10^{23}$ cm$^{-2}$.
Roughly 80\% of the systems have measured
\ion{H}{1} column densities above $2\times10^{20}$
cm$^{-2}$, which is the classical definition of a DLA system 
\citep[e.g.,][]{wolfe:araa}. There is evidence that the true distribution
may extend to somewhat higher column densities, namely the fact that the three
bursts with the highest column densities are dark bursts. The upper range
of the distribution may also reflect the transition where the bulk of the
hydrogen is in molecular form \citep{schaye01,krumholz09}.
We will return to this point in Sect.~\ref{bias}.
The distribution of \ion{H}{1} column densities for GRB absorbers has
recently been successfully reproduced in a high resolution simulation of 
galaxy formation simulation \citep{pontzen09}.

Given that the majority if the bursts have column densities above 10$^{20}$
cm$^{-2}$ it is natural primarily to compare GRB absorbers with the QSO-DLAs.

\begin{deluxetable}{@{}llc@{}}
\tablecaption{\ion{H}{1} column densities from the spectroscopic sample
\label{HItab}}
\tablewidth{0pt}
\tablehead{
GRB & $z$ & $\log(N_\mathrm{HI} / \mathrm{cm}^{-2})$ 
}
\startdata
050319 & 3.240 & 20.90$\pm$0.20 \\
050401 & 2.899 & 22.60$\pm$0.30 \\
050730 & 3.968 & 22.10$\pm$0.10 \\
050820A & 2.612 & 21.10$\pm$0.10 \\
050908 & 3.344 & 17.60$\pm$0.10 \\ 
050922C & 2.198 & 21.55$\pm$0.10 \\
060115 & 3.533 & 21.50$\pm$0.10 \\
060124& 2.30  & $18.5 \pm 0.5$ \\ 
060206 & 4.048 & 20.85$\pm$0.10 \\
060210 & 3.913 & 21.55$\pm$0.15 \\
060526 & 3.221 & 20.00$\pm$0.15 \\
060607A & 3.075 & 16.95$\pm$0.03 \\
060707 & 3.425 & 21.00$\pm$0.20 \\
060714 & 2.711 & 21.80$\pm$0.10 \\
060906 & 3.686 & 21.85$\pm$0.10 \\
060926 & 3.206 & 22.60$\pm$0.15 \\ 
060927 & 5.464 & 22.50$\pm$0.15 \\ 
061110B & 3.433 & 22.35$\pm$0.10 \\
070110 & 2.351 & 21.70$\pm$0.10 \\
070411 & 2.954 & 19.30$\pm$0.30 \\
070506 & 2.308 & 22.00$\pm$0.30 \\ 
070611 & 2.041 & 21.30$\pm$0.20 \\
070721B & 3.628 & 21.50$\pm$0.20 \\
070802 & 2.455 & 21.50$\pm$0.20 \\
071020 & 2.145 & $<$20.30 \\
071031 & 2.692 & 22.15$\pm$0.05 \\
080210 & 2.641 & 21.90$\pm$0.10 \\
080310 & 2.427 & 18.70$\pm$0.10 \\
080413A & 2.433 & 21.85$\pm$0.15 \\
080603B & 2.690 & 21.85$\pm$0.05 \\
080607  & 3.037 & 22.70$\pm$0.15 \\
080721 & 2.591 & 21.60$\pm$0.10 \\
080804 & 2.205 & 21.30$\pm$0.15 
\enddata
\end{deluxetable}

\begin{figure}
\psfig{figure=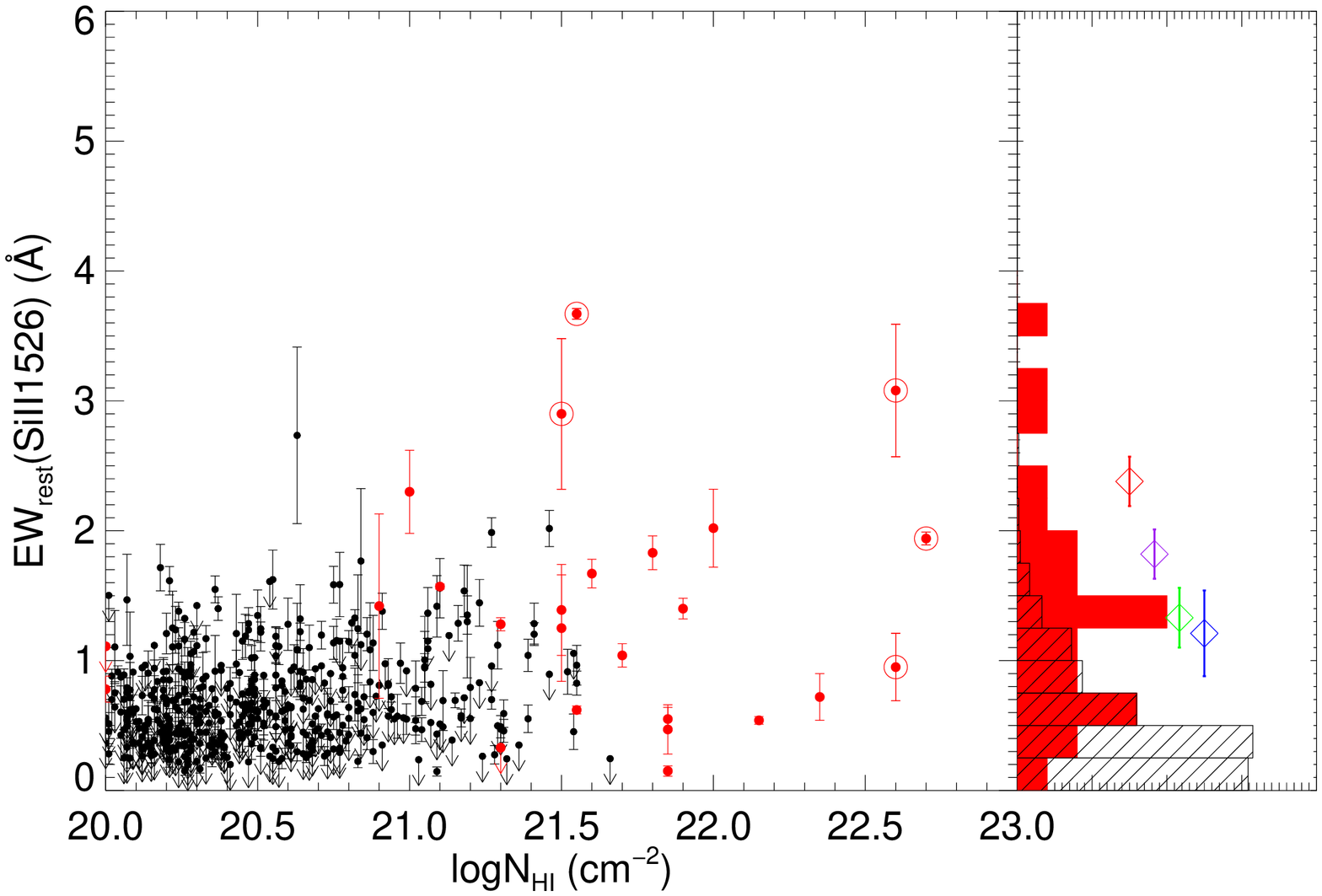,width=15cm}
\caption[]{Black: QSO-DLAs measurements. Red: GRB absorbers from
the spectroscopic sample with $\log{N}>20$.
Encircled points are dark bursts. The histograms are normalized to
have the same area. Open diamonds: LBGs. The LBG points are displaced 
horizontally for visibility reasons. The LBG measurements represent mean values
of 4 composite spectra representing the quartiles of increasing Ly$\alpha$ EW
from blue to red.
}
\label{SiII1526}
\end{figure}

\begin{figure}
\psfig{figure=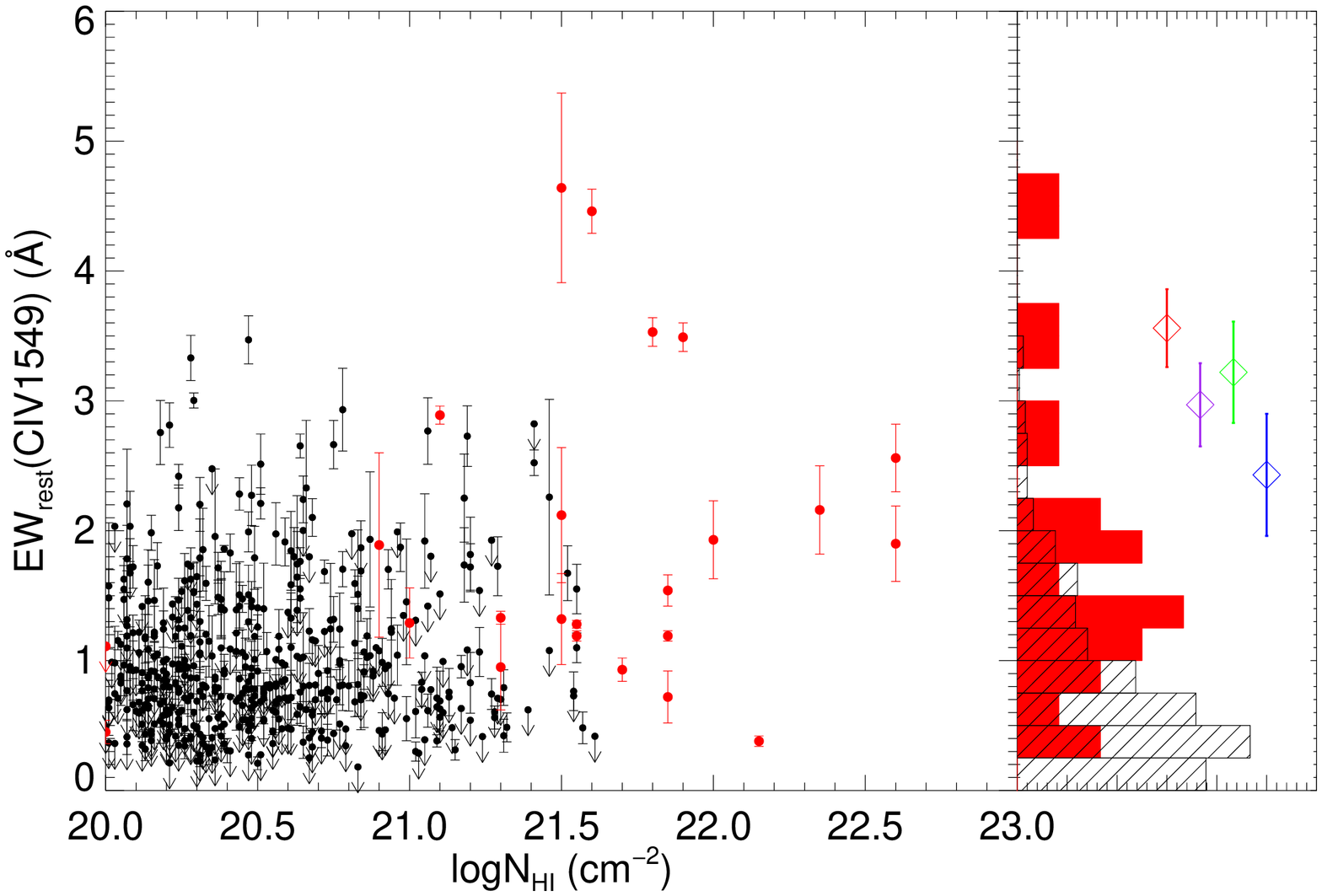,width=15cm}
\caption[]{Black: QSO-DLAs measurements. Red: GRB absorbers from
the spectroscopic sample with $\log{N}>20$.
The histograms are normalized to
have the same area. Open diamonds: LBGs. The LBG points are displaced
horizontally for visibility reasons. The LBG measurements represent mean values
of 4 composite spectra representing the quartiles of increasing Ly$\alpha$ EW
from blue to red.
}
\label{CIV1549}
\end{figure}

In Figs.~\ref{HI},\ref{SiII1526}, and \ref{CIV1549} we compare the distributions
of \ion{H}{1} and metal-line strengths for QSO-DLAs and GRB absorbers in the
spectroscopic sample (the
histogram for QSO absorbers is renormalized to have the same number of systems
as GRB absorbers with $\log{N_\mathrm{HI}} > 20.0$).  In the two latter plots
we only include GRB absorbers with $\log{N_\mathrm{HI}} > 20.0$ where we have
comparison data from Sloan. 
We have used the original QSO-DLA sample from Noterdaeme et al.\ (2009) based
on the SDSS-DR6 database of QSO spectra. These authors automatically
searched for DLA lines, refining their Ly$\alpha$ fits whenever metal lines
are detected redward of the Ly-alpha forest. We have selected all systems from
their list with $\log N($H\,{\sc i}$)\ge 20$ and redshifts in the range
$2.2<z_{\rm abs}\le 3.2$, located at least 5000 km s$^{-1}$ from a
background QSO with $R<21$.

The EWS of two metal transition lines, namely, Si\,{\sc ii}~$\lambda 1526$ 
and C\,{\sc iv}~$\lambda 1549$ (i.e., the blend of the 1548 and 1550
\AA\ lines), were measured. These lines are strong so that they are
detected in most cases and the risk of overestimating their EWs due to 
blending with unrelated absorption is minimal. For the measurements 
themselves, intervals of restframe widths 2 \AA\ (3.5 \AA)
centered on the Si\,{\sc ii}~$\lambda 1526$ (C\,{\sc iv}~$\lambda 1549$) were 
used. However, those systems where these metal lines lie inside the 
Ly$\alpha$ forest were not considered any further.

Among all EW measurements, only those satisfying the following
two criteria were used: $\chi^2_\mathrm{r}<1.5$ (where $\chi^2_\mathrm{r}$ 
is the reduced $\chi^2$) and err$_\mathrm{EW}<0.3$ \AA. 
These thresholds
were determined from the distribution of data points in the $\chi^2_r$
vs. err$_\mathrm{EW}$ plane in order to reject unreliable measurements. Among
the remaining measurements, we provide 3$\sigma$ upper limits when the S/N
ratio is below 3.

Fig.~\ref{SiII1526}-\ref{CIV1549} compare the rest-frame EWs of \ion{Si}{2} and
\ion{C}{4} for the GRB-DLAs and QSO-DLAs. We also show measurements of the
strength of these interstellar lines from Lyman-break galaxies (LBGs) from
\citet{shapley03}. These measurements represent mean values of 4 composite
spectra representing the quartiles of increasing Ly$\alpha$ EW from blue
to red. Note that the \ion{C}{4} EWs for LBGs also contain a stellar
  contribution so they should be considered upper limits.  For GRB-DLAs
and QSO-DLAs, we plot these values against the \nhi\ estimates which separate
the GRB-DLAs from the majority of QSO-DLAs.  In all of these comparisons, there
is significant overlap between the QSO-DLA and GRB-DLA EW distributions but the
GRB-DLA measurements extend to significantly higher EW values than what is seen
in the much larger QSO-DLA sample.  The large dataset presented here confirm
the results of systematically larger Si\,II and CIV EWs in GRB-DLAs reported by
\cite{X:vfields}. Similar results were found by \citet{savaglio04} in a
comparison of \ion{Fe}{2}, \ion{Si}{2} and \ion{Mn}{2} lines between GRB and
QSO-DLA absorbers.  The LBG measurements fall at the high end of the of GRB-DLA
distribution. The LBG systems with the largest line strengths (both \ion{Si}{2}
and \ion{C}{4}) are those that have the strongest Ly$\alpha$ absorption. These
systems also have the highest dust contents \citep{shapley03,noll04}.


To understand the difference between the GRB-DLA and QSO-DLA distributions it
is natural to focus on the different ways the two absorber classes probe their
host galaxies \citep[see also][]{vreeswijk04,prochaska07,fynbo08a}. GRB-DLAs
probe the sight-line to the location of a massive star with a random
orientation relative to the geometry of the galaxy (ignoring for now the issue
of bias).  QSO-DLAs probe \ion{H}{1} cross-section selected random sight-lines
through their host galaxies. The distribution of orientations of the host
galaxy relative to the sight-line is not random, but weighted with the
selection function. If the QSO-DLAs have a flattened geometry this
will tend to produce a shorter sight-line through the host than for a random
distribution of orientations. On the other hand, the GRB absorbers do not probe
the full sight-line through their hosts, in particular for some systems located
in the outskirts of their hosts towards us only a small fraction of the
absorbing material will be probed. An additional piece of evidence is the
distance between the location of the GRB and the bulk of the absorbing material
which has been inferred for a few GRBs based on modeling of fine-structure line
variability \citep{mirka06,vreeswijk07,delia09,delia09b,cedric:050730}. 
These studies have
found that the bulk of the absorbing material probed by the variable
fine-structure lines are at distances of 50--100 pc, 1.7 kpc, 0.7--6 kpc, and
280 pc for the GRBs 020813, 060418, 080319B, and 080330 respectively.  The
nearly ubiquitous detection of strong \ion{Mg}{1} absorption also indicates
that a substantial quantity of neutral gas in GRB-DLAs is located at distances
exceeding 100\,pc \citep{pcb06}. Finally, also vibrationally extited H$_2$
lines have been used to infer a distance in the range 230--940 pc between
GRB\,080607 and the bulk of the absorbing material \citep{scheffer09}.

For strong and saturated lines like those presented in Figs.~\ref{SiII1526} and
\ref{CIV1549}, the EW is weakly sensitive to the abundance of the ion.
Instead, the EW values trace the kinematics of these ions along the sight-line.
\cite{X:vfields} have argued, based on comparisons of resonance and
fine-structure transitions of SiII and FeII, that the large EWs in GRB-DLAs
reflect motions in the "halos'' of these galaxies, i.e.\ in gas at distances
exceeding several kpc.  The data presented here indicates that such motions are
a generic property of gas in the galactic environment of GRBs.  Of principal
interest is whether these motions reflect gravitational dynamics \citep[as
argued by][]{X:vfields} or galactic-scale outflows driven by star-formation,
AGN, etc.  Independent of the mechanism, the fact that both the 
\ion{Si}{2} and CIV histograms extend significantly beyond the range
spanned by QSO-DLA (Figs \ref{SiII1526} and \ref{CIV1549}) indicates the gas is either
predominantly ionized or multi-phase. In other words, if \ion{C}{4} absorption
was mainly produced in a roughly spherical, hot halo whereas the low-ionization
lines where produced in a central, colder component then we would expect the
\ion{C}{4} plot to be different than the \ion{Si}{2} plot. Hence, \ion{C}{4}
seems to probe the same volume as the low-ionization lines.  The same
conclusion has been reached by \citet{X:vfields}.

%

For QSO-DLAs the \ion{Si}{2} 1526 line can be used as a proxy for metallicity
through the relation [M/H]~=~$-0.92$ + $1.41$ $\log$(EW/{\rm\AA})
\citep{X:vfields}. The underlying physical reason for this is believed to be a
velocity width vs. metallicity relation \citep{ledoux:vfield,X:vfields}.
Roughly, a rest-frame EW of 1 \AA \ corresponds to a metallicity of 0.1 Solar.
We do not know yet if GRB absorbers follow this correlation, but there is some
indication that they do \citep{X:vfields}. The metallicities have been measured
for GRBs 050401, 050730, 050820A, 050922C, 070802, 071031, 080310, 080413A and
080607 in our spectroscopic sample
\citep{darach:050401,fox08,ardis:070802,X:080607,cedric:050730}, and these
bursts seem to follow the correlation although possibly with a somewhat steeper
slope.

As seen in Fig.~\ref{SiII1526}, slightly
more than half of the GRB absorbers have EW$_\mathrm{rest}$(\ion{Si}{2}1526)$>$1
\AA \ suggesting that about half of GRBs with afterglow spectroscopy at these 
redshifts have metallicity above 0.1 Solar.  This is consistent with  
the metallicity distribution of GRB-DLAs based on direct estimates of \nhi\
and a metal column density \citep[e.g.][]{prochaska07}.
The true fraction could well be
higher due to the bias against burst with high X-ray absorbing columns in 
the sample of GRBs with OA spectroscopy.

\subsection{Implication for the Escape Fraction of Ionizing Photons in Star-Forming Galaxies at $z>2$}

An important utility of the spectroscopic sample of {\it Swift} GRBs in Table 3
is to constrain the escape fraction of ionizing photons $\fesc$ in distant
star-forming galaxies (Chen et al.\ 2007; Gnedin et al.\ 2008), which specifies
the fraction of stellar-origin ionizing photons ($h\nu > 1$ Ryd) that escape
from star-forming regions into the intergalactic medium (IGM). A traditional
approach to obtain empirical constraints of \fesc\ is to search for high-energy
photons detected at wavelengths below the Lyman limit transition of a distant
galaxy. However, such measurements are subject to a number of systematic
uncertainties that are difficult to quantify, including background subtraction,
intrinsic spectral shape at ultraviolet wavelengths of star-forming galaxies,
and line-of-sight variations of IGM \lya\ absorption (see Chen et al.
2007 for a brief review). 

Chen et al.\ (2007) introduced a new approach that determines \fesc\ in
high-redshift star-forming galaxies based on the range of neutral
hydrogen column density \nhi\ found in the hosts of long-duration
GRBs. The observed \nhi\ in the host of each GRB from early-time
afterglow spectra represents a measure of the integrated optical depth
of Lyman limit photons along the line of sight away from the parent
star-forming region where the progenitor resides.  Considering the
\nhi\ distribution function of an ensemble of GRBs together therefore
yields an estimate of the mean \fesc\ averaged over random lines of
sight.  Adopting a sample of 28 GRBs at $z\apg 2$, Chen et al.\ (2007)
found $\langle\fesc\rangle=0.02\pm 0.02$ with a 95\% c.l. upper limit
$\langle\fesc\rangle \le 0.075$.  Although the best-estimated
$\langle\fesc\rangle$ is among the lowest value reported for $z>2$
star-forming galaxies, the subluminous nature of the majority of GRB
host galaxies (e.g., Jakobsson et al.\ 2005; Chen et al.\ 2009) in
combination with a steep faint-end slope of the distant galaxy
population (e.g.\ Reddy et al.\ 2008) indicates that sub-$L*$ galaxies
with $\langle\fesc\rangle=1-2$ \% can already contribute a comparable
amount of ionizing photons as QSOs to the ultraviolet background
radiation at $z\sim 3$.

\begin{figure}
\begin{center}
\includegraphics[scale=0.5, angle=270]{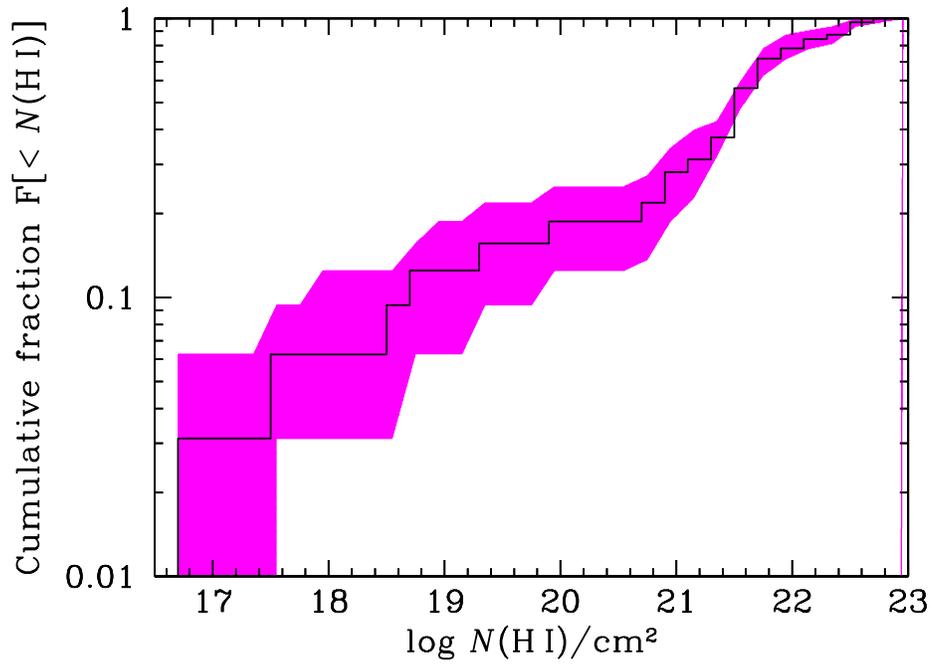}
\caption{Cumulative distribution of neutral hydrogen column density
${\cal F}[<\nhi]$ observed in the host galaxies of 32 long-duration
GRBs discovered at $z\ge 2$ by {\it Swift} (solid histogram).  The
shaded area shows the 1-$\sigma$ uncertainties evaluated using a
bootstrap re-sampling method that accounts for both \nhi\ measurement
uncertainties and sampling errors.
\label{escape}}
\end{center}
\end{figure}

While the new approach of constraining \fesc\ based on the observed
\nhi\ distribution is not affected by the same systematic
uncertainties that bias the \fesc\ measurements from the traditional
method (see Chen et al.\ 2007 for related discussion), the accuracy of
$\langle\fesc\rangle$ depends on an unbiased sample of GRB hosts.
Specifically, if some fraction of optically-thin sightlines are missed
due to a lack of $S/N$ in the afterglow spectra for identifying weak
absorption features, then the estimated $\langle\fesc\rangle$ would be
biased toward a lower value.  Chen et al.\ (2007) considered a sample
of 28 GRBs at $z\apg 2$ with available early-time afterglow spectra
for constraining the \nhi, eight of which are from the pre-{\it Swift}
era when rapid localizations of the optical transients were
challenging.  It is not clear how significant the bias is due to
missed optical transients.

The 33 GRBs discussed in this paper represent a uniform spectroscopic
sample of {\it Swift} bursts with rapid localizations and allow us to
obtain an accurate measurement of $\langle\fesc\rangle$.  Fig.~\ref{escape}
presents the cumulative \nhi\ distribution, ${\cal F}[<\nhi]$ from the
sample of 32 GRB host galaxies (excluding GRB\,071020 due to the
uncertain \nhi), together with the 1-$\sigma$
uncertainties determined based on a bootstrap re-sampling method (see
Chen et al.\ 2007 for further illustrations).  Adopting this sample of
32 GRBs at $z\apg 2$, we find $\langle\fesc\rangle=0.02\pm 0.02$ with
a 95\% c.l. upper limit $\langle\fesc\rangle \le 0.07$, in excellent
agreement with the finding of Chen et al.\  We note that while the two
samples are comparable in size, only roughly half of the new sample
overlaps with those considered by Chen et al.\  In addition, including
GRB\,071020 and assuming a host \nhi\ of $\log\,\nhi=16.85$ (matching
the lowest \nhi\ found so far for a GRB host) would double the
estimated value to $\langle\fesc\rangle=0.04$.  Our new measurement
confirms that the mean escape fraction of ionizing photons in distant
star-forming galaxies is small.

\subsection{Intervening Absorption Systems}
\label{intervening}
The issue of intervening absorbers is interesting for two main reasons. The
first is the puzzling result that there are more intervening \ion{Mg}{2}
absorbers along GRB sight-lines than along QSO sight-lines \citep{prochter,
vergani09,cucchiara09}. On the other hand there is no excess of \ion{C}{4}
systems along GRB sightlines \citep{Sudilovsky07,Tejos07}.
The other reason is that the search for galaxy counterparts of the 
absorbers is much easier along GRB sight-lines as the OA
quickly fades away \citep{vreeswijk03}.

In our sample we detect 
several very strong intervening \ion{Mg}{2} absorbers, i.e. 12
systems with EW$>2$ \AA. For a full statistical analysis of the excess we refer
to Vergani et al.\ (in preparation). 

We also detect a few strong intervening Ly$\alpha$ absorbers, e.g.\ along the
sight-lines to GRB\,050730, GRB\,050908, and GRB\,070721B (see also Schulze et
al.\ and Milvang-Jensen et al.\ in preparation). In is interesting to note that
the $z=2.62$ sub-DLA towards GRB\,050908 also has remarkably strong metal lines
(e.g., a EW$_\mathrm{rest}$ of 2.2\AA \ for the \ion{Si}{2},1526 line). Also,
the $z=3.09$ sub-DLA towards GRB\,070721B has very strong lines (rest EW 1.7\AA
\ for the \ion{Si}{2},1526 line). QSO-DLAs with EW$_\mathrm{rest}$ larger than
1.5 \AA \ for the \ion{Si}{2},1526 line are quite rare  (see Prochaska et al.\
2008b and Fig.~\ref{SiII1526} below). Unfortunately, our sample is too small to
judge if the intervening Ly$\alpha$ absorbers towards GRB sight-lines are also
statistically different from those along QSO sight-lines.


\section{Conclusions}
\label{conclusion}

GRB afterglow spectroscopy provides a detailed analysis of the ISM of actively
star-forming galaxies. In this work we have presented a large sample of GRBs
detected by {\it Swift} for which primarily low-resolution spectroscopic
observations have been secured. The majority ($\sim80$\%) of the GRB host
absorbers bear strong resemblance to high-column density QSO absorbers like
DLAs and sub-DLAs. However, the GRB absorbers are characterized by a broader
range of line strengths extending to significantly stronger lines (both for
\ion{H}{1}, \ion{Si}{2} and \ion{C}{4}) than what is seen for high-column
density QSO absorbers. 

We find that the sample of GRBs for which optical spectroscopy is secured is a
significantly biased subsample. In the statistical sample only about 50\% have
OA spectroscopy and this sub-sample have substantially smaller
X-ray excess absorption and a substantially smaller fraction of dark bursts (by
the Jakobsson et al.\ 2004 definition) than the other half of the sample.

The implication of this work is that the metallicity distribution we derive
from optical afterglow spectroscopy likely is skewed towards low metallicities
compared to the true underlying distribution \citep[see also][]{cedric:050730}. 
This is important to take into
account, e.g., when establishing how well GRBs trace star formation and when
inferring to which extent GRB progenitors may be restricted to massive stars
with low metallicities.

This work also shows that if we can secure optical and/or nearIR spectroscopy
for a much larger fraction of an X-ray selected sample it is likely that we
will pick up sight-lines with significantly higher dust contents, molecular
fractions, and higher metallicities. A similar conclusion was reached by 
\citet{cedric:050730}. Such studies will be ideal to study
extinction curves and properties of star-forming regions, in particular at high
redshifts. In the current sample dusty bursts like GRBs 070802, 080605, and
080607 were observed under fortunate conditions, i.e. very soon after the
bursts and in good observing conditions. It is plausible that bursts for which
the current search strategy and instrumentation cannot secure a
redshift are similar to or even more dust obscured than these bursts.  

Currently, most GRB error circles are only observed in optical bands. For 
further progress, systematic nearIR follow-up of an X-ray selected sample 
will be important. Such systematic nearIR follow-up is currently carried 
out by a few teams using, e.g., GROND, UKIRT and PAIRITEL
\citep[e.g.][]{greiner:080913,kruehler:070802,tanvir:08,X:080607}. For the
spectroscopic follow-up, instruments like the newly commissioned X-shooter
spectrograph that cover the full nearUV to nearIR range in a single shot
\citep{kaper} will be important, but it is likely that we will have to wait for
the advent of $30$--$40$ m telescopes with similar instrumentation before we 
will reach a spectroscopic completeness of, e.g., 90\%.

\acknowledgements
We thank our referee for a thorough and constructive report. JPUF thanks
A. Shapley for helpful discussions.
JPUF thanks the Centre for Astrophysics and Cosmology at the University of
Iceland for hospitality during the writing of most of this work.  We thank the
{\it Swift} team for carrying such a wonderful mission.  The DARK centre is
funded by the DNRF. PJ acknowledges support by a Marie Curie European
Re-integration Grant within the 7th European Community Framework Program under
contract number PERG03-GA-2008-226653, and a Grant of Excellence from the
Icelandic Research Fund. H.-W.C. acknowledges support from NASA grant
NNG\,06GC36G and an NSF grant AST-0607510.

\appendix
\section{Notes on individual objects}
\label{notes}

\subsection{GRB\,050319 ($z=3.2425)$}
The data presented here have previously been published in \citet{palli:NH}.
The spectrum has a low signal-to-noise (SN) ratio, but due to the presence
of the a strong Hydrogen Ly$\alpha$ line the redshift is secure.

\subsection{GRB\,050401 ($z=2.8983$)}
The data presented here have previously been published in \citet{darach:050401}.
This burst has one of the highest \ion{H}{1} column densities measured along
any sightline. The redshift is based on numerous metal lines including 
fine-structure lines. The spectrum also shows evidence for substantial SMC-type
reddening and classifies as a dark burst by the definition of \citet{palli:dark}.

\subsection{GRB\,050408 ($z=1.2356$)}
The data presented here have previously been published in \citet{foley:050408}.
The redshift is based on numerous metal lines (including fine-structure lines).

\subsection{GRB\,050730 ($z=3.9693$)}
The data presented here have previously only been published in the GCN circulars
\citep{delia:050730}. Spectra of this burst were also secured
by Chen et al.\ (2005), Starling et al.\ (2005), D'Elia et al.\ (2007) and
Ledoux et al.\ (2009). In the 2d spectrum we detect a foreground QSO ($z=3.02$)
at an impact parameter of 20 arcsec (153 kpc). This QSO is possibly associated
with the $z=3.022$ absorption system seen in the spectrum of GRB\,050730.

\subsection{GRB\,050801 ($z=1.38$)}
The data presented here have not been published earlier. No standard star was
observed and therefore no flux calibration has been attempted for this spectrum
(the plotted spectrum is normalized to 1 in the continuum). We only detect a
single unidentified line in the spectrum. The upper limit on the redshift based
on the absence of a Ly$\alpha$ forest is about 2.3. Based on UVOT colors a
photometric redshift of 1.38 has been determined for this burst
\citep{oates09}.

\subsection{GRB\,050802 ($z=1.7102$)}
The data presented here have previously only been published in the GCN circulars
\citep{johan:050802}.  No standard star was observed and therefore no flux
calibration has been attempted for this spectrum (the plotted spectrum is
normalized to 1 in the continuum). The redshift is based on several weak
metal lines. The upper limit on the redshift based on the absence of the 
Ly$\alpha$ forest is about 2.3.

\subsection{GRB\,050820A ($z=2.6147$)}
The UVES data for this burst are discussed in \citet{cedric:050820} and 
Fox et al.\ (2008). 

\subsection{GRB\,050824 ($z=0.8278$)}
GRB\,050824 is an X-Ray Flash (XRF).  The data presented here have previously
been published in \citet{jesper:050824}. That work also established evidence for
an associated SN and detected the host galaxy of the burst. Here the data
have been re-reduced and reanalyzed and more lines have been identified.
The redshift is based on both metal absorption lines and emission lines
from the underlying host galaxy.

\subsection{GRB\,050908 ($z=3.3467$)}
The data presented here have previously only been published in the GCN circulars
\citep{fugazza:050908}. The spectra will be discussed in more detail in Smette
et al.\ (in preparation) where a \ion{H}{1} column density of $\log{N_\mathrm{HI}}
=17.60\pm0.10$ is derived. This system is
interesting due to its low \ion{H}{1} column density and the detection of
escaping ionizing radiation along its line-of-sight. Furthermore, there is
an intervening sub-DLA at $z=2.62$ with very strong metal lines.

\subsection{GRB\,050922C ($z=2.1995$)}
The data presented here have previously been published in \citet{palli:NH}.
For this burst high resolution spectra have been presented in
\citet{piranomonte08}. The plotted spectrum is normalized.

\subsection{GRB\,060115 ($z=3.5328$)}
The data presented here have previously only been published in the GCN circulars
\citep{piranomonte:060115}. The redshift is based on both Ly$\alpha$ 
and fine-structure lines and is hence secure.

\subsection{GRB\,060124 ($z=2.3000$)}
The data presented here have previously only been published in the GCN circulars
\citep{X:060124}. The GRB system is characterized by a low \ion{H}{1} column
density and the absence of low-ionization lines. 
A detailed discussion of the prompt emission and afterglow of this burst
can be found in \citet{Romano:060124}.

\subsection{GRB\,060206 ($z=4.0559$)}
The data presented here have previously been published in \citet{johan:060206}
and \citet{thoene:060206}. Spectroscopic observations of this afterglow
have also been published in \citet{hao07} who claimed variable absorption
lines in the intervening system. This was subsequently falsified by
\citet{Aoki08} and \citet{thoene:060206}. There is tentative evidence
for molecular absorption in this spectrum (Fynbo et al.\ 2006b, see also
Prochaska et al.\ 2009). No standard star was observed at the night of 
the observation so we here show a normalized spectrum. The spectrum
is strongly affected by fringing redwards of 7000 \AA.

\subsection{GRB\,060210 ($z=3.9122$)}
The data presented here have previously only been published in the GCN circulars
\citep{cucchiara:060210}. We do not confirm their identification of an 
intervening system at $z=1.47$. The burst has very strong low-ionization
lines and classifies as a dark burst by the \citet{palli:dark} definition.
A detailed discussion of the prompt emission and afterglow of this burst
can be found in \citet{curran:060210} who also infer substantial reddening
based on broadband modeling of the afterglow.

\subsection{GRB\,060502A ($z=1.5026$)}
The data presented here have previously only been published in the GCN circulars
\citep{cucchiara:060502A}. The redshift is based on a rich spectrum of low
ionization (including fine-structure) absorption lines.

\subsection{GRB\,060512 ($z=2.1$?)}
GRB\,060512 is an XRF. The data presented here have previously only been
published in the GCN circulars \citep{starling:060512a}. For this burst there
is uncertainty about the redshift in the GCN reports. An afterglow spectrum
taken at the TNG indicates a redshift of around 2.1 \citep{starling:060512b},
which is supported by a spectral break detected by UVOT
\citep{DePasquale:060512,oates09} and which is supported by modeling of the 
broad band spectral energy distribution of the burst \citep{schady07}.
\citet{Bloom:060512} measure a redshift of 0.44
from a galaxy close to the afterglow position. This galaxy is offset by about
0.7 arcsec from the position of the afterglow. Based on a binning of the
spectrum shown here we confirm the presence of a broad absorption line around
3750 \AA \ consistent with a redshift of around 2.1 if interpreted as
Ly$\alpha$. This suggest that the $z=0.44$ galaxy close to the line of sight is
a foreground object. We do not detect significant metal lines in the afterglow
spectrum.  

\subsection{GRB\,060526 ($z=3.2213$)}
The data presented here have previously been published in \citet{palli:NH}.
Additional medium resolution spectra are discussed in \citet{ct:060526}. 
The redshift is based on Ly$\alpha$ and metal absorption lines.

\subsection{GRB\,060604 ($z\lesssim3$)}
The data presented here have previously only been published in circulars
(Castro-Tirado et al.\ 2006). In the
spectrum we detect a single absorption line at 5219 \AA. The origin of the
line is unclear. The line appears too narrow to be \ion{Mg}{2} at $z=0.864$. A
possibility is \ion{Al}{2} at $z=2.124$. This would be compatible with the
tentative break detected by UVOT (Blustin \& Page 2006). From the absence of
the onset of the Ly$\alpha$ forest we infer an approximate upper limit on
the redshift of $\sim$3. We do not confirm the tentative redshift of 
$z=2.68$ proposed by Castro-Tirado et al.\ (2006).

\subsection{GRB\,060607A ($z=3.0749$)}
For this burst UVES spectra were obtained starting only about 7.5 minutes after
the burst \citep{cedric:060607A}.  For a full analysis of the UVES data see Fox
et al.\ (2008), Prochaska et al.\ (2008a), and Smette et al.\ (in preparation). 

\subsection{GRB\,060614 ($z=0.1257$)}
The data presented here have previously been published in Della Valle et al.\
(2006). This is the lowest redshift burst in our spectroscopic sample. However,
from the present spectrum no lines (either absorption or emission) are
significantly detected. The upper limit on the redshift based on the absence of
a Ly$\alpha$ forest is about 2.6. The redshift was later found to be 
$z=0.1257$ based on emission lines from the underlying host galaxy (Della
Valle et al.\ 2006). GRB\,060614 is remarkable in being a long
burst without an associated (bright) supernova (Fynbo et al.\ 2006c, Della
Valle et al.\ 2006, Gal-Yam et al.\ 2006).

\subsection{GRB\,060707 ($z=3.4240$)}
The data presented here have previously been published in \citet{palli:NH}.
The burst is remarkable in having very strong metal lines while showing no 
sign of extinction.

\subsection{GRB\,060708 ($z=1.92$)}
The data presented here have previously only been published in the GCN circulars
\citep{palli:060708}. The spectrum does not display any significant absorption
or emission lines. An upper limit of $z\lesssim2.8$ can be placed on the
redshift of GRB\,060708 from the lack of significant Ly$\alpha$ forest lines in the
spectrum of the afterglow redwards of 4600 \AA. Bluewards of 4600 \AA \ there
is very marginal evidence for broad absorption lines in the spectrum, but the
spectrum is too noisy to establish a precise redshift. The {\it Swift} team has
established a photometric redshift of 1.92 for this burst based on the UVOT
imaging of the OA \citep{uvot:060708}.

\subsection{GRB\,060714 ($z=2.7108$)}
The data presented here have previously been published in \citet{palli:NH}.  In
the trough of the DLA line of the host absorption system Ly$\alpha$ emission
is detected, presumably from the underlying host galaxy.

\subsection{GRB\,060719 ($z\lesssim4.6$)}
The data presented here have not been published earlier. This afterglow was
extremely red with a R$-$K color of about 4.5 \citep{malesani:060719}
and the burst was considered a candidate high redshift burst. Our spectrum,
taken with the 600I grating,
shows a featureless continuum in the range 6800--9400 \AA \ thereby placing 
an upper limit of $\sim$4.6 on the redshift. 
Hence, this is most likely a dust obscured burst.

\subsection{GRB\,060729 ($z=0.5428$)}
The data presented here have previously only been published in the GCN circulars
\citep{ct:060729}. The broad
undulations seen in the blue end of the spectrum are caused by systematic errors
in the flux calibration due to the high airmass of the observation.
All of the observed strong, spectral lines can be associated
with a single absorption system.  We associate the GRB redshift with
this gas.
The upper limit on the redshift based on the absence of a Ly$\alpha$ forest
is about 2.1.

\subsection{GRB\,060807 ($z\lesssim3.4$)}
The data presented here have not been published elsewhere. We retrieved the
data from the ESO archive. There is a single marginal line at
5200 \AA \ in the spectrum. If due to Ly$\alpha$ the redshift of the burst
is 3.28. From the absence of a spectral break we set a conservative upper
limit of $z<3.4$. For this burst there is evidence for emission lines in the
X-ray spectrum (Butler 2006).

\subsection{GRB\,060904B ($z=0.7029$)}
The data presented here have previously only been published in the GCN circulars
\citep{fugazza:060904}. 
The burst is not included in the statistical sample as it has too high
foreground extinction (A$_\mathrm{V}$ = 0.57, Schlegel et al.\ 1998). The broad
undulations seen in the spectrum are due to errors in the flux calibration.
All of the observed strong, spectral lines can be associated
with a single absorption system.  We associate the GRB redshift with
this gas.
The upper limit on the redshift based on the absence of a Ly$\alpha$ forest
is about 2.5.

\subsection{GRB\,060906 ($z=3.6856$)}
The data presented here have previously been published in Jakobsson et al.\ 
(2006b). The spectrum was obtained in twilight just before sun rise.
We do not include this burst
in the statistical sample due to its too high foreground extinction of
A$_\mathrm{V}$ = 0.54 (Schlegel et al.\ 1998). The redshift is based on
both Ly$\alpha$ and metal absorption lines.

\subsection{GRB\,060908 ($z=1.8836$)}
The data presented here have previously only been published in the GCN
circulars \citep{rol:060908}. The redshift is based on a single \ion{C}{4}
doublet and is confirmed by the detection of the Ly$\alpha$ line from the
underlying host galaxy (Milvang-Jensen et al.\ in preparation). We hence do not
confirm the tentative redshift of $z=2.43$ reported by \citet{rol:060908}.  The
redshift we infer is also consistent with the UVOT constraints
\citep{morgan:060908}.

\subsection{GRB\,060926 ($z=3.2086$)}
The data presented here have previously been published in
\citet{piranomonte:060926}.  We do not include this burst in the statistical
sample due to its too high foreground extinction of A$_\mathrm{V}$ = 0.52. In
the trough of the DLA line of the host absorption system Ly$\alpha$ emission
from the underlying host galaxy is detected. The \ion{H}{1} column density is
among the highest detected and there is evidence for substantial reddening.

\subsection{GRB\,060927 ($z=5.4636$)}
The data presented here have previously been published in \citet{alma:060927}.
The precise redshift is based on a single low S/N \ion{Si}{2}
line. Based alone on the onset of the Ly$\alpha$ forest the redshift can
be estimated to be in the range 5.5--5.6, the 2nd highest in the
spectroscopic sample.

\subsection{GRB\,061007 ($z=1.2622$)}
The data presented here have previously only been published in circulars
\citep{palli:061007}. This burst is one of the brightest in $\gamma$-rays
observed by {\it Swift} and its afterglow was also very luminous both in the
X-ray and optical bands \citep{schady:061007}. The redshift is based on both
metal absorption lines and emission lines from the underlying host galaxy.

\subsection{GRB\,061021 ($z=0.3463$)}
The data presented here have previously only been published in circulars
\citep{ct:061021}.  In the GCN circular no redshift measurement was reported.
It was only after measuring the redshift from the underlying host galaxy
(Hjorth et al.\ in preparation) that we identified \ion{Mg}{2} absorption lines
in the very blue end of the afterglow spectrum. This is the 2nd lowest redshift
in our spectroscopic sample.

\subsection{GRB\,061110A ($z=0.7578$)}
The data presented here have previously only been published in circulars
\citep{johan:061110A}. The redshift is mainly based  on emission lines from the 
underlying host galaxy. We also identify a broad absorption feature consistent
with the \ion{Mg}{2} doublet at the same redshift.

\subsection{GRB\,061110B ($z=3.4344$)}
The data presented here have previously only been published in circulars
\citep{johan:061110B}. The redshift is based on both a very strong
Ly$\alpha$ line and metal absorption lines, including fine structure lines.

\subsection{GRB\,061121 ($z=1.3145$)}
The data presented here have previously only been published in the GCN circulars
\citep{bloom:061121}. The redshift is based on a rich spectrum of low
ionization (including fine-structure) absorption lines.

\subsection{GRB\,070110 ($z=2.3521$)}
The data presented here have previously only been published in circulars
\citep{andreas:070110}. In the trough of the DLA line of the host absorption
system Ly$\alpha$ emission from the underlying host galaxy is detected.

\subsection{GRB\,070125 ($z=1.5471$)}
The data presented here have not previously been published. The redshift of the
burst was determined from Gemini data (Cenko et al.\ 2008). The burst is peculiar in
being very bright and having very weak absorption lines in its afterglow
spectrum. For a typical fainter OA the redshift for such an
absorption system would be very difficult to establish. The burst is not
included in the statistical sample as the XRT position was distributed later
than 12 hr after the burst.  The upper limit on the redshift based on the
absence of Ly$\alpha$ forest lines is about 2.3.
The broad undulations in the blue end of the spectrum are
due to systematics in the flux calibration.

\subsection{GRB\,070129 ($z\lesssim3.4$)}
The data presented here have not previously been published. 
A redshift could not be established as no significant lines are
detected in the spectrum.  We place a conservative upper limit of
$z\lesssim3.4$ from the upper limit on wavelength of the onset of the
Ly$\alpha$ forest.

\subsection{GRB\,070306 ($z=1.4965$)}
The data presented here have previously been published in
\citet{andreas:070306}. The redshift is based on a single oxygen emission
line. The identification with [OII] is secure as the doublet is resolved in a
higher resolution spectrum \citep{andreas:070306}. The burst is remarkable in
having a highly extinguished afterglow only detected in the $K$ and $H$ bands
while being located in a blue host galaxy \citep{andreas:070306}. 

\subsection{GRB\,070318 ($z=0.8397$)}
The data presented here have previously only been published in circulars
\citep{andreas:070318}.  The afterglow spectrum displays a spectral break
around 5000 \AA \ (or about 2700 \AA \ in the rest-frame), the nature of which
is currently not understood. 
All of the observed strong, spectral lines can be associated with a single
absorption system.  We associate the GRB redshift with this gas.  The upper
limit on the redshift based on the absence of Ly$\alpha$-frost features is
about 2.1.

\subsection{GRB\,070411 ($z=2.9538$)}
The data presented here have previously only been published in circulars
\citep{palli:070411}. The spectra were obtained with the two 
medium resolution gratings 1400V and 1200R. The burst is not included in our
statistical sample as the foreground Galactic extinction is too high 
(A$_\mathrm{V}$=0.76, Schlegel et al.\ 1998). The GRB absorption system 
has one of the lower \ion{H}{1} column densities in our sample 
(see Table~\ref{HItab}).

\subsection{GRB\,070419 ($z=0.9705$)}
The data presented here have not been published before. 
All of the observed strong, spectral lines can be associated
with a single absorption system.  We associate the GRB redshift with
this gas.

\subsection{GRB\,070506 ($z=2.3090$)}
The data presented here have previously only been published in circulars
\citep{ct:070506}. Redwards of 7000 \AA \ the spectrum is strongly affected by
fringing. The redshift is based on both Ly$\alpha$ and metal absorption lines.

\subsection{GRB\,070508 ($z\lesssim3.0$)}
The data presented here have previously only been published in circulars
\citep{palli:070508}. No lines are detected in the spectrum and an upper
limit of $z\lesssim3$ is placed on the redshift from the upper limit on
wavelength of the onset of the Ly$\alpha$ forest.  The burst is not included
in the statistical sample as the declination is too low (-78$^o$). The burst
classifies as a dark burst by the definition of \citet{palli:dark}.

\subsection{GRB\,070611 ($z=2.0394$)}
The data presented here have previously only been published in circulars
\citep{ct:070611}. The redshift is based on both Ly$\alpha$ and metal
absorption lines. Also detected in the spectrum are two intervening
\ion{Mg}{2} systems.

\subsection{GRB\,070721B ($z=3.6298$)}
The data presented here have previously only been published in circulars
\citep{daniele:070721B}. A bright Ly$\alpha$ emitter is located 20 arcsec from
the GRB position at the the same redshift as the GRB. A $z=3.09$ intervening
sub-DLA is detected in both the spectrum of the afterglow and in the spectrum
of the neighbor Ly$\alpha$ emitter. A possible galaxy counterpart of the
absorber is detected in emission 2 arcsec from the afterglow position. The
spectrum will be discussed in more detail in Milvang-Jensen et al.\ (in
preparation).

\subsection{GRB\,070802 ($z=2.4541$)}
The data presented here have previously been published in \citet{ardis:070802}.
The spectrum is remarkable is showing a strong 2175 \AA \ dust extinction
feature at the GRB redshift of $z=2.45$ \citep[see also][]{kruehler:070802}.
The burst classifies as a dark burst
by the definition of \citet{palli:dark}.  There are also two intervening
\ion{Mg}{2} systems in the spectrum, one of which is very strong. 

\subsection{GRB\,071020 ($z=2.1462$)}
The data presented here have previously only been published in circulars
\citep{palli:071020}. The spectrum will be discussed in more detail in Nardini
et al.\ (in preparation). 
All of the observed strong, spectral lines can be associated
with a single absorption system. We associate the GRB redshift with
this gas.

The upper limit based on the absence of a Ly$\alpha$ forest is about 3.

\subsection{GRB\,071025 ($z=5.2$?)}
The data presented here have not previously been published. The data were
obtained with the high resolution HIRES spectrograph. Weak, continuum flux is
detected from the afterglow at wavelengths $\lambda > 7500$\AA, but no emission
is detected at shorter wavelengths.  This flux `decrement' may be associated
with the IGM at $z \approx 5.2$ but could also be associated with a significant
reddening of the afterglow.  If one presumes the latter, the burst is still
very likely to be at high redshift ($z>2.5$) because a sharp decrement from
dust is only expected at ultraviolet wavelengths.  This redshift of $z \approx
5.2$ is consistent with the photometry which displays a break between the $R$
and $I$ bands \citep[e.g.][]{rykoff07}. We do not show a figure of this 
spectrum.

\subsection{GRB\,071031 ($z=2.6918$)}
The data presented here have previously only been published in circulars
\citep{cedric:071031}.  In addition to the 300V spectrum shown here spectra
were secured with a range of higher resolution gratings and with the
high-resolution UVES spectrograph (e.g., Fox et al.\ 2008). In the trough of
the DLA line at the host redshift Ly$\alpha$ emission from the underlying host
galaxy is detected. In the medium and high resolution spectra a large number
of \ion{Fe}{2} fine structure and metastable lines are detected.

\subsection{GRB\,071112C ($z=0.8227$)}
The data presented here have previously only been published in circulars
\citep{palli:071112C}.  The redshift is based both on absorption lines from the
GRB host system as well as emission lines from the underlying host galaxy.

\subsection{GRB\,071117 ($z=1.3308$?)}
The data presented here have previously only been published in circulars
\citep{palli:071117}. The redshift is based on a single emission line at 8687
\AA, which we interpret as \ion{O}{2} at $z=1.331$. The spectrum is
affected by fringing redwards of about 7000 \AA \ and the line is only seen 
after subtracting two of the individual exposures where the source has been
offset along the slit from each other. Hence, we consider the significance of 
this redshift measurement to be marginal.

\subsection{GRB\,080210 ($z=2.6419$)}
The data presented here have previously only been published in circulars
\citep{palli:080210}. The spectrum
will be discussed in more detail in De Cia et al.\ (in preparation). There
is evidence for substantial reddening. The redshift is based on both
Ly$\alpha$ and metal absorption lines.

\subsection{GRB\,080310 ($z=2.4274$)}
This burst was observed both with the VLT/UVES spectrograph and the Kast
spectrograph. Here we show the Kast spectrum previously published in 
\citet{X:080310}.  The GRB absorber is
characterized by a low \ion{H}{1} column density and several high ionization
lines (see also Fox et al.\ 2008; Ledoux et al.\ 2009; De Cia et al.\ in preparation).

\subsection{GRB\,080319B ($z=0.9382$)}
This is the famous ``naked eye burst'' \citep{racusin08,wozniak09,bloom09}. The FORS2 spectrum
presented here was obtained quite late, 26 hr after the burst. We retrieved the
data from the ESO archive. 

\subsection{GRB\,080319C ($z=1.9492$)}
The data presented here have previously only been published in circulars
\citep{wiersema:080319C}. The data were obtained in poor conditions 
with cloud cover. The plotted spectrum is normalized. The redshift is based
on the highest redshift metal line system in the spectrum. We also detect
an intervening \ion{Mg}{2} system. The upper limit on the redshift based
on the absence of significant Ly$\alpha$ forest lines is about 2.7.
The burst classifies as a dark burst by the \citet{palli:dark} definition.

\subsection{GRB\,080330 ($z=1.5119$)}
GRB\,080330 is an XRF. The data presented here have previously been published
in Guidorzi et al.\ (2009). The burst was also observed with the VLT-UVES
spectrograph \citep{delia09b}. All of the observed strong, spectral lines can
be associated with a single absorption system. We associate the GRB redshift
with this gas.

\subsection{GRB\,080411 ($z=1.0301$)}
The data presented here have previously only been published in circulars
\citep{thoene:080411}. 
The burst is not included in the statistical sample as it has too high
foreground extinction (A$_\mathrm{V}$ = 0.54, Schlegel et al.\ 1998).
The burst is remarkable in having a bright X-ray afterglow visible for
more than three weeks after the burst \citep{marshall:gcn7664}. The 
OA spectrum has a fairly strong unidentified line at 
4439\AA.

\subsection{GRB\,080413A ($z=2.4330$)}
This burst was observed with VLT/UVES.
The burst is not included in the statistical sample as it has too high
foreground extinction (A$_\mathrm{V}$ = 0.52, Schlegel et al.\ 1998).

\subsection{GRB\,080413B ($z=1.1014$)}
The data presented here have previously only been published in circulars
\citep{vreeswijk:080413B}. Spectra were also secured in a number of higher
resolution gratings. The broad undulations in the blue end of the spectrum are
due to systematics in the flux calibration.

\subsection{GRB\,080520 ($z=1.5457$)}
The data presented here have previously only been published in circulars
\citep{palli:080520}.  The afterglow was very faint at the time of observation,
but the redshift is based on both host emission lines and afterglow metal
absorption lines and is hence secure.

\subsection{GRB\,080523 ($z\lesssim3.0$)}
The data presented here have previously only been published in circulars
\citep{fynbo:080523}. The afterglow was very faint at the
time of observation and due to the lack of significant lines or breaks
we can only place an upper limit of about 3.0 on the redshift.

\subsection{GRB\,080603B ($z=2.6892$)}
The data presented here have previously only been published in circulars
\citep{fynbo:080603B}. The redshift is on both Ly$\alpha$ and metal (including
fine-structure) absorption lines. We also detect an intervening \ion{Mg}{2}
system in the spectrum.

\subsection{GRB\,080604 ($z=1.4171$)}
The data presented here have previously only been published in circulars
\citep{wiersema:080604}. The plotted spectrum is normalized.  All of the
observed strong, spectral lines can be associated with a single absorption
system.  We associate the GRB redshift with this gas.  The upper limit on the
redshift is about 3.1 based on the absence of a Ly$\alpha$ forest in the
spectrum.

\subsection{GRB\,080605 ($z=1.6403$)}
The data presented here have previously only been published in circulars
\citep{palli:080605}.  Spectra were also secured with a number of higher
resolution gratings.  This spectrum like the spectrum of GRB\,070802 shows both
\ion{C}{1} absorption and evidence for the 2175 \AA \ extinction feature.
Spectroscopy of this afterglow will be further discussed in Xu et al.\ in
preparation.

\subsection{GRB\,080607 ($z=3.0368$)}
The data presented here have previously been published in \citet{X:080607}.
The spectrum is characterized by a very strong Ly$\alpha$ line and 
by the detection of both H$_2$ and CO molecular lines. The spectrum shows
evidence for substantial reddening. Based on the broad
band photometry there is also evidence for the presence of the 2175 \AA \
extinction bump \citep{X:080607}. The burst classifies
as a dark burst by the \citet{palli:dark} definition. In the figure of this
spectrum we only show the R400 spectrum. For the blue grating spectrum
covering Ly$\alpha$ we refer to \citet{X:080607}.

\subsection{GRB\,080707 ($z=1.2322$)}
The data presented here have previously only been published in circulars
\citep{fynbo:080707}. The spectrum was obtained under quite poor conditions with
bad seeing and at high airmass. A spectrum was also secured with a higher
resolution grating. All of the observed strong, spectral lines can be associated
with a single absorption system.  We associate the GRB redshift with
this gas.
The upper limit on the redshift is about 2.2 based on the
absence of a Ly$\alpha$ forest.

\subsection{GRB\,080710 ($z=0.8454$)}
The data presented here have previously only been published in circulars
\citep{perley:080710}. 
All of the observed strong, spectral lines can be associated
with a single absorption system.  
The system also shows fine structure lines.
We associate the GRB redshift with
this gas. The plotted spectrum has been normalized.

\subsection{GRB\,080721 ($z=2.5914$)}
The data presented here have previously been published in \citet{starling:080721}.
Spectra were also secured with a number of higher
resolution gratings. The spectrum is affected by fringing redwards of about 7000
\AA. The redshift is on both Ly$\alpha$ and metal (including fine-structure)
absorption lines.

\subsection{GRB\,080804 ($z=2.2045$)}
The data presented here have previously only been published in circulars
\citep{thoene:080804}. The spectrum was obtained with the high resolution UVES 
spectrograph.

\subsection{GRB\,080805 ($z=1.5042$)}
The data presented here have previously only been published in circulars
\citep{palli:080805}. The strongest line in the spectrum is from a very strong
intervening \ion{Mg}{2} system. The proposed GRB redshift of 1.504 is based on
a higher redshift weaker metal absorption line system.  There is evidence for
the presence of the 2175 \AA \ extinction feature in the shape of the spectrum
at the proposed GRB redshift. The burst classifies as a dark burst according
the Jakobsson et al.\ (2004) definition.

\subsection{GRB\,080810 ($z=3.3604$)}
The data presented here have previously only been published in circulars
\citep{antonio:080810}.  A higher resolution spectrum was secured at the Keck
telescope and this can be inspected in Page et al.\ (2009). The host absorber
is remarkable in having a relatively low \ion{H}{1} column density. There is
evidence for Ly$\alpha$ emission presumably from the underlying host galaxy.

\subsection{GRB\,080905B ($z=2.3739$)}
The data presented here have previously only been published in circulars
\citep{vreeswijk:080905B}.  The light form the afterglow is blended with light
from another object on the slit.  Spectra were also secured with a number of
higher resolution gratings. The redshift is based on metal absorption lines
including fine structure lines.

\subsection{GRB\,080913 ($z=6.7$)}
The data presented here have previously been published in Greiner et al.\ (2009).
The redshift is based on the detection of the onset of the Gunn-Peterson trough
at 9400 \AA. No absorption or emission lines are detected in the spectrum. This 
is the most distant GRB for which the redshift has been determined in our
spectroscopic sample. 
The burst classifies as a dark burst by the \citet{palli:dark} definition.

\subsection{GRB\,080916A ($z=0.6887$)}
The data presented here have previously only been published in circulars
\citep{fynbo:080916A}.  The redshift is based on both absorption lines and
emission lines from the underlying host galaxy.

\subsection{GRB\,080928 ($z=1.6919$)}
The data presented here have previously only been published in circulars
\citep{vreeswijk:080928}.  This is another good example of a spectrum where the
lines from the intervening absorber are significantly stronger than in the 
likely GRB absorber. The proposed GRB system is the highest redshift, but
weaker, of two metal line systems detected in the spectrum. The upper limit
on the redshift based on the absence of Ly$\alpha$ forest features is about 2.1.

\section{Linelists and spectra of GRB afterglows}
\label{tablesandspectra}

\begin{deluxetable}{@{}llllc@{}}
\tablecaption{GRB\,050319 ($z=3.2425$) 
\label{050319}}
\tablewidth{0pt}
\tablehead{
$\lambda_{\textrm{obs}}$ [\AA] &
$\lambda_{\textrm{rest}}$ [\AA] &
\hspace*{2mm} $z$ &
Feature &
EW$_{\textrm{obs}}$ [\AA]
}
\startdata

5157. & 1215.7         &        & Ly$\alpha$             &  \\
5348.2 & 1260.4        & 3.2425  & \ion{Si}{2}         &  $8\pm3$  \\
\multirow{2}{*}{5530.5} & 1302.2  & 3.2425  & \ion{O}{1} &  \multirow{2}{*}{9$\pm$3}  \\
     & 1304.4        & 3.2425  & \ion{Si}{2}     &    \\
5663.8 & 1334.5        & 3.2425  & \ion{C}{2}          &  $9\pm3$  \\
5913.2 & 1393.7        & 3.2425  & \ion{Si}{4}         &  $5\pm3$   \\
5949.9 & 1402.7        & 3.2425  & \ion{Si}{4}         &  $9\pm3$  \\
6472.8 & 1526.7        & 3.2425  & \ion{Si}{2}         &  $6\pm3$  \\
\multirow{2}{*}{6578.9} & 1548.2        & 3.2425  & \ion{C}{4}          &  \multirow{2}{*}{$8\pm3$}  \\
       & 1550.8        & 3.2425  & \ion{C}{4}         &    \\
\enddata
\end{deluxetable}

\begin{deluxetable}{@{}llllc@{}}
\tablecaption{GRB\,050401 ($z=2.8983$)}
\tablewidth{0pt}
\tablehead{
$\lambda_{\textrm{obs}}$ [\AA] &
$\lambda_{\textrm{rest}}$ [\AA] &
\hspace*{2mm} $z$ &
Feature &
EW$_{\textrm{obs}}$ [\AA]
}
\startdata
4739.         & 1215.7  &  2.8983        &  Ly$\alpha$  &  \\
\multirow{2}{*}{5074.4} &  1302.2        &  2.8983 & \ion{O}{1}      & \multirow{2}{*}{14$\pm$4} \\
     &  1304.4        &  2.8983 & \ion{Si}{2}    &  \\
5180.0 &              &         &           & 6$\pm$3  \\
5194.7 &                &         &         & 14$\pm$4 \\
\multirow{2}{*}{5205.0} & 1334.5 & 2.8983 & \ion{C}{2} & \multirow{2}{*}{12$\pm$3} \\  
     & 1335.7 & 2.8983 & \ion{C}{2}* & \\  
\hdashline
5422.2 &  1548.2        &  2.4982 & \ion{C}{4}     & \multirow{3}{*}{14$\pm$2} \\
\enddata
\tablenotetext{1}{Likely blended with another line.}
\end{deluxetable}

\clearpage
\begin{deluxetable}{lcccc}
\tablecaption{GRB 050408 ($z=1.2356$)\label{tab:050408}}
\tablewidth{0pt}
\tablehead{\colhead{$\lambda_{\textrm{obs}}$ [\AA]} &
\colhead{$\lambda_{\textrm{rest}}$ [\AA]} &
\colhead{\hspace*{2mm} $z$} &
\colhead{Feature} &
\colhead{EW$_{\textrm{obs}}$ [\AA]} }
\startdata
\cutinhead{GMOS/R150}
4610.4&2056.25&1.2356&CrII&$  6.17\pm1.51$\\
&2062.2&1.2356&CrII&\\
&2062.7&1.2356&ZnII&\\
&2066.2&1.2356&CrII&\\
5241.4&2333.5&1.2356&FeII*&$  5.49\pm0.77$\\
&2338.7&1.2356&FeII*&\\
&2344.2&1.2356&FeII&\\
&2345.0&1.2356&FeII*&\\
&2349.0&1.2356&FeII*&\\
5310.0&2365.6&1.2356&FeII*&$  3.00\pm0.67$\\
&2374.5&1.2356&FeII&\\
5328.7&2381.5&1.2356&FeII*&$  4.22\pm0.65$\\
&2382.8&1.2356&FeII&\\
&2383.8&1.2356&FeII*&\\
&2389.4&1.2356&FeII*&\\
5782.9&2586.7&1.2356&FeII&$  3.55\pm0.43$\\
5812.7&2594.5&1.2356&MnII&$  7.83\pm0.55$\\
&2599.2&1.2356&FeII*&\\
&2600.2&1.2356&FeII&\\
&2606.5&1.2356&MnII&\\
&2607.9&1.2356&FeII*&\\
&2612.7&1.2356&FeII*&\\
6261.7&2796.4&1.2356&MgII&$ 10.90\pm0.66$\\
&2803.5&1.2356&MgII&\\
6377.9&2853.0&1.2356&MgI&$  3.48\pm0.39$\\
7246.3&3240.0&1.2356&TiII*&$  2.24\pm0.48$\\
&3242.9&1.2356&TiII&\\
\cutinhead{GMOS/R831}
6378.1&2853.0&1.2356&MgI&$  3.02\pm0.29$\\
6752.7&3021.5&1.2356&FeI&$  0.69\pm0.19$\\
6875.1&3073.9&1.2356&TiII&$  2.91\pm0.30$\\
7566.9&3384.7&1.2356&TiII&$  1.11\pm0.21$\\
\enddata
\end{deluxetable}

\begin{deluxetable}{@{}llllc@{}}
\tablecaption{GRB\,050730 ($z=3.9693$)}
\tablewidth{0pt}
\tablehead{
$\lambda_{\textrm{obs}}$ [\AA] &
$\lambda_{\textrm{rest}}$ [\AA] &
\hspace*{2mm} $z$ &
Feature &
EW$_{\textrm{obs}}$ [\AA]
}
\startdata

4889.3 & 1215.7 & 3.022  & Ly$\alpha$  &   \\ 
5549.2 & 1215.7 & 3.5680  & Ly$\alpha$  &   \\ 
6040.  & 1215.7 & 3.9693 & Ly$\alpha$  &   \\ 
6230.6 & 1253.8 & 3.9693 & \ion{S}{2} & 1.38$\pm$0.16 \\
\hdashline
\multirow{2}{*}{6261.9} & 1259.5 & 3.9693 & \ion{S}{2} & \multirow{3}{*}{5.5$\pm$0.3} \\
     & 1260.4 & 3.9693 & \ion{Si}{2} &  \\
6291.1 & 1264.7 & 3.9693 & \ion{Si}{2}* &  \\
\hdashline
6400.8 & 1402.7 & 3.5680 & \ion{Si}{4} & 0.65$\pm$0.13 \\
\enddata
\end{deluxetable}

\clearpage
\begin{deluxetable}{lcccc}
\tablecaption{GRB 050801\label{tab:050801}}
\tablewidth{0pt}
\tablehead{\colhead{$\lambda_{\textrm{obs}}$ [\AA]} &
\colhead{$\lambda_{\textrm{rest}}$ [\AA]} &
\colhead{\hspace*{2mm} $z$} &
\colhead{Feature} &
\colhead{EW$_{\textrm{obs}}$ [\AA]} }
\startdata
\cutinhead{LRISb/B600}
\cutinhead{LRISr/R600}
6040.7&&&&$ 2.32\pm0.76$\\
\enddata
\end{deluxetable}

\begin{deluxetable}{@{}llllc@{}}
\tablecaption{GRB\,050802 ($z=1.7102$)}
\tablewidth{0pt}
\tablehead{
$\lambda_{\textrm{obs}}$ [\AA] &
$\lambda_{\textrm{rest}}$ [\AA] &
\hspace*{2mm} $z$ &
Feature &
EW$_{\textrm{obs}}$ [\AA]
}
\startdata
4199.4 & 1548.2        &  1.7102  & \ion{C}{4}  & \multirow{2}{*}{7.4$\pm$1.5} \\
4199.4 & 1550.8        &  1.7102  & \ion{C}{4}  &  \\
6353.5 & 2344.2        &  1.7102  & \ion{Fe}{2} & 7.1$\pm$0.9 \\
6458.7 & 2382.8        &  1.7102  & \ion{Fe}{2} & 7.3$\pm$0.9 \\
7009.8 & 2586.7        &  1.7102  & \ion{Fe}{2} & 5.0$\pm$1.1 \\
\enddata
\end{deluxetable}

\clearpage
\begin{deluxetable}{lcccc}
\tablecaption{GRB 050820A ($z=2.6147$)\label{tab:050820A}}
\tablewidth{0pt}
\tablehead{\colhead{$\lambda_{\textrm{obs}}$ [\AA]} &
\colhead{$\lambda_{\textrm{rest}}$ [\AA]} &
\colhead{\hspace*{2mm} $z$} &
\colhead{Feature} &
\colhead{EW$_{\textrm{obs}}$ [\AA]} }
\startdata
5518.6&1526.7&2.6147&SiII&$  5.68\pm0.05$\\
5596.3&1548.2&2.6147&CIV&5.44$\pm$0.04\\
5605.7&1550.8&2.6147&CIV&5.01$\pm$0.05\\
\enddata
\end{deluxetable}

\begin{deluxetable}{@{}llllc@{}}
\tablecaption{GRB\,050824 ($z=0.8278$)}
\tablewidth{0pt}
\tablehead{
$\lambda_{\textrm{obs}}$ [\AA] &
$\lambda_{\textrm{rest}}$ [\AA] &
\hspace*{2mm} $z$ &
Feature &
EW$_{\textrm{obs}}$ [\AA]
}
\startdata
4284.7 & 2344.2 & 0.8278 & \mbox{Fe\,{\sc ii}} & 1.8$\pm$0.5 \\
4337.9 & 2374.5 & 0.8278 & \mbox{Fe\,{\sc ii}} & \multirow{2}{*}{4.5$\pm$0.8} \\
4354.2 & 2382.8 & 0.8278 & \mbox{Fe\,{\sc ii}} &  \\
4728.6 & 2586.7 & 0.8278 & \mbox{Fe\,{\sc ii}} & 1.5$\pm$0.4 \\
5108.6 & 2796.3 & 0.8278 & \mbox{Mg\,{\sc ii}} & 1.7$\pm$0.3 \\ 
5124.5 & 2803.5 & 0.8278 & \mbox{Mg\,{\sc ii}} & 2.3$\pm$0.3 \\ 
6814.5 & 3727.7 & 0.8278 & [\mbox{O\,{\sc ii}}] & emission \\
7074.4 & 3869.8 & 0.8278 & [\mbox{Ne\,{\sc iii}}] & emission \\
7192.4 & 3934.8 & 0.8278 & \ion{Ca}{2} & 3.0$\pm$0.5 \\ 
7752.9 &        &        &                     & 1.2$\pm$0.4 \\
7938.4 & 4341.7 & 0.8278 & H$\gamma$ & emission    \\
8889.2 & 4862.7 & 0.8278 & H$\beta$  & emission    \\
9067.6 & 4960.3 & 0.8278 & [\mbox{O\,{\sc iii}}]  & emission    \\
9155.0 & 5008.2 & 0.8278 & [\mbox{O\,{\sc iii}}]  & emission    \\
\enddata

\end{deluxetable}

\begin{deluxetable}{@{}llllc@{}}
\tablecaption{GRB\,050908 ($z=3.3467$)}
\tablewidth{0pt}
\tablehead{
$\lambda_{\textrm{obs}}$ [\AA] &
$\lambda_{\textrm{rest}}$ [\AA] &
\hspace*{2mm} $z$ &
Feature &
EW$_{\textrm{obs}}$ [\AA]
}
\startdata
4402.7 & 1215.7 & 2.6208 & Ly$\alpha$ & --  \\
5280.6 & 1215.7 & 3.3467 & Ly$\alpha$ & --  \\
5527.8 & 1526.7  & 2.6208 & \ion{Si}{2}       & 8.1$\pm$0.2 \\
\multirow{2}{*}{5607.4} & 1548.2  & 2.6208 & \ion{C}{4} & \multirow{2}{*}{3.2$\pm$0.2} \\
& 1550.8  & 2.2613 & \ion{C}{4} &  \\
5627.8 & 1393.7 & 3.0383 & \ion{Si}{4} & 1.9$\pm$0.2 \\
5664.9 & 1402.7 & 3.0383 & \ion{Si}{4} & 0.80$\pm$0.15\\
\multirow{2}{*}{5683.8} & 1302.2 & 3.3467 & \ion{O}{1} & \multirow{2}{*}{1.3$\pm$0.2} \\
\enddata
\end{deluxetable}

\begin{deluxetable}{@{}llllc@{}}
\tablecaption{GRB\,050922C ($z=2.1995$)}
\tablewidth{0pt}
\tablehead{
$\lambda_{\textrm{obs}}$ [\AA] &
$\lambda_{\textrm{rest}}$ [\AA] &
\hspace*{2mm} $z$ &
Feature &
EW$_{\textrm{obs}}$ [\AA]
}
\startdata
3630.0 & 1215.7 & 1.9904 & Ly$\alpha$ & \\
3654.6 & 1215.7 & 2.0008 & Ly$\alpha$ & \\
3740.0 & 1215.7 & 2.0768 & Ly$\alpha$ & \\
3893.0 & 1215.7 & 2.1995 & Ly$\alpha$ & \\
3975.5 & 1242.8 & 2.1995 & NV          & 2.5$\pm$0.2 \\
\hdashline
\multirow{3}{*}{4004.1} & 1302.2 & 2.0768 & \ion{O}{1}        & \multirow{3}{*}{0.8$\pm$0.2} \\
       & 1304.4 & 2.0768 & \ion{Si}{2}         & \\
       & 1250.6 & 2.1995 & \ion{S}{2}          &  \\
\enddata

\end{deluxetable}

\begin{deluxetable}{@{}llllc@{}}
\tablecaption{GRB\,060115 ($z=3.5328$)}
\tablewidth{0pt}
\tablehead{
$\lambda_{\textrm{obs}}$ [\AA] &
$\lambda_{\textrm{rest}}$ [\AA] &
\hspace*{2mm} $z$ &
Feature &
EW$_{\textrm{obs}}$ [\AA]
}
\startdata
5507.0 & 1215.7 & 3.5328 & Ly$\alpha$ &  \\
5634.3 & 1242.8 & 3.5328 & NV & 4$\pm$2 \\
5713.1 & 1260.4 & 3.5328 & \ion{Si}{2} & 8$\pm$3 \\
5735.2 & 1264.7 & 3.5328 & \ion{Si}{2}* & 5$\pm$2 \\
\multirow{2}{*}{5909.3} & 1302.2 & 3.5328 & \ion{O}{1} & \multirow{2}{*}{11.5$\pm$1.8} \\
& 1304.4 & 3.5328 & \ion{Si}{2} &  \\
6030.0 &        &      &        & 2.8$\pm$1.4 \\
6051.0 & 1334.5 & 3.5328 & \ion{C}{2} & 8.6$\pm$1.8 \\
6918.5 & 1526.7 & 3.5328 & \ion{Si}{2} & 6.3$\pm$1.6 \\
\multirow{2}{*}{7022.5} & 1548.8 & 3.5328 & \ion{C}{4} & \multirow{2}{*}{6.0$\pm$1.6} \\
& 1550.2 & 3.5328 & \ion{C}{4} &  \\
7572.8 & 1670.8 & 3.5328 & \ion{Al}{2} & 8.1$\pm$1.5 \\ 
\enddata

\end{deluxetable}

\clearpage
\begin{deluxetable}{lcccc}
\tablecaption{GRB 060124 ($z=2.3000$)\label{tab:060124}}
\tablewidth{0pt}
\tablehead{\colhead{$\lambda_{\textrm{obs}}$ [\AA]} &
\colhead{$\lambda_{\textrm{rest}}$ [\AA]} &
\colhead{\hspace*{2mm} $z$} &
\colhead{Feature} &
\colhead{EW$_{\textrm{obs}}$ [\AA]} }
\startdata
\cutinhead{LRIS/R300}
4011.7&1215.7&2.3000&Ly$\alpha$&$ 13.12\pm2.44$\\
4231.2&&&&$10.03\pm1.70$\\
4598.5&1393.8&2.3000&SiIV&$  4.73\pm1.08$\\
4846.6&&&&$ 5.73\pm0.84$\\
4862.6&&&&$ 2.43\pm0.71$\\
5108.8&1548.2&2.3000&CIV&$  9.68\pm0.74$\\
&1550.8&2.3000&CIV&\\
6873.6&&&&$ 2.56\pm0.70$\\
\enddata
\end{deluxetable}

\begin{deluxetable}{@{}llllc@{}}
\tablecaption{GRB\,060206 ($z=4.0559$)}
\tablewidth{0pt}
\tablehead{
$\lambda_{\textrm{obs}}$ [\AA] &
$\lambda_{\textrm{rest}}$ [\AA] &
\hspace*{2mm} $z$ &
Feature &
EW$_{\textrm{obs}}$ [\AA]
}

\startdata

6147.9 & 1215.7 & 4.0559 & Ly$\alpha$ &  \\
6370.8 & 1260.4 & 4.0559 & \ion{Si}{2} & \multirow{2}{*}{11.99$\pm$0.09} \\
6395.2 & 1264.7 & 4.0559 & \ion{Si}{2}* & \\
6421.5 & 2586.7 & 1.4822 & \ion{Fe}{2} & 0.29$\pm$0.09 \\
6453.7 & 2600.2 & 1.4822 & \ion{Fe}{2} & 0.68$\pm$0.09 \\
\multirow{2}{*}{6589.0} & 1302.2 & 4.0559 & \ion{O}{1} & \multirow{2}{*}{14.04$\pm$0.09} \\
     & 1304.4 & 4.0559 & \ion{Si}{2} & \\
6746.0 & 1334.5 & 4.0559 & \ion{C}{2}  & 11.45$\pm$0.14 \\  
\enddata


\end{deluxetable}

\clearpage
\begin{deluxetable}{lcccc}
\tablecaption{GRB 060210 ($z=3.9122$)\label{tab:060210}}
\tablewidth{0pt}
\tablehead{\colhead{$\lambda_{\textrm{obs}}$ [\AA]} &
\colhead{$\lambda_{\textrm{rest}}$ [\AA]} &
\colhead{\hspace*{2mm} $z$} &
\colhead{Feature} &
\colhead{EW$_{\textrm{obs}}$ [\AA]} }
\startdata
\cutinhead{GMOS/R400}
5963.5&2026.1&1.9253&ZnII&$174.46\pm1.59$\\
&1215.7&3.9133&Ly$\alpha$&\\
6088.3&1238.8&3.9133&NV&$  5.02\pm0.53$\\
&1239.9&3.9133&MgII&\\
&1240.4&3.9133&MgII&\\
6106.9&1242.8&3.9133&NV&$  4.93\pm0.46$\\
6154.4&1250.6&3.9133&SII&$  8.32\pm0.49$\\
&1253.8&3.9133&SII&\\
\enddata
\end{deluxetable}

\clearpage
\begin{deluxetable}{lcccc}
\tablecaption{GRB 060502A ($z=1.5025$)\label{tab:060502A}}
\tablewidth{0pt}
\tablehead{\colhead{$\lambda_{\textrm{obs}}$ [\AA]} &
\colhead{$\lambda_{\textrm{rest}}$ [\AA]} &
\colhead{\hspace*{2mm} $z$} &
\colhead{Feature} &
\colhead{EW$_{\textrm{obs}}$ [\AA]} }
\startdata
\cutinhead{GMOS/R400}
5688.2&2796.4&1.0323&MgII&$  5.11\pm0.47$\\
&2803.5&1.0323&MgII&\\
5727.5&&&&$ 3.53\pm0.48$\\
5860.6&2338.7&1.5026&FeII*&$  1.31\pm0.43$\\
5897.1&&&&$ 1.96\pm0.40$\\
5936.4&2374.5&1.5026&FeII&$  2.73\pm0.30$\\
5957.8&2381.5&1.5026&FeII*&$  2.82\pm0.26$\\
5970.9&2382.8&1.5026&FeII&$  6.01\pm0.28$\\
      &2383.8&1.5026&FeII*&\\
6013.0&2396.2&1.5026&FeII*&$ 12.57\pm0.52$\\
      &2396.4&1.5026&FeII*&\\
      &2400.0&1.5026&FeII*&\\
      &2405.2&1.5026&FeII*&\\
      &2405.6&1.5026&FeII*&\\
      &2407.4&1.5026&FeII*&\\
      &2411.3&1.5026&FeII*&\\
      &2411.8&1.5026&FeII*&\\
      &2414.1&1.5026&FeII*&\\
6090.6&&&&$ 1.13\pm0.23$\\
6445.6&2576.9&1.5026&MnII&$  0.83\pm0.21$\\
6470.2&2586.7&1.5026&FeII&$  2.08\pm0.25$\\
6504.7&2599.2&1.5026&FeII*&$  3.91\pm0.23$\\
      &2600.2&1.5026&FeII&\\
6521.0&2606.5&1.5026&MnII&$  0.77\pm0.22$\\
      &2607.9&1.5026&FeII*&\\
6535.6&2612.7&1.5026&FeII*&$  1.10\pm0.25$\\
6839.2&&&&$ 2.38\pm0.33$\\
6884.7&2740.4&1.5026&FeII*&$  7.74\pm0.46$\\
      &2747.3&1.5026&FeII*&\\
      &2747.9&1.5026&FeII*&\\
      &2750.2&1.5026&FeII*&\\
      &2756.6&1.5026&FeII*&\\
6998.0&2796.4&1.5026&MgII&$  4.73\pm0.18$\\
7016.0&2803.5&1.5026&MgII&$  4.26\pm0.21$\\
7033.5&&&&$ 3.34\pm0.24$\\
7139.9&2853.0&1.5026&MgI&$  2.26\pm0.20$\\
7208.9&&&&$ 1.08\pm0.30$\\
\enddata
\end{deluxetable}

\begin{deluxetable}{@{}llllc@{}}
\tablecaption{GRB\,060512 ($z=2.1$)}
\tablewidth{0pt}
\tablehead{
$\lambda_{\textrm{obs}}$ [\AA] &
$\lambda_{\textrm{rest}}$ [\AA] &
\hspace*{2mm} $z$ &
Feature &
EW$_{\textrm{obs}}$ [\AA]
}

\startdata
3750 & 1215.7 & 2.1 & Ly$\alpha$ & - \\
\enddata
\end{deluxetable}

\begin{deluxetable}{@{}llllc@{}}
\tablecaption{GRB\,060526 ($z=3.2213$)}
\tablewidth{0pt}
\tablehead{
$\lambda_{\textrm{obs}}$ [\AA] &
$\lambda_{\textrm{rest}}$ [\AA] &
\hspace*{2mm} $z$ &
Feature &
EW$_{\textrm{obs}}$ [\AA]
}

\startdata
5131.0 & 1215.7 & 3.2213 & Ly$\alpha$ &  \\
5248.0 & 1242.8 & 3.2213 & NV & 1.1$\pm$0.3 \\
5319.8 & 1260.4 & 3.2213 & \ion{Si}{2} & 6.0$\pm$0.3 \\
\multirow{2}{*}{5500.7} & 1302.2 & 3.2213 & \ion{O}{1} & \multirow{2}{*}{7.5$\pm$0.4} \\
& 1304.4 & 3.2213 & \ion{Si}{2} &  \\
5633.4 & 1334.4 & 3.2213 & \ion{C}{2} & 7.5$\pm$0.4 \\
6443.9 & 1526.7 & 3.2213 & \ion{Si}{2} & 3.3$\pm$0.4 \\
\multirow{2}{*}{6539.0} & 1548.2 & 3.2213 & \ion{C}{4} & \multirow{2}{*}{1.9$\pm$0.4} \\
& 1550.8 & 3.2213 & \ion{C}{4} & \\
6788.7 & 1608.4 & 3.2213 & \ion{Fe}{2} & 1.9$\pm$0.4 \\
7053.2 & 1670.7 & 3.2213 & \ion{Al}{2} & 3.6$\pm$0.4 \\

\enddata


\end{deluxetable}

\begin{deluxetable}{@{}llllc@{}}
\tablecaption{GRB\,060604}
\tablewidth{0pt}
\tablehead{
$\lambda_{\textrm{obs}}$ [\AA] &
$\lambda_{\textrm{rest}}$ [\AA] &
\hspace*{2mm} $z$ &
Feature &
EW$_{\textrm{obs}}$ [\AA]
}

\startdata

5217.0  &  & & & 13$\pm$4 \\

\enddata


\end{deluxetable}

\clearpage
\begin{deluxetable}{lcccc}
\tablecaption{GRB 060607A ($z=3.0749$)\label{tab:060607A}}
\tablewidth{0pt}
\tablehead{\colhead{$\lambda_{\textrm{obs}}$ [\AA]} &
\colhead{$\lambda_{\textrm{rest}}$ [\AA]} &
\colhead{\hspace*{2mm} $z$} &
\colhead{Feature} &
\colhead{EW$_{\textrm{obs}}$ [\AA]} }
\startdata
6308.8&1548.2&3.0746&CIV&$  1.13\pm0.05$\\
6319.4&1550.8&3.0746&CIV&$  0.83\pm0.05$\\
\enddata
\end{deluxetable}

\begin{deluxetable}{@{}llllc@{}}
\tablecaption{GRB\,060707 ($z=3.4240$)}
\tablewidth{0pt}
\tablehead{
$\lambda_{\textrm{obs}}$ [\AA] &
$\lambda_{\textrm{rest}}$ [\AA] &
\hspace*{2mm} $z$ &
Feature &
EW$_{\textrm{obs}}$ [\AA]
}

\startdata

5342.6 & 1215.7 & 3.4240  & Ly$\alpha$           &      \\
5538.8 & 1250.6 & 3.4240 & \ion{S}{2} & 6.7$\pm$1.3   \\
5724.4 &    &    &     & 4.5$\pm$1.2   \\
5760.8 & 1302.2 & 3.4240 & \mbox{O\,{\sc i}}    & 9.2$\pm$1.1 \\
5768.8 & 1304.4 & 3.4240 & \mbox{Si\,{\sc ii}}  & 6.0$\pm$2.0 \\
5902.1 & 1334.5 & 3.4240 & \mbox{C\,{\sc ii}}   & \multirow{2}{*}{10.4$\pm$1.3} \\
5909.0 & 1335.7 & 3.4240 & \mbox{C\,{\sc ii}}*  &  \\
6755.7 & 1526.7 & 3.4240 & \mbox{Si\,{\sc ii}}  & 10.2$\pm$1.4 \\
6851.6 & 1548.2 & 3.4240 & \mbox{C\,{\sc iv}}   & \multirow{2}{*}{5.7$\pm$1.2} \\
6858.3 & 1550.8 & 3.4240 & \mbox{C\,{\sc iv}}   &   \\
7391.4 & 1670.8 & 3.4240 & \mbox{Al\,{\sc ii}}  & 9.4$\pm$1.6 \\
8002.1 & 1808.0 & 3.4240 & \mbox{Si\,{\sc ii}}  & 3.8$\pm$1.9 \\

\enddata


\end{deluxetable}

\begin{deluxetable}{@{}llllc@{}}
\tablecaption{GRB\,060714 ($z=2.7108$)}
\tablewidth{0pt}
\tablehead{
$\lambda_{\textrm{obs}}$ [\AA] &
$\lambda_{\textrm{rest}}$ [\AA] &
\hspace*{2mm} $z$ &
Feature &
EW$_{\textrm{obs}}$ [\AA]
}

\startdata

4503.8 & 1215.7 & 2.7108 & Ly$\alpha$             &      \\
4640.2 & 1250.6 & 2.7108 & \mbox{S\,{\sc ii}}     &  3.3$\pm$0.7 \\
4651.6 & 1253.8 & 2.7108 & \mbox{S\,{\sc ii}}     &  3.5$\pm$0.6 \\
4676.5 & 1260.4 & 2.7108 & \mbox{Si\,{\sc ii}}    &  7.2$\pm$0.6 \\
4693.8 & 1264.7 & 2.7108 & \mbox{Si\,{\sc ii}}*   &  3.7$\pm$0.6 \\
\multirow{2}{*}{4837.3} & 1302.2 & 2.7108 & \mbox{O\,{\sc i}} & \multirow{2}{*}{12.7$\pm$0.6}  \\
                & 1304.4 & 2.7108 & \mbox{Si\,{\sc ii}} &  \\
4859.7 & 1309.3 & 2.7108 & \mbox{Si\,{\sc ii}}*   &  1.0$\pm$0.4 \\
\enddata


\end{deluxetable}

\begin{deluxetable}{@{}llllc@{}}
\tablecaption{GRB\,060729 ($z=0.5428$)}
\tablewidth{0pt}
\tablehead{
$\lambda_{\textrm{obs}}$ [\AA] &
$\lambda_{\textrm{rest}}$ [\AA] &
\hspace*{2mm} $z$ &
Feature &
EW$_{\textrm{obs}}$ [\AA]
}

\startdata

4010.7 & 2600.2 & 0.5428 & \mbox{Fe\,{\sc ii}}     & 0.7$\pm$0.2  \\
4314.7 & 2796.3 & 0.5428 & \mbox{Mg\,{\sc ii}}     & \multirow{2}{*}{2.80$\pm$0.06} \\
4325.5 & 2803.5 & 0.5428 & \mbox{Mg\,{\sc ii}}     &      \\
4402.1 & 2853.0 & 0.5428 & \mbox{Mg\,{\sc i}}      & 0.91$\pm$0.06 \\
6070.1 & 3934.8 & 0.5428 & \mbox{Ca\,{\sc ii}} (K) & 1.16$\pm$0.04 \\
6123.7 & 3969.6 & 0.5428 & \mbox{Ca\,{\sc ii}} (H) & 0.85$\pm$0.04 \\

\enddata


\end{deluxetable}

\begin{deluxetable}{@{}llllc@{}}
\tablecaption{GRB\,060807}
\tablewidth{0pt}
\tablehead{
$\lambda_{\textrm{obs}}$ [\AA] &
$\lambda_{\textrm{rest}}$ [\AA] &
\hspace*{2mm} $z$ &
Feature &
EW$_{\textrm{obs}}$ [\AA]
}

\startdata

5200.0 &     &    &    & 0.6$\pm$0.3 \\

\enddata


\end{deluxetable}

\begin{deluxetable}{@{}llllc@{}}
\tablecaption{GRB\,060904B ($z=0.7029$)}
\tablewidth{0pt}
\tablehead{
$\lambda_{\textrm{obs}}$ [\AA] &
$\lambda_{\textrm{rest}}$ [\AA] &
\hspace*{2mm} $z$ &
Feature &
EW$_{\textrm{obs}}$ [\AA]
}

\startdata

4383.5 & 2576.9 & 0.7029 & \mbox{Mn\,{\sc ii}} & 0.6$\pm$0.3 \\
4404.0 & 2586.7 & 0.7029 & \mbox{Fe\,{\sc ii}} & 1.6$\pm$0.3 \\
4427.3 & 2600.2 & 0.7029 & \mbox{Fe\,{\sc ii}} & 3.1$\pm$0.3 \\
4763.7 & 2796.3 & 0.7029 & \mbox{Mg\,{\sc ii}} & \multirow{2}{*}{7.9$\pm$0.2} \\
4772.6 & 2803.5 & 0.7029 & \mbox{Mg\,{\sc ii}} & \\
4858.4 & 2853.0 & 0.7029 & \mbox{Mg\,{\sc i}} & 1.7$\pm$0.2 \\
6352.6 & 3727.7 & 0.7029 & [\mbox{O\,{\sc ii}}] & emission \\
6701.7 & 3934.8 & 0.7029 & \mbox{Ca\,{\sc ii}} (K) & 2.3$\pm$0.2 \\
6760.6 & 3969.6 & 0.7029 & \mbox{Ca\,{\sc ii}} (H) & 1.8$\pm$0.2 \\
8531.7 & 5008.2 & 0.7029 & [\mbox{O\,{\sc iii}}] & emission \\

\enddata


\end{deluxetable}

\clearpage

\begin{deluxetable}{@{}llllc@{}}
\tablecaption{GRB\,060906 ($z=3.6856$)}
\tablewidth{0pt}
\tablehead{
$\lambda_{\textrm{obs}}$ [\AA] &
$\lambda_{\textrm{rest}}$ [\AA] &
\hspace*{2mm} $z$ &
Feature &
EW$_{\textrm{obs}}$ [\AA]
}

\startdata

5689.8 & 1215.7 & 3.6856 & Ly$\alpha$             &      \\
5904.0 & 1260.4 & 3.6856 & \mbox{Si\,{\sc ii}}    & 4.0$\pm$1.0 \\
\multirow{2}{*}{6108.2} & 1302.2 & 3.6856 & \mbox{O\,{\sc i}} & \multirow{2}{*}{5.5$\pm$1.1} \\
& 1304.4 & 3.6856 & \mbox{Si\,{\sc ii}} &  \\
6254.6 & 1334.5 & 3.6856 & \mbox{C\,{\sc ii}}  & 6.1$\pm$0.9 \\
6335.9 & 2796.3 & 1.2659 & \mbox{Mg\,{\sc ii}} & \multirow{2}{*}{3.7$\pm$0.7} \\
6352.5 & 2803.5 & 1.2659 & \mbox{Mg\,{\sc ii}} & \\
6529.2 & 1393.8 & 3.6856 & \mbox{Si\,{\sc iv}}    & 3.2$\pm$0.8 \\
7153.6 & 1526.7 & 3.6856 & \mbox{Si\,{\sc ii}}    & 2.2$\pm$0.9 \\
\multirow{2}{*}{7258.1}  & 1548.2 & 3.6856& \ion{C}{4} & \multirow{2}{*}{3.4$\pm$0.9} \\
                         & 1550.8 & 3.6856& \ion{C}{4} & \\ 

\enddata


\end{deluxetable}

\begin{deluxetable}{@{}llllc@{}}
\tablecaption{GRB\,060908 ($z=1.8836$)}
\tablewidth{0pt}
\tablehead{
$\lambda_{\textrm{obs}}$ [\AA] &
$\lambda_{\textrm{rest}}$ [\AA] &
\hspace*{2mm} $z$ &
Feature &
EW$_{\textrm{obs}}$ [\AA]
}

\startdata

4464.8 & 1548.2 & 1.8836 & \ion{C}{4} & \multirow{2}{*}{9.3$\pm$1.5} \\
4471.5 & 1550.8 & 1.8836 & \ion{C}{4} &  \\

\enddata


\end{deluxetable}

\begin{deluxetable}{@{}llllc@{}}
\tablecaption{GRB\,060926 ($z=3.2086$)}
\tablewidth{0pt}
\tablehead{
$\lambda_{\textrm{obs}}$ [\AA] &
$\lambda_{\textrm{rest}}$ [\AA] &
\hspace*{2mm} $z$ &
Feature &
EW$_{\textrm{obs}}$ [\AA]
}

\startdata

5116.2 & 1215.7 & 3.2086 & Ly$\alpha$           &      \\
5382.1 & 2796.3 & 0.9242 & \mbox{Mg\,{\sc ii}}  & \multirow{2}{*}{4.8$\pm$1.8} \\
5393.1 & 2803.5 & 0.9242 & \mbox{Mg\,{\sc ii}}  &      \\
5614.3 & 1334.5 & 3.2086 & \mbox{C\,{\sc ii}}   & \multirow{2}{*}{5.5$\pm$1.4} \\
5625.1 & 1335.7 & 3.2086 & \mbox{C\,{\sc ii}}*  &      \\
5863.7 & 1393.8 & 3.2086 & \ion{Si}{4} & 3.9$\pm$1.4 \\
6426.2 & 1526.7 & 3.2086 & \mbox{Si\,{\sc ii}}  & 4.0$\pm$1.1 \\
\multirow{2}{*}{6517.7} & 1548.2 & 3.2086 & \mbox{C\,{\sc iv}} & \multirow{2}{*}{8.0$\pm$1.2} \\
& 1550.8 & 3.2086 & \mbox{C\,{\sc iv}} &  \\
6637.5 & 2374.5 & 1.7954 & \mbox{Fe\,{\sc ii}}  & 4.8$\pm$1.0 \\
6660.6 & 2382.8 & 1.7954 & \mbox{Fe\,{\sc ii}}  & 4.7$\pm$1.4 \\
7232.3 & 2586.7 & 1.7954 & \mbox{Fe\,{\sc ii}}  & 7.3$\pm$1.5 \\
7267.9 & 2600.2 & 1.7954 & \mbox{Fe\,{\sc ii}}  & 4.6$\pm$1.5 \\
7816.2 & 2796.3 & 1.7954 & \mbox{Mg\,{\sc ii}}  & 7.5$\pm$2.0 \\
7837.1 & 2803.5 & 1.7954 & \mbox{Mg\,{\sc ii}}  & 7.8$\pm$2.0 \\

\enddata


\end{deluxetable}


\begin{deluxetable}{@{}llllc@{}}
\tablecaption{GRB\,060927 ($z=5.4636$)}
\tablewidth{0pt}
\tablehead{
$\lambda_{\textrm{obs}}$ [\AA] &
$\lambda_{\textrm{rest}}$ [\AA] &
\hspace*{2mm} $z$ &
Feature &
EW$_{\textrm{obs}}$ [\AA]
}

\startdata

7865.  & 1215.7& 5.4636 & Ly$\alpha$ &  \\
8146.7 & 1260.4 & 5.4636 & \mbox{Si\,{\sc ii}}  & 9.0$\pm$4.0 \\

\enddata


\end{deluxetable}

\begin{deluxetable}{@{}llllc@{}}
\tablecaption{GRB\,061007 ($z=1.2622$)}
\tablewidth{0pt}
\tablehead{
$\lambda_{\textrm{obs}}$ [\AA] &
$\lambda_{\textrm{rest}}$ [\AA] &
\hspace*{2mm} $z$ &
Feature &
EW$_{\textrm{obs}}$ [\AA]
}

\startdata

5302.8 & 2344.2 & 1.2622 & \mbox{Fe\,{\sc ii}}  & 6.4$\pm$1.7 \\
5370.4 & 2374.5 & 1.2622 & \mbox{Fe\,{\sc ii}}  & 8.0$\pm$2.0 \\
5389.5 & 2382.8 & 1.2622 & \mbox{Fe\,{\sc ii}}  & 5.1$\pm$1.5 \\
5777.2 & 2796.3 & 1.0660 & \mbox{Mg\,{\sc ii}}  & \multirow{2}{*}{13.9$\pm$1.8} \\
5788.2 & 2803.5 & 1.0660 & \mbox{Mg\,{\sc ii}}  &  \\
5827.7 & 2576.9 & 1.2622 & \mbox{Mn\,{\sc ii}}  & 2.8$\pm$1.3 \\
5852.0 & 2586.7 & 1.2622 & \mbox{Fe\,{\sc ii}}  & 7.3$\pm$1.3 \\
5884.5 & 2600.2 & 1.2622 & \mbox{Fe\,{\sc ii}}  & 5.9$\pm$1.4 \\
6324.5 & 2796.3 & 1.2622 & \mbox{Mg\,{\sc ii}}  & 5.8$\pm$1.1 \\
6341.9 & 2803.5 & 1.2622 & \mbox{Mg\,{\sc ii}}  & 6.4$\pm$1.1 \\
8134.3 & 3934.8 & 1.0660 & \ion{Ca}{2} & 3.5$\pm$1.1 \\
8429.3 & 3726.0 & 1.2622 & [\mbox{O\,{\sc ii}}] & emission \\
8440.3 & 3728.8 & 1.2622 & [\mbox{O\,{\sc ii}}] & emission \\

\enddata


\end{deluxetable}

\begin{deluxetable}{@{}llllc@{}}
\tablecaption{GRB\,061021 ($z=0.3463$)}
\tablewidth{0pt}
\tablehead{
$\lambda_{\textrm{obs}}$ [\AA] &
$\lambda_{\textrm{rest}}$ [\AA] &
\hspace*{2mm} $z$ &
Feature &
EW$_{\textrm{obs}}$ [\AA]
}

\startdata
\multirow{2}{*}{3767.9} & 2796.4 & 0.3463 & \mbox{Mg\,{\sc ii}} & \multirow{2}{*}{$2.5 \pm 0.7$} \\
                      & 2803.5 & 0.3463 & \mbox{Mg\,{\sc ii}} & \\

\enddata


\end{deluxetable}

\begin{deluxetable}{@{}llllc@{}}
\tablecaption{GRB\,061110A ($z=0.7578$)}
\tablewidth{0pt}
\tablehead{
$\lambda_{\textrm{obs}}$ [\AA] &
$\lambda_{\textrm{rest}}$ [\AA] &
\hspace*{2mm} $z$ &
Feature &
EW$_{\textrm{obs}}$ [\AA]
}

\startdata

\multirow{2}{*}{4914.} & 2796.3 & 0.7578 & \ion{Mg}{2} & \multirow{2}{*}{6.4$\pm$1.4} \\
                       & 2803.5 & 0.7578 & \ion{Mg}{2} & \\
6551.8 & 3727.7 & 0.7578 & H$\beta$  & emission    \\
8719.7 & 4960.3 & 0.7578 & [\mbox{O\,{\sc iii}}]  & emission    \\
8803.3 & 5008.2 & 0.7578 & [\mbox{O\,{\sc iii}}]  & emission    \\

\enddata


\end{deluxetable}

\begin{deluxetable}{@{}llllc@{}}
\tablecaption{GRB\,061110B ($z=3.4344$)}
\tablewidth{0pt}
\tablehead{
$\lambda_{\textrm{obs}}$ [\AA] &
$\lambda_{\textrm{rest}}$ [\AA] &
\hspace*{2mm} $z$ &
Feature &
EW$_{\textrm{obs}}$ [\AA]
}

\startdata

5350.7 & 1215.7 & 3.4344 & Ly$\alpha$           &      \\
5774.2 & 1302.2 & 3.4344 & \mbox{O\,{\sc i}}    & $4.7 \pm 1.0$  \\
5785.8 & 1304.4 & 3.4344 & \mbox{Si\,{\sc ii}}  & $6.9 \pm 1.1$  \\
\multirow{2}{*}{5919.3} & 1334.5 & 3.4344 & \mbox{C\,{\sc ii}} & \multirow{2}{*}{$6.7 \pm 1.4$} \\
     & 1335.7 & 3.4344 & \mbox{C\,{\sc ii}}*  &      \\
6182.1 & 1393.8 & 3.4344 & \mbox{Si\,{\sc iv}}  & $6.6 \pm 1.2$  \\
6221.4 & 1402.8 & 3.4344 & \mbox{Si\,{\sc iv}}  & $5.9 \pm 1.3$  \\
6767.8 & 1526.7 & 3.4344 & \mbox{Si\,{\sc ii}}  & $3.2 \pm 0.8$  \\
6802.2 & 1533.4 & 3.4344 & \mbox{Si\,{\sc ii}}* & $3.5 \pm 0.9$  \\
6862.5 & 1548.2 & 3.4344 & \ion{C}{4}           & \multirow{2}{*}{9.6$\pm$1.5} \\
6875.3 & 1550.8 & 3.4344 & \ion{C}{4} & \\
7409.6 & 1670.8 & 3.4344 & \mbox{Al\,{\sc ii}}  & $4.1 \pm 1.3$  \\

\enddata


\end{deluxetable}

\clearpage
\begin{deluxetable}{lcccc}
\tablecaption{GRB 061121 ($z=1.3145$)\label{tab:061121}}
\tablewidth{0pt}
\tablehead{\colhead{$\lambda_{\textrm{obs}}$ [\AA]} &
\colhead{$\lambda_{\textrm{rest}}$ [\AA]} &
\colhead{\hspace*{2mm} $z$} &
\colhead{Feature} &
\colhead{EW$_{\textrm{obs}}$ [\AA]} }
\startdata
\cutinhead{LRISb/B600}
3534.1&1526.7&1.3145&SiII&$  4.74\pm0.13$\\
3549.0&1532.5&1.3145&PII&$  1.75\pm0.13$\\
      &1533.4&1.3145&SiII*&\\
3587.2&1548.2&1.3145&CIV&$ 10.20\pm0.14$\\
      &1550.8&1.3145&CIV&\\
3609.2&1560.3&1.3145&CI&$  1.14\pm0.10$\\
3722.7&1608.5&1.3145&FeII&$  4.26\pm0.08$\\
3732.4&1611.2&1.3145&FeII&$  1.16\pm0.09$\\
\enddata
\end{deluxetable}

\begin{deluxetable}{@{}llllc@{}}
\tablecaption{GRB\,070110 ($z=2.3521$)}
\tablewidth{0pt}
\tablehead{
$\lambda_{\textrm{obs}}$ [\AA] &
$\lambda_{\textrm{rest}}$ [\AA] &
\hspace*{2mm} $z$ &
Feature &
EW$_{\textrm{obs}}$ [\AA]
}

\startdata

4067.7 & 1215.7 & 2.3521 & Ly$\alpha$           &      \\
4224.2 & 1260.4 & 2.3521 & \mbox{Si\,{\sc ii}}  & $4.5 \pm 0.4$ \\
4240.7 & 1264.7 & 2.3521 & \mbox{Si\,{\sc ii}}* & $2.4 \pm 0.4$ \\
\multirow{2}{*}{4368.6} & 1302.2 & 2.3521 & \mbox{O\,{\sc i}}   & \multirow{2}{*}{$7.1 \pm 0.4$} \\
                      & 1304.4 & 2.3521 & \mbox{Si\,{\sc ii}} & \\
\multirow{2}{*}{4425.1} & 1548.2 & 1.856  & \mbox{C\,{\sc iv}} & \multirow{2}{*}{8.2$\pm$0.4} \\
                        & 1550.8 & 1.856  & \mbox{C\,{\sc iv}} & \\
\multirow{2}{*}{4474.1} & 1334.5 & 2.3521 & \mbox{C\,{\sc ii}}  & \multirow{2}{*}{$3.8 \pm 0.3$} \\
\enddata


\end{deluxetable}

\begin{deluxetable}{@{}llllc@{}}
\tablecaption{GRB\,070125 ($z=1.5471$)}
\tablewidth{0pt}
\tablehead{
$\lambda_{\textrm{obs}}$ [\AA] &
$\lambda_{\textrm{rest}}$ [\AA] &
\hspace*{2mm} $z$ &
Feature &
EW$_{\textrm{obs}}$ [\AA]
}

\startdata
\multirow{2}{*}{3950.3} & 1548.2 & 1.5471 & \mbox{C\,{\sc iv}} & \multirow{2}{*}{$1.7 \pm 0.3$} \\
                      & 1550.8 & 1.5471 & \mbox{C\,{\sc iv}} & \\
7122.4 & 2796.3 & 1.5471 & \mbox{Mg\,{\sc ii}}  & $0.5 \pm 0.1$ \\
7140.8 & 2803.5 & 1.5471 & \mbox{Mg\,{\sc ii}}  & $0.4 \pm 0.1$ \\

\enddata


\end{deluxetable}

\begin{deluxetable}{@{}llllc@{}}
\tablecaption{GRB\,070306 ($z=1.4965$)}
\tablewidth{0pt}
\tablehead{
$\lambda_{\textrm{obs}}$ [\AA] &
$\lambda_{\textrm{rest}}$ [\AA] &
\hspace*{2mm} $z$ &
Feature &
EW$_{\textrm{obs}}$ [\AA]
}

\startdata

9306.3 & 3727.7 & 1.4965 & [\mbox{O\,{\sc ii}}] & emission \\

\enddata


\end{deluxetable}

\begin{deluxetable}{@{}llllc@{}}
\tablecaption{GRB\,070318 ($z=0.8397$)}
\tablewidth{0pt}
\tablehead{
$\lambda_{\textrm{obs}}$ [\AA] &
$\lambda_{\textrm{rest}}$ [\AA] &
\hspace*{2mm} $z$ &
Feature &
EW$_{\textrm{obs}}$ [\AA]
}

\startdata

4313.1 & 2344.2 & 0.8397 & \mbox{Fe\,{\sc ii}} & $1.4 \pm 0.3$ \\
4365.2 & 2374.5 & 0.8397 & \mbox{Fe\,{\sc ii}} & $0.8 \pm 0.2$ \\
4381.7 & 2382.8 & 0.8397 & \mbox{Fe\,{\sc ii}} & $2.2 \pm 0.3$ \\
\hdashline
4743.5 & 2576.9 & 0.8397 & \mbox{Mn\,{\sc ii}} & \multirow{5}{*}{$6.7 \pm 0.4$} \\
4758.1 & 2586.7 & 0.8397 & \mbox{Fe\,{\sc ii}} & \\
4772.6 & 2594.5 & 0.8397 & \mbox{Mn\,{\sc ii}} & \\
4783.6 & 2600.2 & 0.8397 & \mbox{Fe\,{\sc ii}} & \\
4795.0 & 2606.5 & 0.8397 & \mbox{Mn\,{\sc ii}} & \\
\hdashline
5145.4 & 2796.4 & 0.8397 & \mbox{Mg\,{\sc ii}} & \multirow{2}{*}{$4.8 \pm 0.3$} \\
5157.9 & 2803.5 & 0.8397 & \mbox{Mg\,{\sc ii}} & \\
5247.8 & 2853.0 & 0.8397 & \mbox{Mg\,{\sc i}}  & $1.2 \pm 0.2$ \\
6228.2 & 3384.7 & 0.8397 & \ion{Ti}{2} & 1.0$\pm$0.3 \\
7238.4 & 3934.8 & 0.8397 & \ion{Ca}{2} & 3.8$\pm$0.4 \\
7305.5 & 3969.6 & 0.8397 & \ion{Ca}{2} & 1.8$\pm$0.4 \\

\enddata


\end{deluxetable}

\begin{deluxetable}{@{}llllc@{}}
\tablecaption{GRB\,070411 ($z=2.9538$)}
\tablewidth{0pt}
\tablehead{
$\lambda_{\textrm{obs}}$ [\AA] &
$\lambda_{\textrm{rest}}$ [\AA] &
\hspace*{2mm} $z$ &
Feature &
EW$_{\textrm{obs}}$ [\AA]
}

\startdata

4806.5 & 1215.7 & 2.9538 & Ly$\alpha$ & \\
4983.9 & 1260.4 & 2.9538 & \mbox{Si\,{\sc ii}}  & $1.1 \pm 0.3$ \\
\multirow{2}{*}{5156.9} & 1302.2 & 2.9538 & \mbox{O\,{\sc i}}   & \multirow{2}{*}{$1.3 \pm 0.3$} \\
                        & 1304.4 & 2.9538 & \mbox{Si\,{\sc ii}} & \\
5276.6 & 1334.5 & 2.9538 & \mbox{C\,{\sc ii}}   & $2.1 \pm 0.3$ \\
5280.6 & 1335.7 & 2.9538 & \mbox{C\,{\sc ii}}*  & $0.7 \pm 0.2$ \\
5509.3 & 1393.8 & 2.9538 & \mbox{Si\,{\sc iv}}  & $3.1 \pm 0.4$ \\
5545.9 & 1402.8 & 2.9538 & \mbox{Si\,{\sc iv}}  & $2.6 \pm 0.4$ \\
6033.8 & 1526.7 & 2.9538 & \mbox{Si\,{\sc ii}}  & $1.7 \pm 0.4$ \\
6120.5 & 1548.2 & 2.9538 & \mbox{C\,{\sc iv}}   & \multirow{2}{*}{$10.1 \pm 0.5$} \\
6131.8 & 1550.8 & 2.9538 & \mbox{C\,{\sc iv}}   & \\
6358.8 & 1608.4 & 2.9538 & \mbox{Fe\,{\sc ii}}  & $0.9 \pm 0.3$ \\
6606.3 & 1670.8 & 2.9538 & \mbox{Al\,{\sc ii}}  & $1.8 \pm 0.3$ \\

\enddata


\end{deluxetable}

\clearpage
\begin{deluxetable}{lcccc}
\tablecaption{GRB 070419A ($z=0.9705$)\label{tab:070419A}}
\tablewidth{0pt}
\tablehead{\colhead{$\lambda_{\textrm{obs}}$ [\AA]} &
\colhead{$\lambda_{\textrm{rest}}$ [\AA]} &
\colhead{\hspace*{2mm} $z$} &
\colhead{Feature} &
\colhead{EW$_{\textrm{obs}}$ [\AA]} }
\startdata
\cutinhead{GMOS/R831}
7753.8&3934.8&0.9705&CaII&$  2.46\pm0.31$\\
7822.2&3969.6&0.9705&CaII&$  1.21\pm0.29$\\
8332.3&4227.9&0.9705&CaI&$  0.84\pm0.18$\\
\enddata
\end{deluxetable}

\begin{deluxetable}{@{}llllc@{}}
\tablecaption{GRB\,070506 ($z=2.3090$)}
\tablewidth{0pt}
\tablehead{
$\lambda_{\textrm{obs}}$ [\AA] &
$\lambda_{\textrm{rest}}$ [\AA] &
\hspace*{2mm} $z$ &
Feature &
EW$_{\textrm{obs}}$ [\AA]
}

\startdata

4017.1 & 1215.7 & 2.3090 & Ly$\alpha$ &   \\
4166.8 & 1260.4 & 2.3090 & \ion{Si}{2}        & 9.0$\pm$2.0 \\
\multirow{2}{*}{4314.5} & 1302.2 & 2.3090 & \ion{O}{1} & \multirow{2}{*}{8.5$\pm$2.0} \\ 
     & 1304.4 & 2.3090 & \ion{Si}{2} & \\
4641.3 & 1402.7 & 2.3090 & \ion{Si}{4} & 4.4$\pm$1.5 \\
\multirow{2}{*}{4758.7} & 1548.2 & 2.0709 & \ion{C}{4} & \multirow{2}{*}{7.5$\pm$1.4} \\
       & 1550.8 & 2.0709 & \ion{C}{4} &  \\
5049.5 & 1526.7 & 2.3090 & \ion{Si}{2} & 6.7$\pm$0.9 \\
5085.2 &        &       &             & 6.4$\pm$0.9 \\
\multirow{2}{*}{5122.} & 1548.2 & 2.3090 & \ion{C}{4} & \multirow{2}{*}{6.4$\pm$1.0} \\
     & 1550.8 & 2.3090 & \ion{C}{4} & \\ 
5324.7 & 1608.4 & 2.3090 & \ion{Fe}{2} & 4.4$\pm$0.7 \\
5528.4 & 1670.7 & 2.3090 & \ion{Al}{2} & 4.7$\pm$0.6 \\
5694.2 & 1854.7 & 2.0709 & \ion{Al}{3} & 3.1$\pm$0.5 \\
5721.4 & 1862.7 & 2.0709 & \ion{Al}{3} & 3.1$\pm$0.5 \\
6140.1 & 1854.7 & 2.3090 & \ion{Al}{3} & 2.6$\pm$0.5 \\
\multirow{2}{*}{6706.8} & 2026.1 & 2.3090 & \ion{Zn}{2} & \multirow{2}{*}{2.8$\pm$0.5} \\
                        & 2026.5 & 2.3090 & \ion{Mg}{1} & \\
\multirow{2}{*}{6824.8} & 2062.2 & 2.3090 & \ion{Cr}{2} & \multirow{2}{*}{2.9$\pm$0.5} \\
                        & 2062.6 & 2.3090 & \ion{Zn}{2} &  \\
\enddata


\end{deluxetable}

\begin{deluxetable}{@{}llllc@{}}
\tablecaption{GRB\,070611 ($z=2.0394$)}
\tablewidth{0pt}
\tablehead{
$\lambda_{\textrm{obs}}$ [\AA] &
$\lambda_{\textrm{rest}}$ [\AA] &
\hspace*{2mm} $z$ &
Feature &
EW$_{\textrm{obs}}$ [\AA]
}

\startdata

3694.8 & 1215.7 & 2.0394 & Ly$\alpha$           &     \\
4359.4 & 2796.3 & 0.5588 & \mbox{Mg\,{\sc ii}}  & 2.5$\pm$1.0 \\
4369.8 & 2803.5 & 0.5588 & \mbox{Mg\,{\sc ii}}  & 2.5$\pm$1.0 \\
\multirow{2}{*}{4708.8} & 1548.2 & 2.0394 & \mbox{C\,{\sc iv}} & \multirow{2}{*}{2.9$\pm$1.0} \\
     & 1550.8 & 2.0394 & \mbox{C\,{\sc iv}} & \\
5472.6 & 2382.8 & 1.2973 & \mbox{Fe\,{\sc ii}}  & 2.2$\pm$0.7 \\
6425.4 & 2796.3 & 1.2973 & \mbox{Mg\,{\sc ii}}  & 3.9$\pm$0.7 \\
6440.4 & 2803.5 & 1.2973 & \mbox{Mg\,{\sc ii}}  & 3.2$\pm$0.7 \\
8497.0 & 2796.3 & 2.0394 & \mbox{Mg\,{\sc ii}}  & 2.8$\pm$1.2 \\ 
8523.2 & 2803.5 & 2.0394 & \mbox{Mg\,{\sc ii}}  & 3.6$\pm$0.9 \\ 

\enddata


\end{deluxetable}

\begin{deluxetable}{@{}llllc@{}}
\tablecaption{GRB\,070721B ($z=3.6298$)}
\tablewidth{0pt}
\tablehead{
$\lambda_{\textrm{obs}}$ [\AA] &
$\lambda_{\textrm{rest}}$ [\AA] &
\hspace*{2mm} $z$ &
Feature &
EW$_{\textrm{obs}}$ [\AA]
}

\startdata

4970.0 & 1215.7 & 3.0939 & Ly$\alpha$           &      \\
5590.0 & 1215.7 & 3.6298 & Ly$\alpha$           &      \\
\multirow{2}{*}{6028.1} & 1302.2 & 3.6298 & \mbox{O\,{\sc i}}  & \multirow{2}{*}{6.5$\pm$1.9} \\
     & 1304.4 & 3.6298 & \mbox{Si\,{\sc ii}} & \\
6179.8 & 1334.5 & 3.6298 & \mbox{C\,{\sc ii}}  & 5.0$\pm$2.0 \\
6248.7 & 1526.7 & 3.0939 & \mbox{Si\,{\sc ii}}  & 7.0$\pm$2.2  \\
6337.3 & 1548.2 & 3.0939 & \mbox{C\,{\sc iv}}   & \multirow{2}{*}{19.0$\pm$3.0} \\
6349.9 & 1550.8 & 3.0939 & \mbox{C\,{\sc iv}}   &  \\
6450.2 & 1393.8 & 3.6298 & \mbox{Si\,{\sc iv}}  & 6.0$\pm$1.9 \\
6492.7 & 1402.8 & 3.6298 & \mbox{Si\,{\sc iv}}  & 7.0$\pm$2.2 \\
6584.2 & 1608.5 & 3.0939 & \mbox{Fe\,{\sc ii}}  & 9.0$\pm$1.9 \\
6842.2 & 1670.8 & 3.0939 & \mbox{Al\,{\sc ii}}  & 7.6$\pm$2.0 \\
7066.2 & 1526.7 & 3.6298 & \mbox{Si\,{\sc ii}}  & 5.8$\pm$1.9 \\
7104.6 & 1533.4 & 3.6298 & \ion{Si}{2}*         & 5.9$\pm$1.8 \\
7167.5 & 1548.2 & 3.6298 & \mbox{C\,{\sc iv}}   & \multirow{2}{*}{14.6$\pm$2.6} \\
7180.8 & 1550.8 & 3.6298 & \mbox{C\,{\sc iv}}   &      \\

\enddata


\end{deluxetable}

\begin{deluxetable}{@{}llllc@{}}
\tablecaption{GRB\,070802 ($z=2.4511$)}
\tablewidth{0pt}
\tablehead{
$\lambda_{\textrm{obs}}$ [\AA] &
$\lambda_{\textrm{rest}}$ [\AA] &
\hspace*{2mm} $z$ &
Feature &
EW$_{\textrm{obs}}$ [\AA]
}

\startdata

4200.0 & 1215.7 & 2.4541& Ly$\alpha$ &  \\
4607.9 & 1334.5 & 2.4541 & \mbox{C\,{\sc ii}}  & 16$\pm$6 \\
5141.4 & 1670.7 & 2.0785 & \mbox{Al\,{\sc ii}} & 6.0$\pm$1.7 \\
5273.9 & 1526.7 & 2.4541 & \mbox{Si\,{\sc ii}} & 10$\pm$2 \\
\multirow{2}{*}{5350.9} & 1548.2 & 2.4541 & \mbox{C\,{\sc iv}} & \multirow{2}{*}{7.3$\pm$1.8} \\
& 1550.8 & 2.4541 & \mbox{C\,{\sc iv}} & \\
5390.7 & 1560.3 & 2.4541 & \mbox{C\,{\sc i}} & 2.4$\pm$1.2 \\
5556.4 & 1608.4 & 2.4541 & \mbox{Fe\,{\sc ii}} & 10.5$\pm$1.6 \\
\enddata


\end{deluxetable}

\begin{deluxetable}{@{}llllc@{}}
\tablecaption{GRB\,071020 ($z=2.1462$)}
\tablewidth{0pt}
\tablehead{
$\lambda_{\textrm{obs}}$ [\AA] &
$\lambda_{\textrm{rest}}$ [\AA] &
\hspace*{2mm} $z$ &
Feature &
EW$_{\textrm{obs}}$ [\AA]
}

\startdata

5254.7 & 1670.8 & 2.1462 & \mbox{Al\,{\sc ii}}  & 6$\pm$2 \\
7376.8 & 2344.2 & 2.1462 & \mbox{Fe\,{\sc ii}}  & 5$\pm$2 \\
7498.3 & 2382.8 & 2.1462 & \mbox{Fe\,{\sc ii}}  & 6$\pm$2 \\
8136.8 & 2586.7 & 2.1462 & \mbox{Fe\,{\sc ii}}  & 7$\pm$2 \\
8181.6 & 2600.2 & 2.1462 & \mbox{Fe\,{\sc ii}}  & 11$\pm$2 \\
8274.5 &        &        &                      &  6$\pm$2 \\
8797.7 & 2796.3 & 2.1462 & \mbox{Mg\,{\sc ii}}  & 14$\pm$2 \\
8820.9 & 2803.5 & 2.1462 & \mbox{Mg\,{\sc ii}}  & 10$\pm$2 \\

\enddata


\end{deluxetable}

\clearpage

\begin{deluxetable}{@{}llllc@{}}
\tablecaption{GRB\,071031 ($z=2.6918$)}
\tablewidth{0pt}
\tablehead{
$\lambda_{\textrm{obs}}$ [\AA] &
$\lambda_{\textrm{rest}}$ [\AA] &
\hspace*{2mm} $z$ &
Feature &
EW$_{\textrm{obs}}$ [\AA]
}

\startdata

4483.0 & 1215.7 & 2.6918 & Ly$\alpha$             &      \\
4652.0 & 1260.4 & 2.6918 & \mbox{Si\,{\sc ii}}    & 2.65$\pm$0.15 \\ 
4670.0 & 1264.7 & 2.6918 & \mbox{Si\,{\sc ii}}*   & 2.00$\pm$0.15 \\ 
\multirow{2}{*}{4815.2} & 1302.2 & 2.6918 & \mbox{O\,{\sc i}} & \multirow{2}{*}{4.91$\pm$0.11} \\
     & 1304.4 & 2.6918 & \mbox{Si\,{\sc ii}} &  \\
4833.5 & 1309.3 & 2.6918 & \mbox{Si\,{\sc ii}}*   & 1.45$\pm$0.13 \\
4862.9 & 1317.2 & 2.6918 & \mbox{Ni\,{\sc ii}}    & 0.46$\pm$0.15 \\
4900.1 & 1854.7 & 1.6421 & \mbox{Al\,{\sc iii}}   & 0.56$\pm$0.12 \\
\enddata


\end{deluxetable}

\begin{deluxetable}{@{}llllc@{}}
\tablecaption{GRB\,071112C ($z=0.8227$)}
\tablewidth{0pt}
\tablehead{
$\lambda_{\textrm{obs}}$ [\AA] &
$\lambda_{\textrm{rest}}$ [\AA] &
\hspace*{2mm} $z$ &
Feature &
EW$_{\textrm{obs}}$ [\AA]
}

\startdata

4273.4 & 2344.2 & 0.8227 & \mbox{Fe\,{\sc ii}}    & 2.0$\pm$0.5 \\
4328.7 & 2374.5 & 0.8227 & \mbox{Fe\,{\sc ii}}    & 1.0$\pm$0.5 \\
4345.0 & 2382.8 & 0.8227 & \mbox{Fe\,{\sc ii}}    & 1.6$\pm$0.5 \\
4715.9 & 2586.7 & 0.8227 & \mbox{Fe\,{\sc ii}}    & 2.0$\pm$0.5 \\
4741.0 & 2600.2 & 0.8227 & \mbox{Fe\,{\sc ii}}    & 2.4$\pm$0.5 \\
5096.4 & 2796.3 & 0.8227 & \mbox{Mg\,{\sc ii}}    & \multirow{2}{*}{4.9$\pm$0.6} \\
5110.1 & 2803.5 & 0.8227 & \mbox{Mg\,{\sc ii}}    &     \\
6795.8 & 3727.7 & 0.8227 & [\mbox{O\,{\sc ii}}]   & emission    \\
7051.2 & 3869.8 & 0.8227 & [\mbox{Ne\,{\sc iii}}] & emission    \\
7170.0 & 3934.8 & 0.8227 & \ion{Ca}{2} & 2.4$\pm$0.5 \\
7233.1 & 3969.6 & 0.8227 & \ion{Ca}{2} & 2.8$\pm$0.8 \\
9038.5 & 4960.3 & 0.8227 & [\mbox{O\,{\sc iii}}]  & emission    \\
9127.7 & 5008.2 & 0.8227 & [\mbox{O\,{\sc iii}}]  & emission    \\

\enddata


\end{deluxetable}

\begin{deluxetable}{@{}llllc@{}}
\tablecaption{GRB\,080210 ($z=2.6419$)}
\tablewidth{0pt}
\tablehead{
$\lambda_{\textrm{obs}}$ [\AA] &
$\lambda_{\textrm{rest}}$ [\AA] &
\hspace*{2mm} $z$ &
Feature &
EW$_{\textrm{obs}}$ [\AA]
}

\startdata

4406.6 & 1215.7 & 2.6419 & Ly$\alpha$             &      \\
4591.5 & 1260.4 & 2.6419 & \mbox{Si\,{\sc ii}}    & 4.2$\pm$0.7 \\
4609.3 & 1264.7 & 2.6419 & \mbox{Si\,{\sc ii}}*   & 2.3$\pm$0.6 \\
4652.8 &        &        &                        & 1.4$\pm$0.4 \\
\multirow{2}{*}{4747.6} & 1302.2 & 2.6419 & \mbox{O\,{\sc i}} & \multirow{2}{*}{9.7$\pm$0.6} \\
     & 1304.4 & 2.6419 & \mbox{Si\,{\sc ii}} &  \\
4768.8 & 1309.3 & 2.6419 & \mbox{Si\,{\sc ii}}*   &  1.3$\pm$0.5  \\
4862.5 & 1334.5 & 2.6419 & \mbox{C\,{\sc ii}}  & 8.7$\pm$0.5 \\
\enddata


\end{deluxetable}%


\begin{figure}
\epsscale{1.00}
\plotone{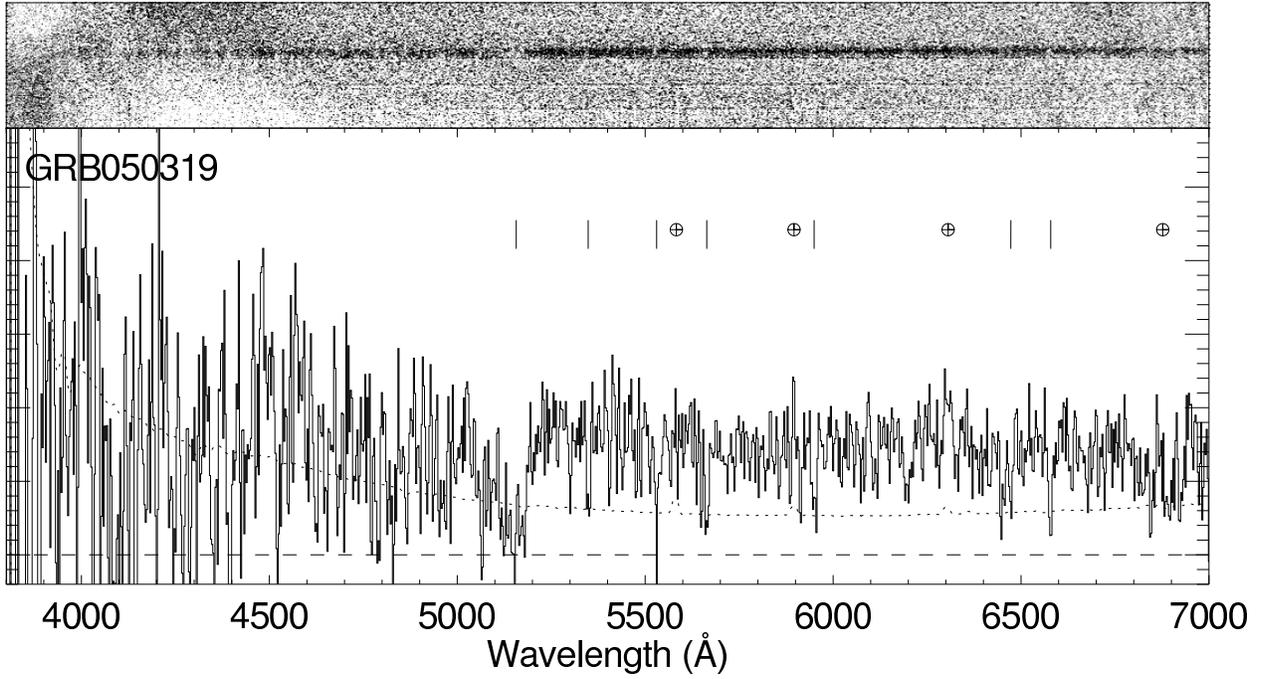}
\caption{Shown are 1- and 2-dimensional spectra for GRBs 050319--080928.  Lines
from the GRB absorption systems are marked with vertical lines whereas
unidentified lines or lines from intervening systems are marked with vertical 
dashed lines. Telluric features are marked with a telluric symbol. The error
spectrum is plotted as a dotted line. When in the spectral range we also plot
the position of the Lyman-limit as a vertical dashed line.
\label{fig050319}}
\end{figure}

\begin{figure}
\epsscale{1.00}
\plotone{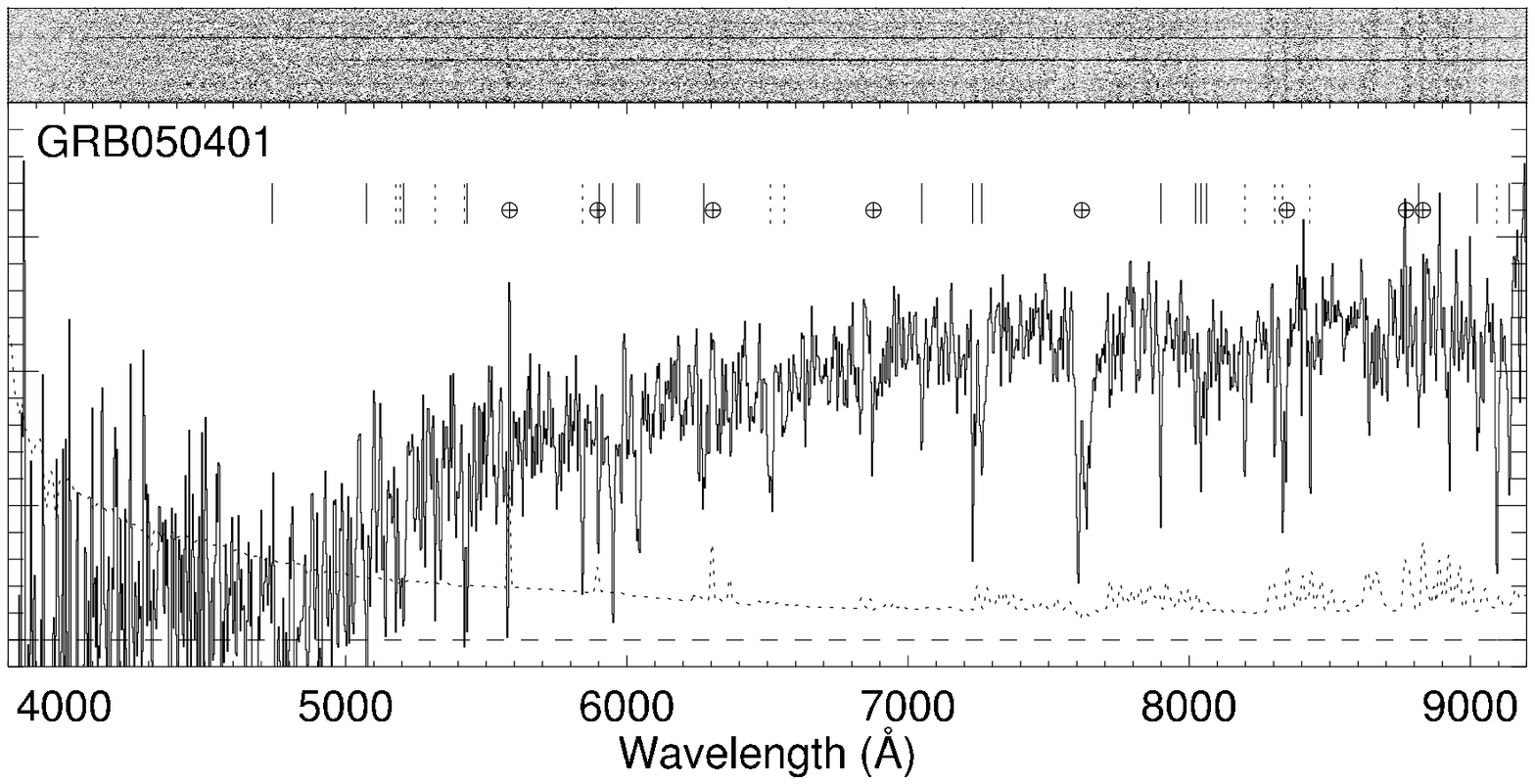}
\end{figure}

\begin{figure}
\epsscale{1.00}
\plotone{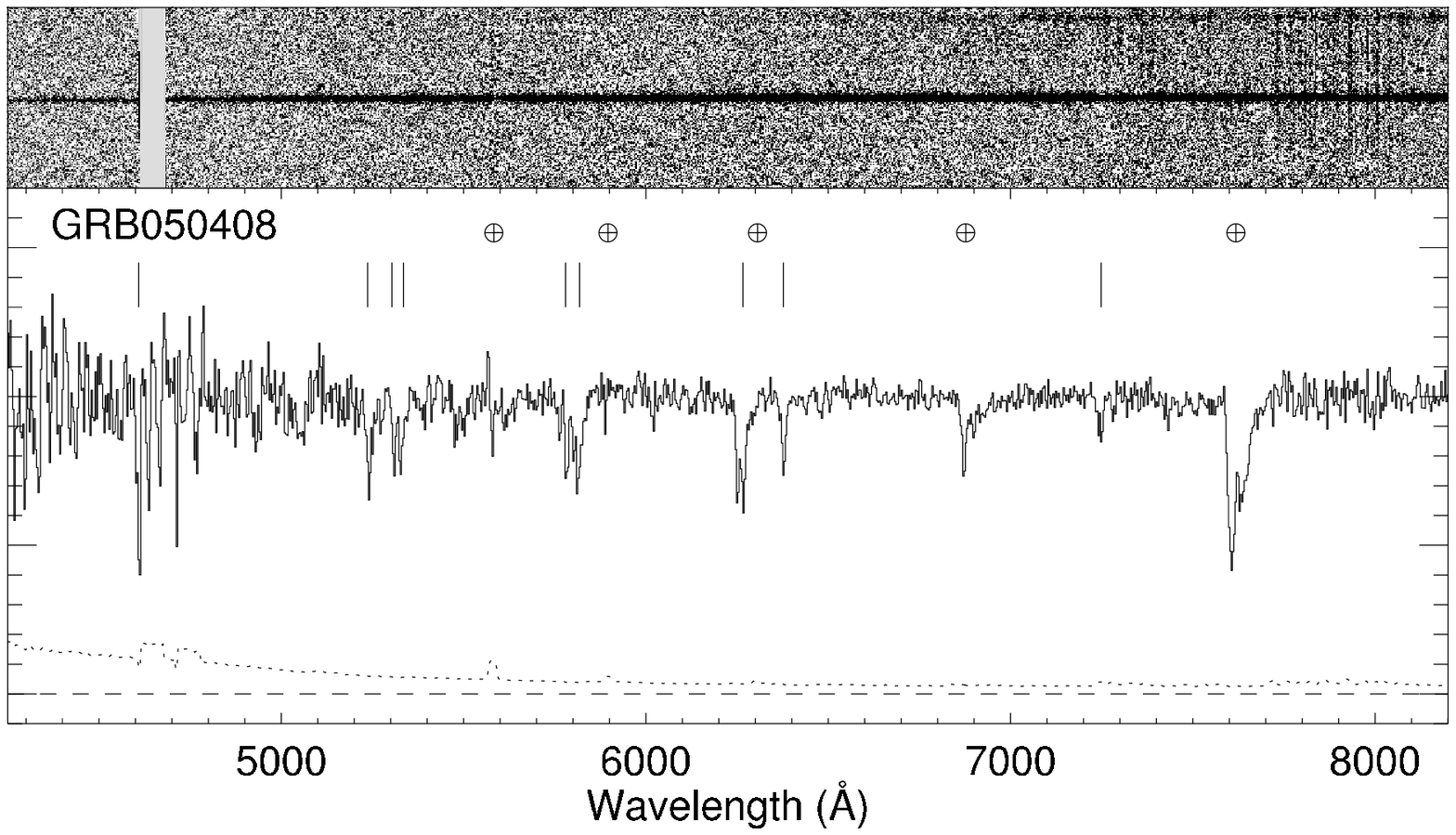}
\end{figure}

\begin{figure}
\epsscale{1.00}
\plotone{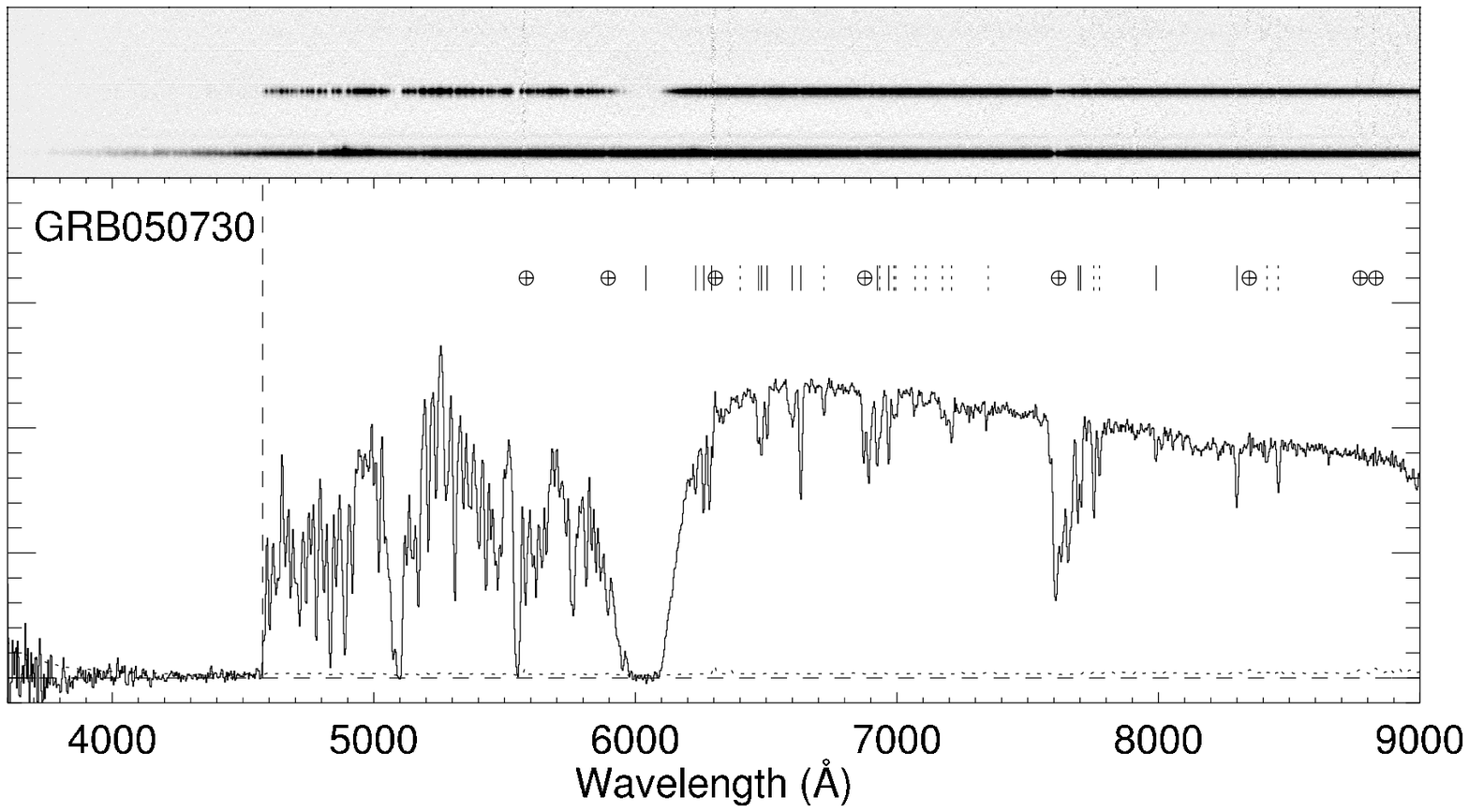}
\end{figure}

\begin{figure}
\epsscale{1.00}
\plotone{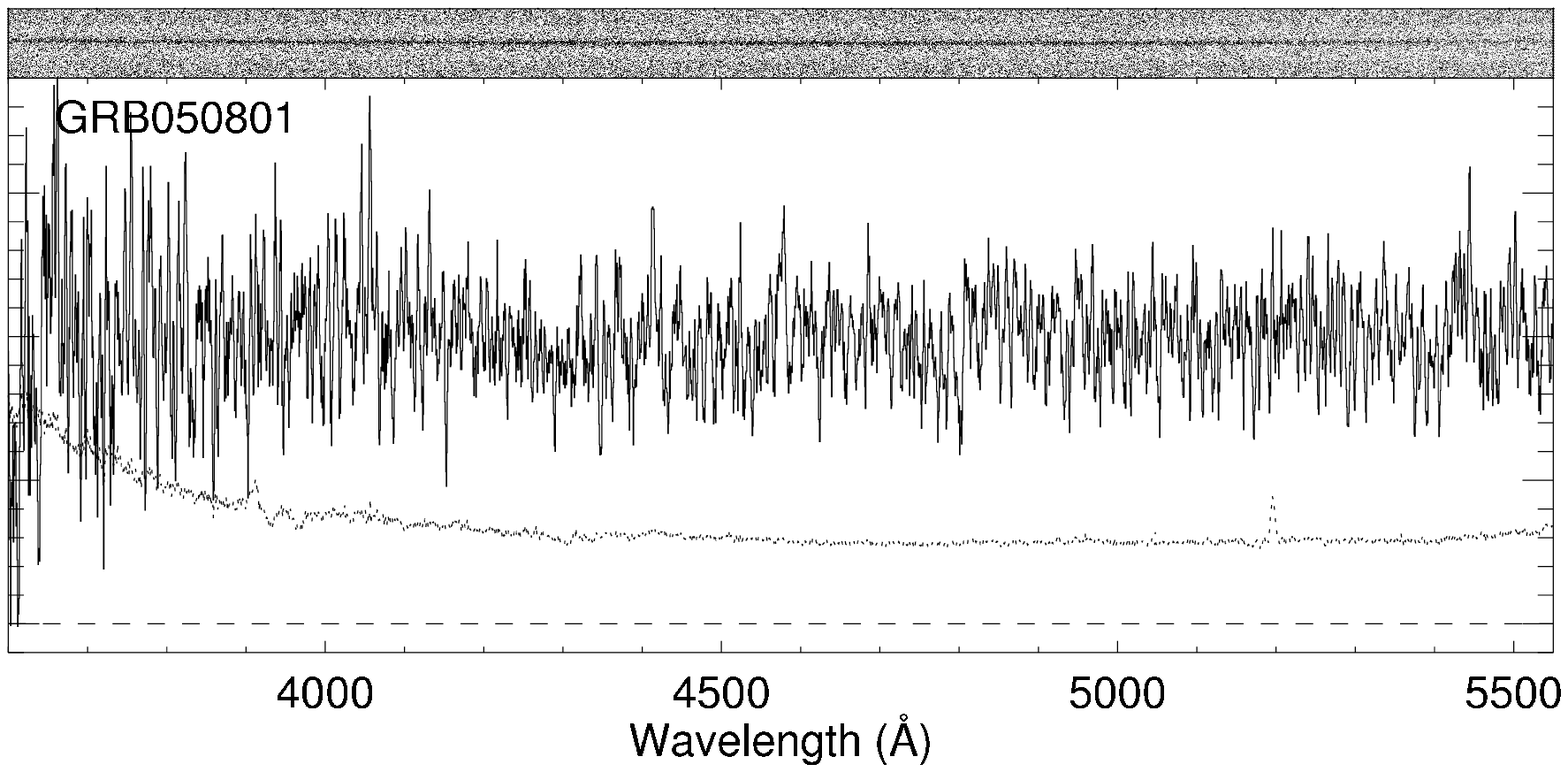}
\end{figure}

\begin{figure}
\epsscale{1.00}
\plotone{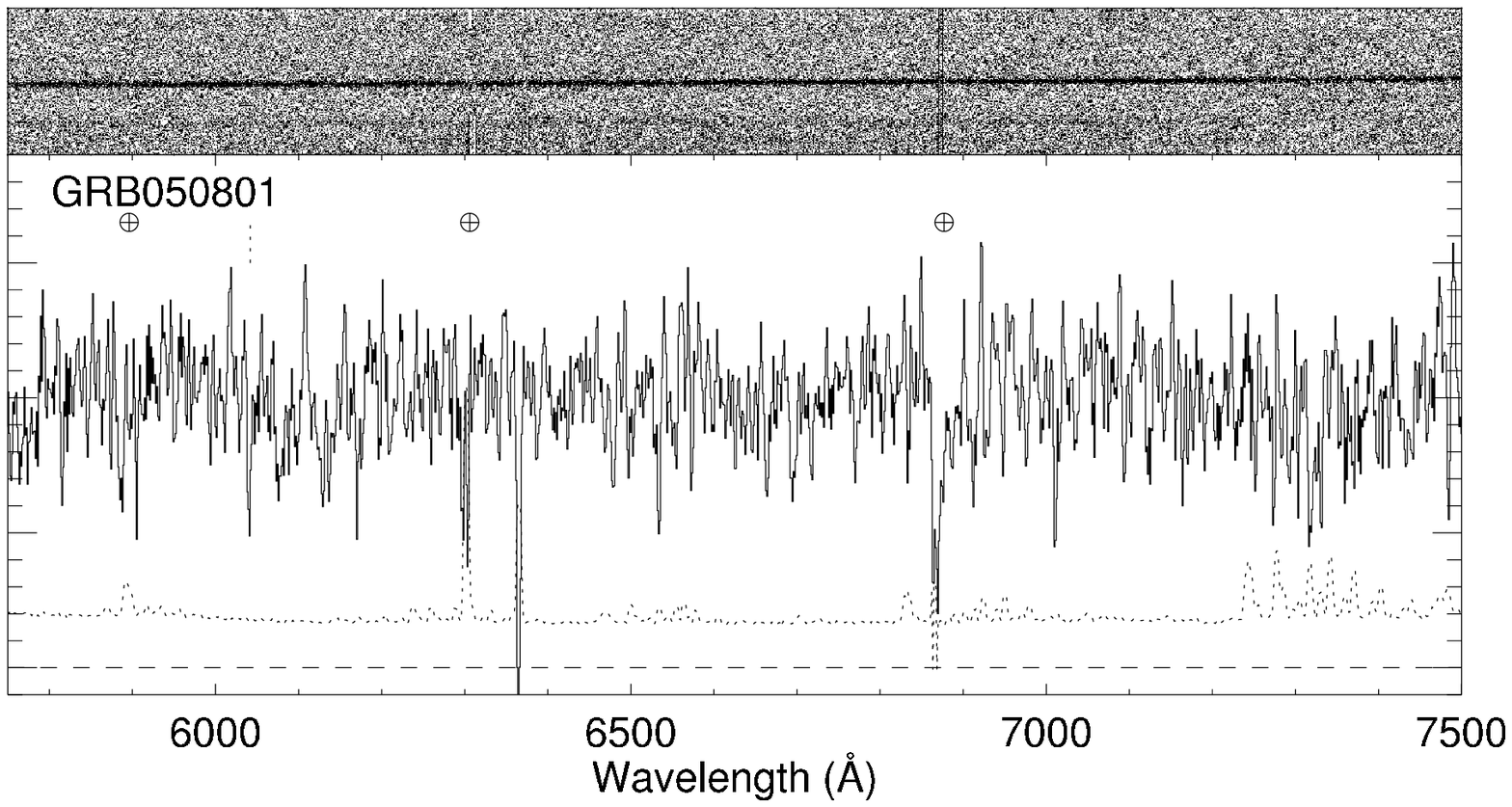}
\end{figure}

\begin{figure}
\epsscale{1.00}
\plotone{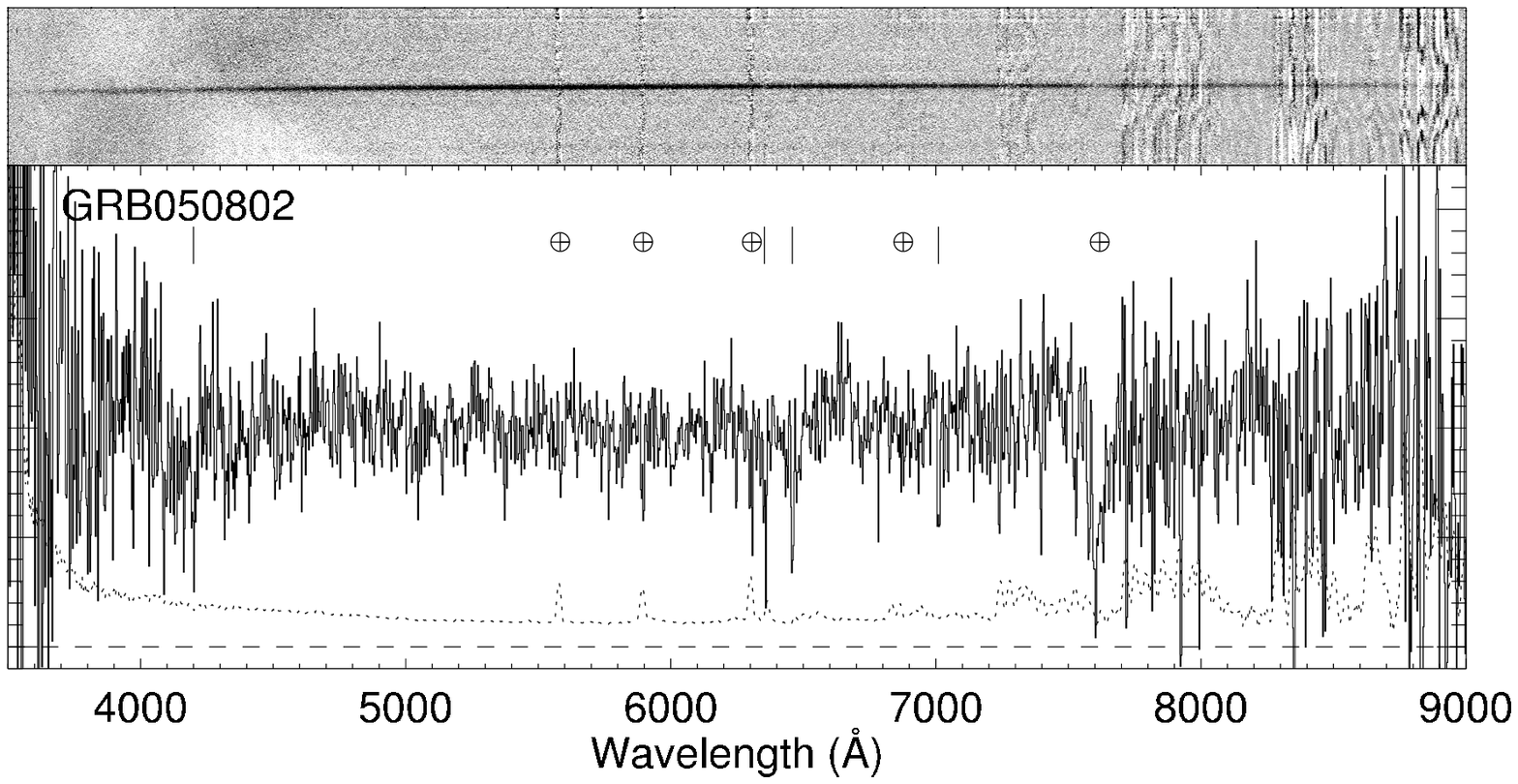}
\end{figure}

\begin{figure}
\epsscale{1.00}
\plotone{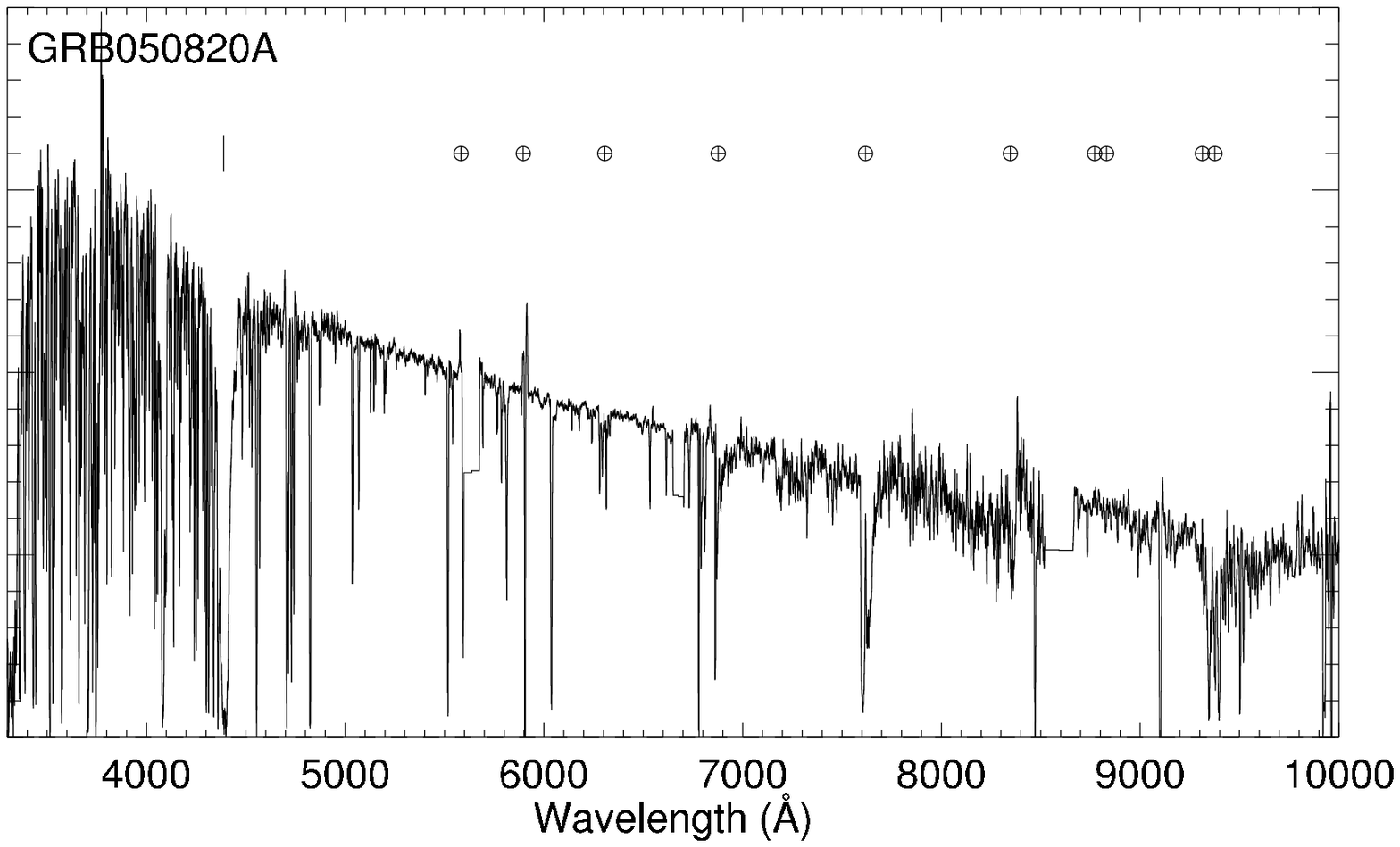}
\end{figure}

\begin{figure}
\epsscale{1.00}
\plotone{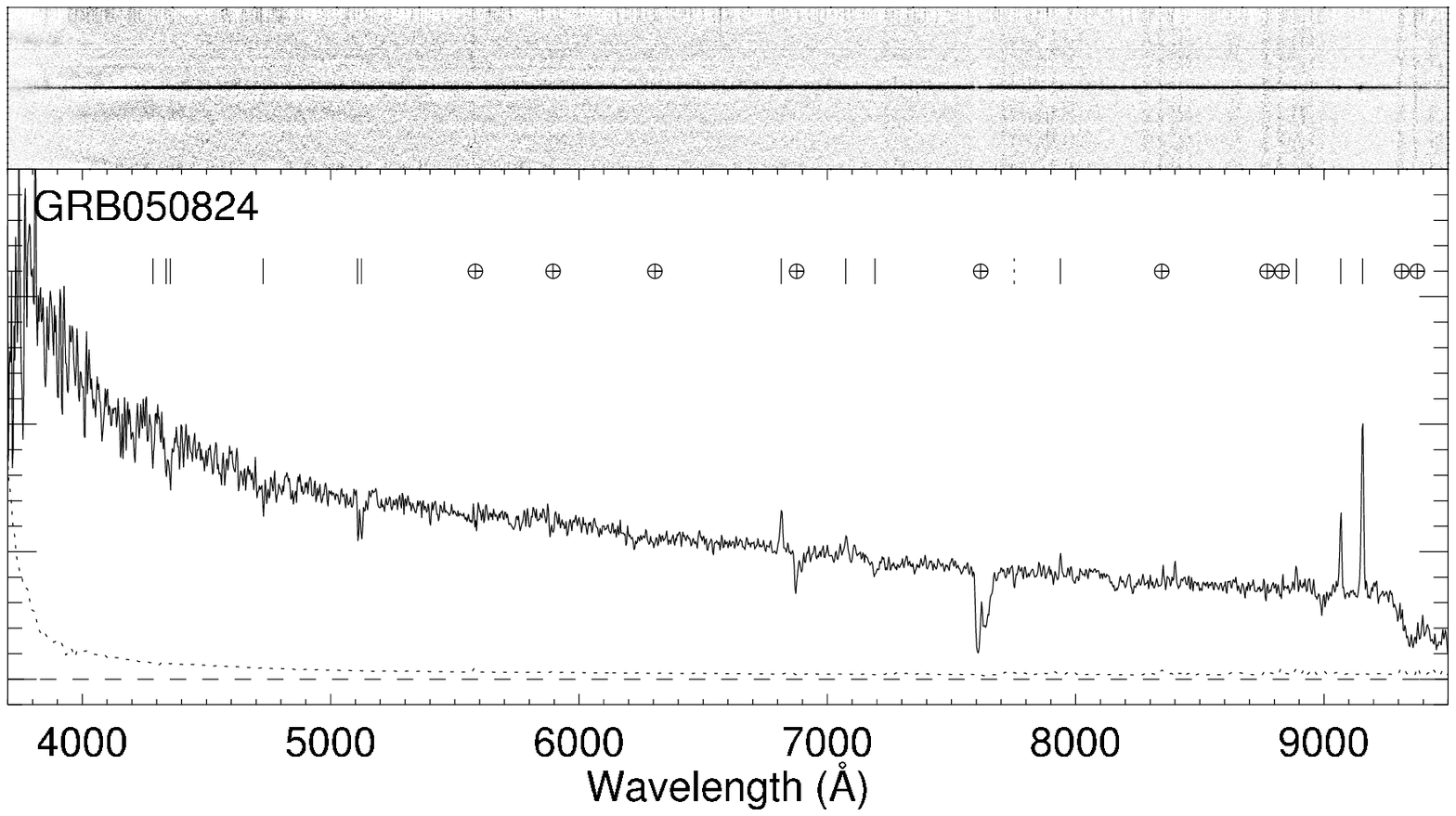}
\end{figure}

\clearpage

\begin{figure}
\epsscale{1.00}
\plotone{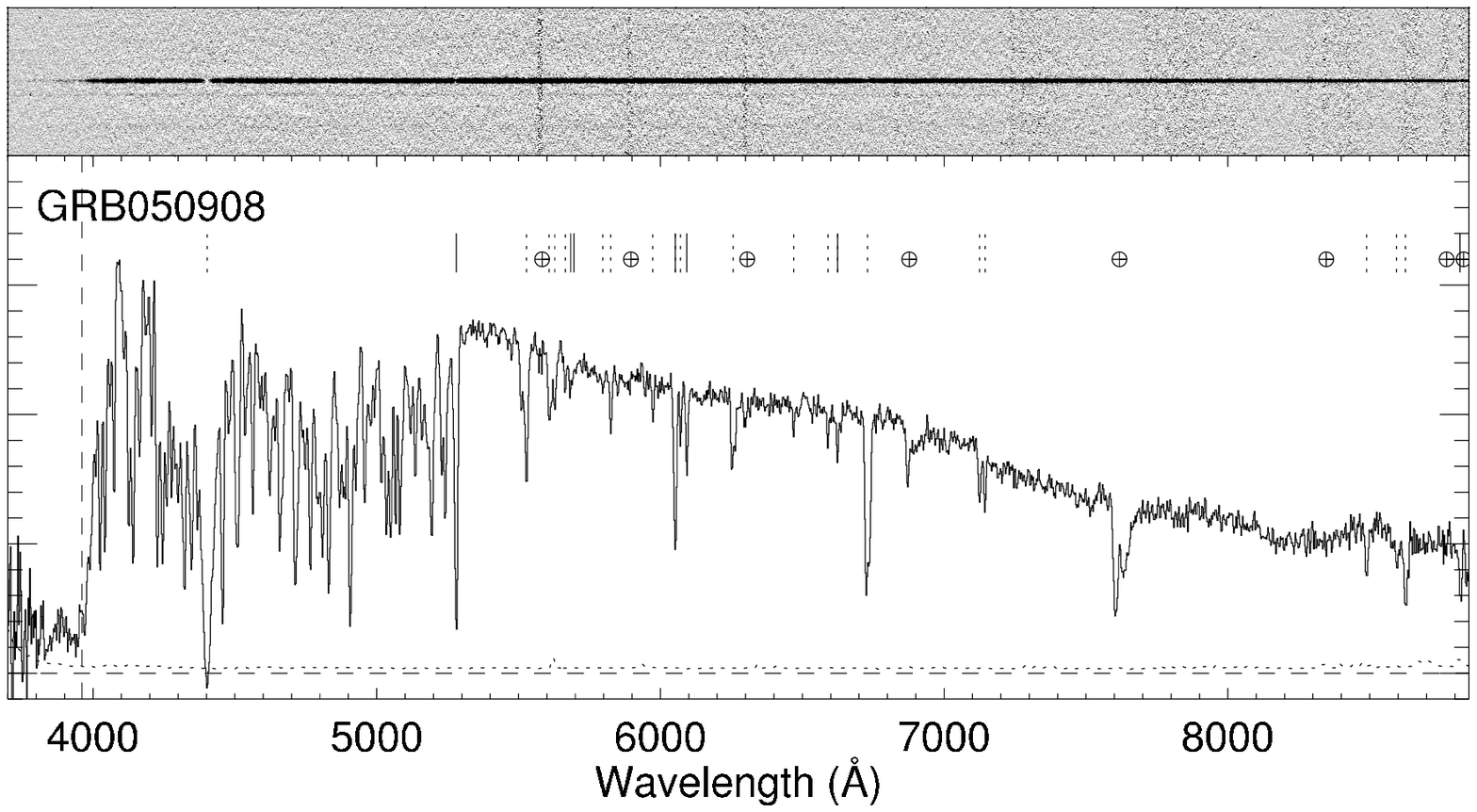}
\end{figure}

\begin{figure}
\epsscale{1.00}
\plotone{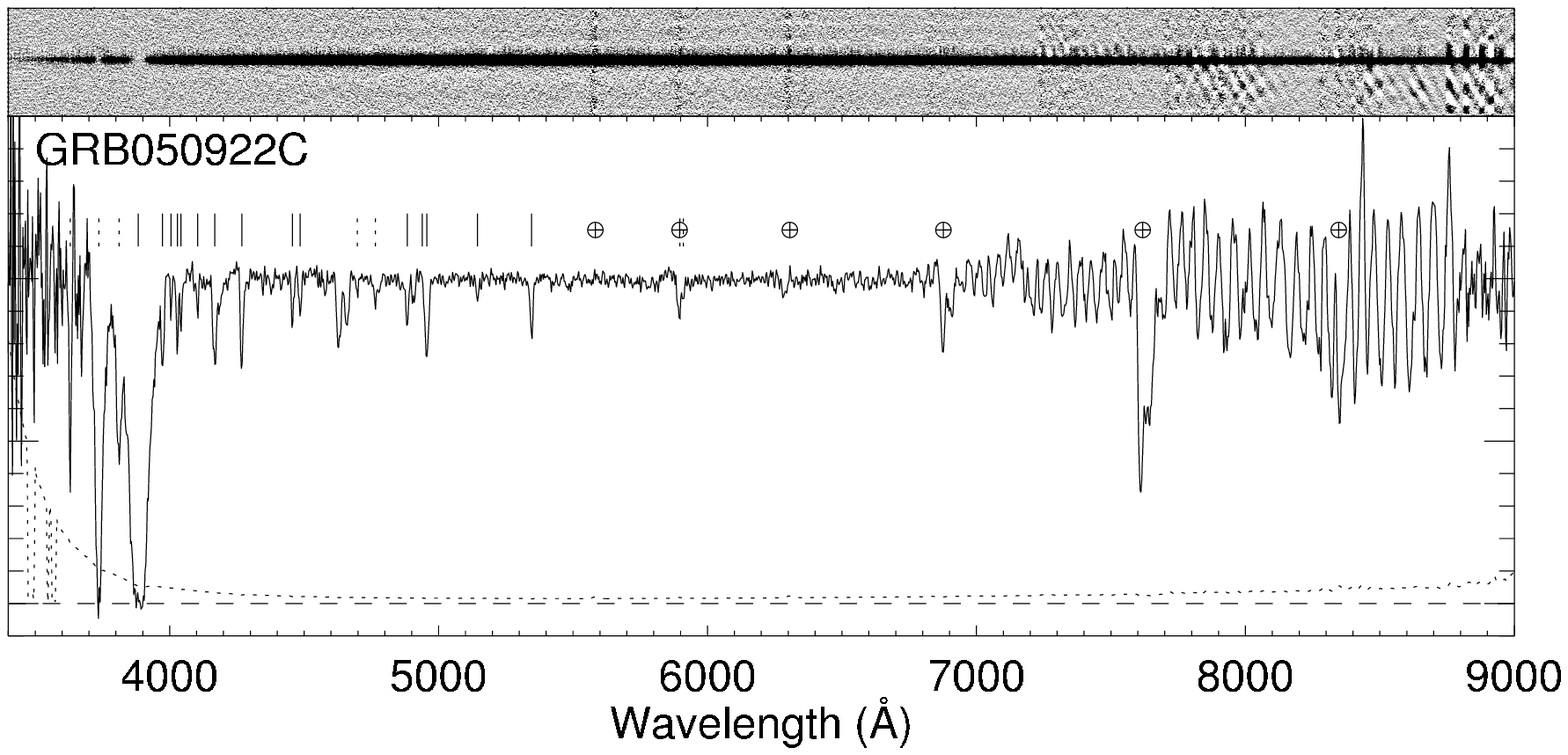}
\end{figure}

\begin{figure}
\epsscale{1.00}
\plotone{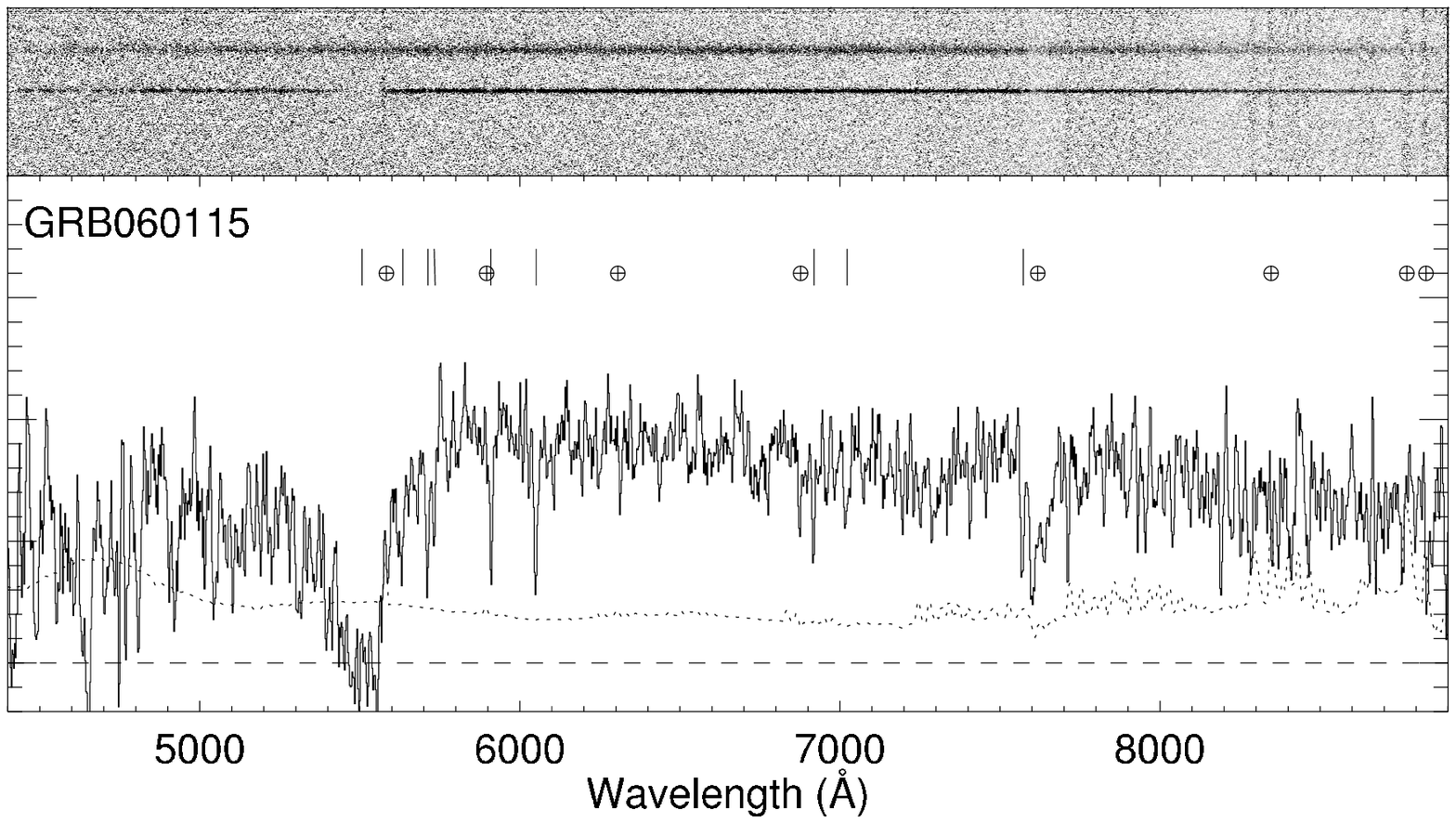}
\end{figure}

\clearpage

\begin{figure}
\epsscale{1.00}
\plotone{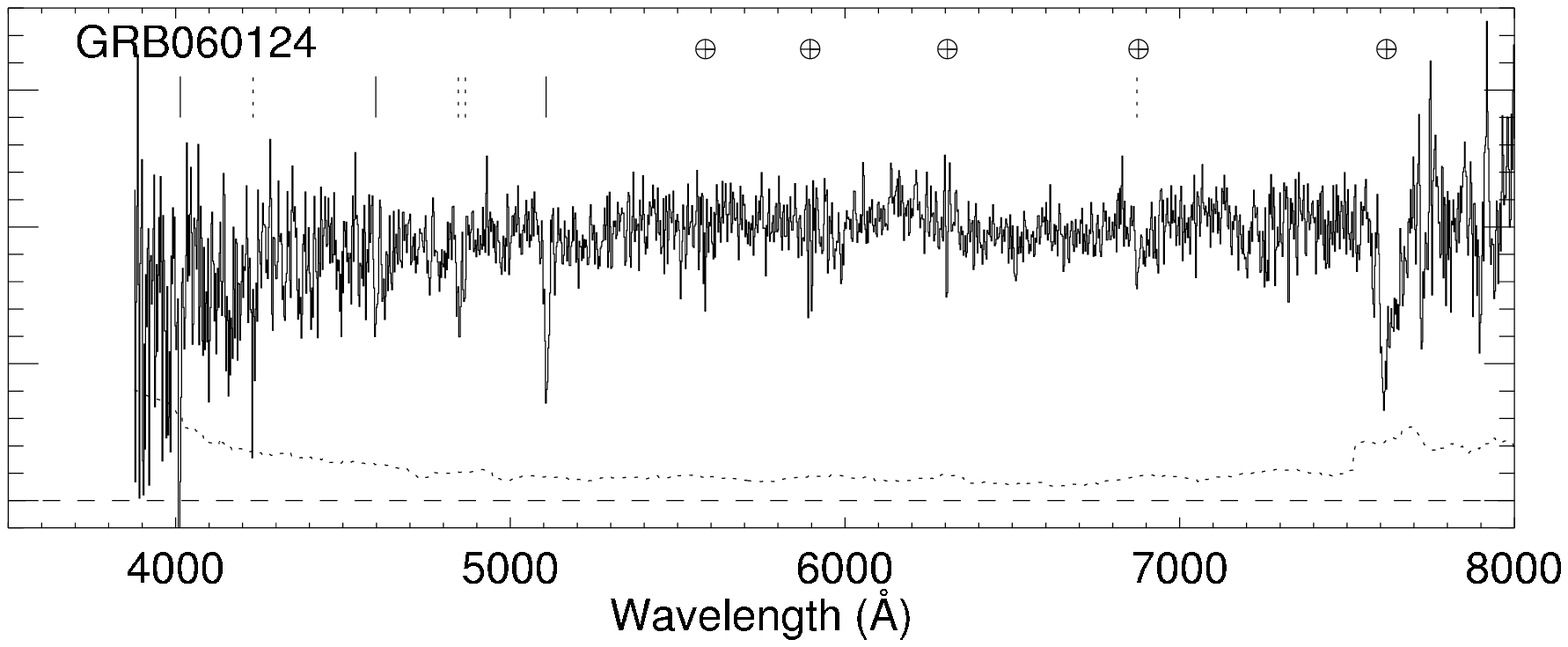}
\end{figure}

\begin{figure}
\epsscale{1.00}
\plotone{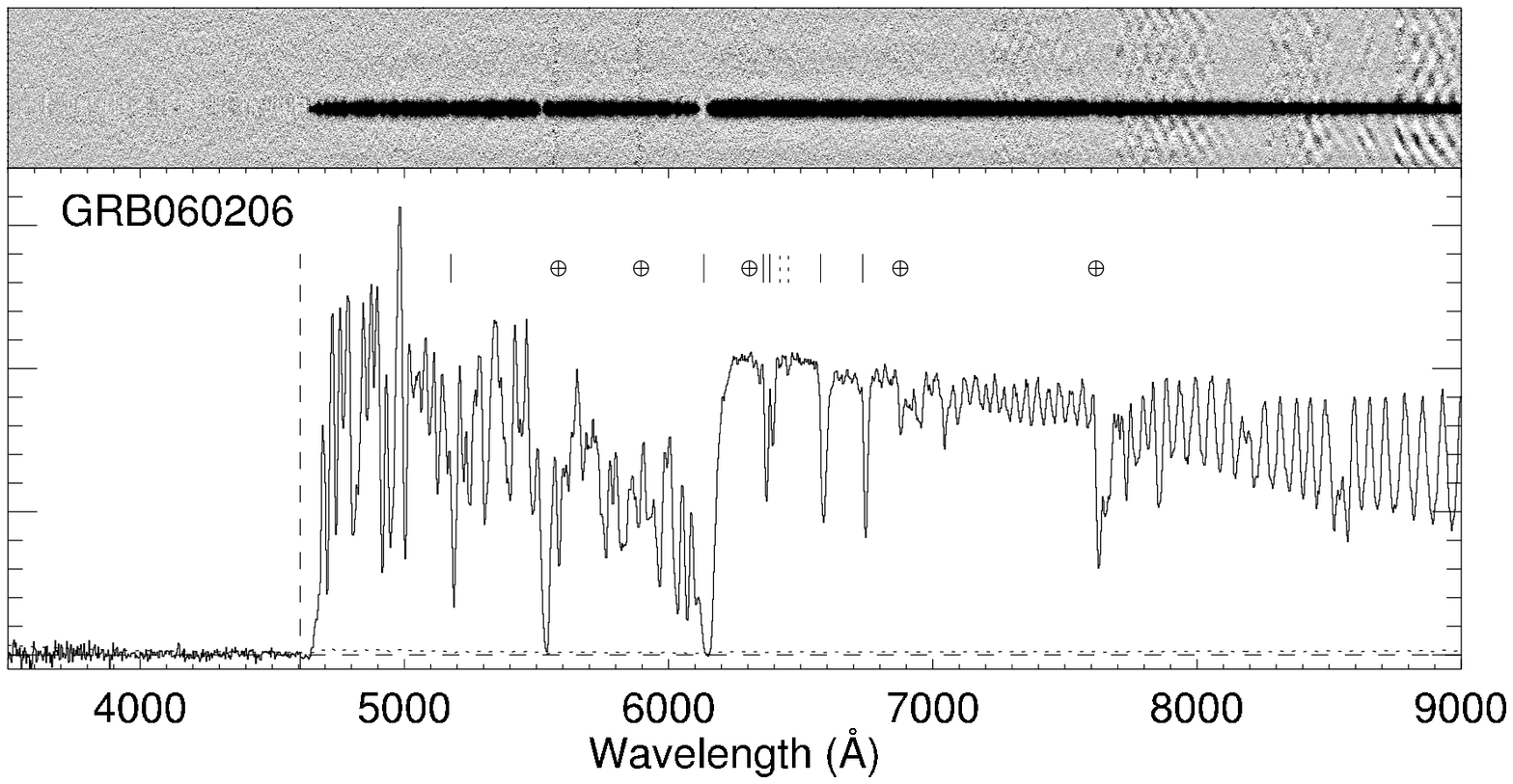}
\end{figure}

\begin{figure}
\epsscale{1.00}
\plotone{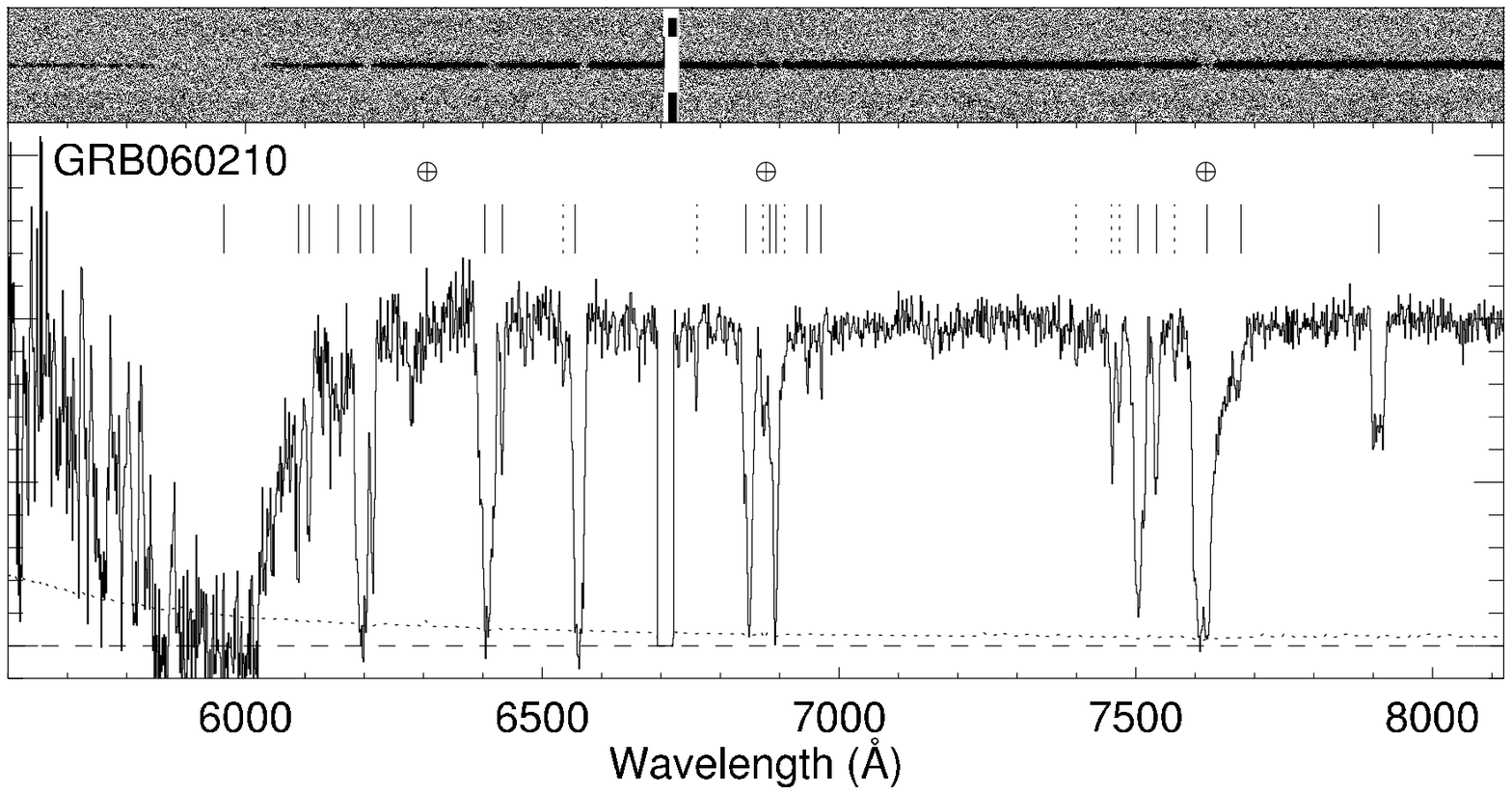}
\end{figure}

\clearpage

\begin{figure}
\epsscale{1.00}
\plotone{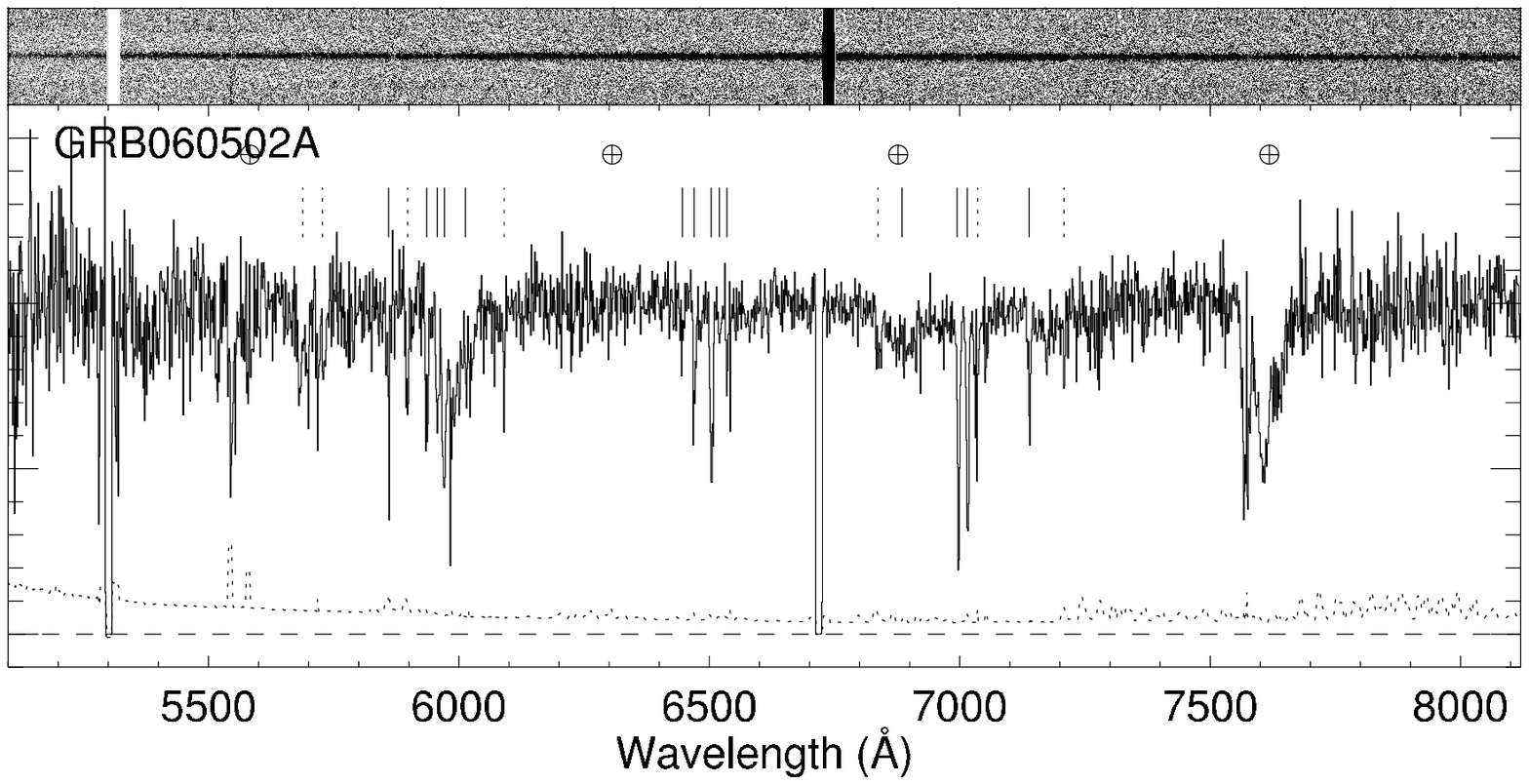}
\end{figure}

\begin{figure}
\epsscale{1.00}
\plotone{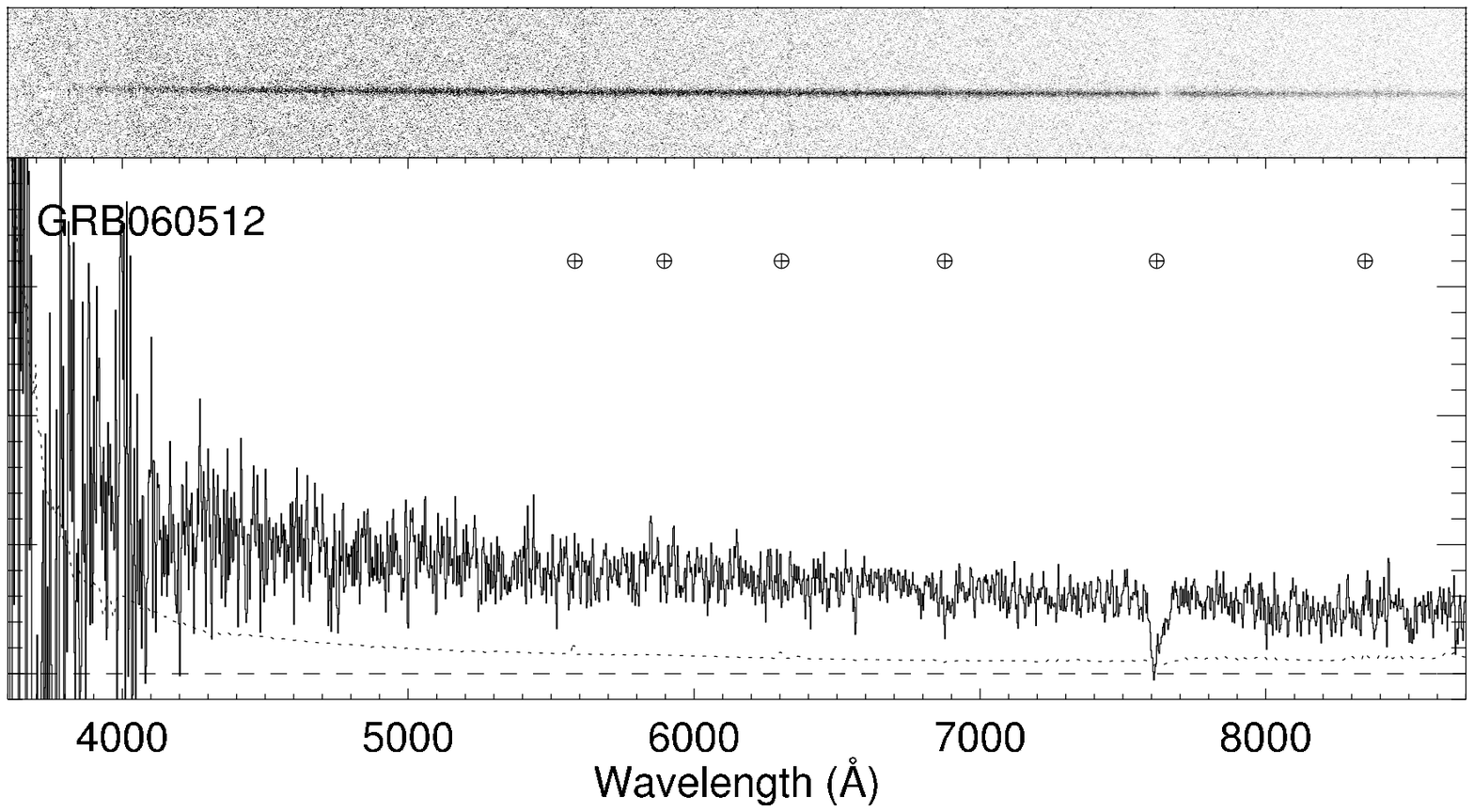}
\end{figure}

\begin{figure}
\epsscale{1.00}
\plotone{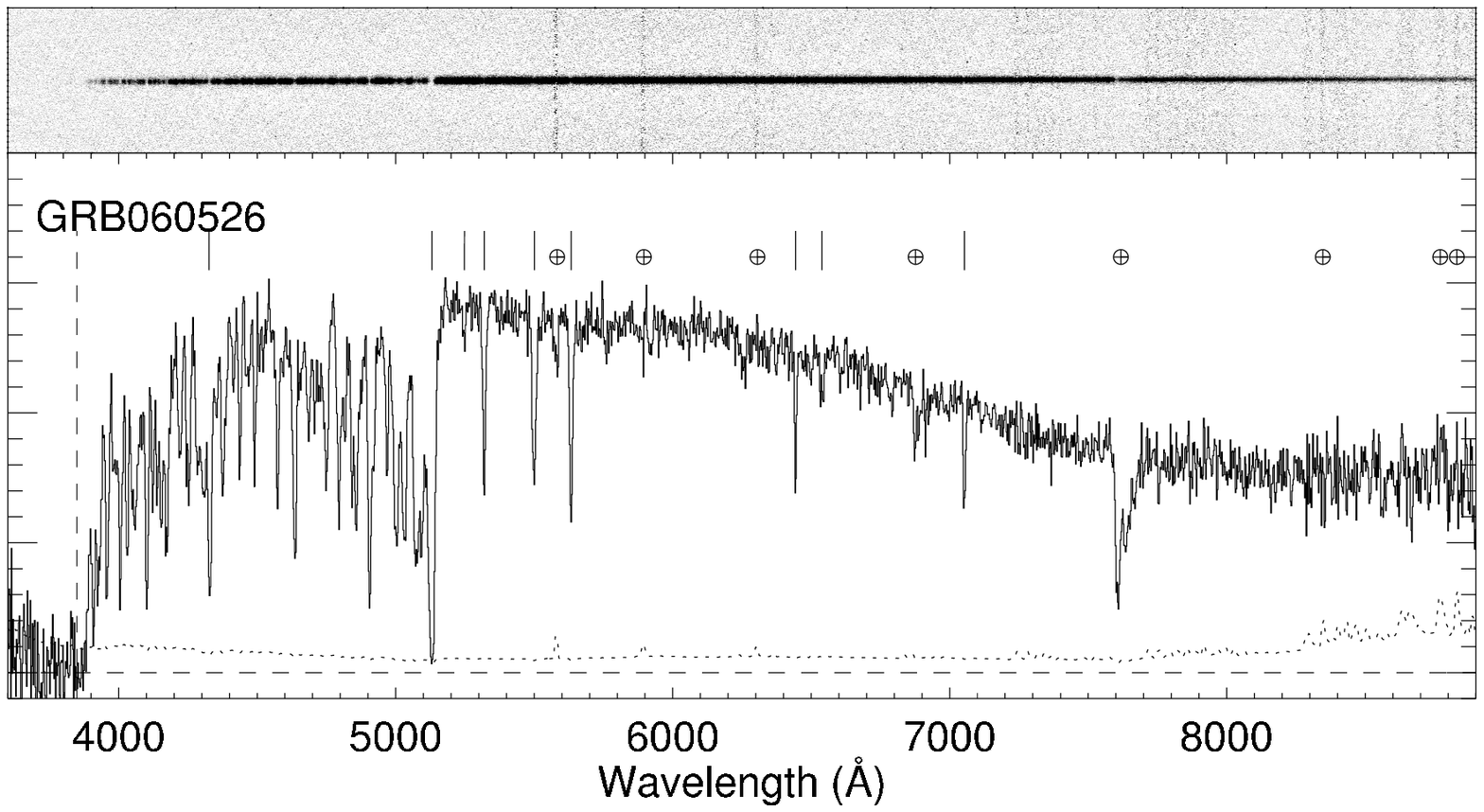}
\end{figure}

\begin{figure}
\epsscale{1.00}
\plotone{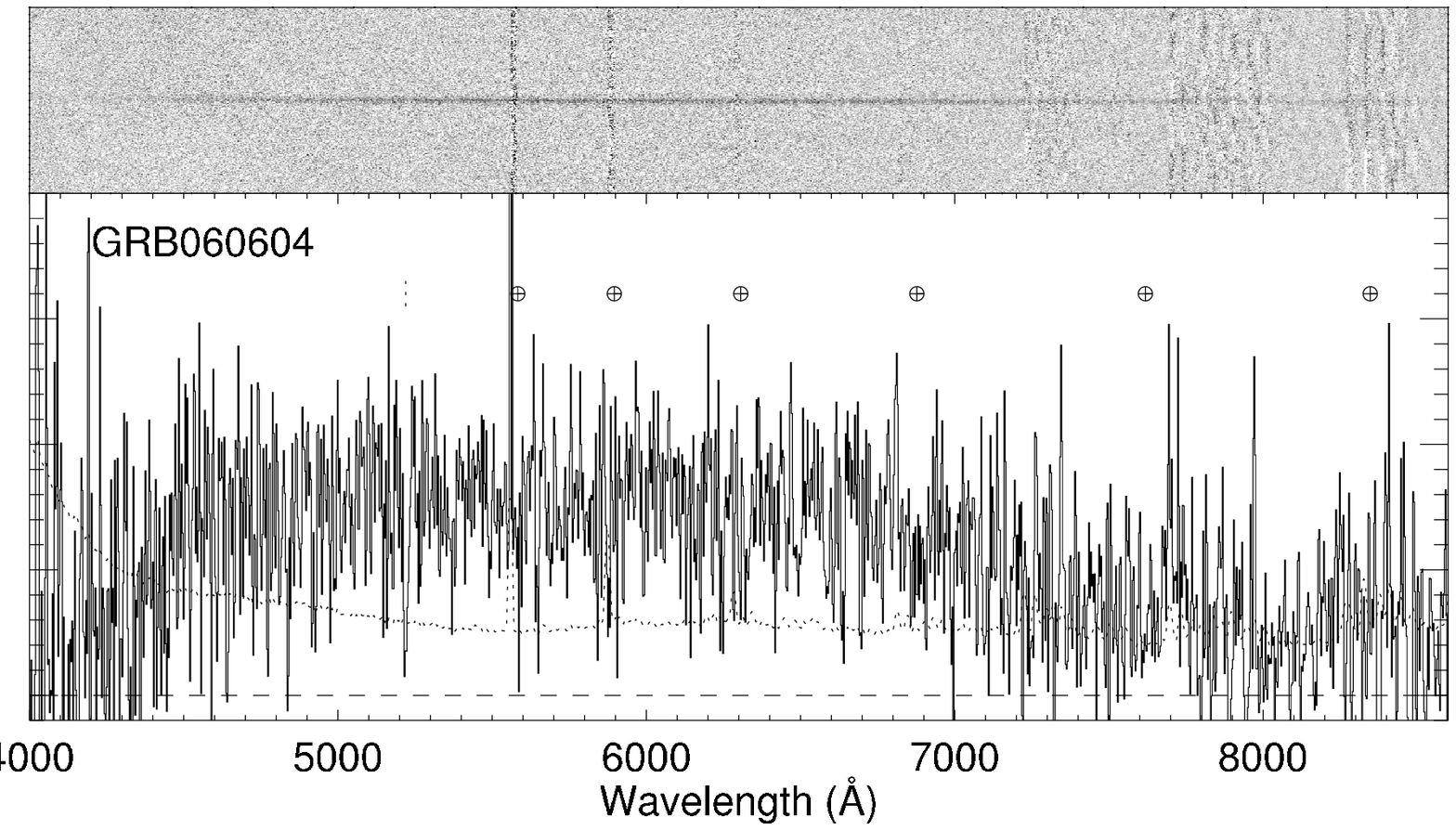}
\end{figure}

\begin{figure}
\epsscale{1.00}
\plotone{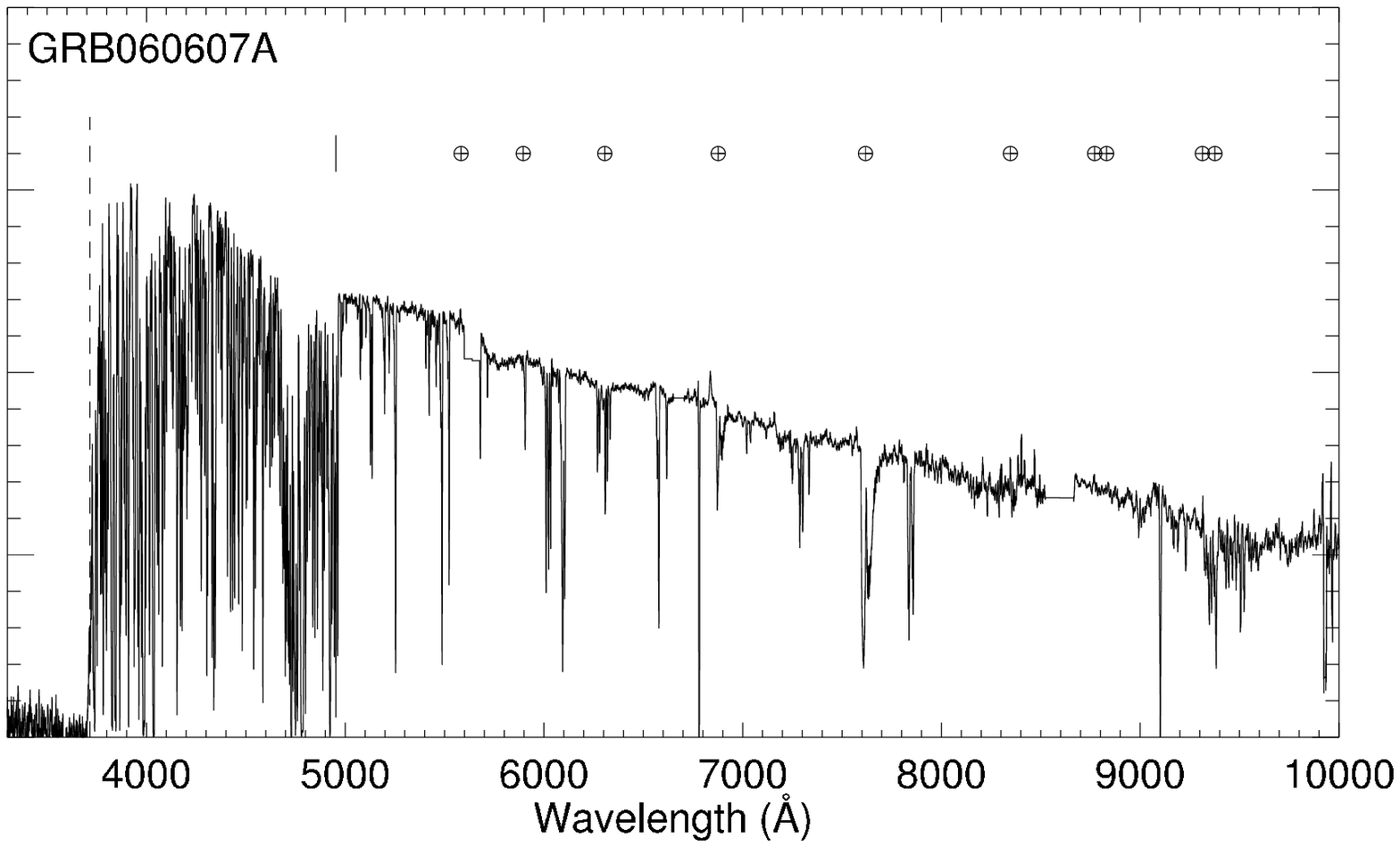}
\end{figure}

\begin{figure}
\epsscale{1.00}
\plotone{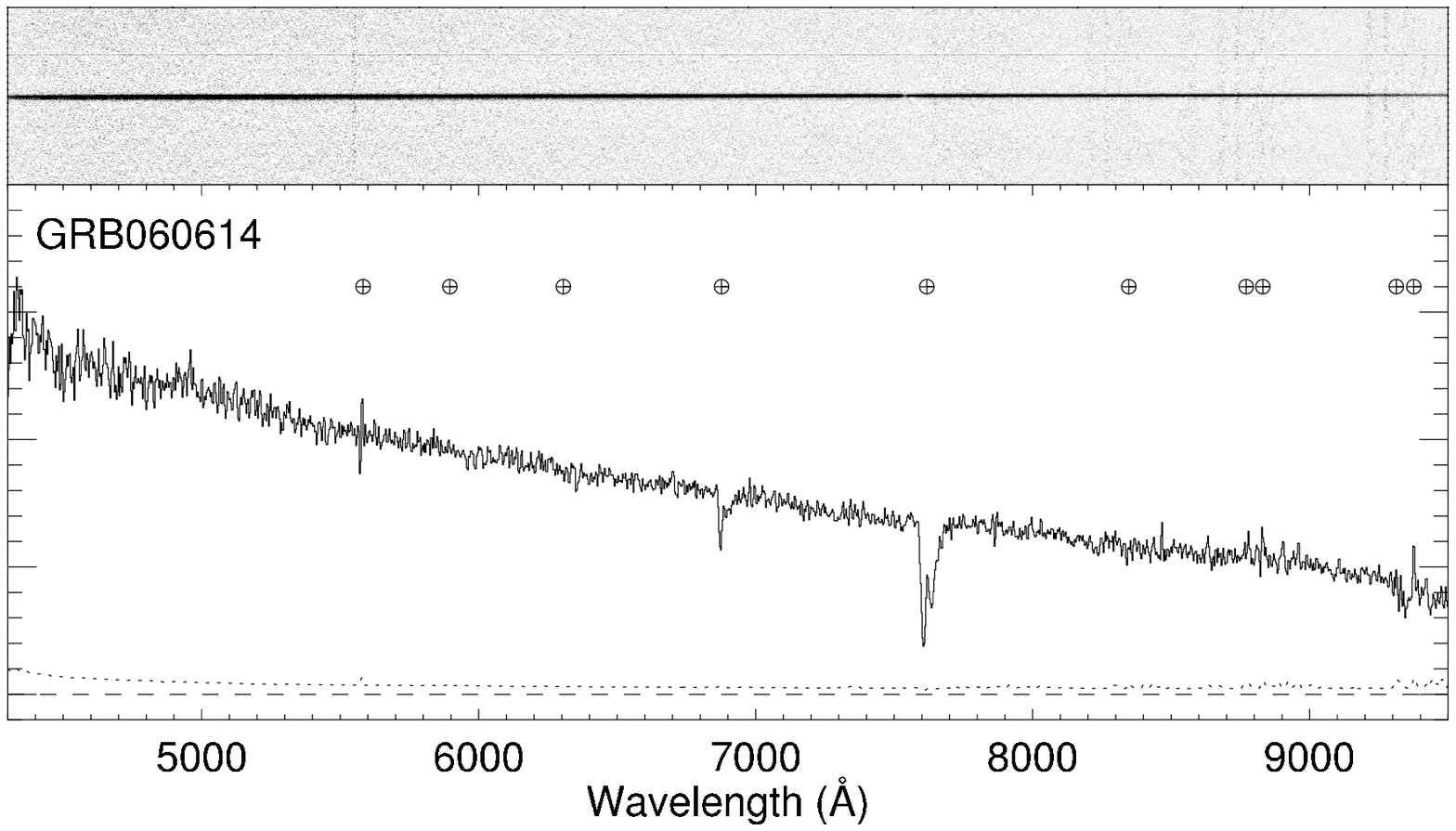}
\end{figure}

\begin{figure}
\epsscale{1.00}
\plotone{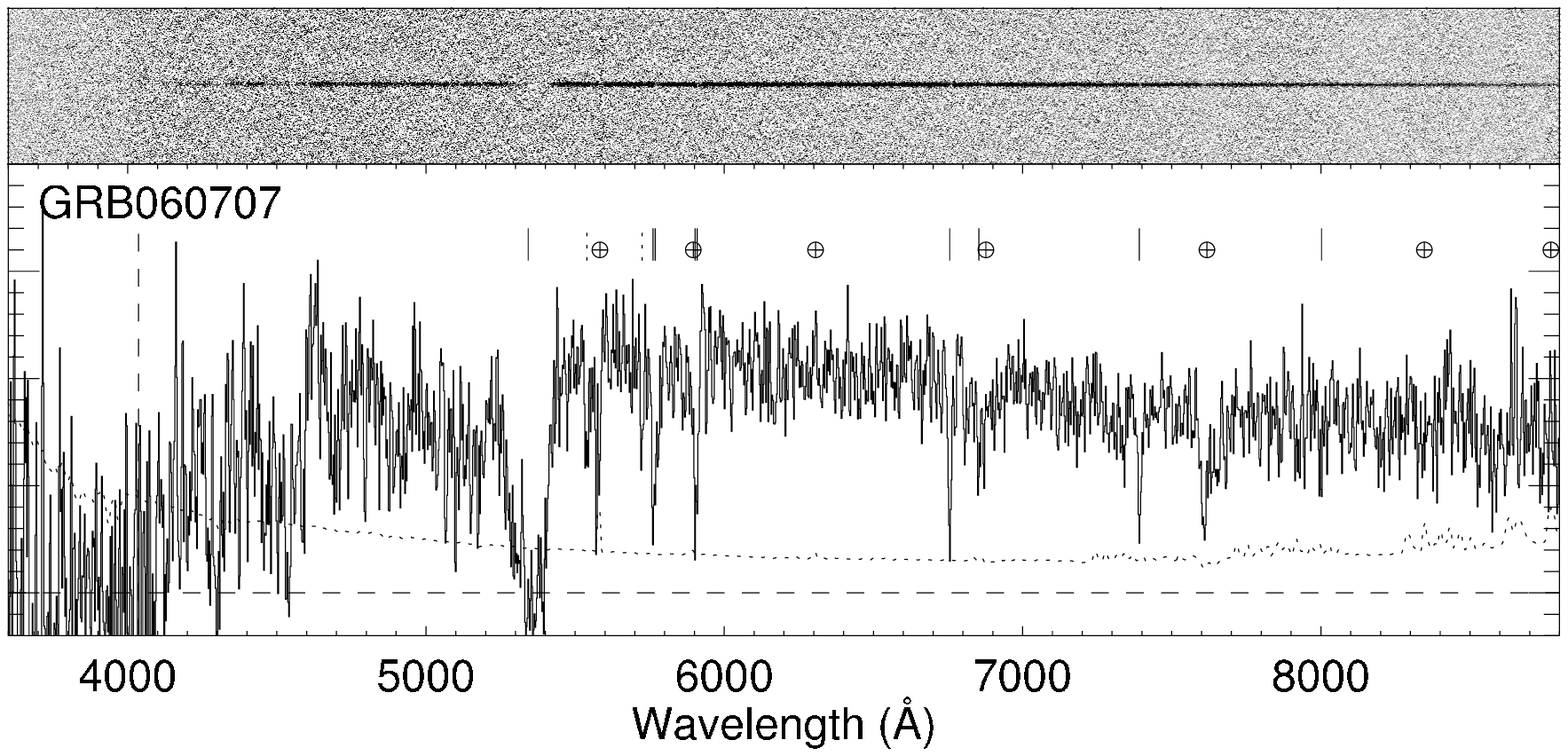}
\end{figure}

\begin{figure}
\epsscale{1.00}
\plotone{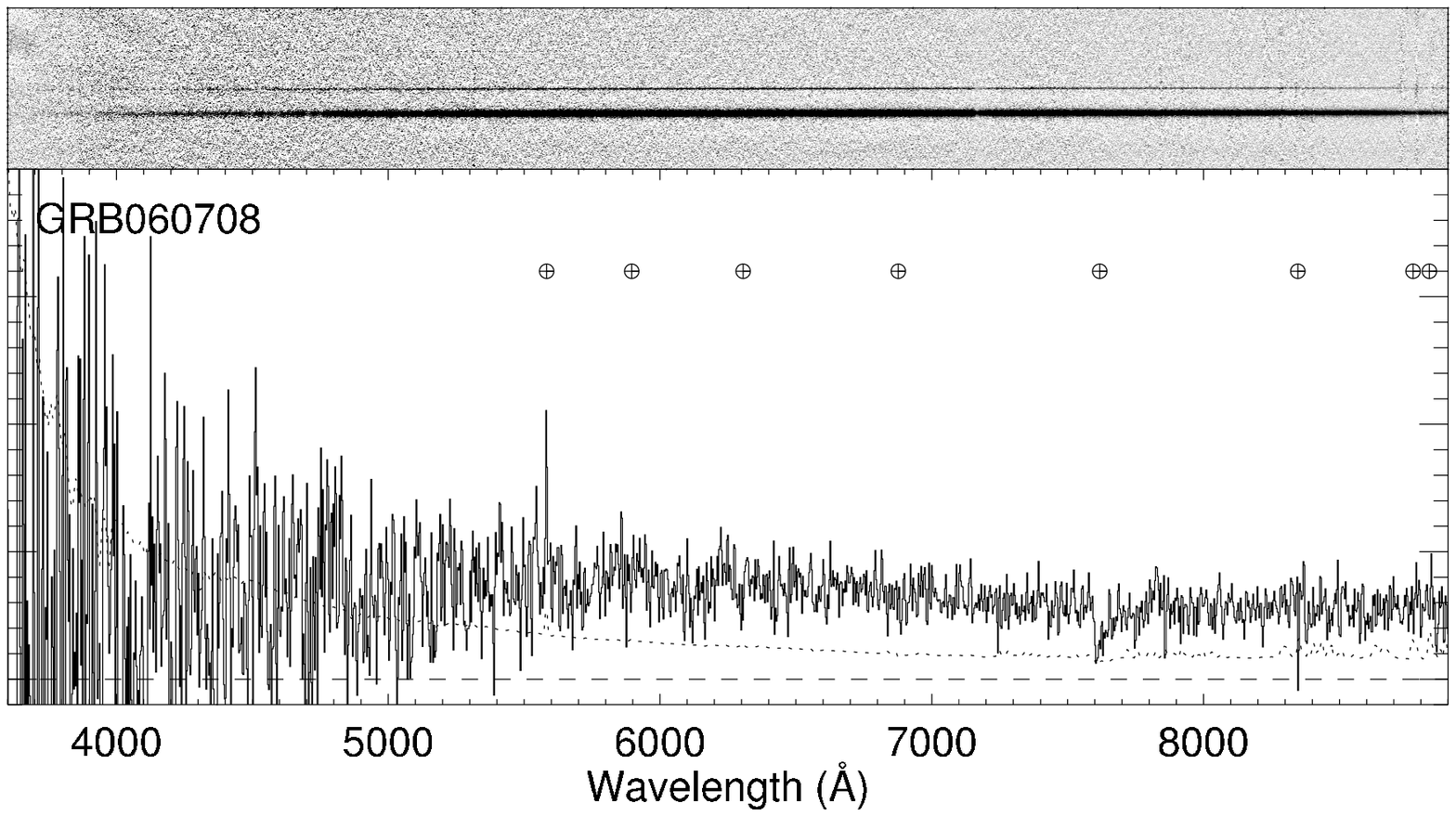}
\end{figure}

\begin{figure}
\epsscale{1.00}
\plotone{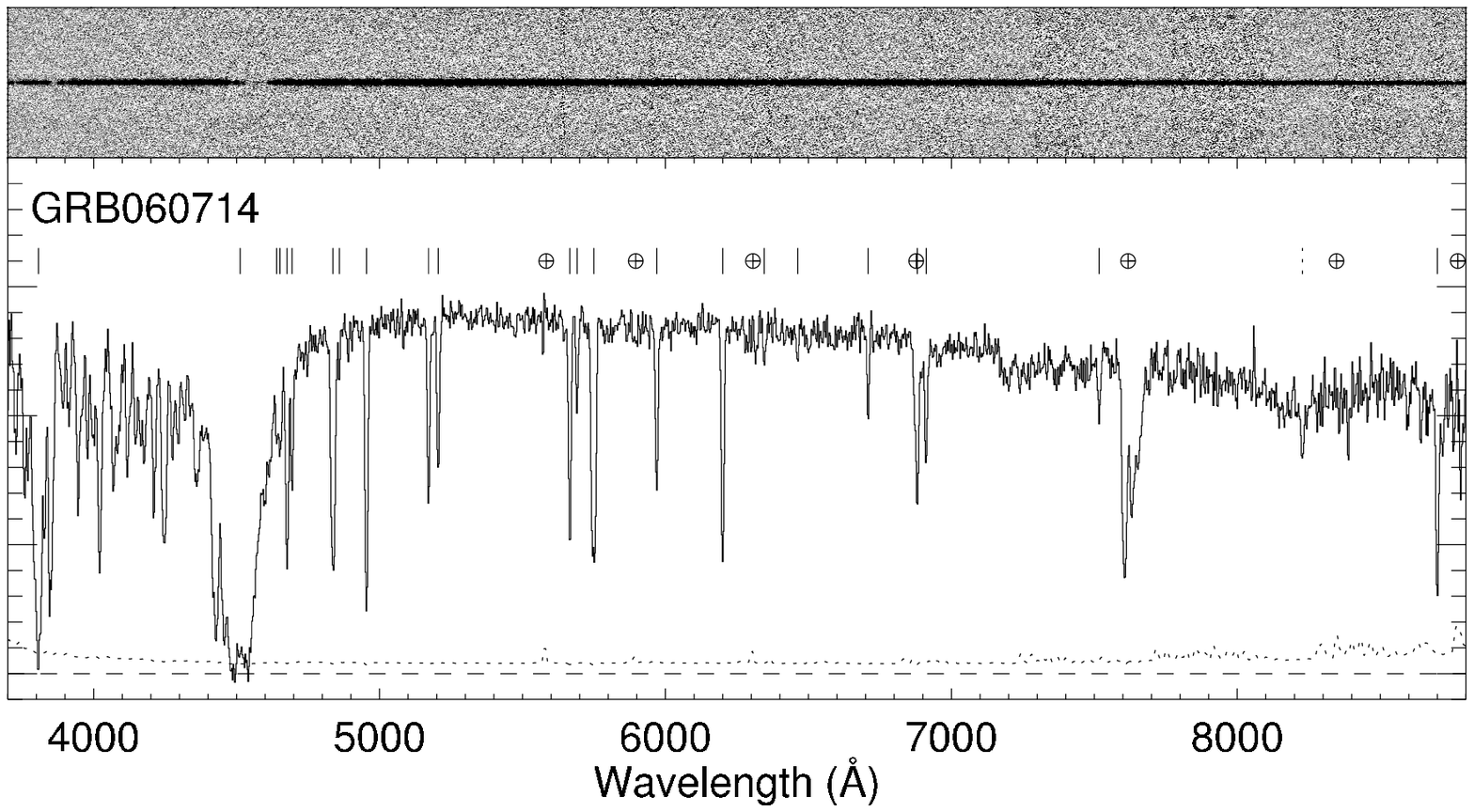}
\end{figure}

\begin{figure}
\epsscale{1.00}
\plotone{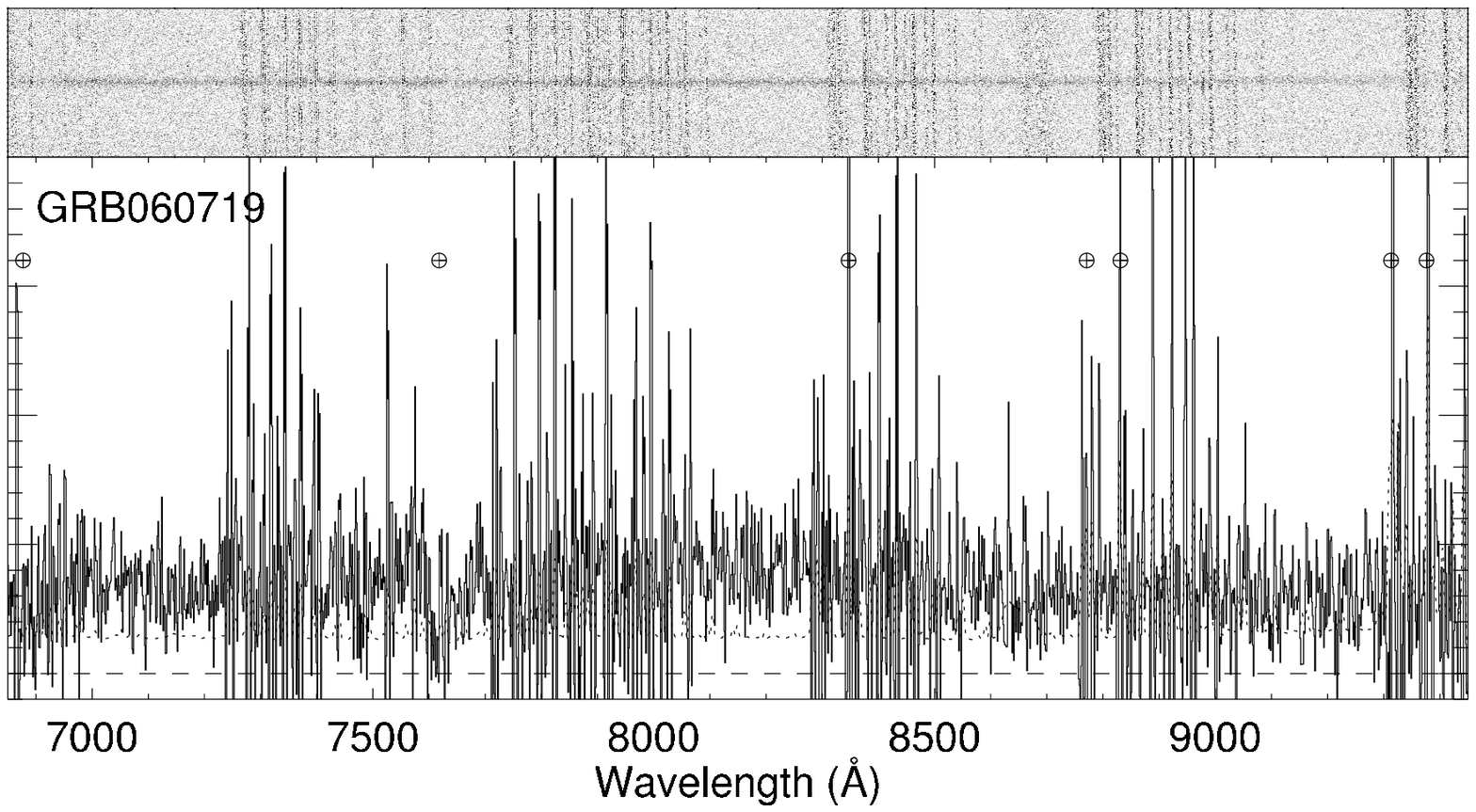}
\end{figure}

\begin{figure}
\epsscale{1.00}
\plotone{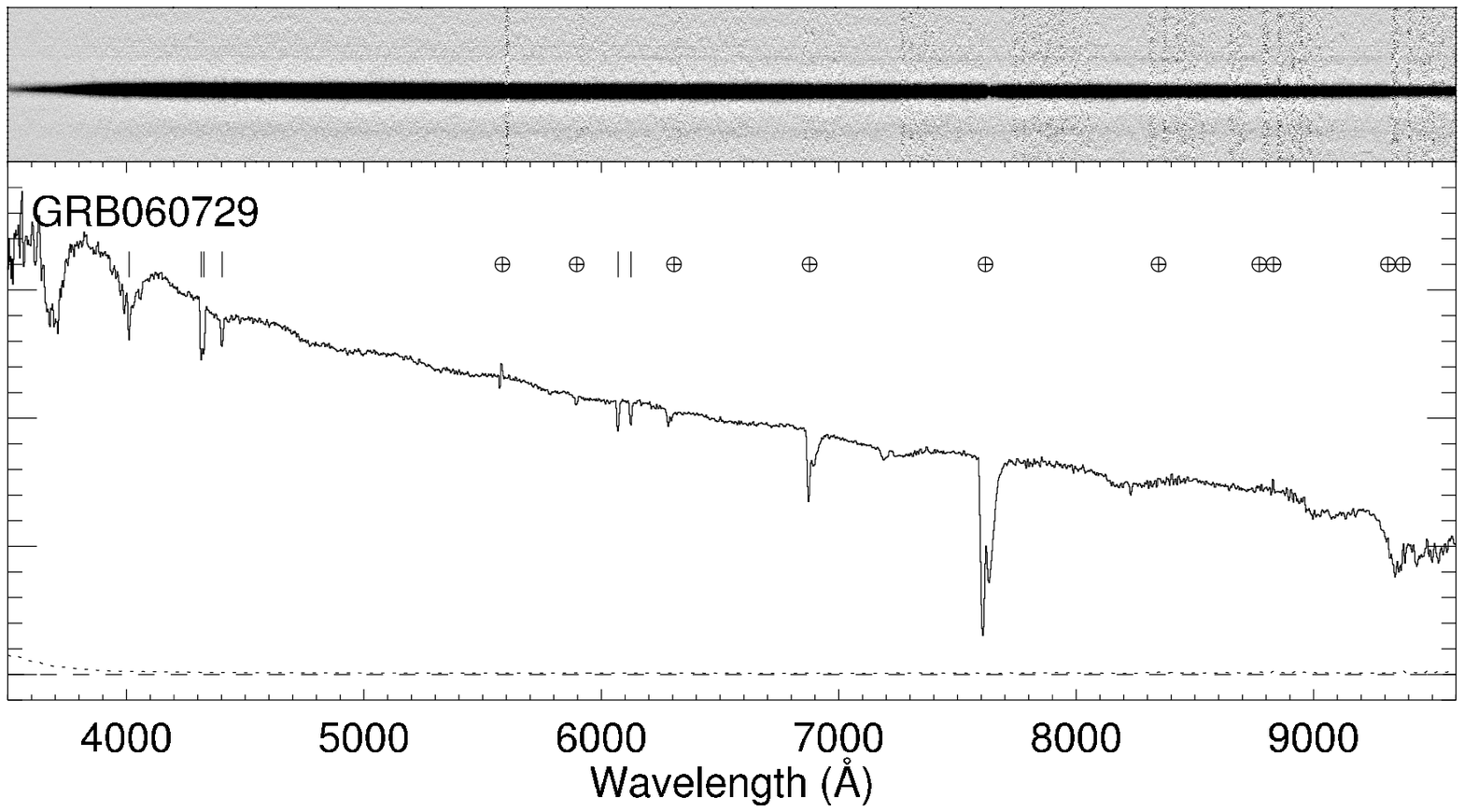}
\end{figure}

\begin{figure}
\epsscale{1.00}
\plotone{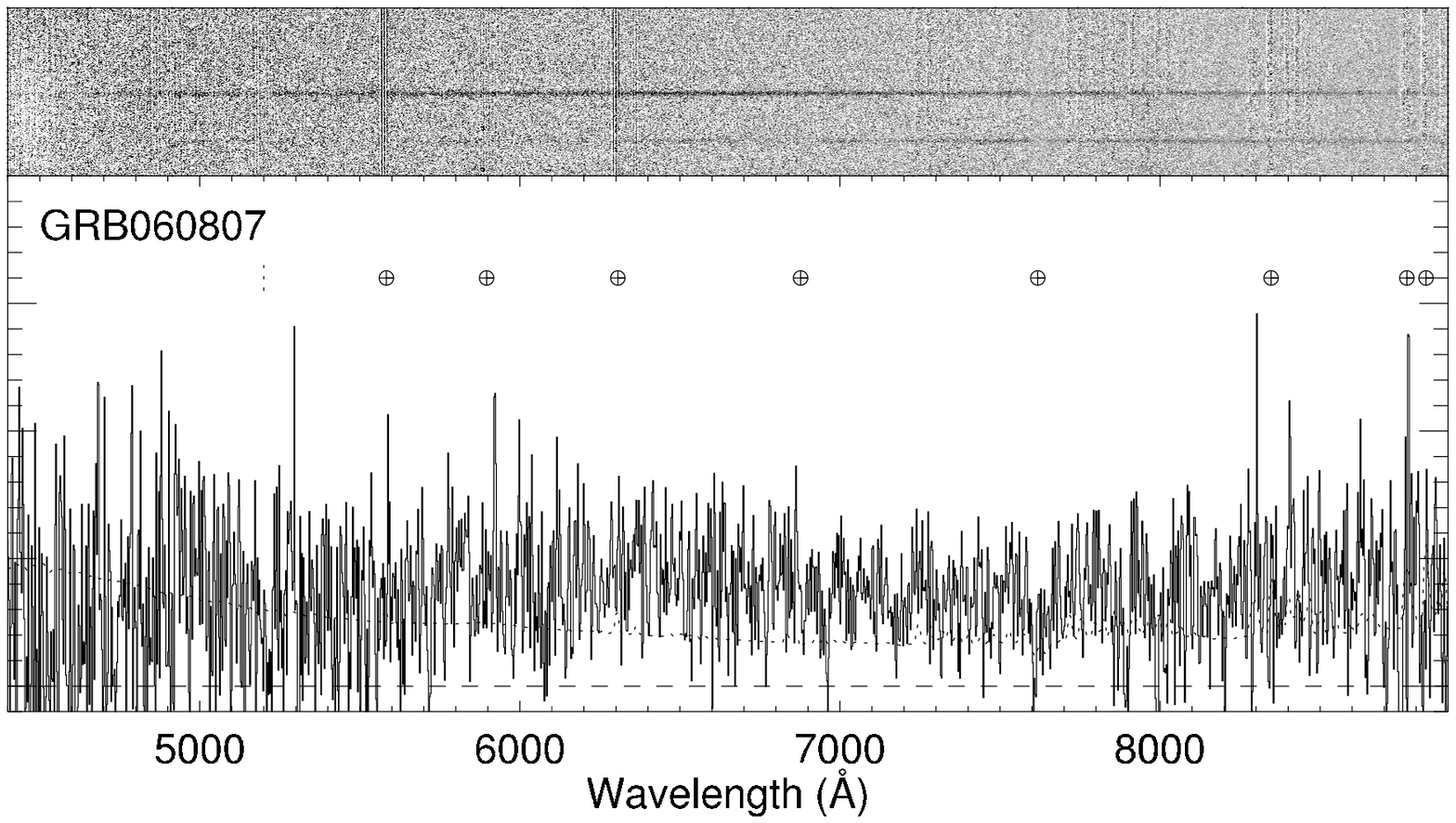}
\end{figure}

\begin{figure}
\epsscale{1.00}
\plotone{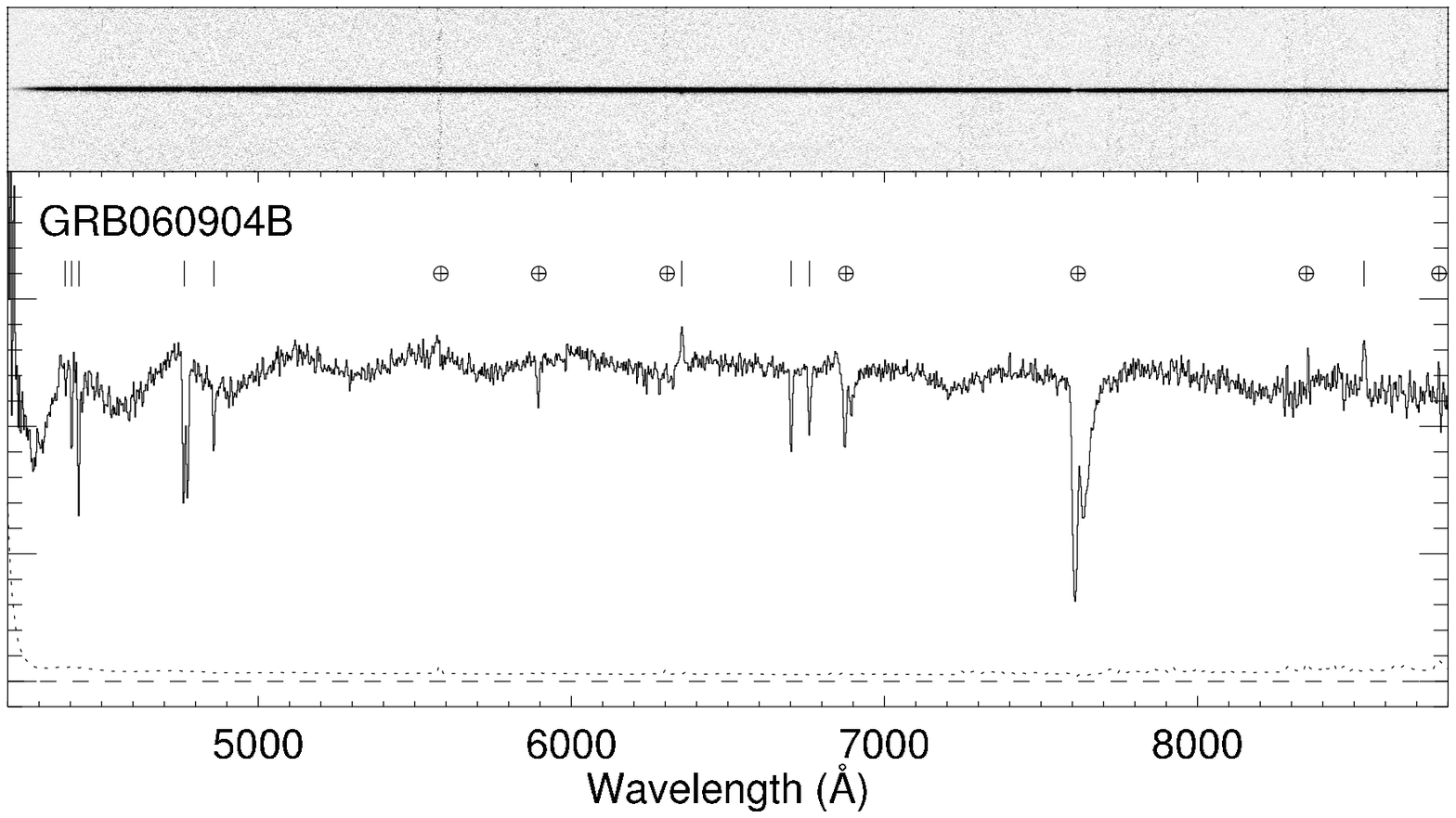}
\end{figure}

\begin{figure}
\epsscale{1.00}
\plotone{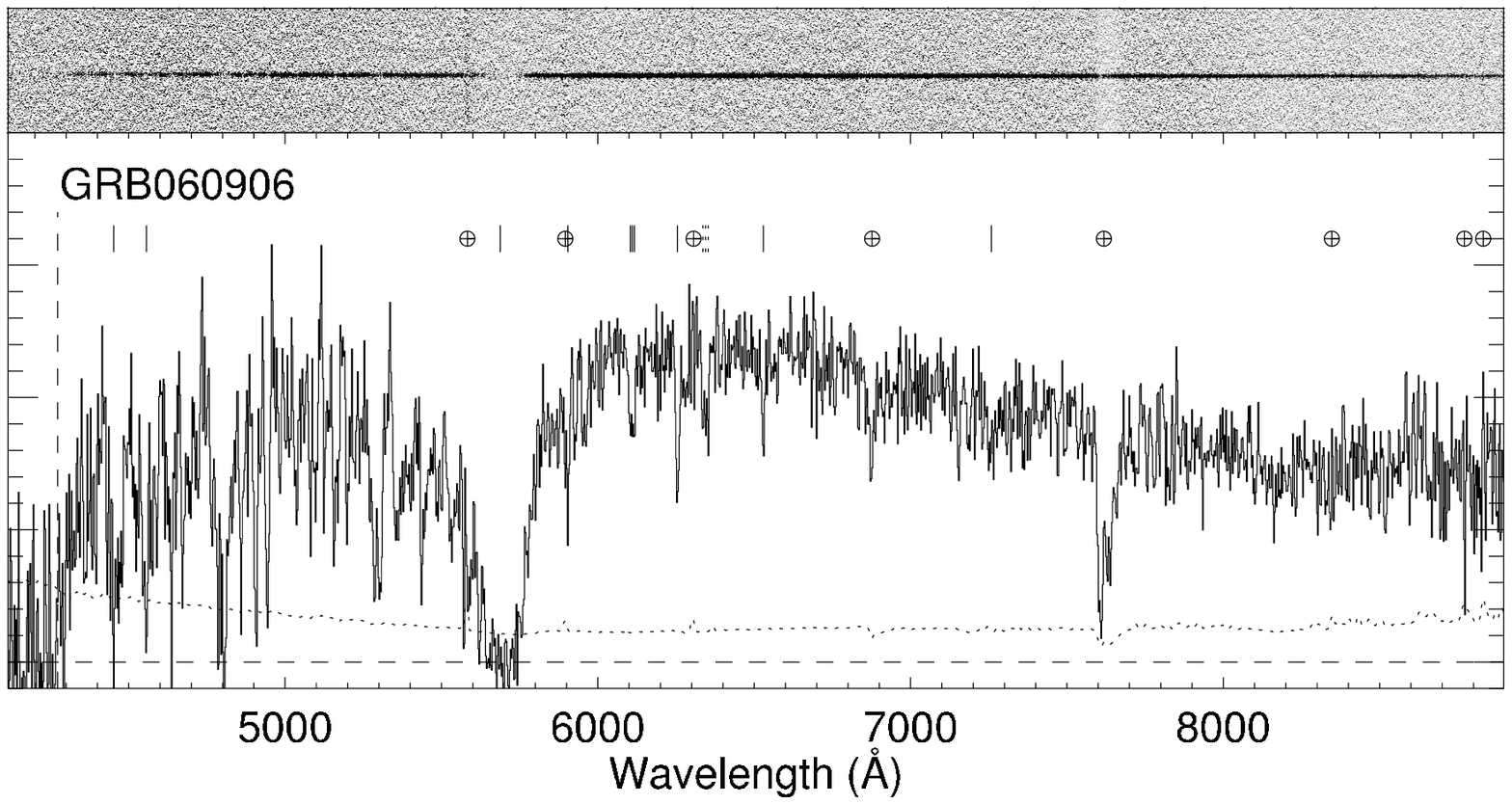}
\end{figure}

\begin{figure}
\epsscale{1.00}
\plotone{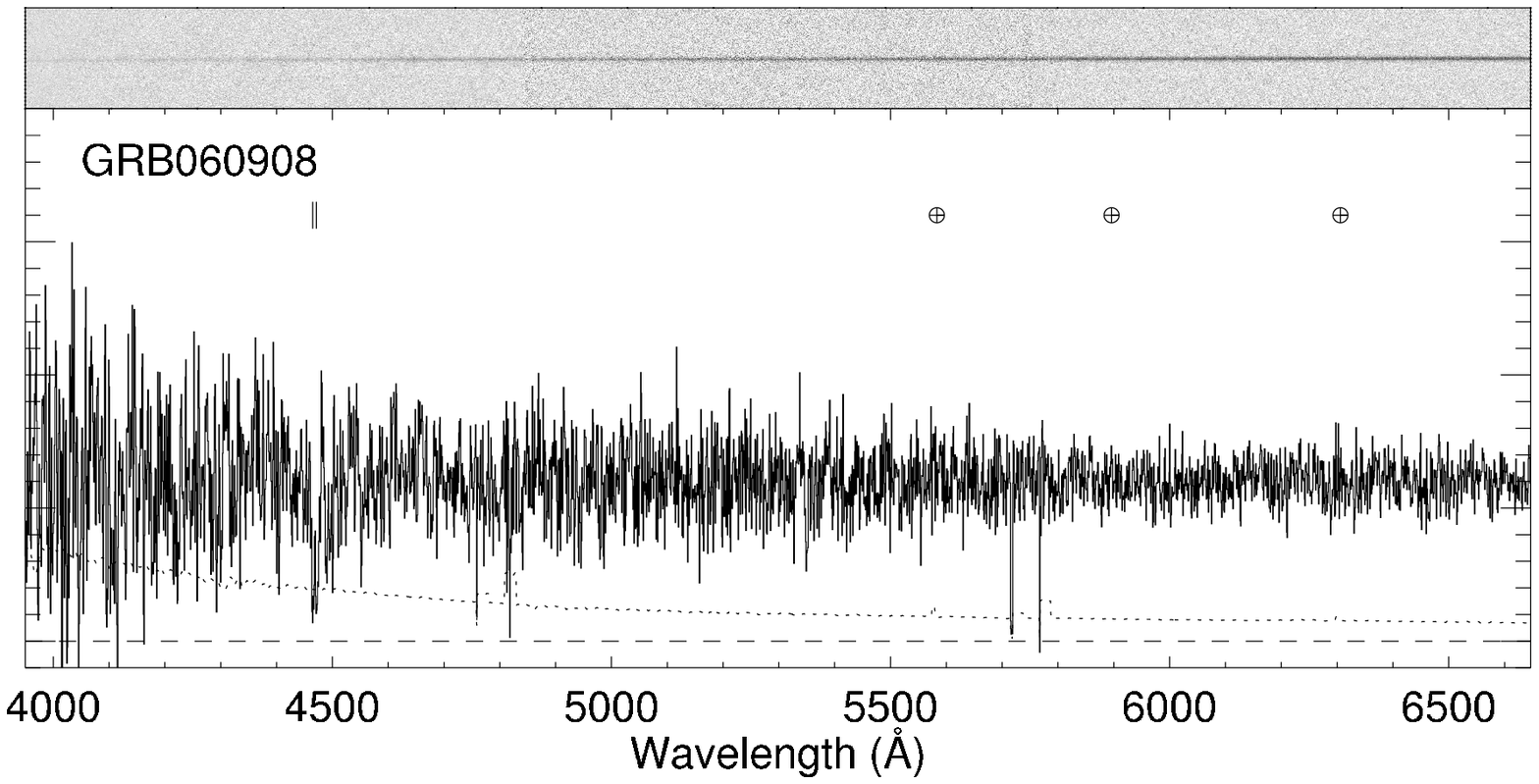}
\end{figure}

\begin{figure}
\epsscale{1.00}
\plotone{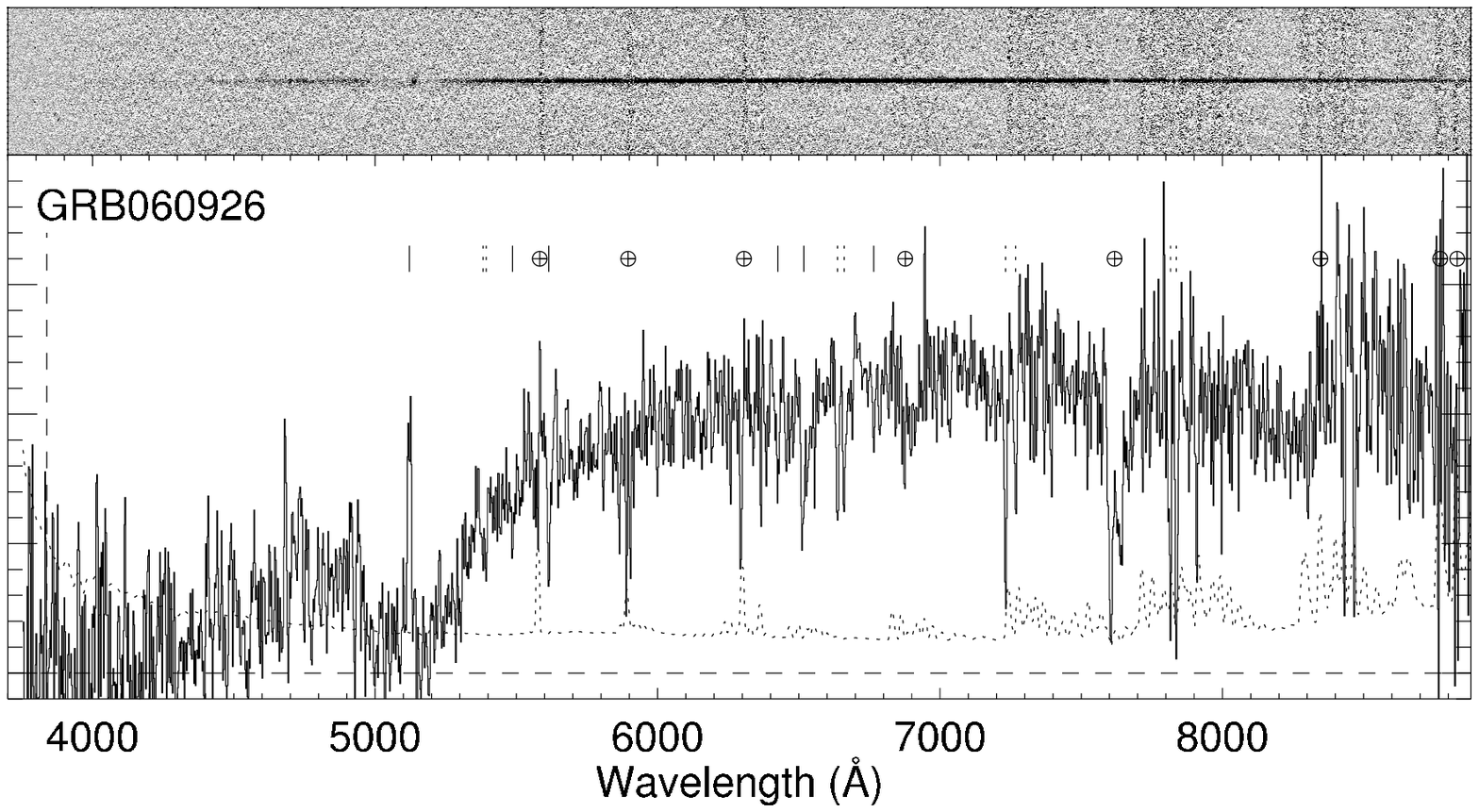}
\end{figure}

\clearpage

\begin{figure}
\epsscale{1.00}
\plotone{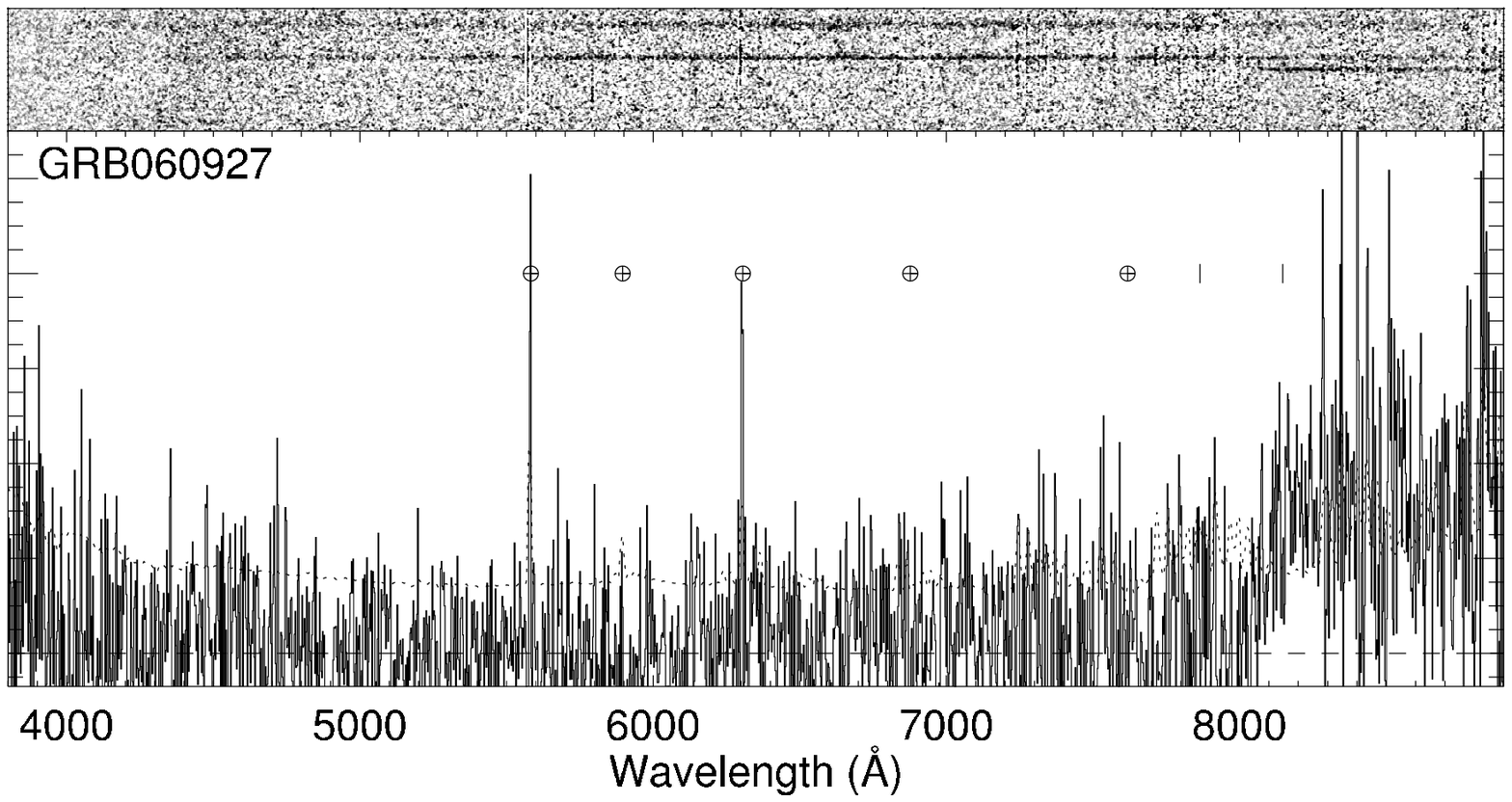}
\end{figure}

\begin{figure}
\epsscale{1.00}
\plotone{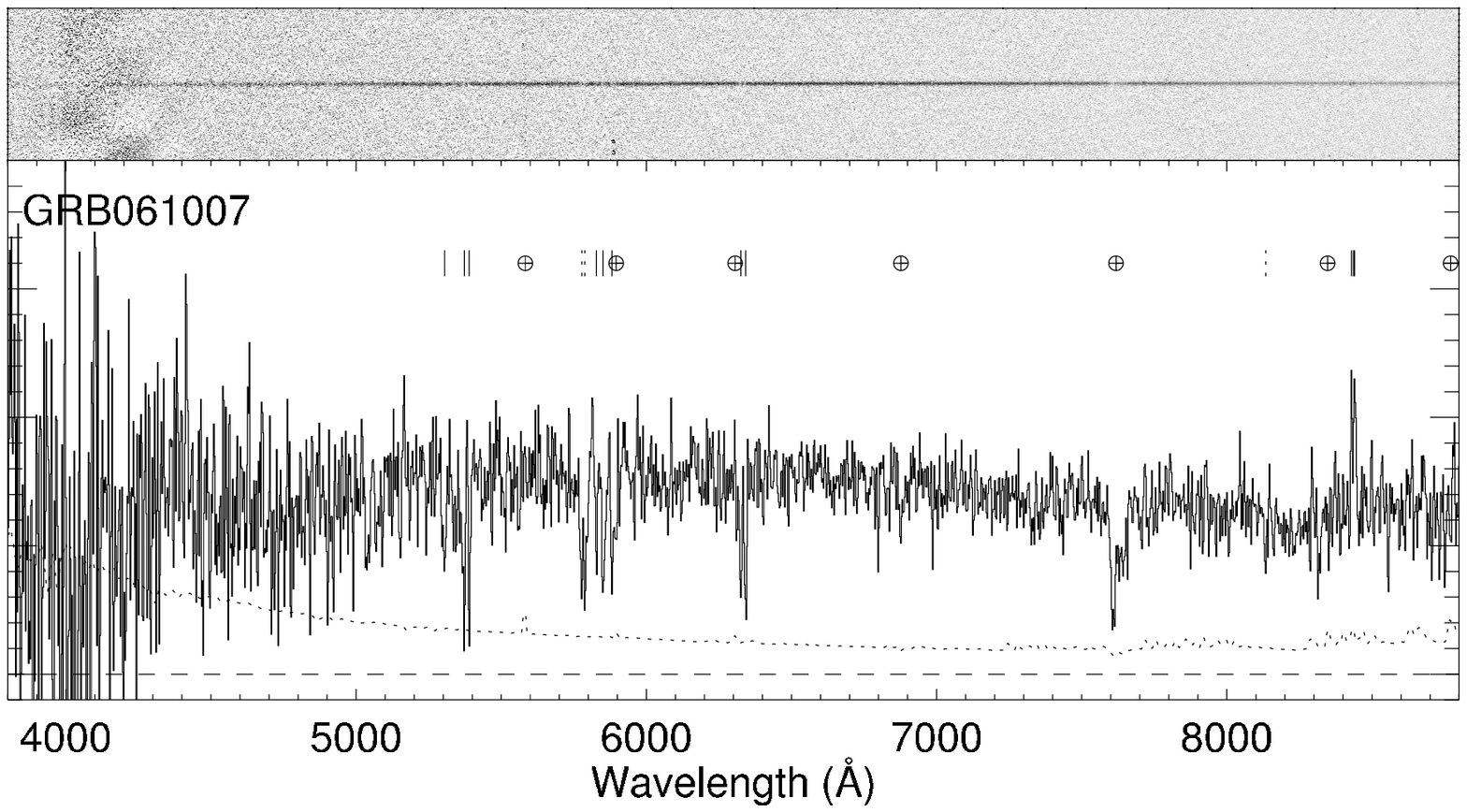}
\end{figure}

\begin{figure}
\epsscale{1.00}
\plotone{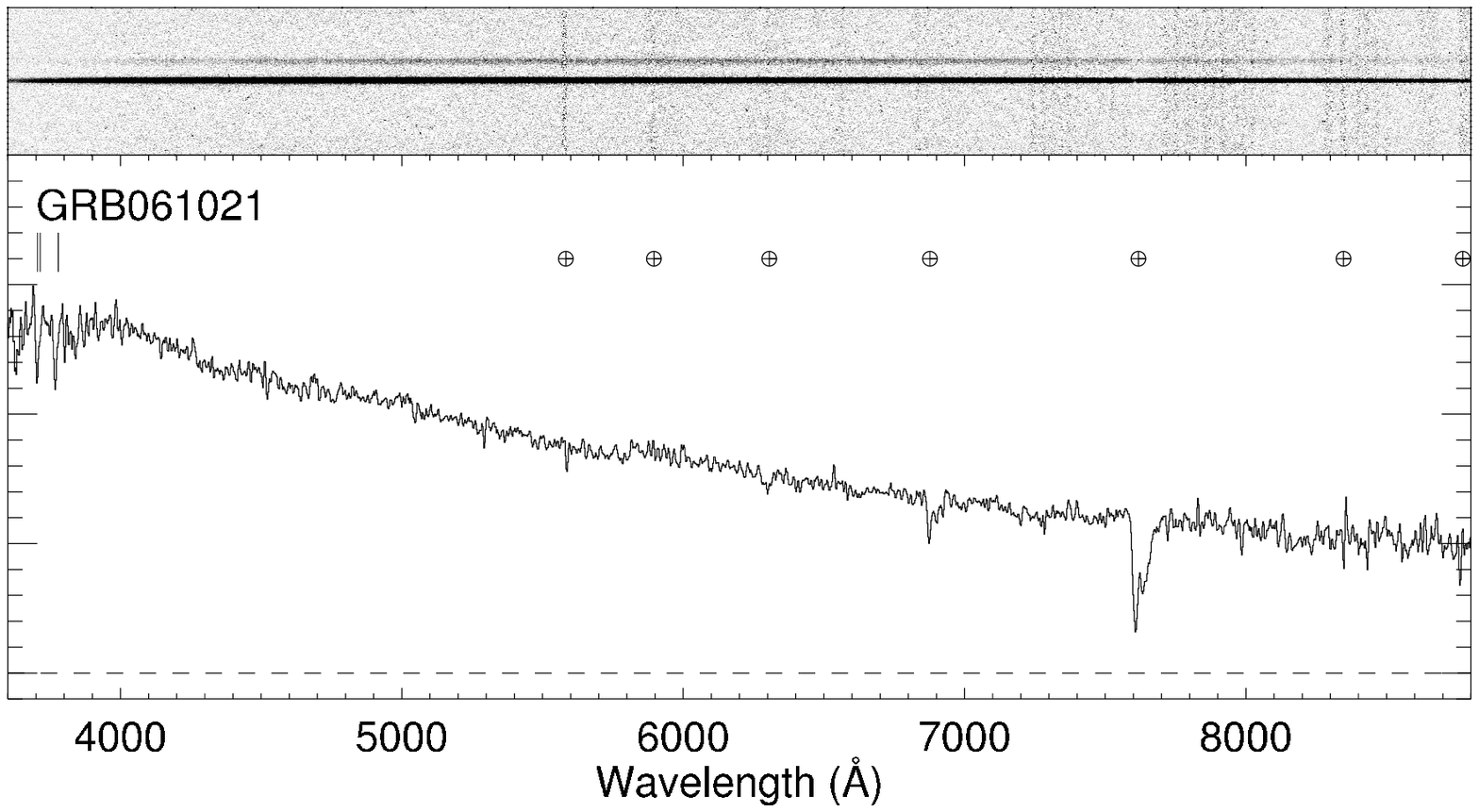}
\end{figure}

\begin{figure}
\epsscale{1.00}
\plotone{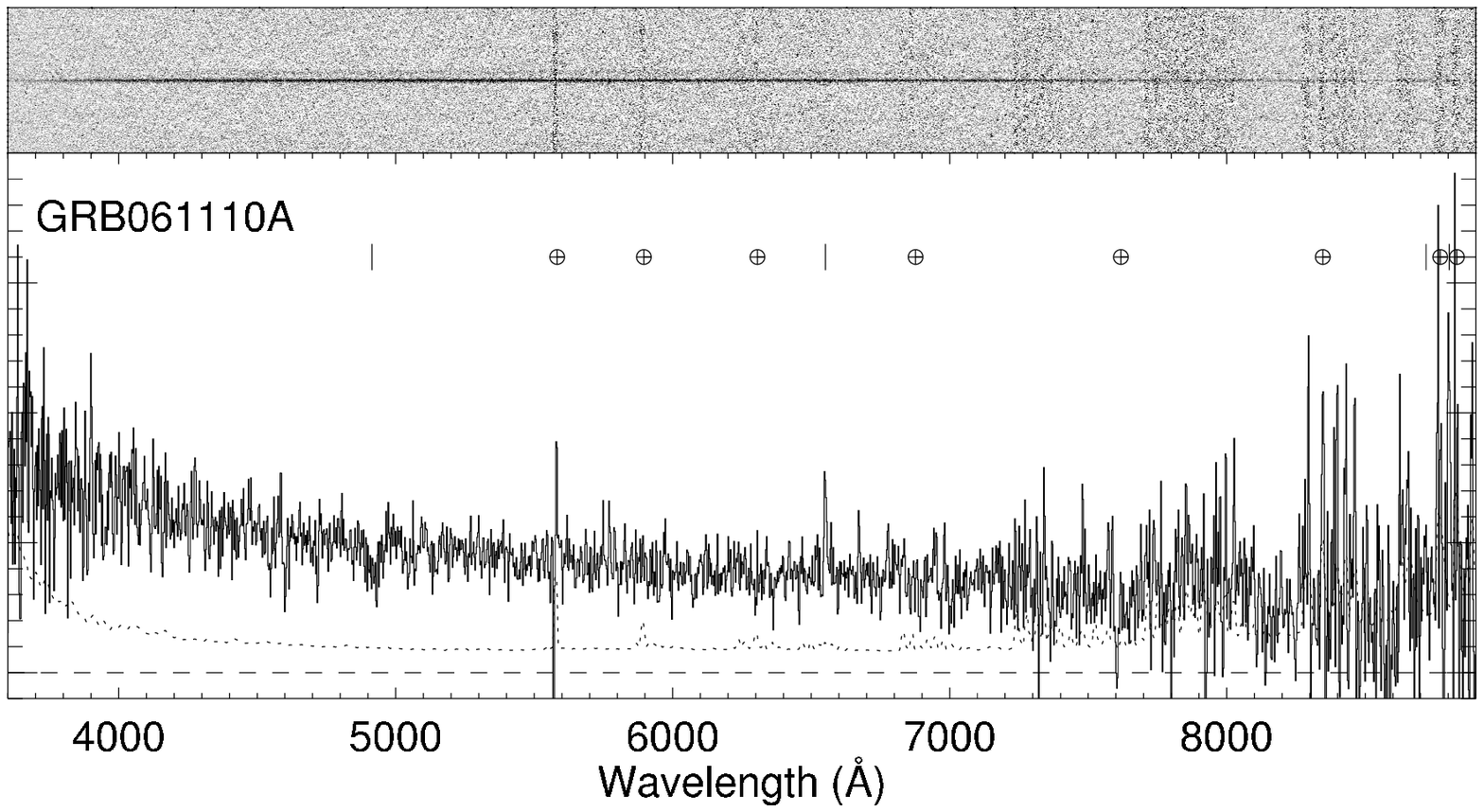}
\end{figure}

\begin{figure}
\epsscale{1.00}
\plotone{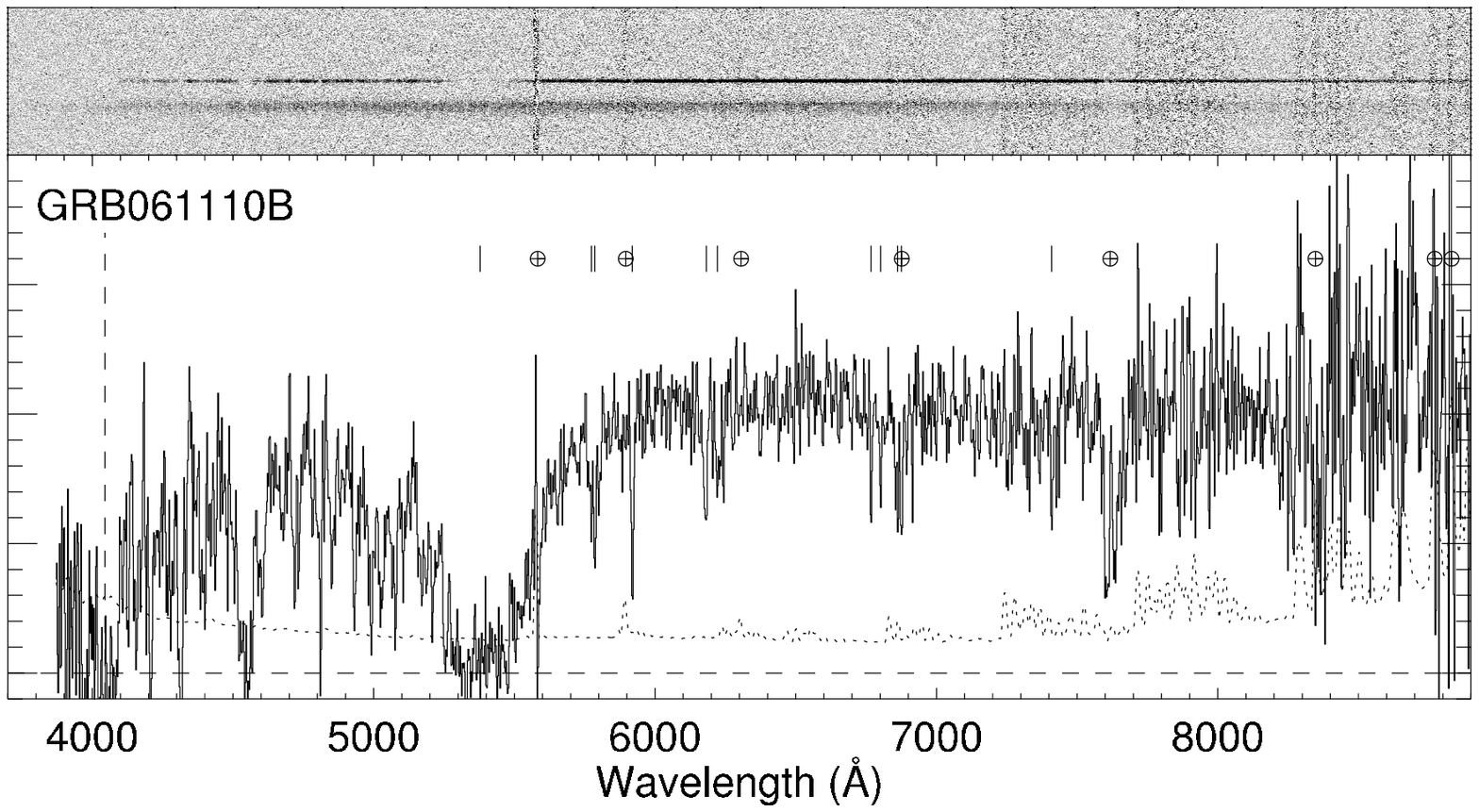}
\end{figure}

\begin{figure}
\epsscale{1.00}
\plotone{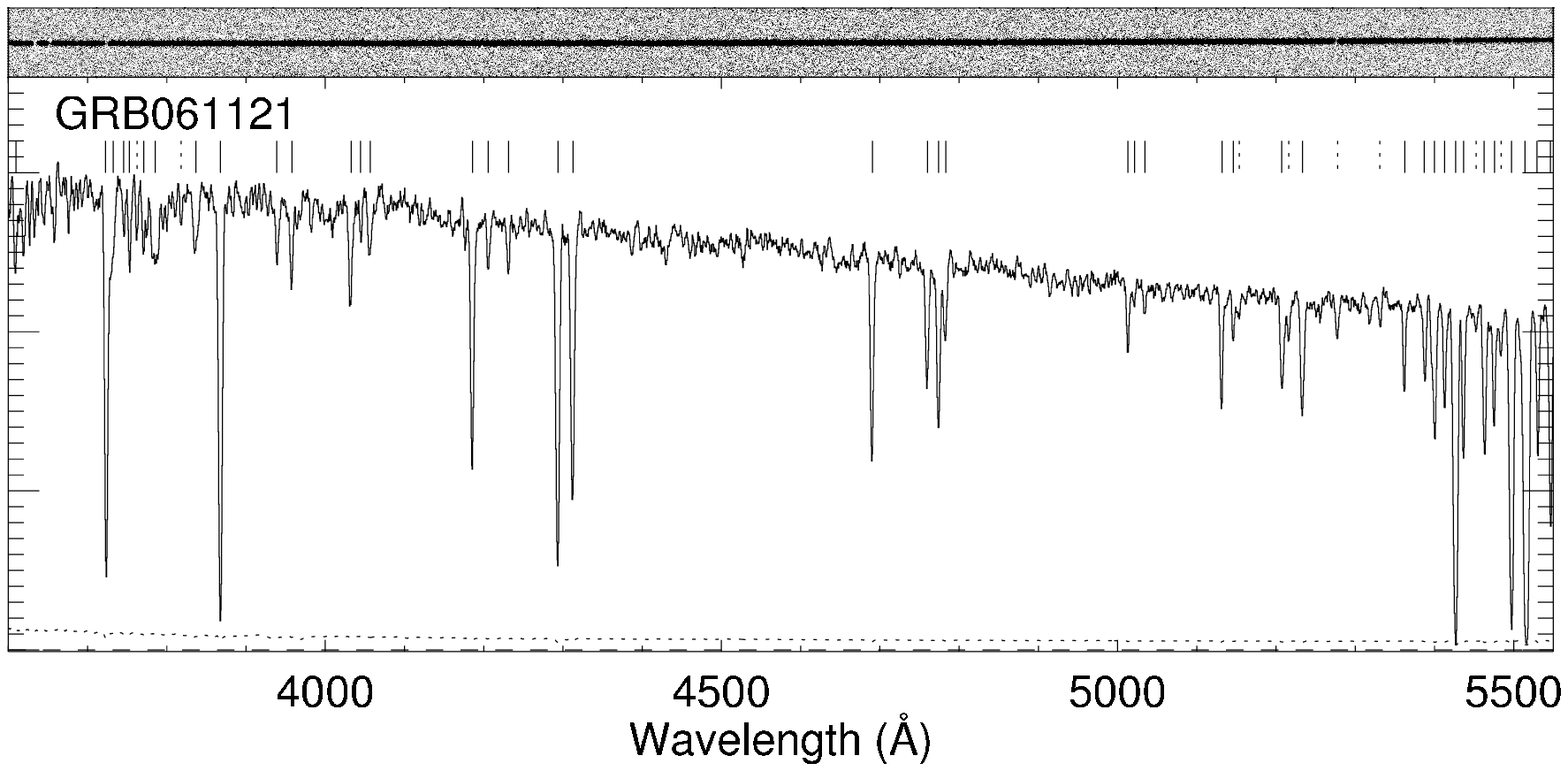}
\end{figure}

\begin{figure}
\epsscale{1.00}
\plotone{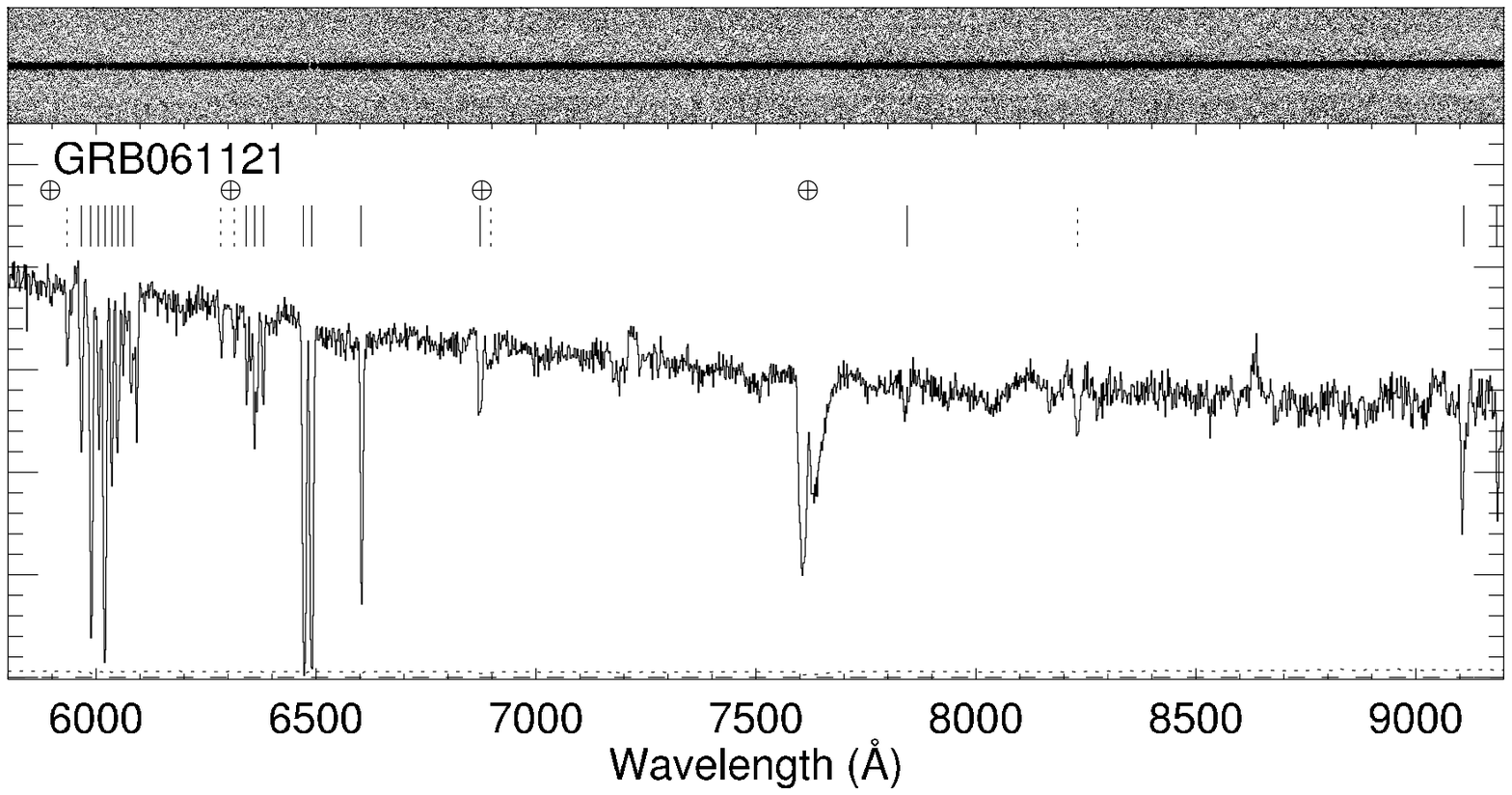}
\end{figure}

\begin{figure}
\epsscale{1.00}
\plotone{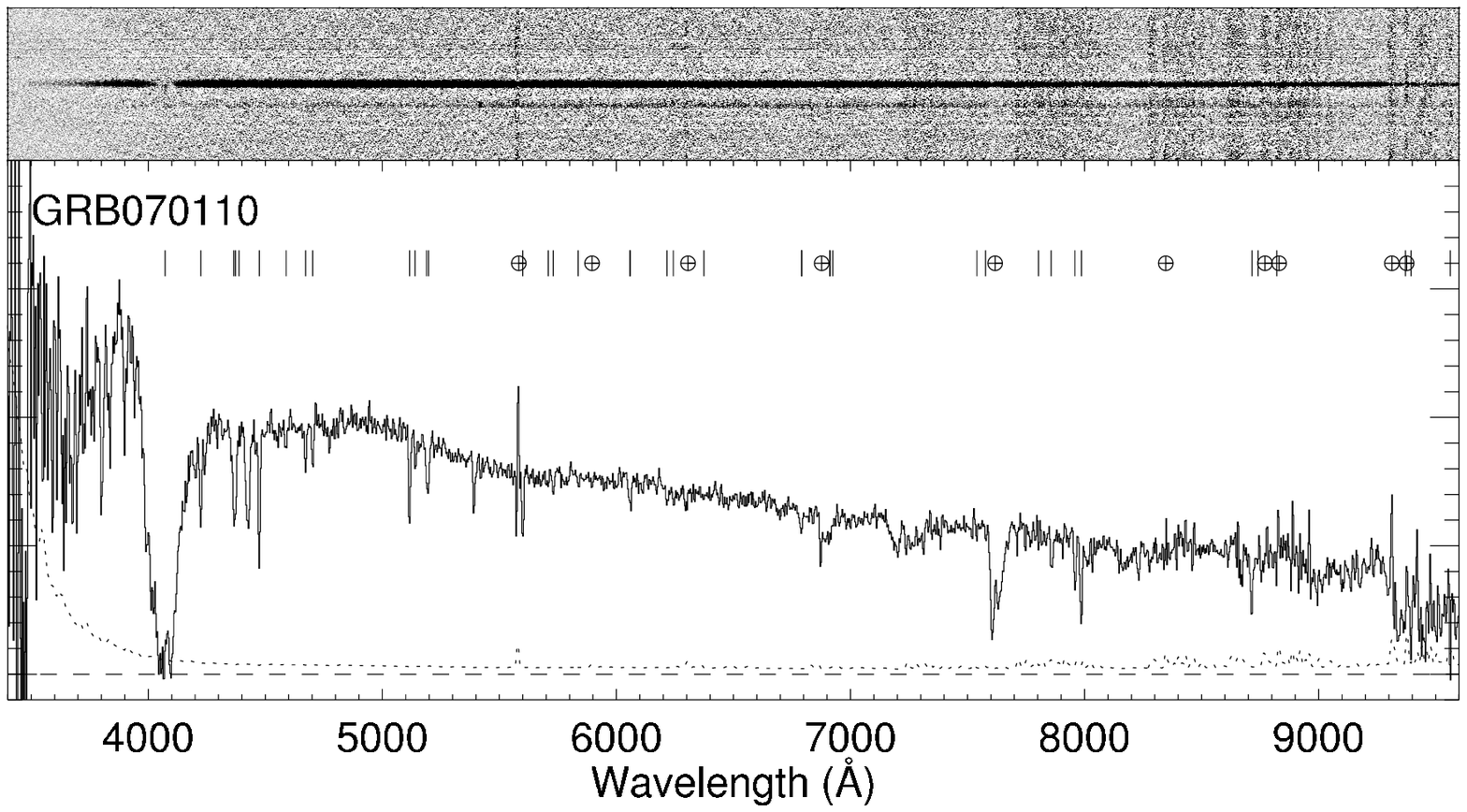}
\end{figure}

\begin{figure}
\epsscale{1.00}
\plotone{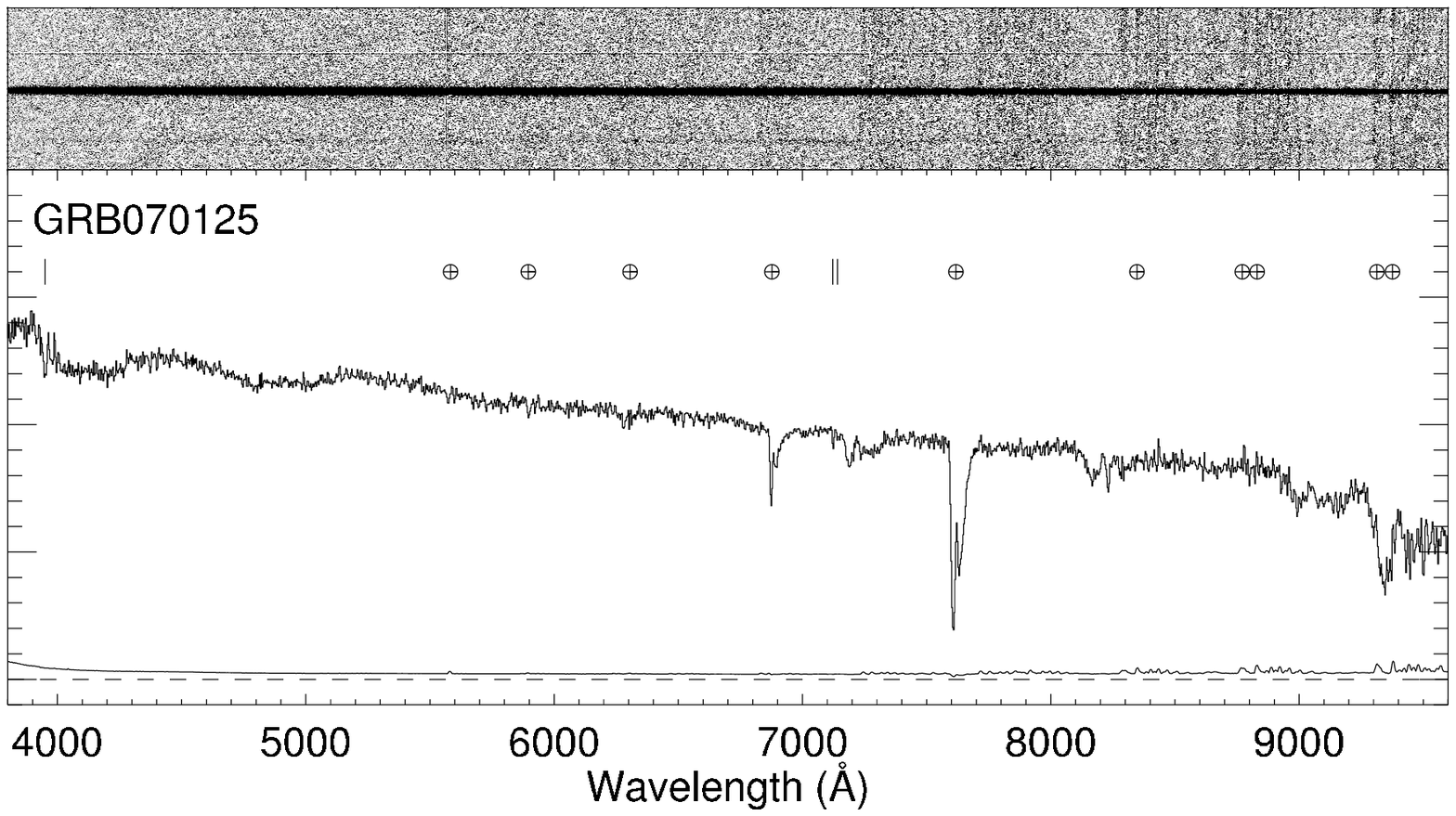}
\end{figure}

\begin{figure}
\epsscale{1.00}
\plotone{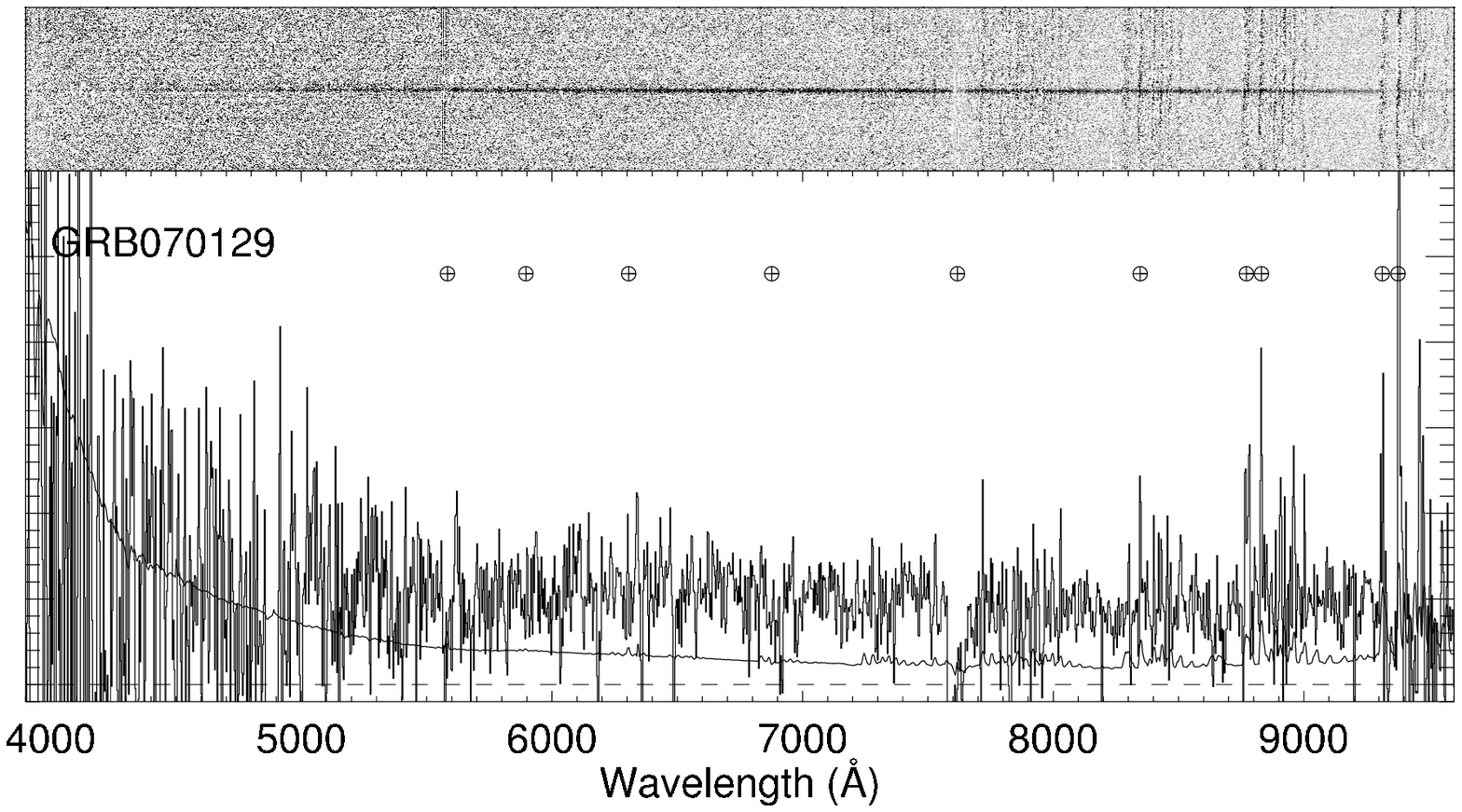}
\end{figure}

\begin{figure}
\epsscale{1.00}
\plotone{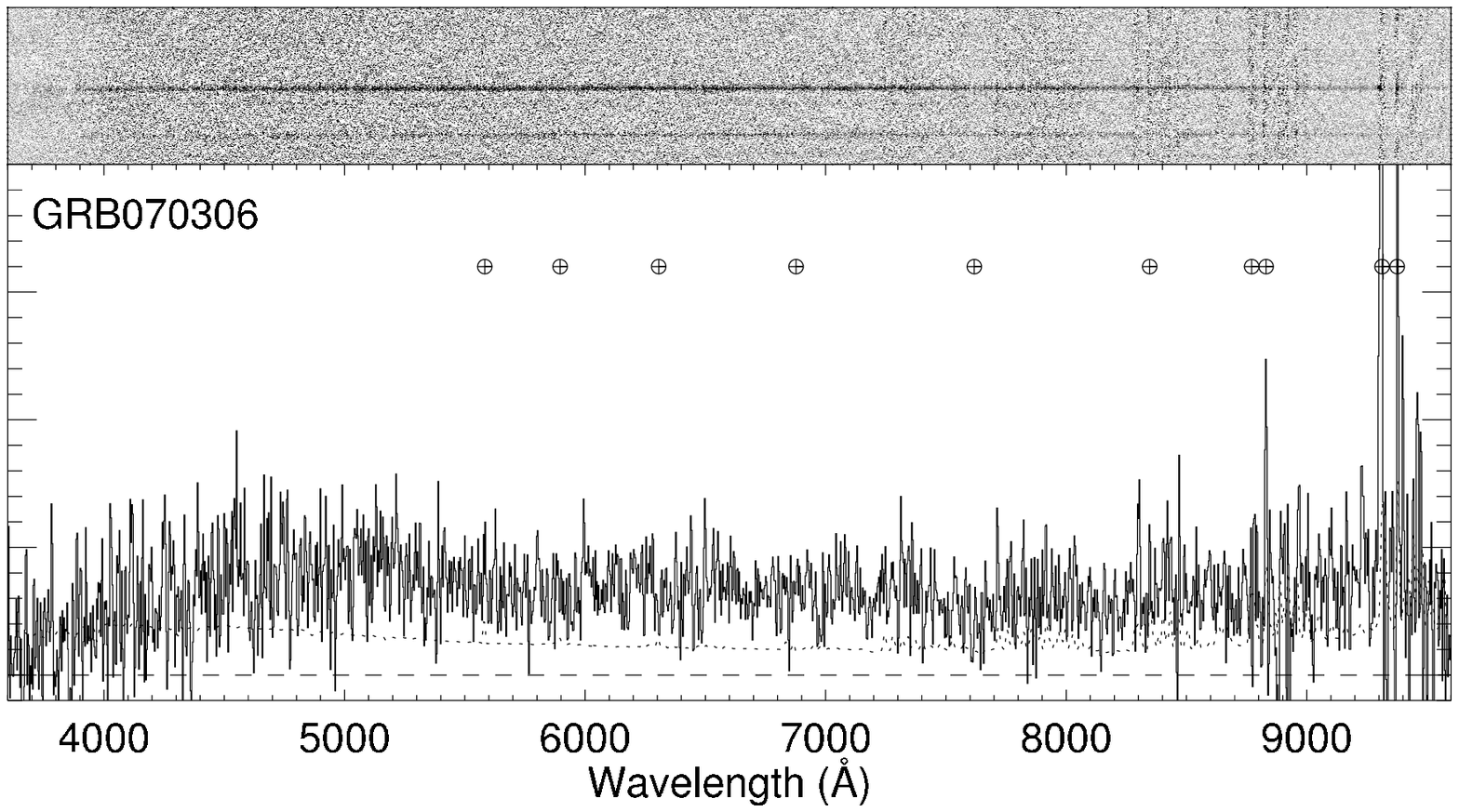}
\end{figure}

\begin{figure}
\epsscale{1.00}
\plotone{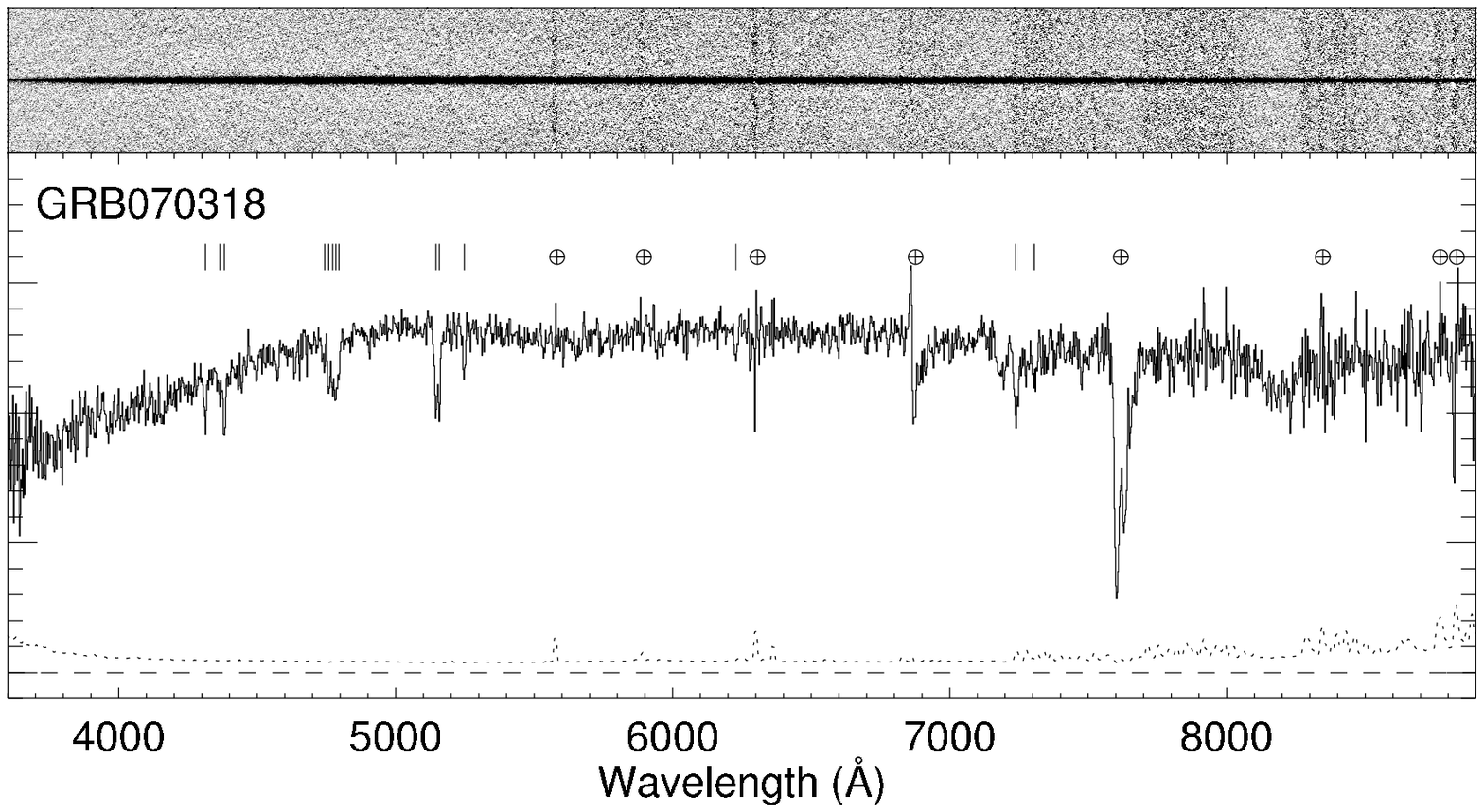}
\end{figure}

\begin{figure}
\epsscale{1.00}
\plotone{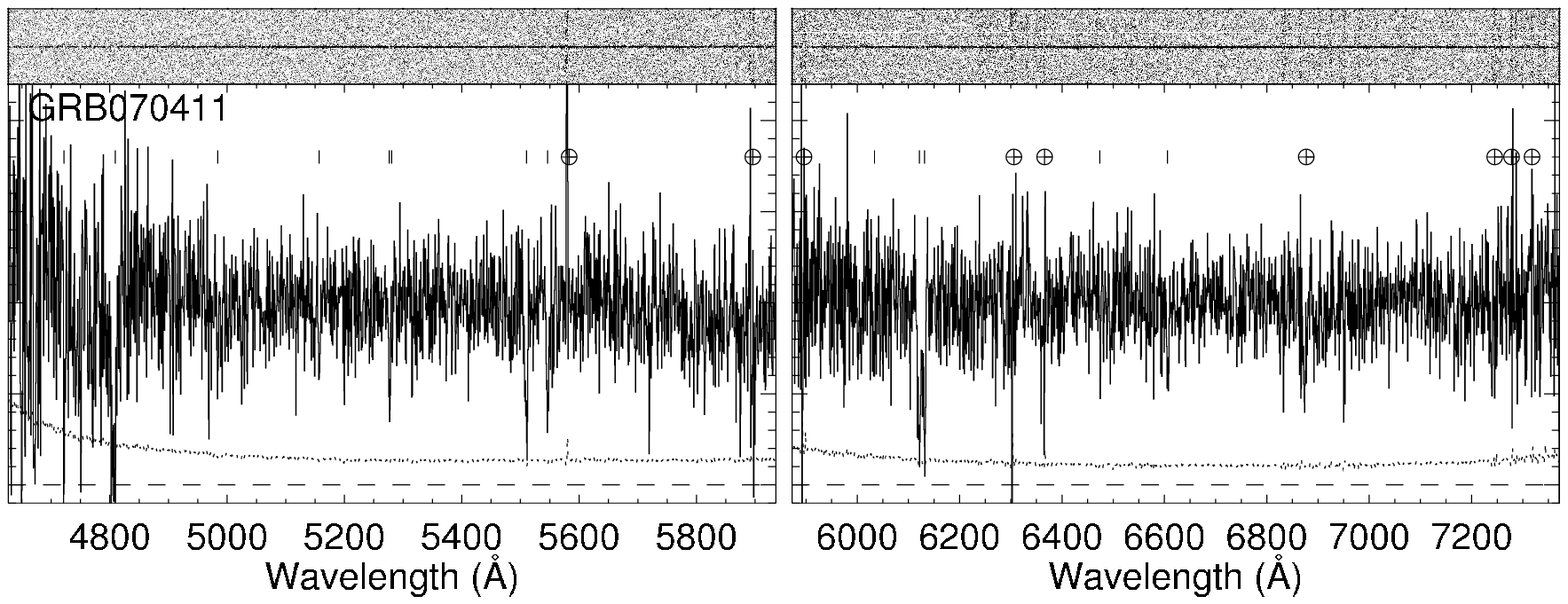}
\end{figure}

\clearpage

\begin{figure}
\epsscale{1.00}
\plotone{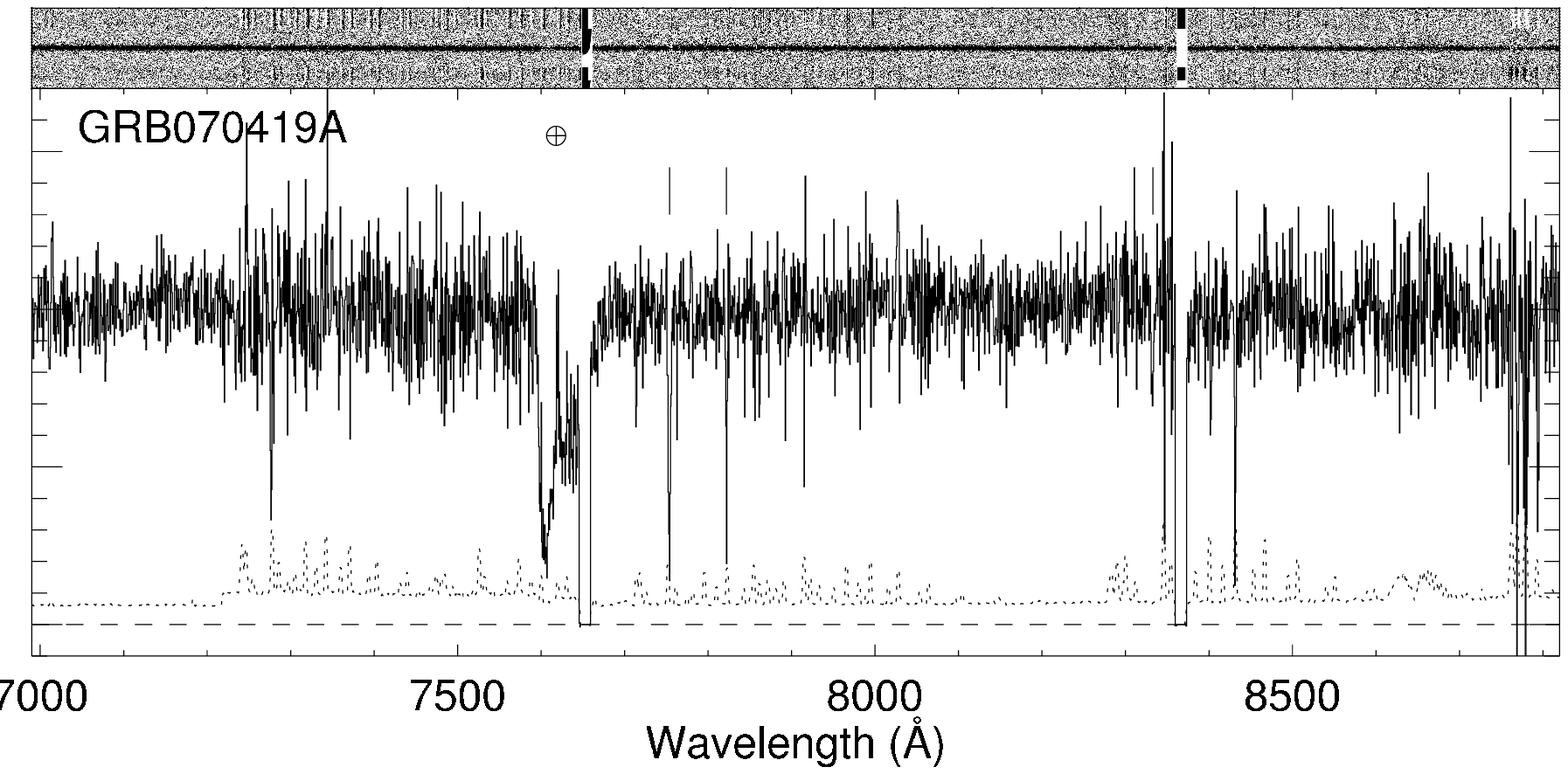}
\end{figure}

\begin{figure}
\epsscale{1.00}
\plotone{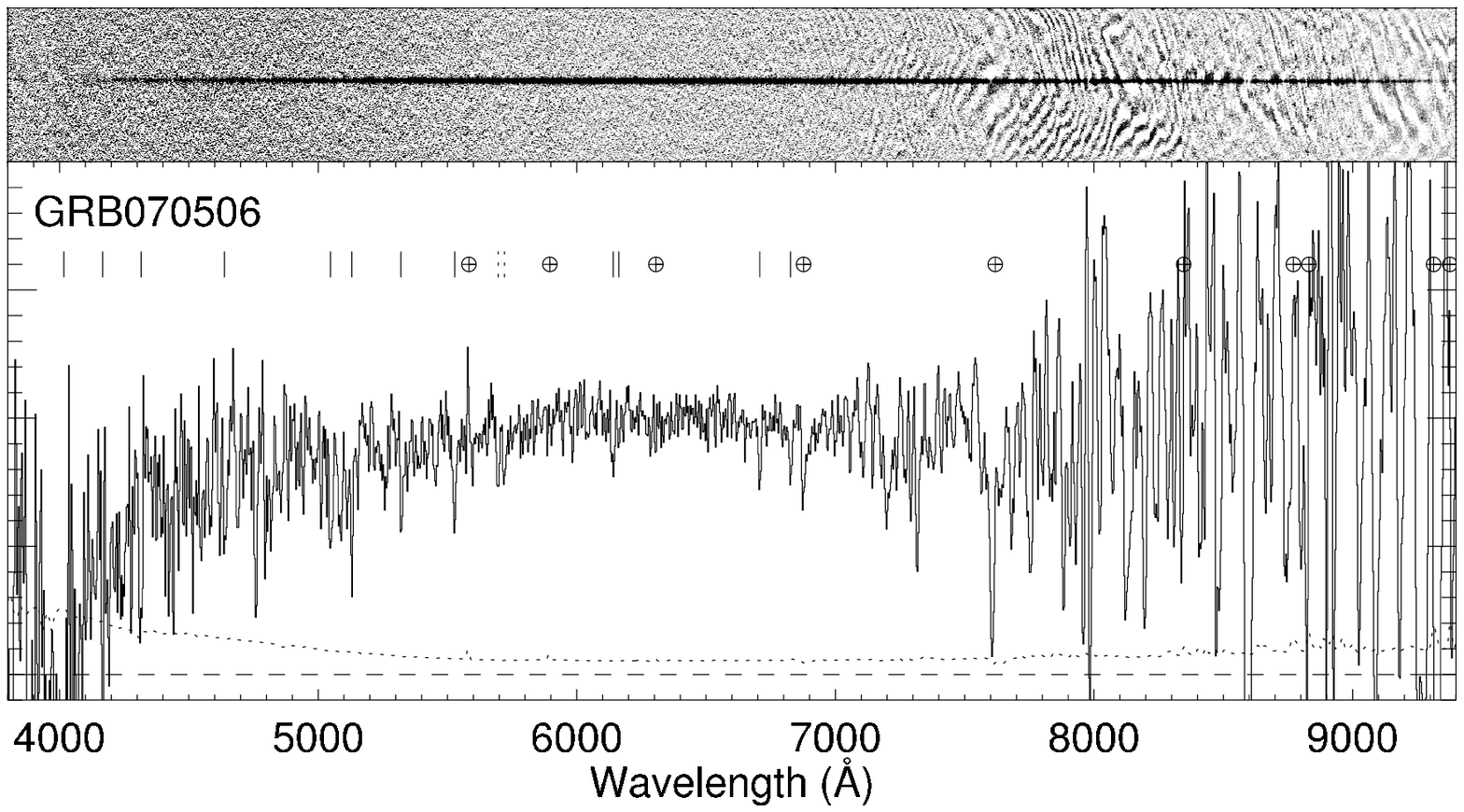}
\end{figure}

\begin{figure}
\epsscale{1.00}
\plotone{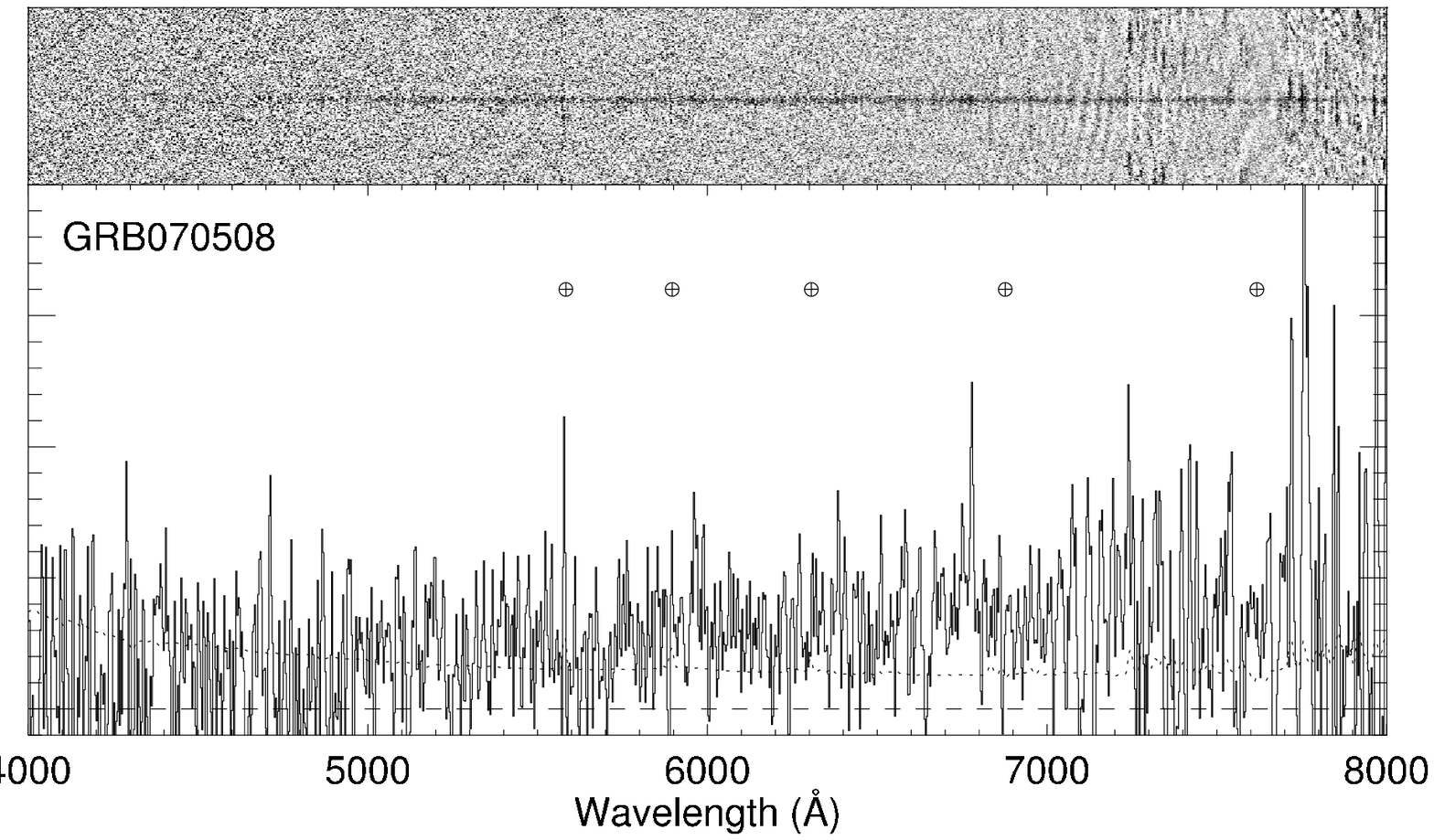}
\end{figure}

\begin{figure}
\epsscale{1.00}
\plotone{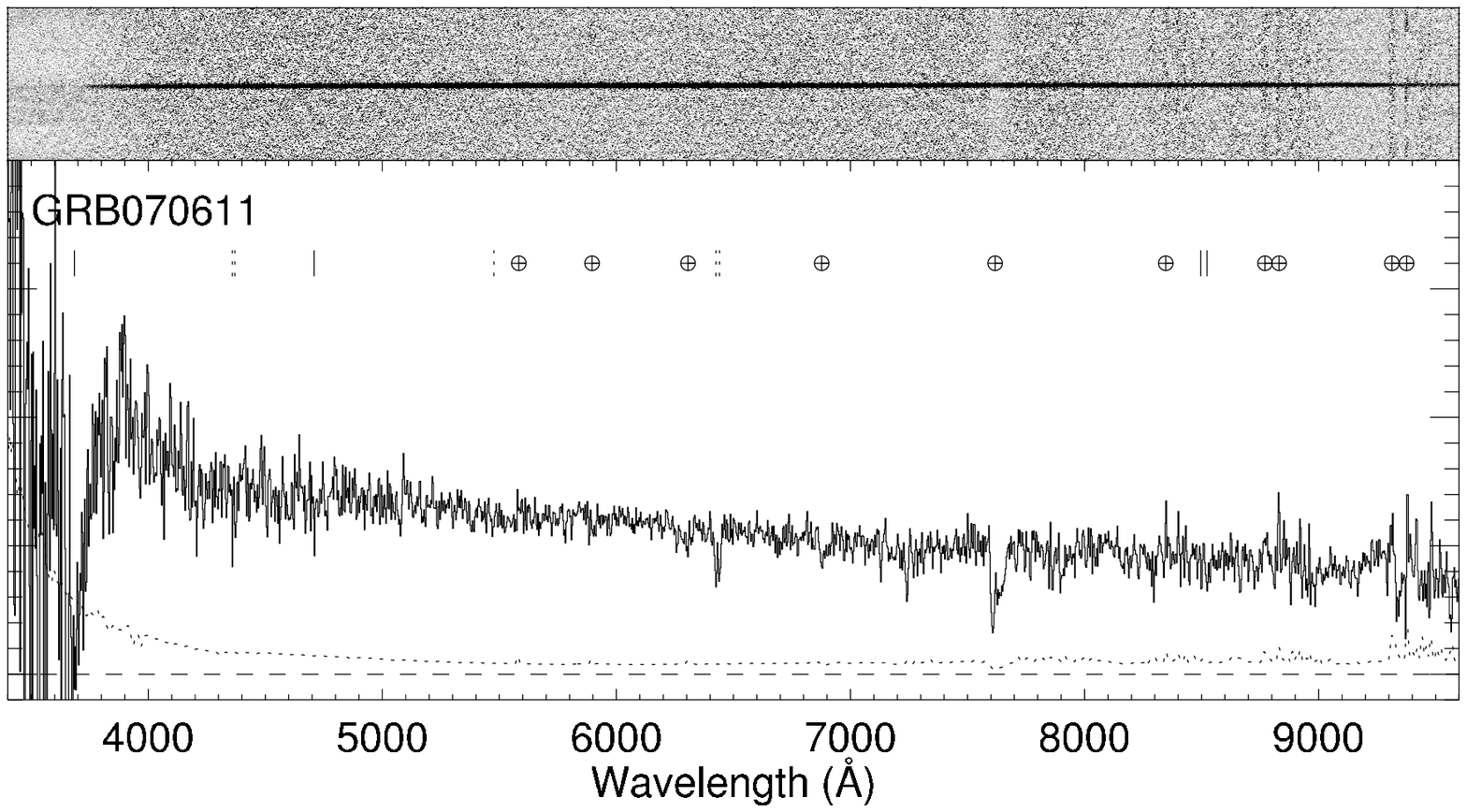}
\end{figure}

\begin{figure}
\epsscale{1.00}
\plotone{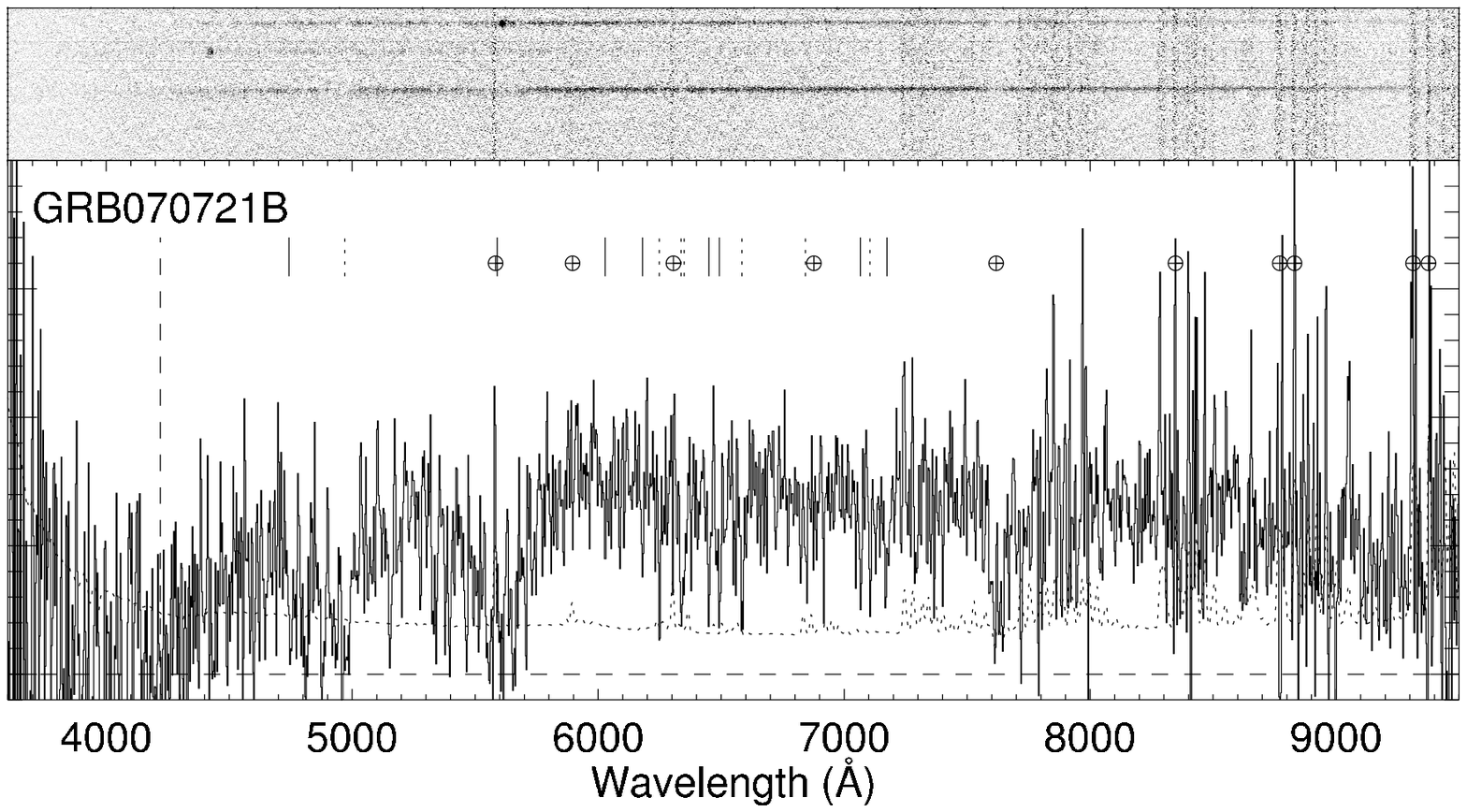}
\end{figure}

\begin{figure}
\epsscale{1.00}
\plotone{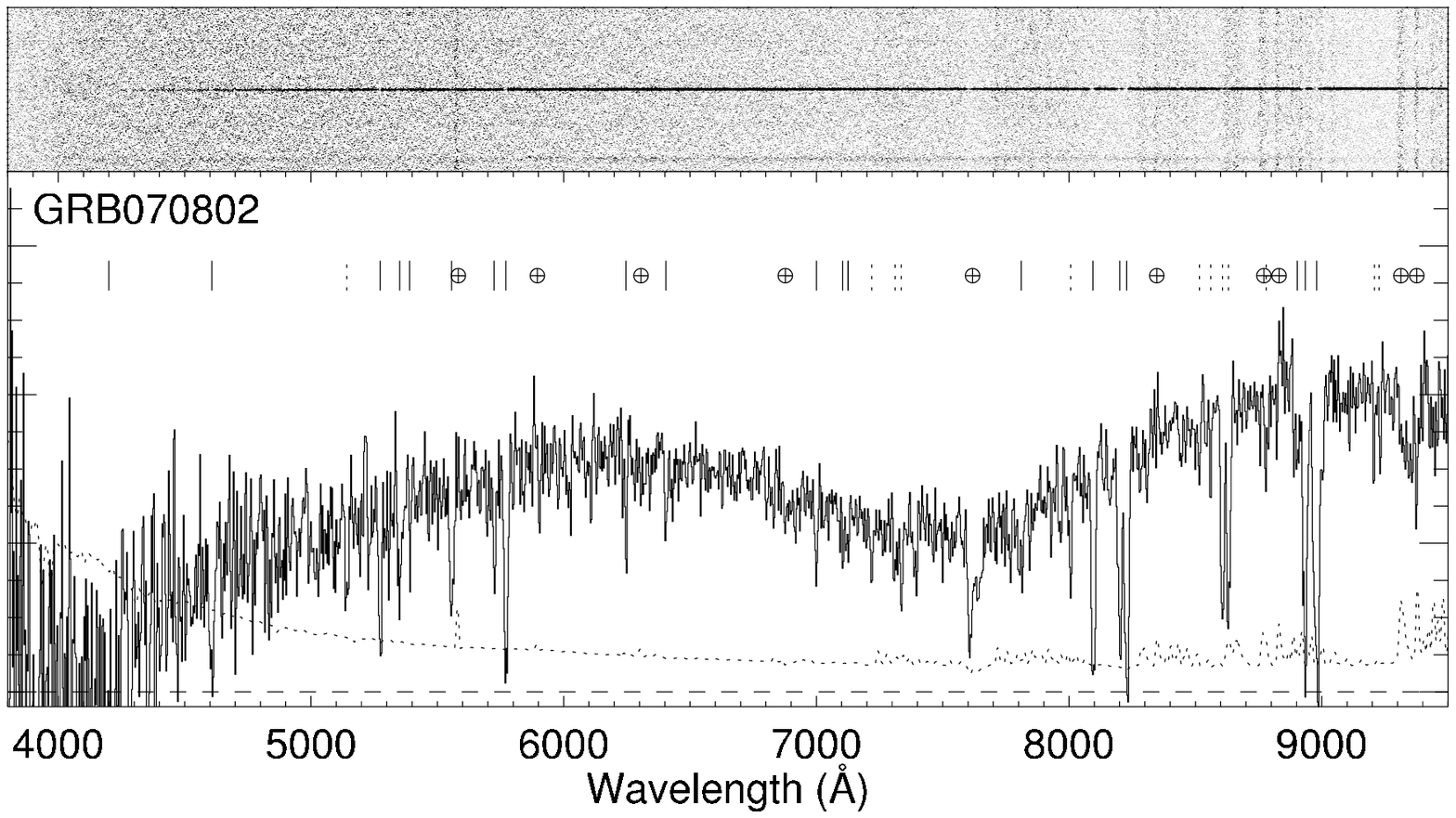}
\end{figure}

\begin{figure}
\epsscale{1.00}
\plotone{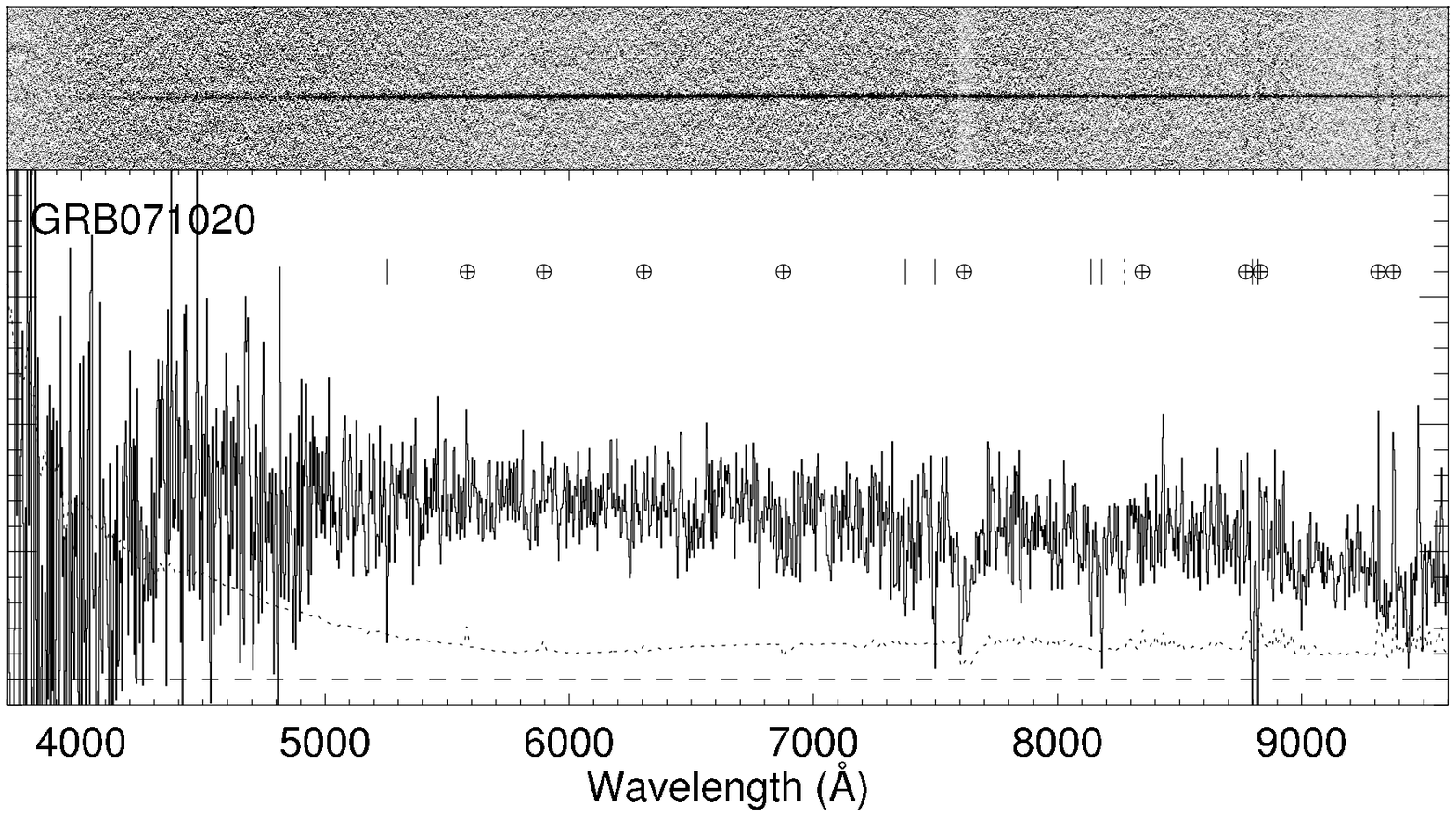}
\end{figure}

\begin{figure}
\epsscale{1.00}
\plotone{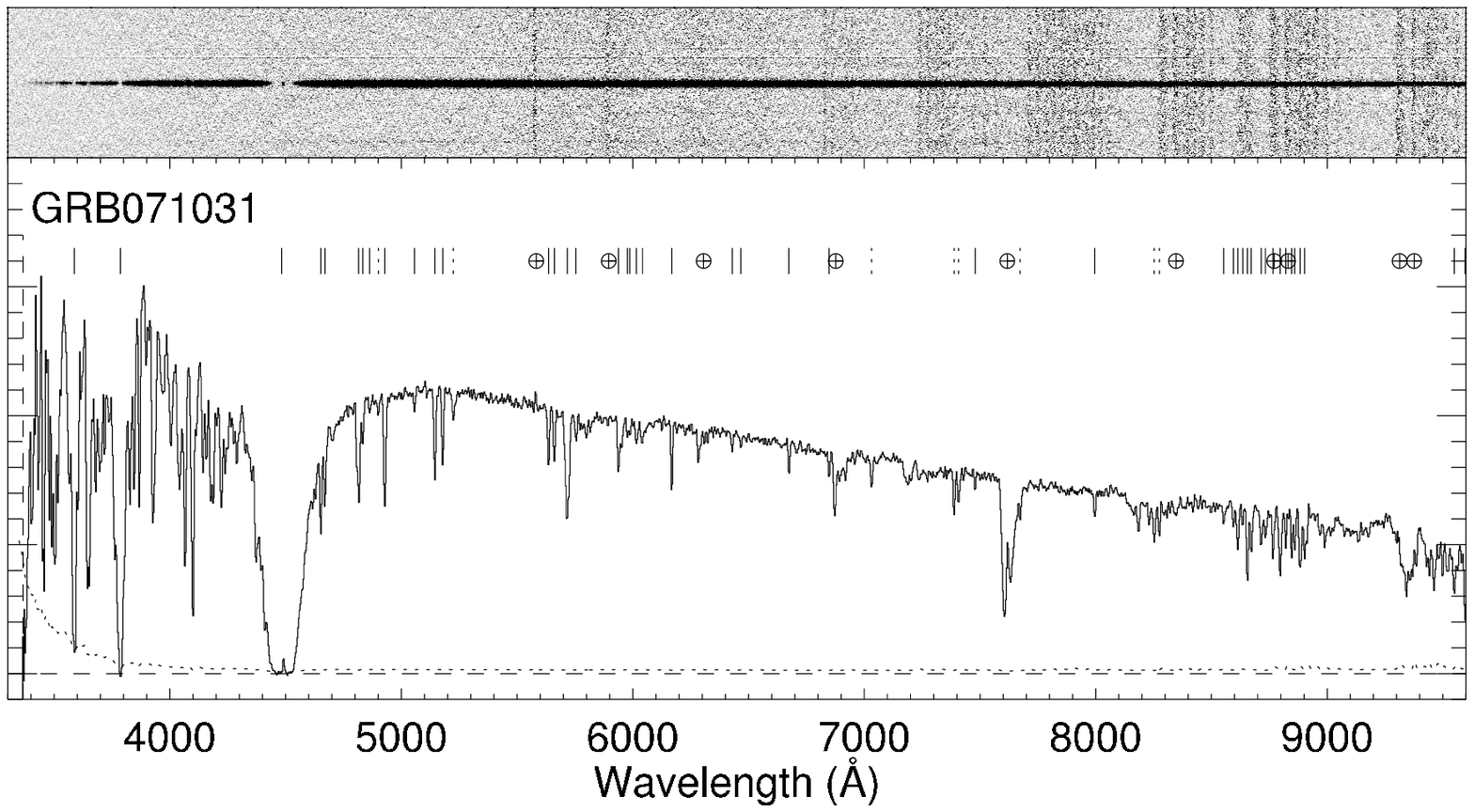}
\end{figure}

\clearpage

\begin{figure}
\epsscale{1.00}
\plotone{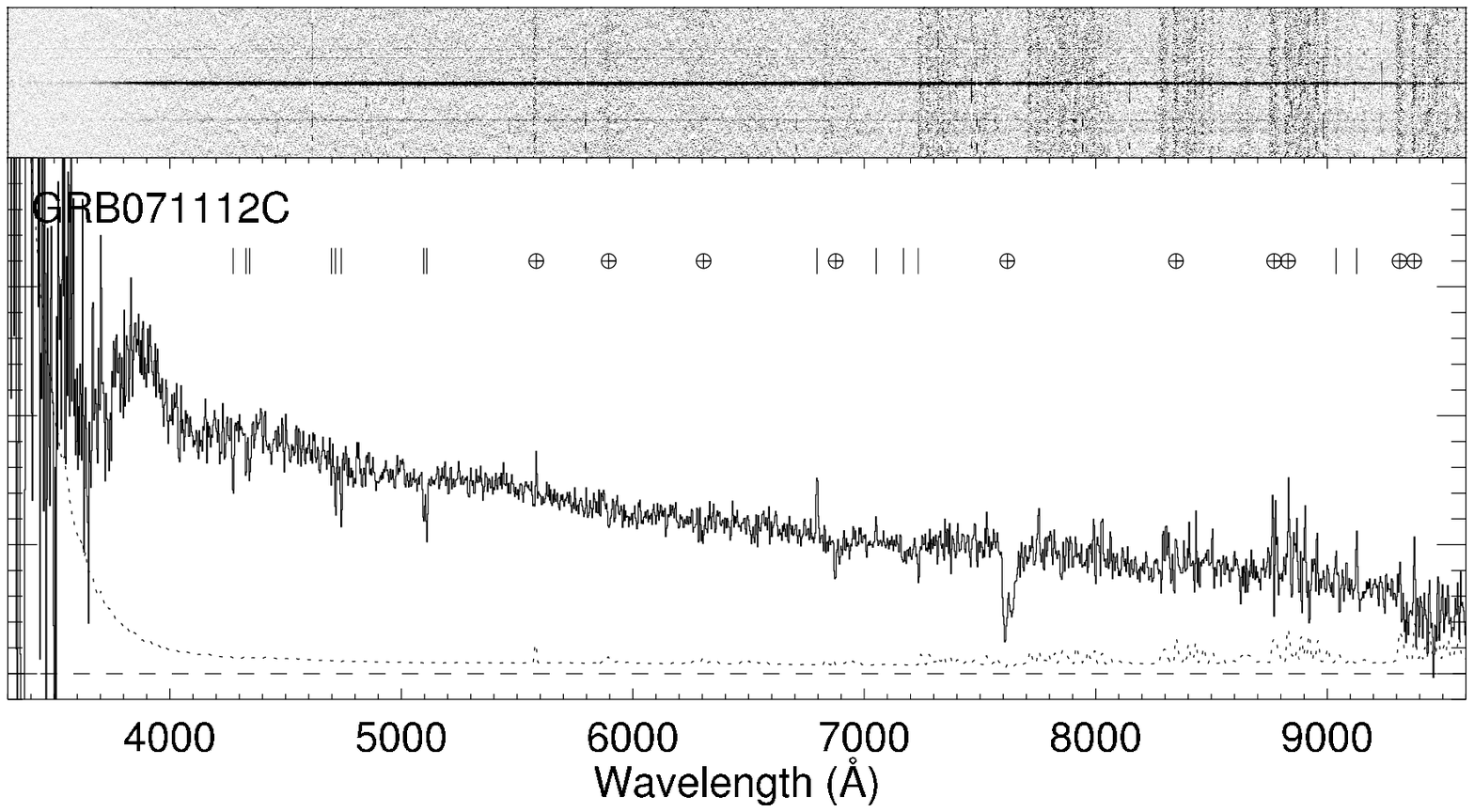}
\end{figure}

\begin{figure}
\epsscale{1.00}
\plotone{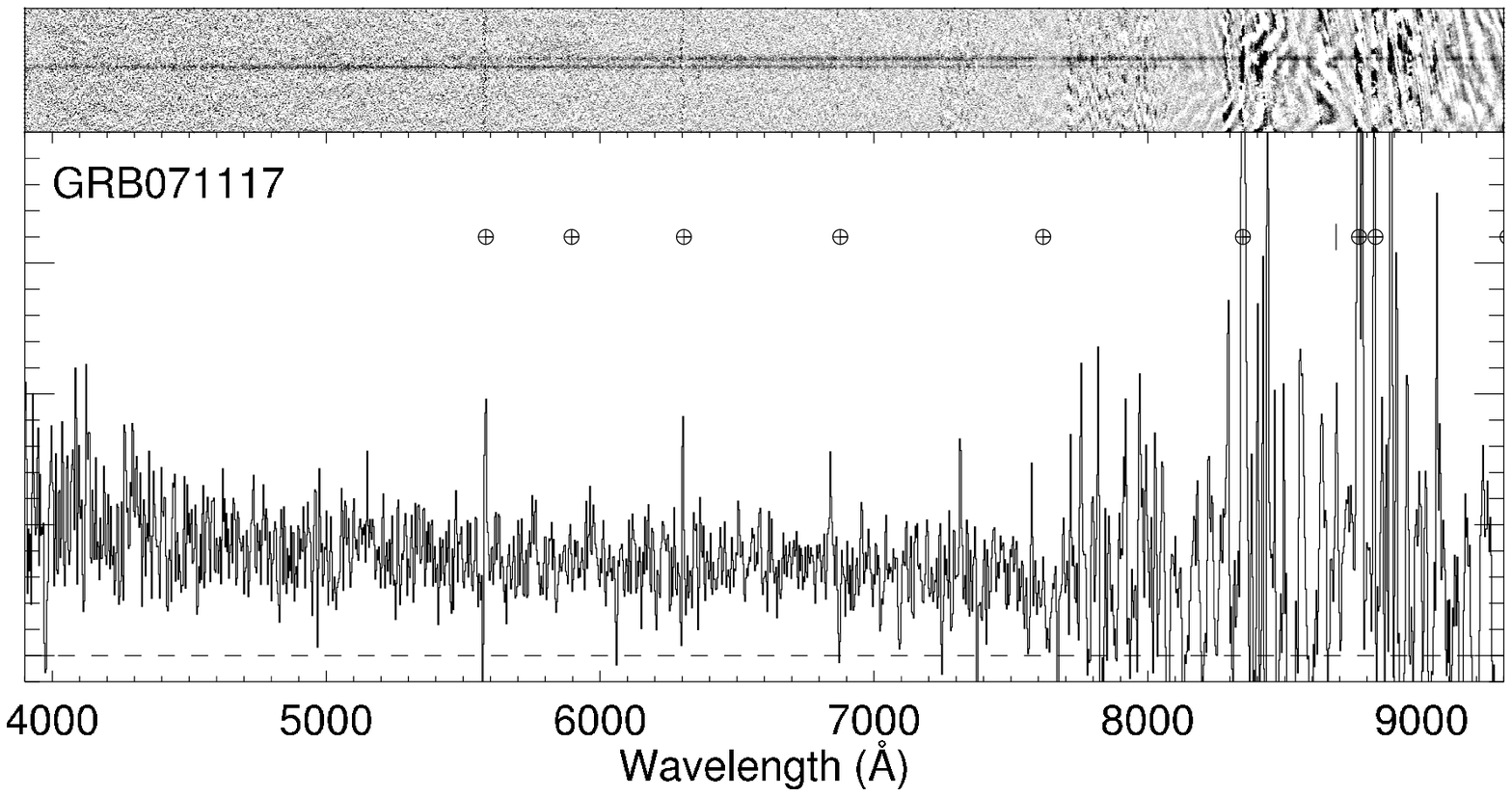}
\end{figure}

\begin{figure}
\epsscale{1.00}
\plotone{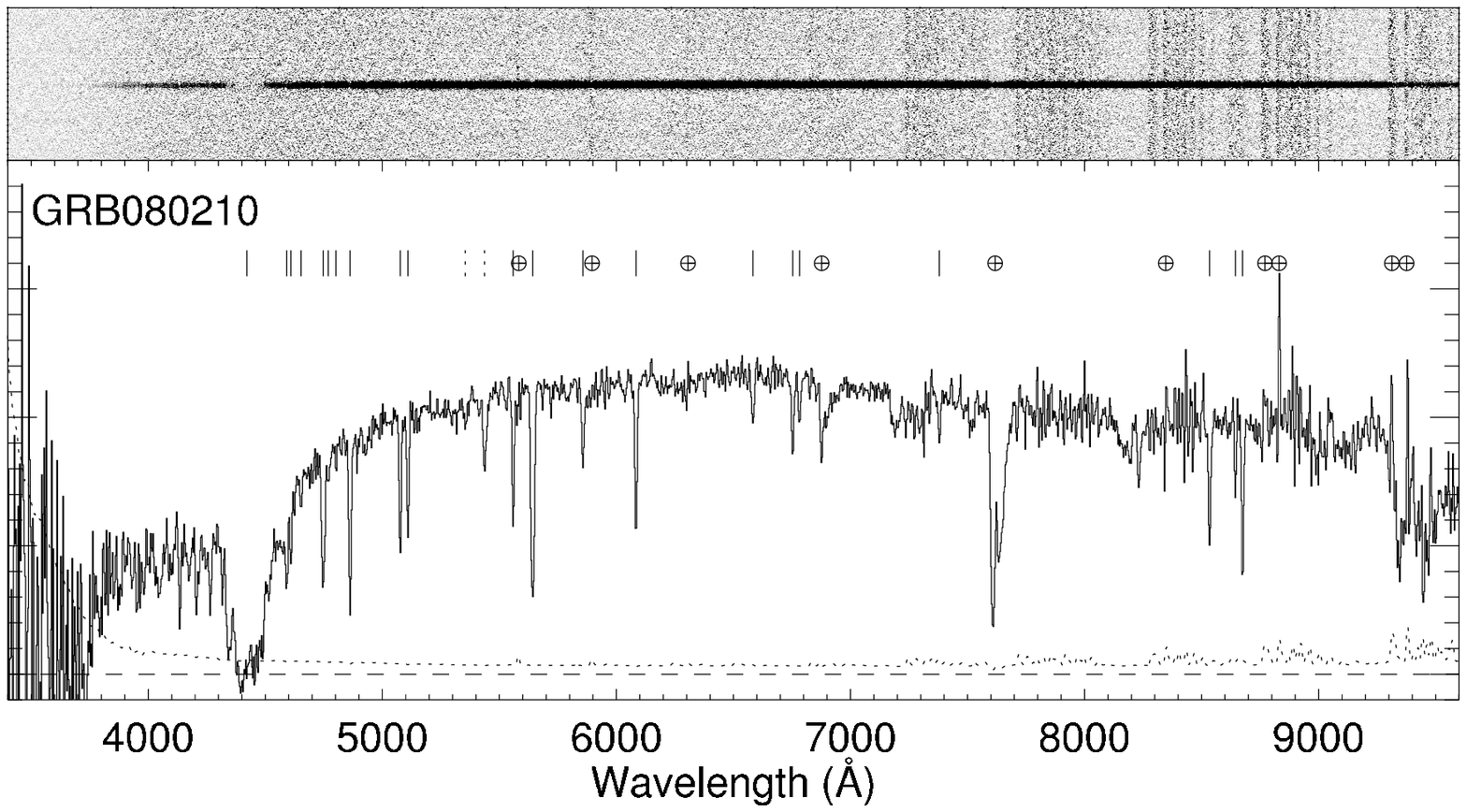}
\end{figure}


\begin{figure}
\epsscale{1.00}
\plotone{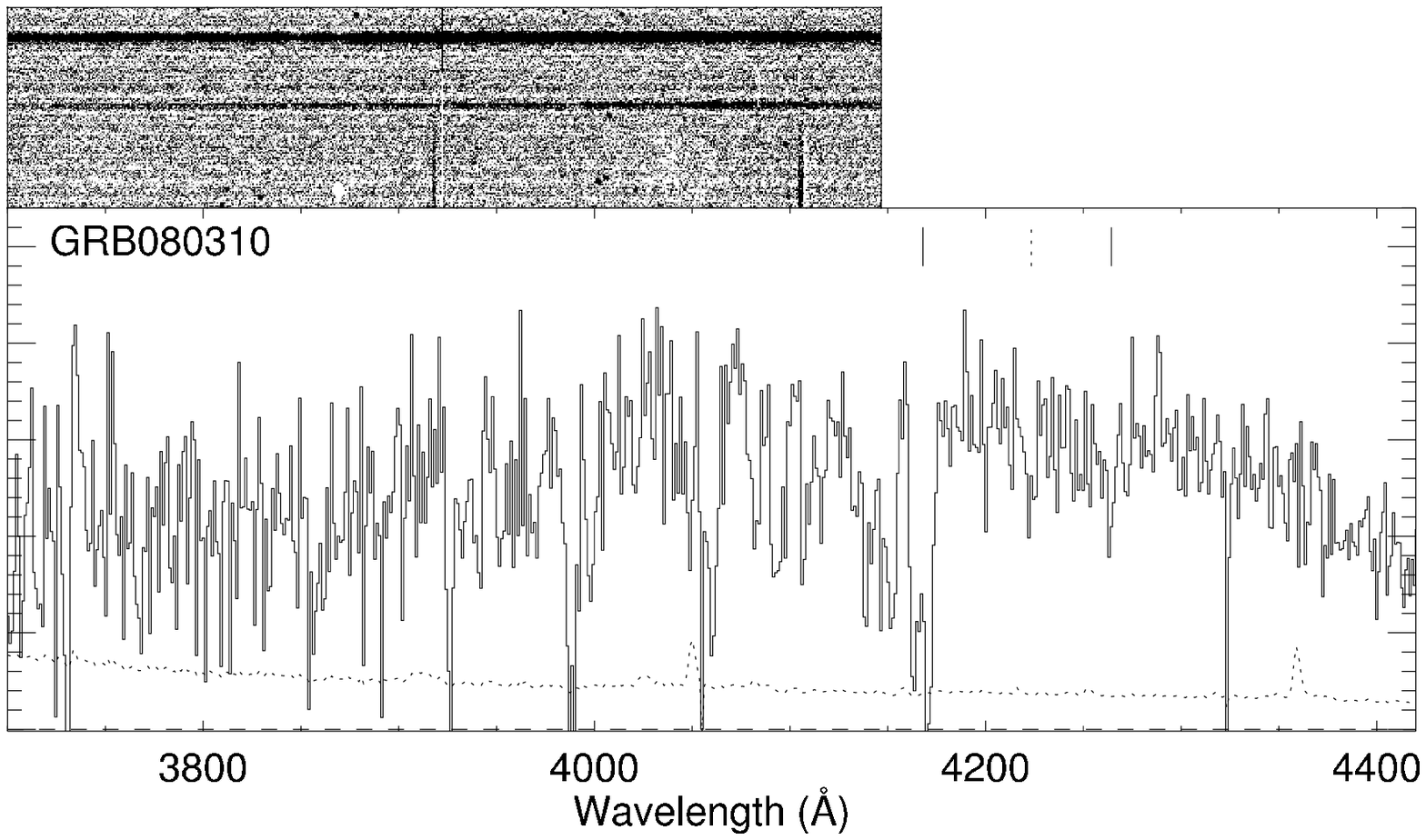}
\end{figure}

\begin{figure}
\epsscale{1.00}
\plotone{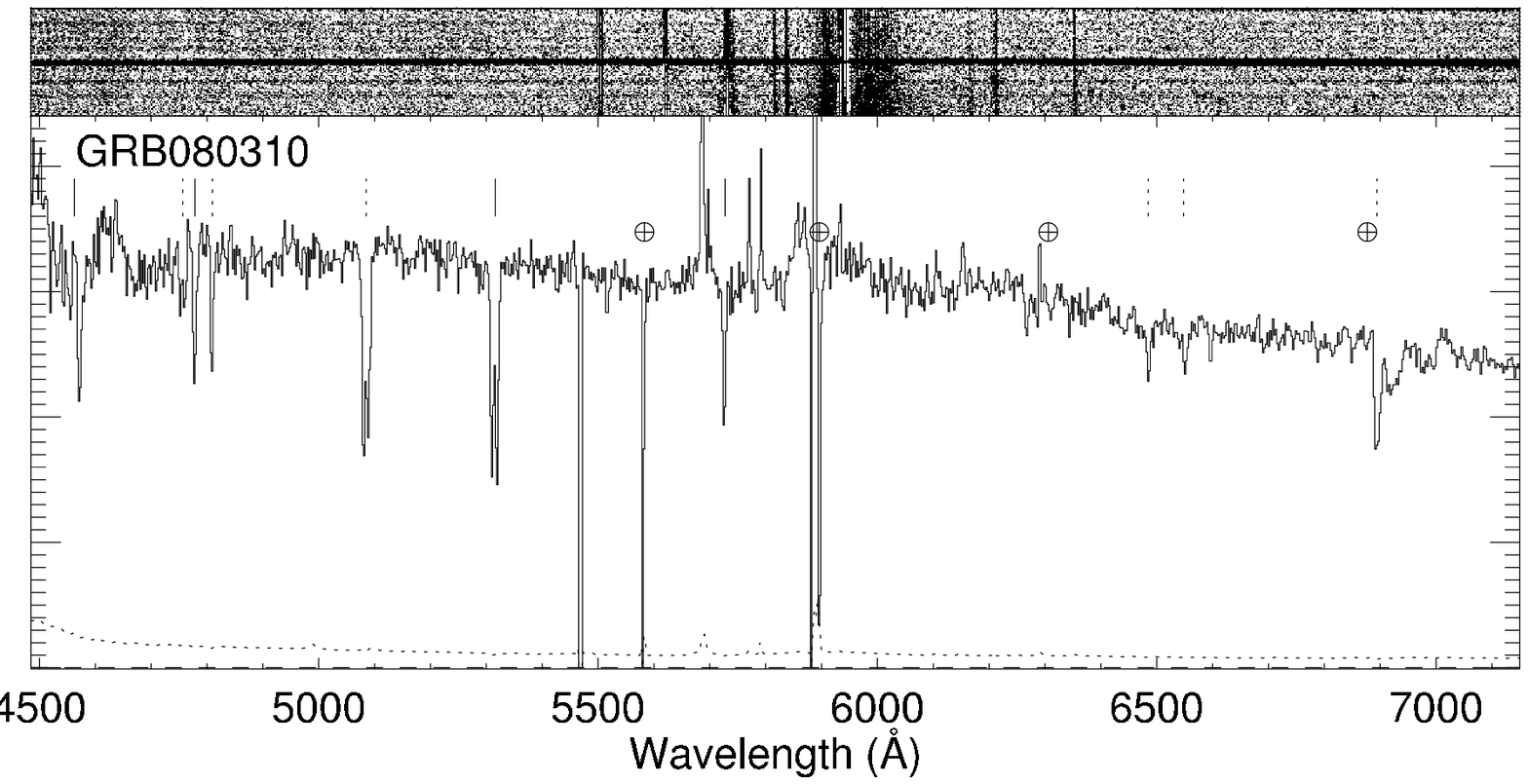}
\end{figure}

\begin{figure}
\epsscale{1.00}
\plotone{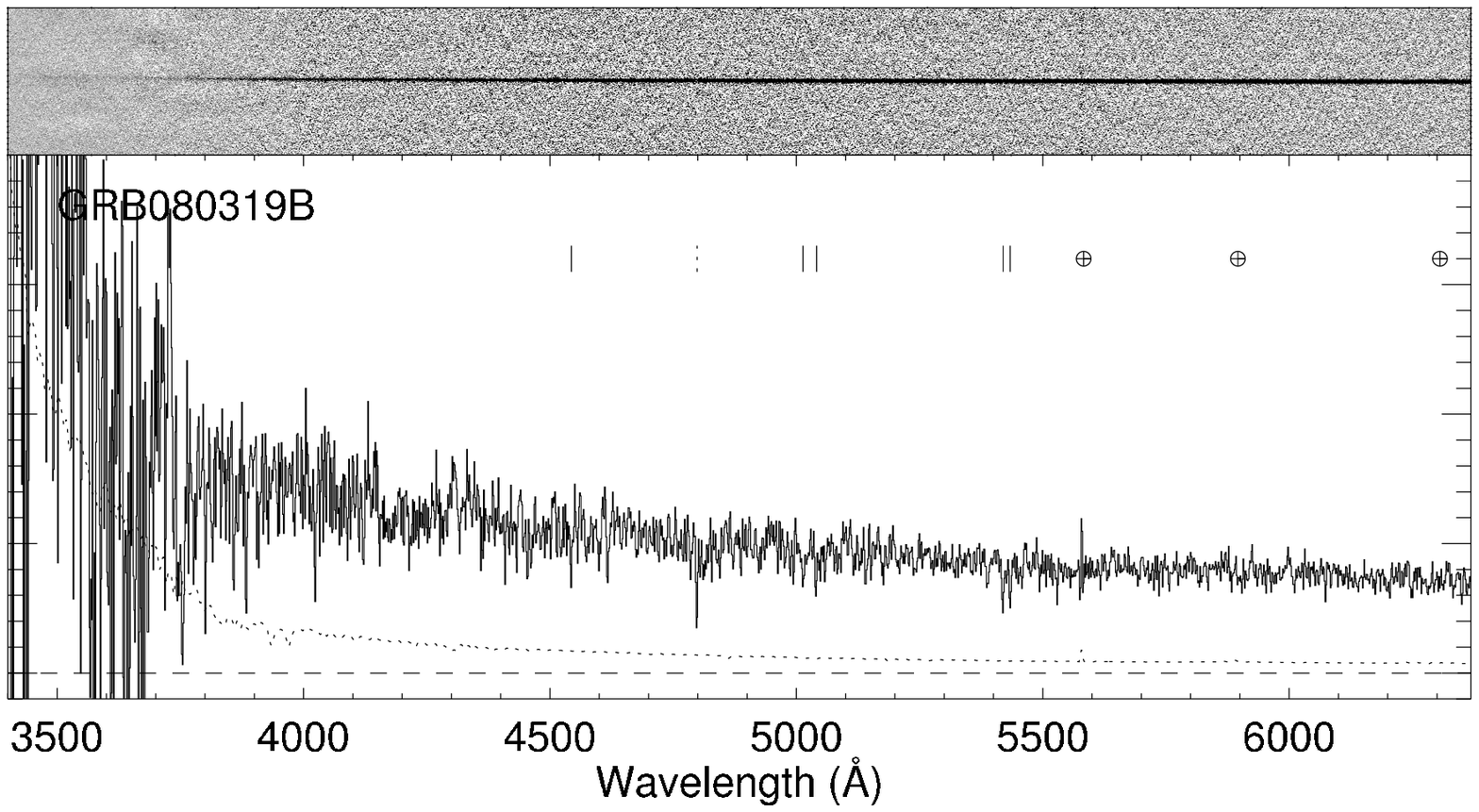}
\end{figure}

\begin{figure}
\epsscale{1.00}
\plotone{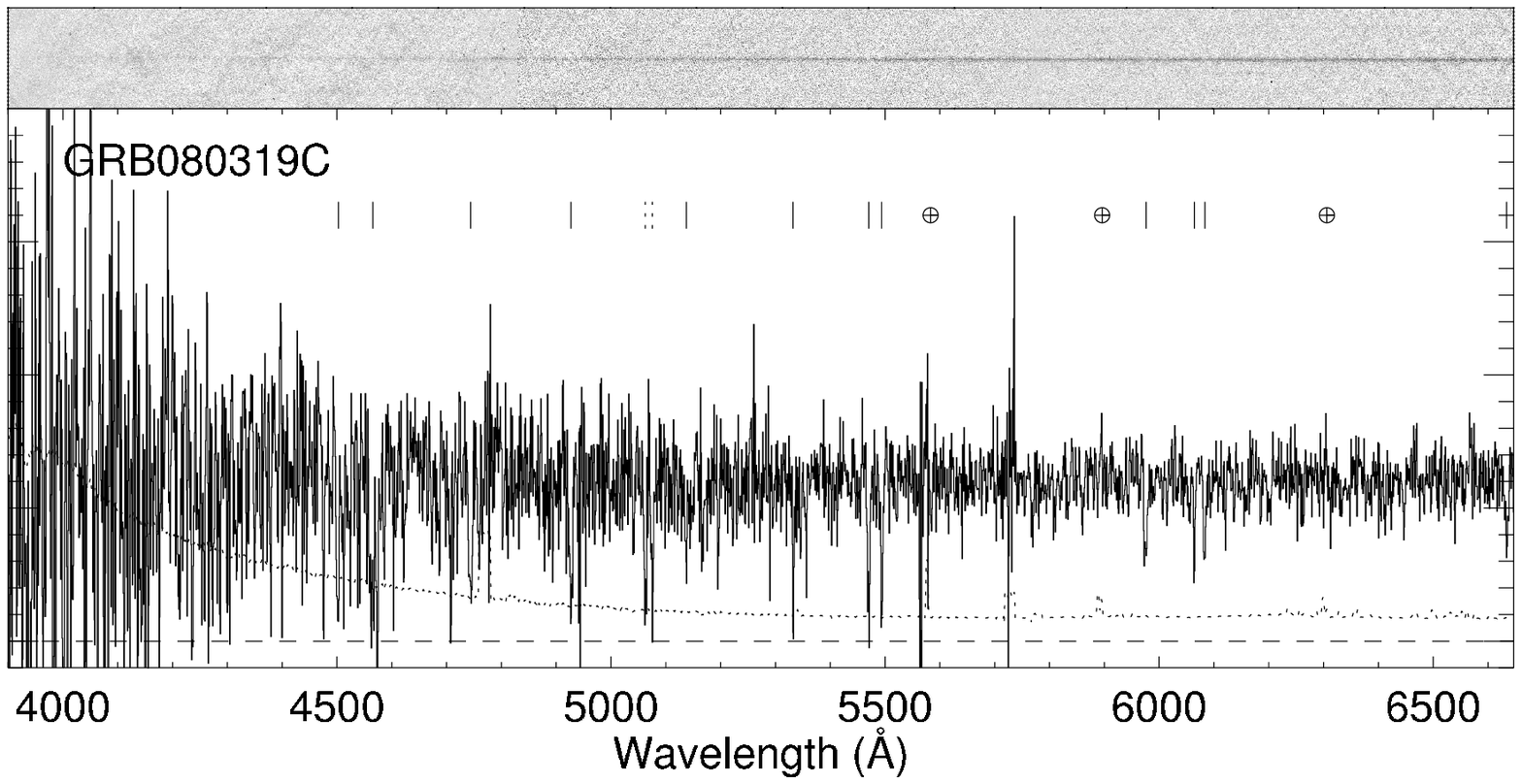}
\end{figure}

\begin{figure}
\epsscale{1.00}
\plotone{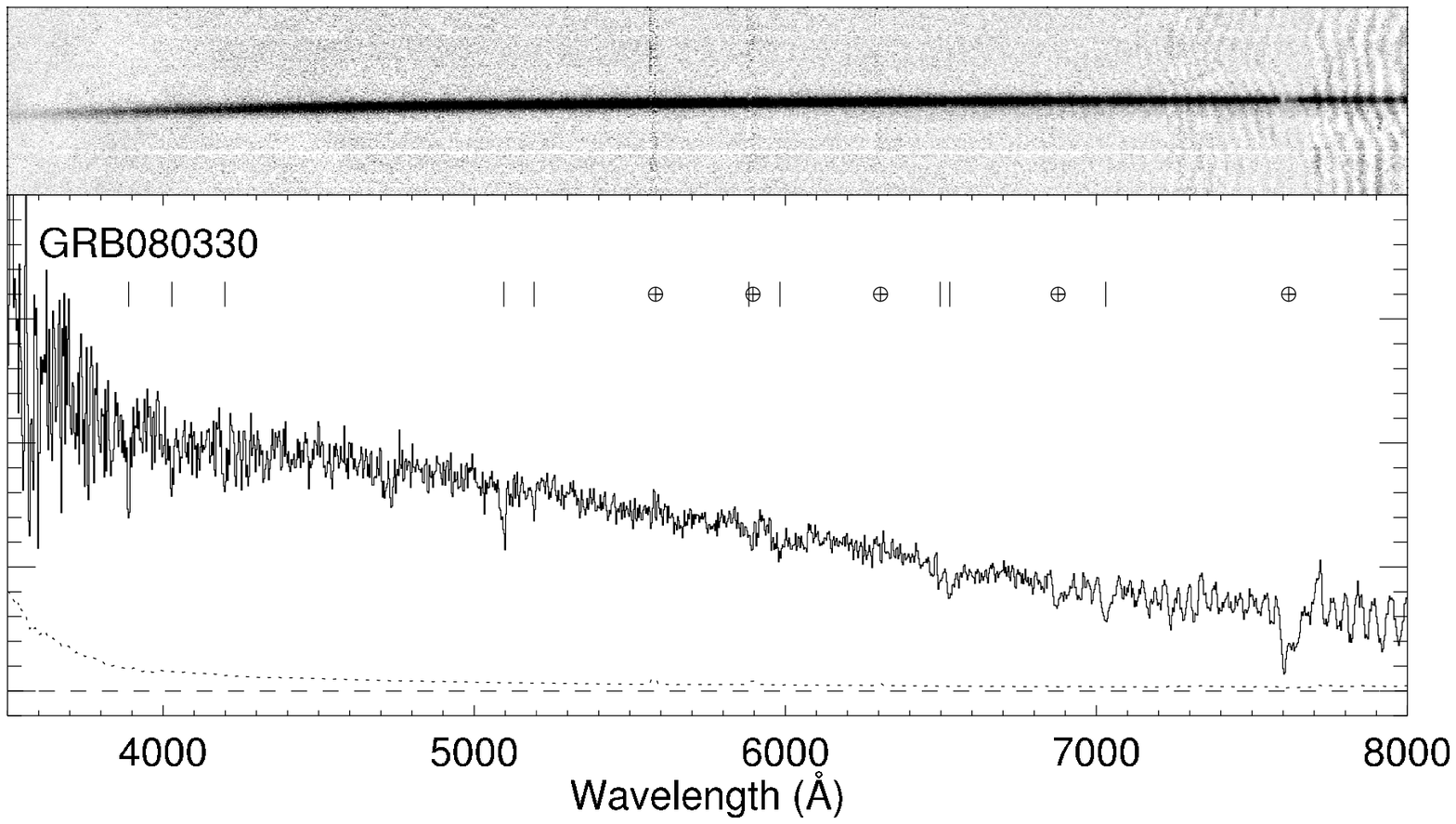}
\end{figure}

\begin{figure}
\epsscale{1.00}
\plotone{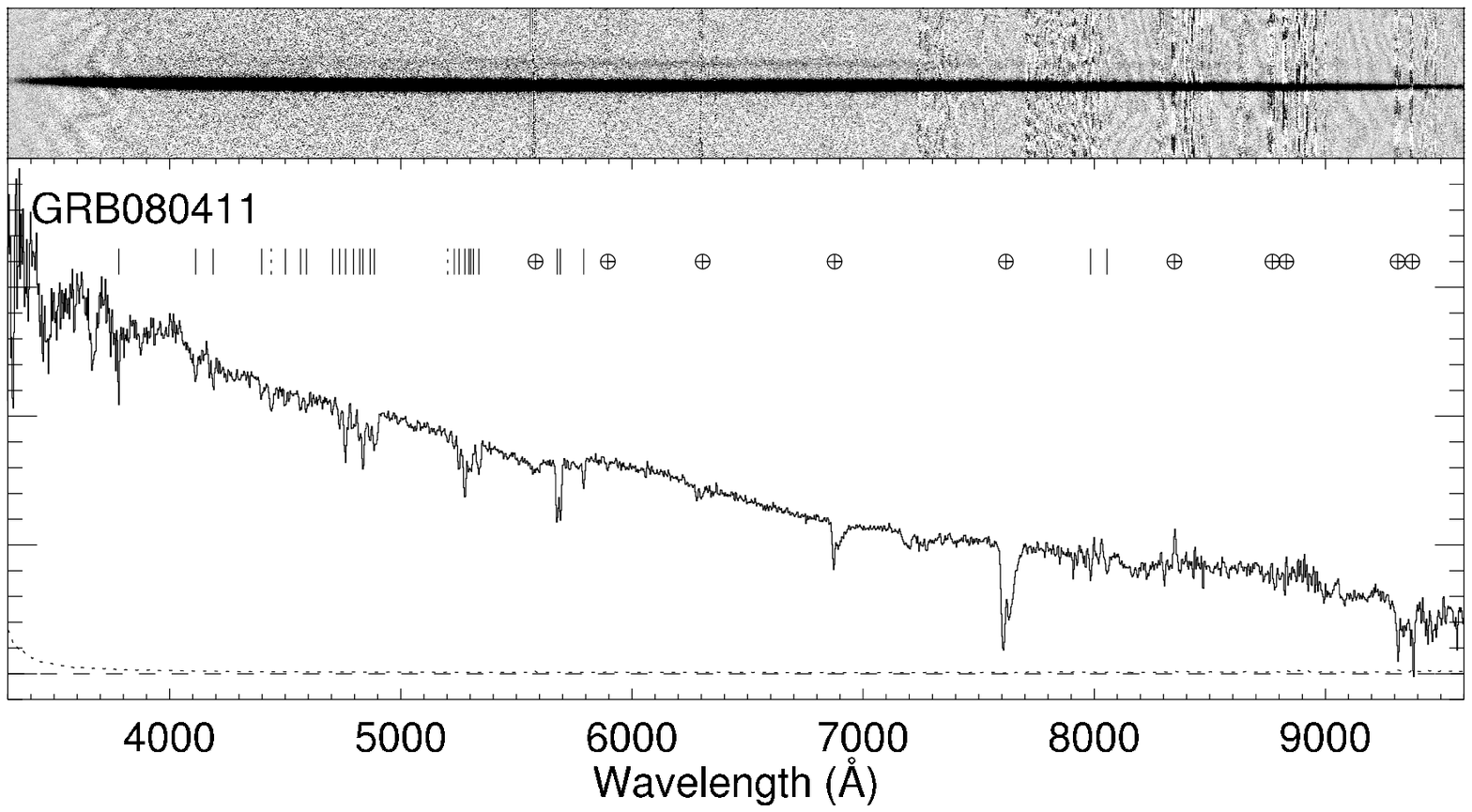}
\end{figure}

\begin{figure}
\epsscale{1.00}
\plotone{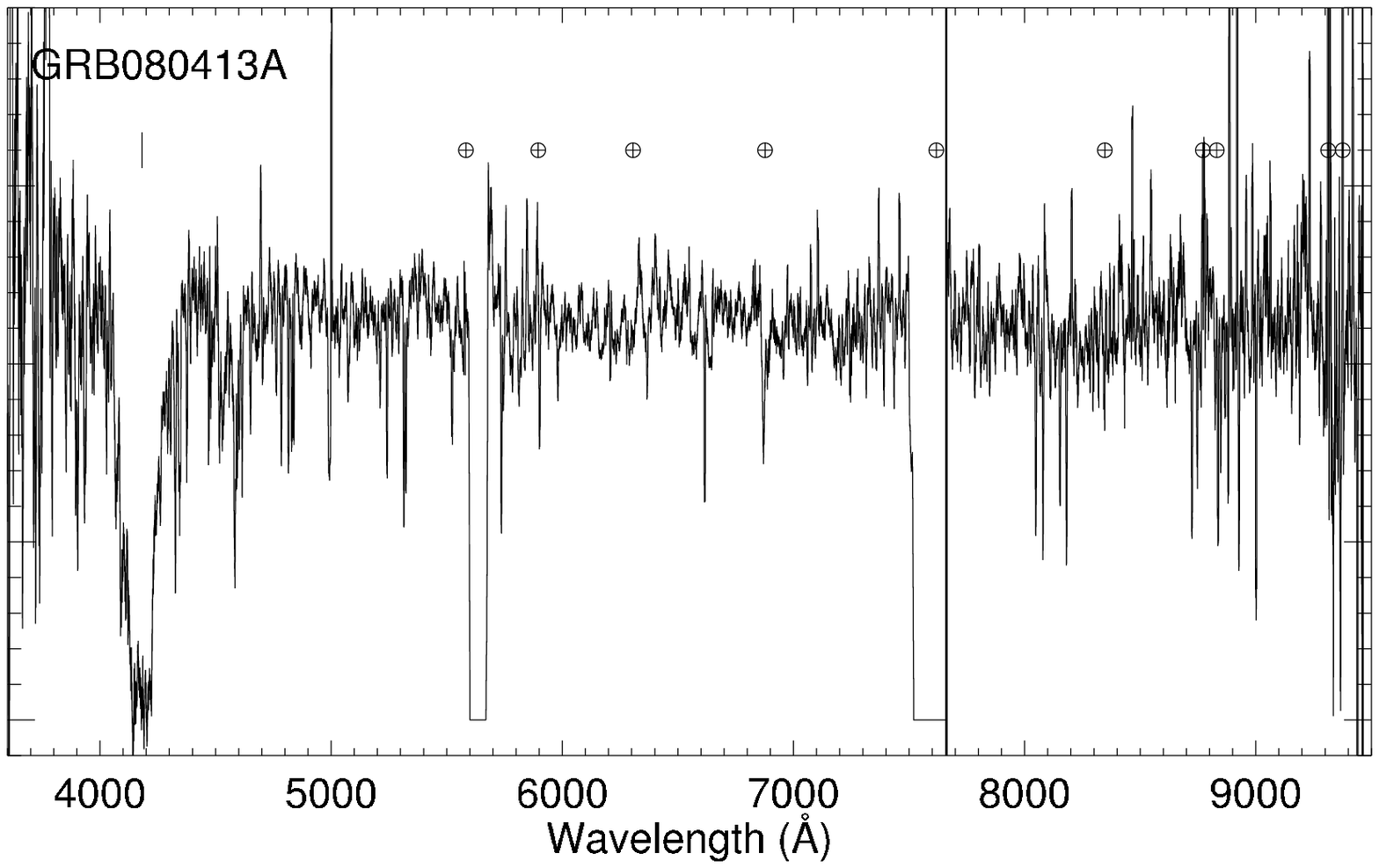}
\end{figure}

\begin{figure}
\epsscale{1.00}
\plotone{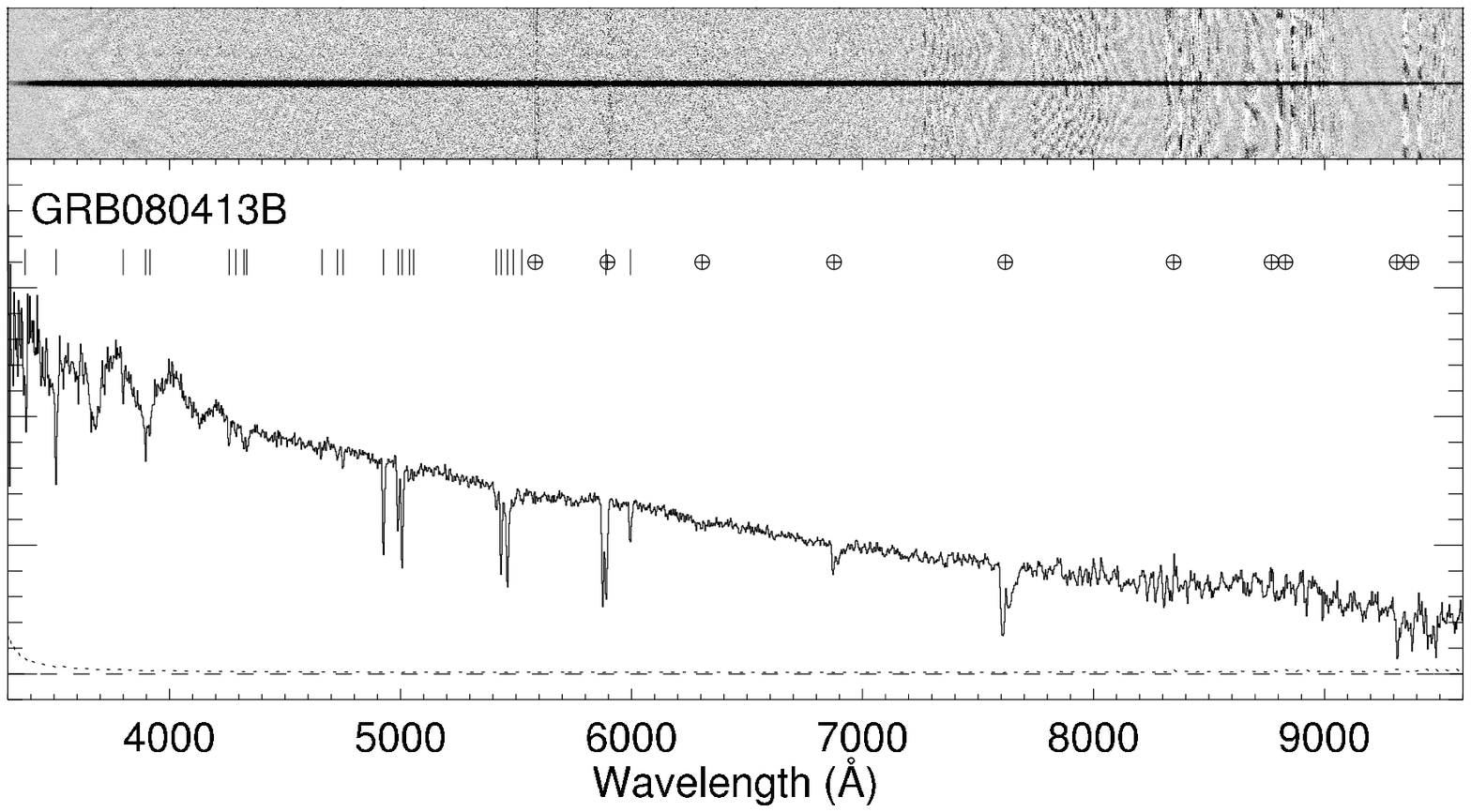}
\end{figure}

\begin{figure}
\epsscale{1.00}
\plotone{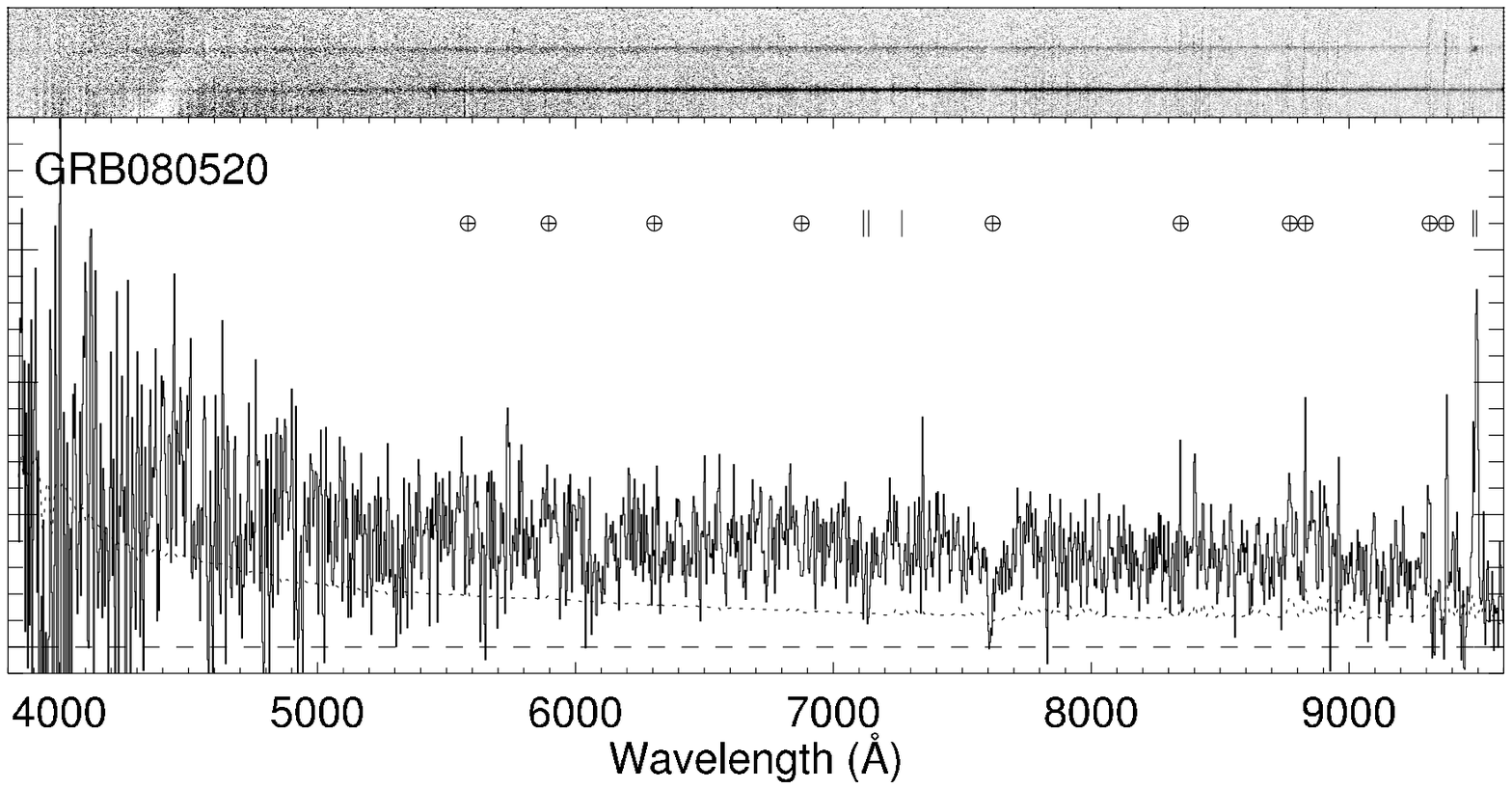}
\end{figure}

\begin{figure}
\epsscale{1.00}
\plotone{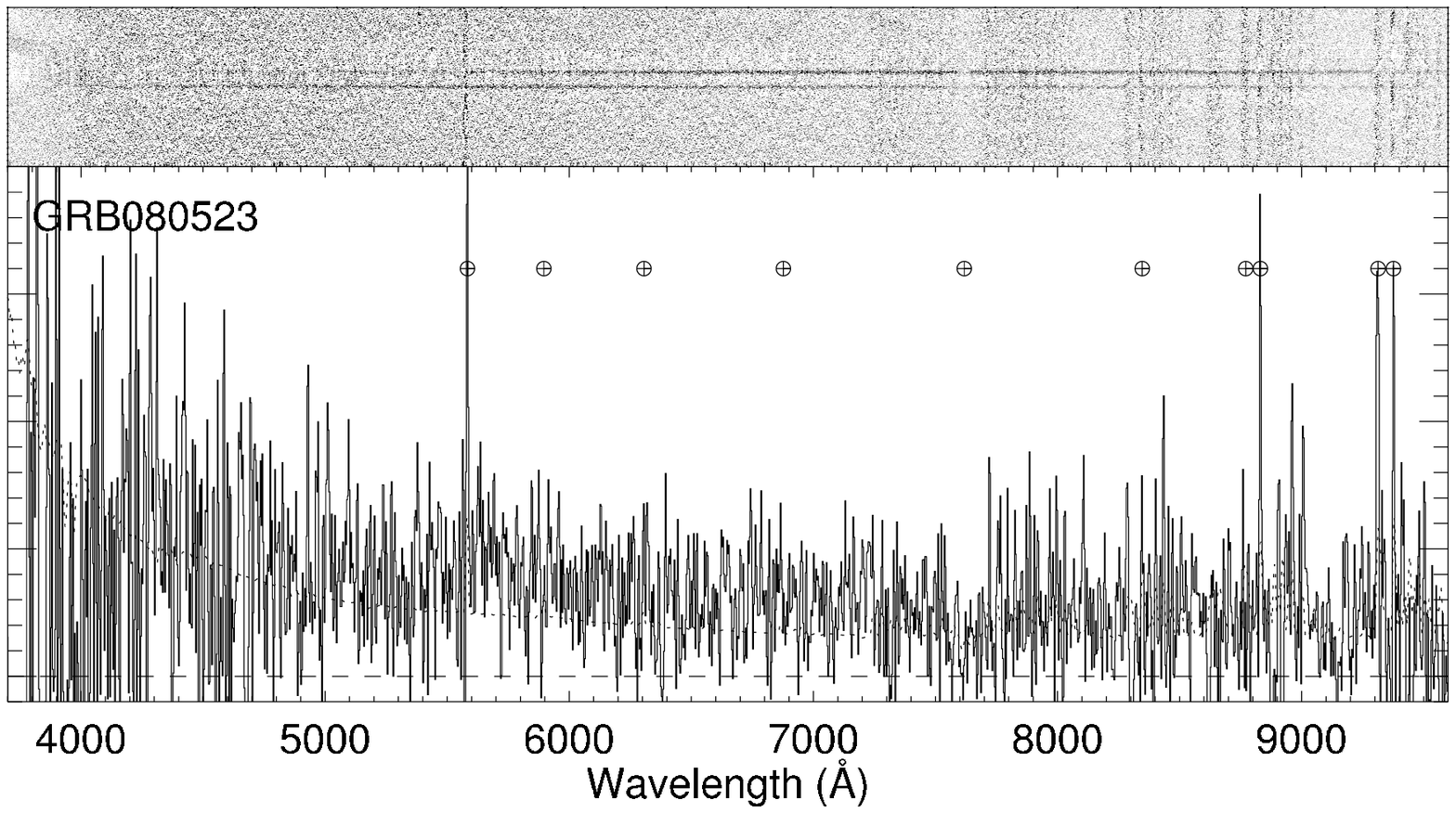}
\end{figure}

\clearpage

\begin{figure}
\epsscale{1.00}
\plotone{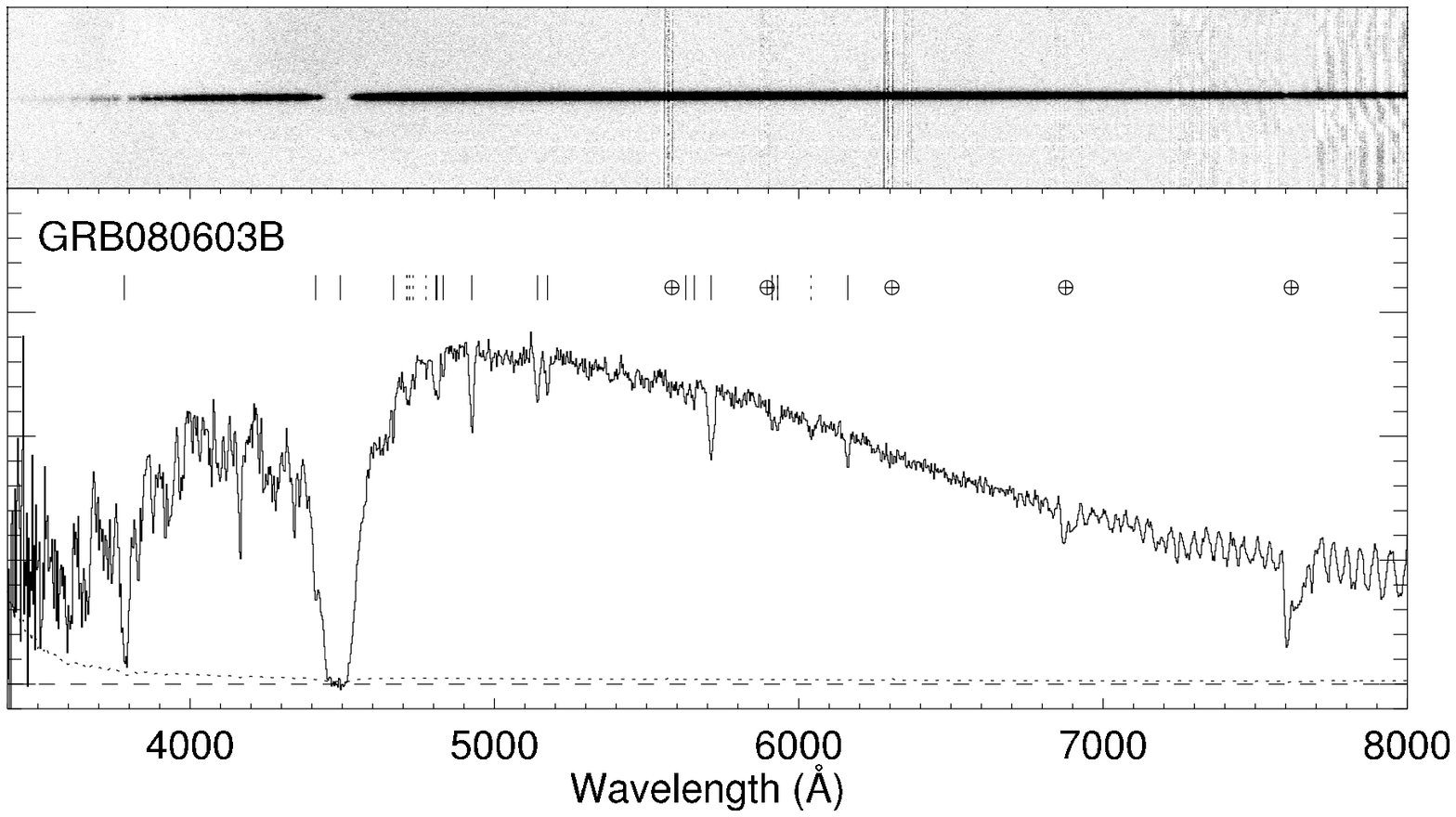}
\end{figure}

\begin{figure}
\epsscale{1.00}
\plotone{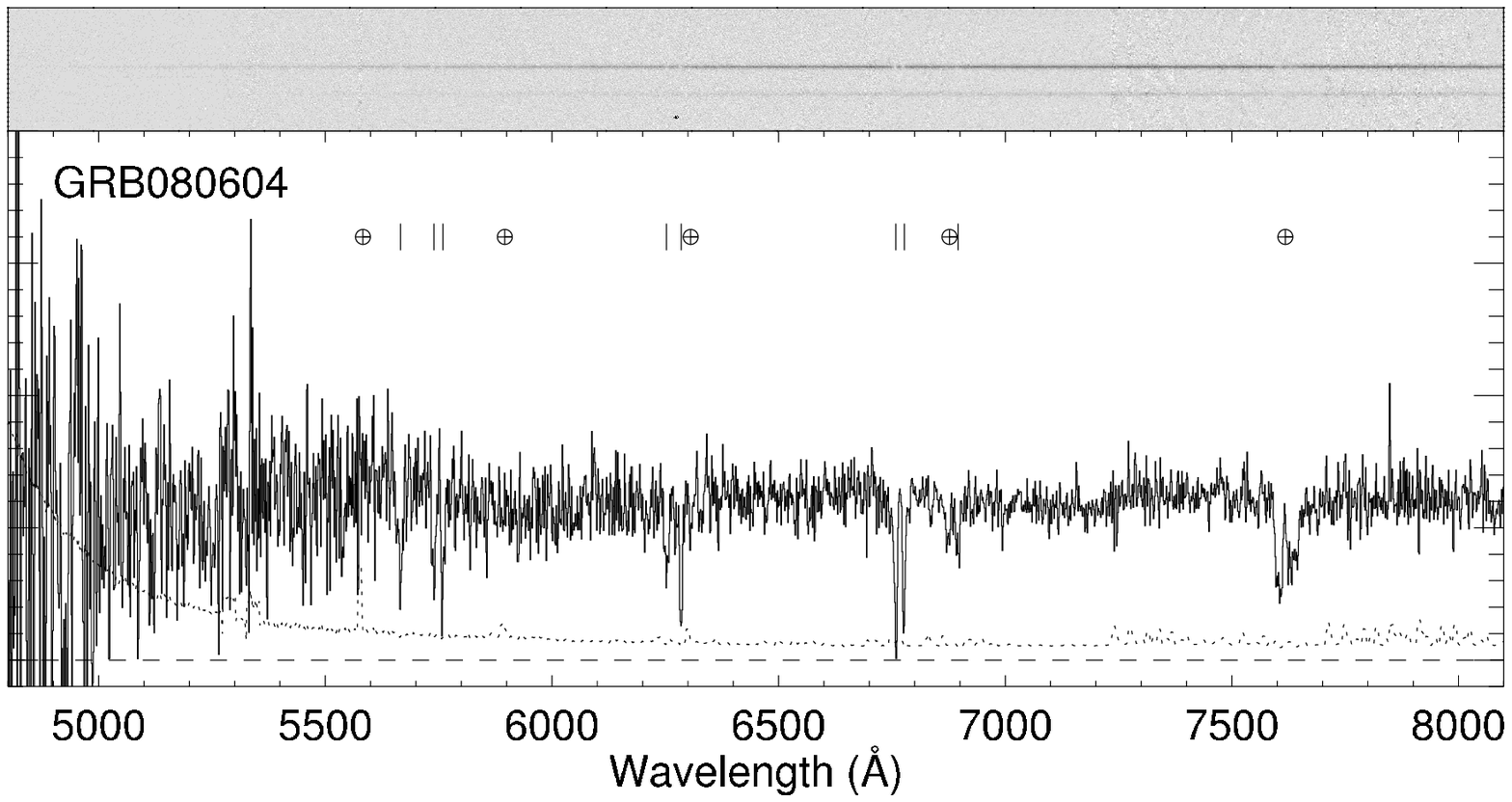}
\end{figure}

\begin{figure}
\epsscale{1.00}
\plotone{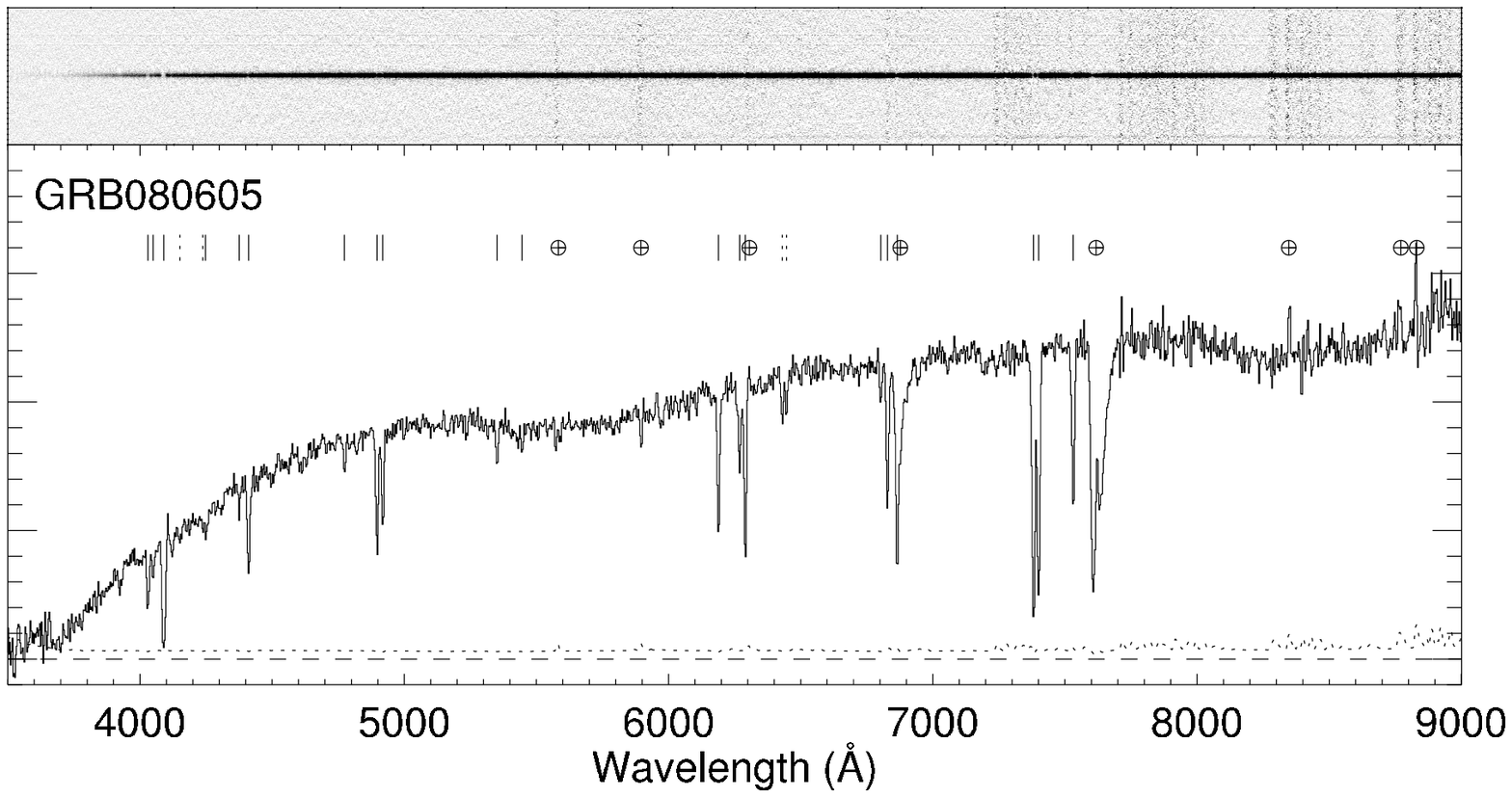}
\end{figure}

\begin{figure}
\epsscale{1.00}
\plotone{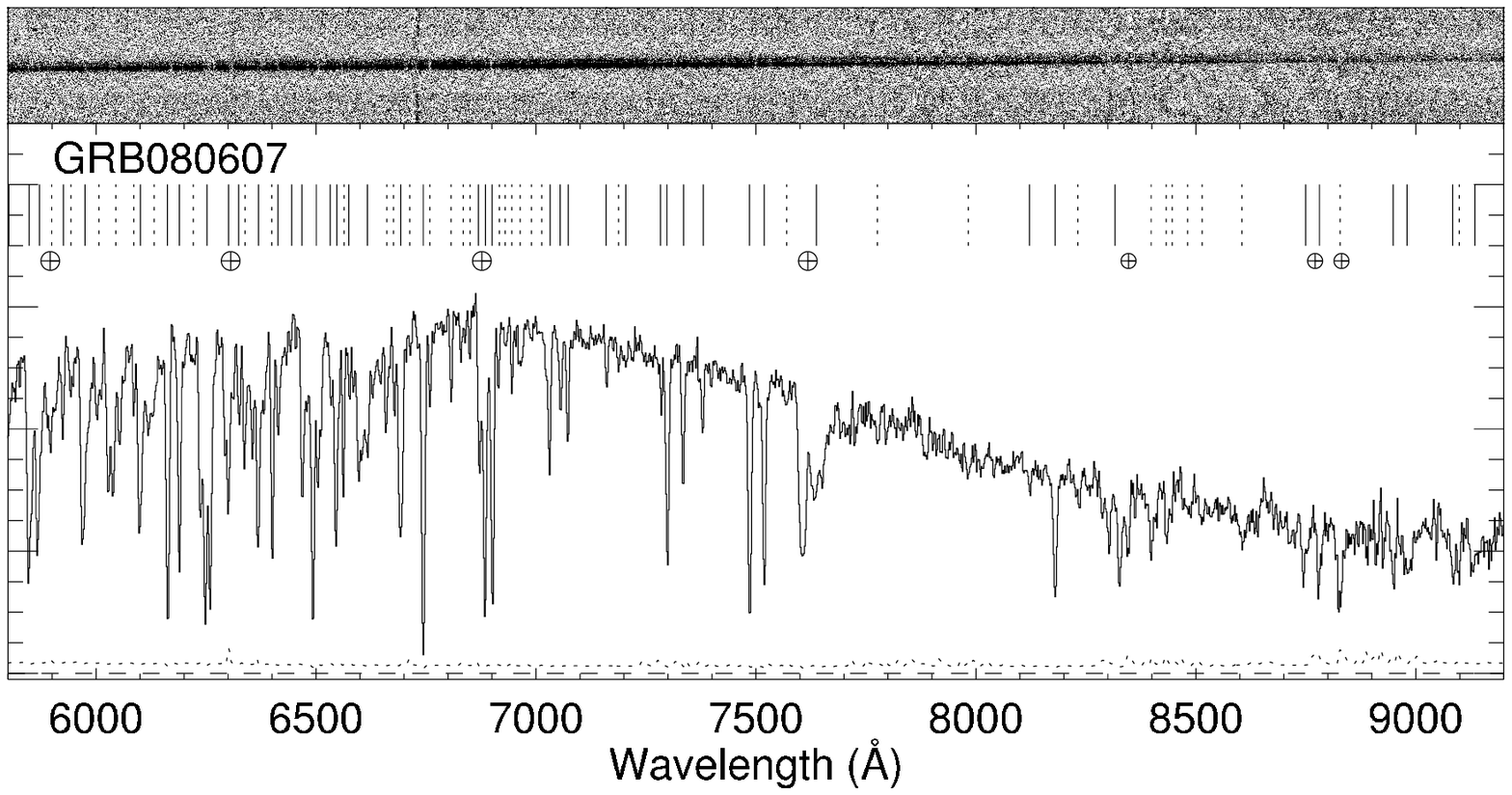}
\end{figure}

\begin{figure}
\epsscale{1.00}
\plotone{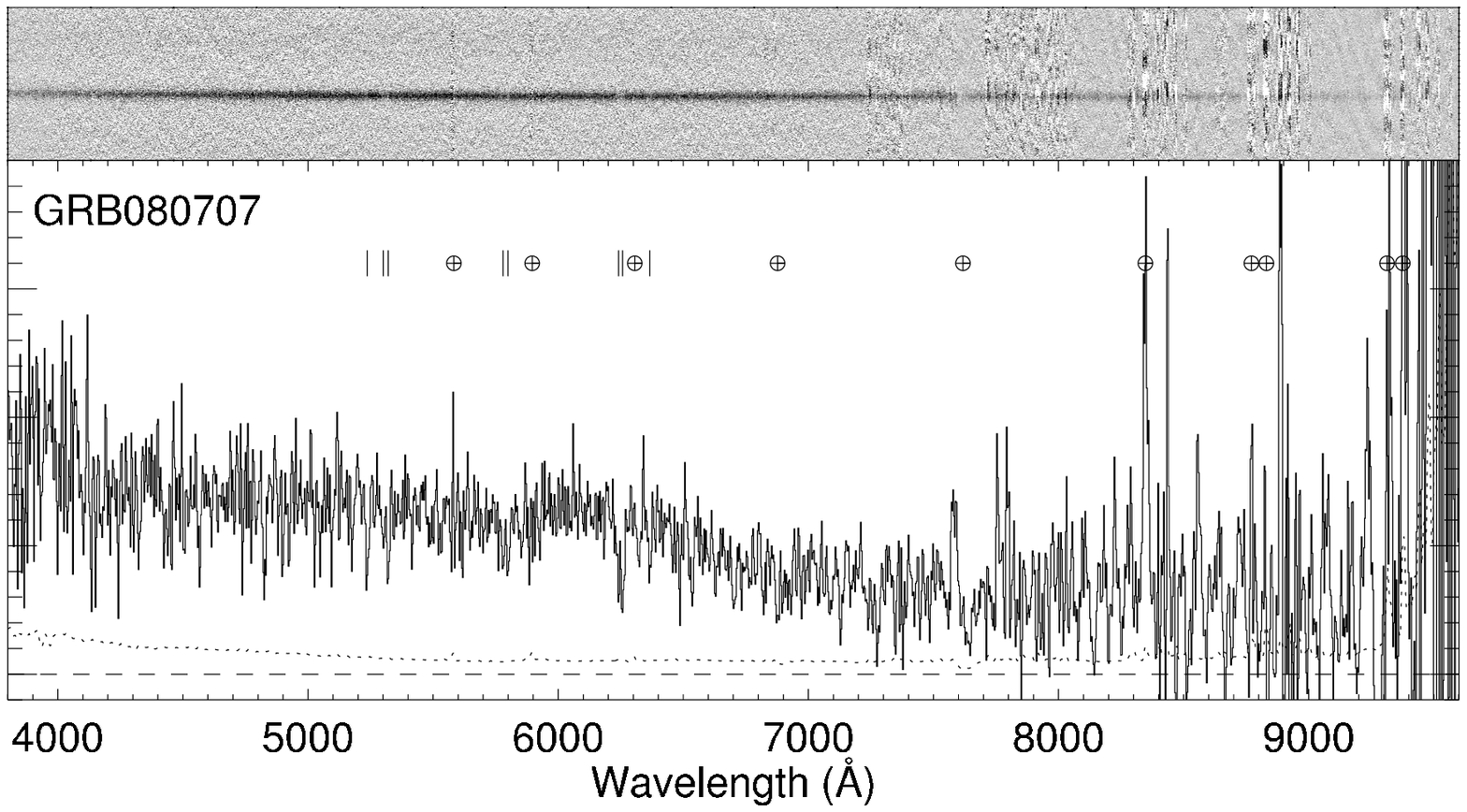}
\end{figure}

\begin{figure}
\epsscale{1.00}
\plotone{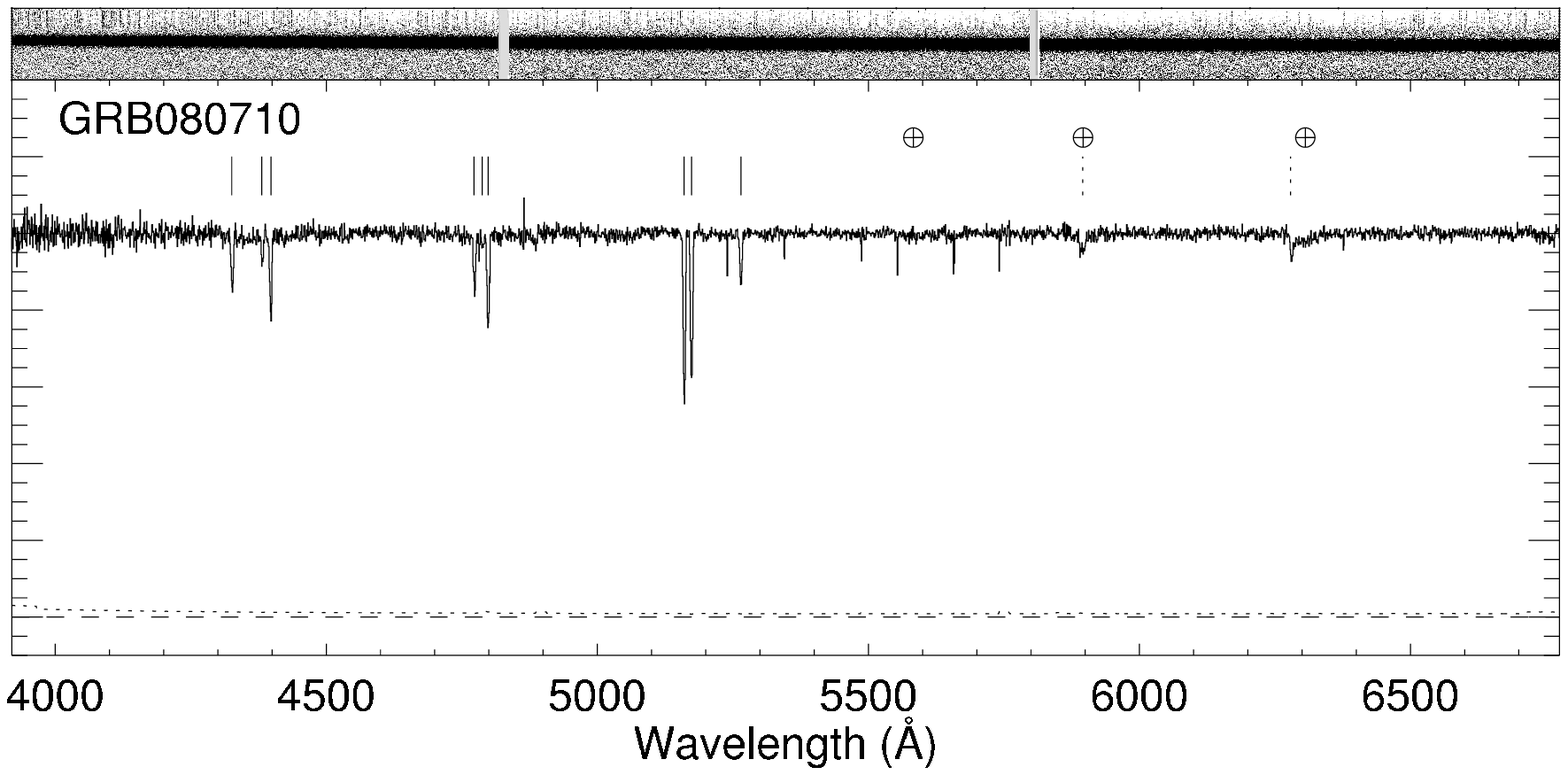}
\end{figure}

\begin{figure}
\epsscale{1.00}
\plotone{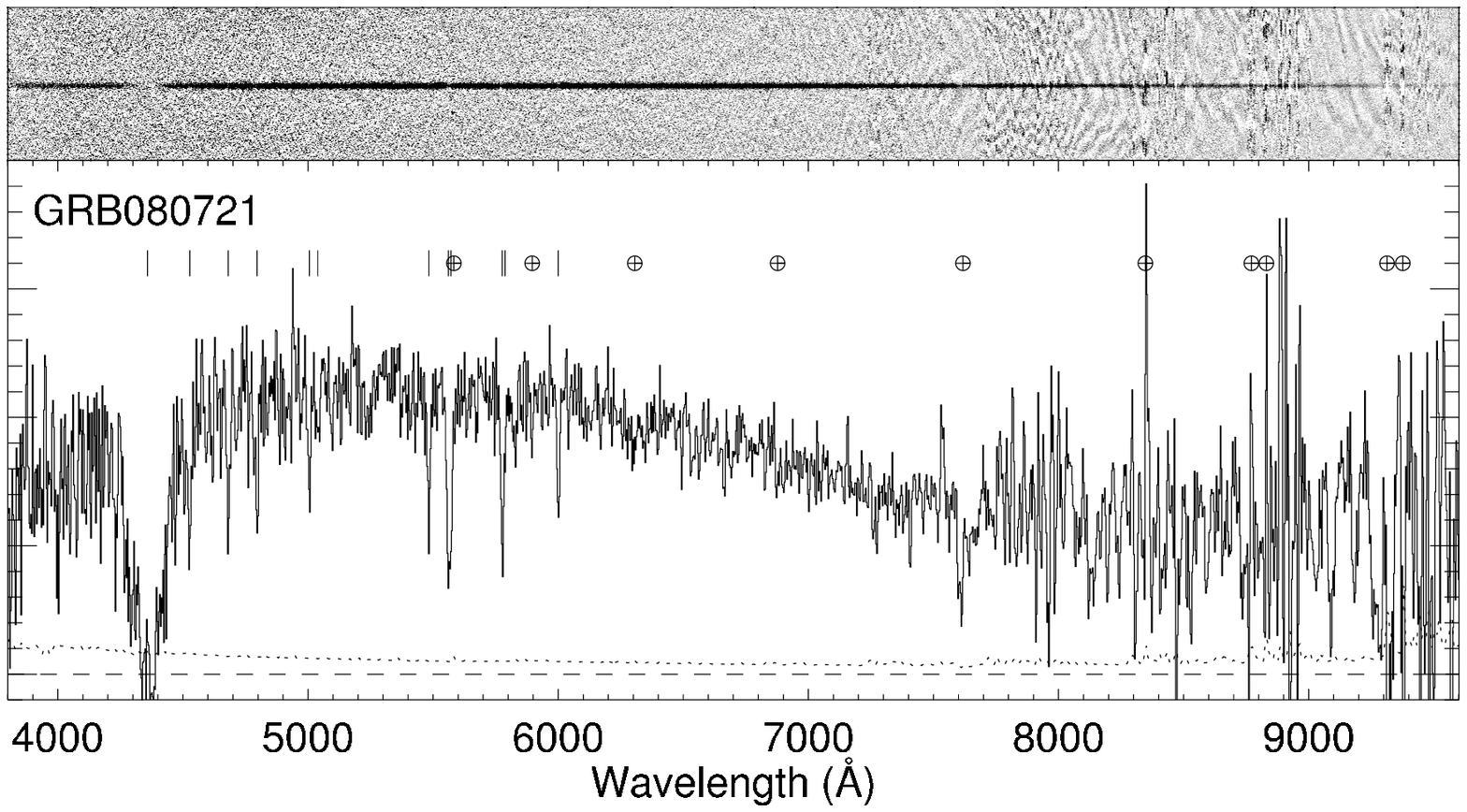}
\end{figure}

\begin{figure}
\epsscale{1.00}
\plotone{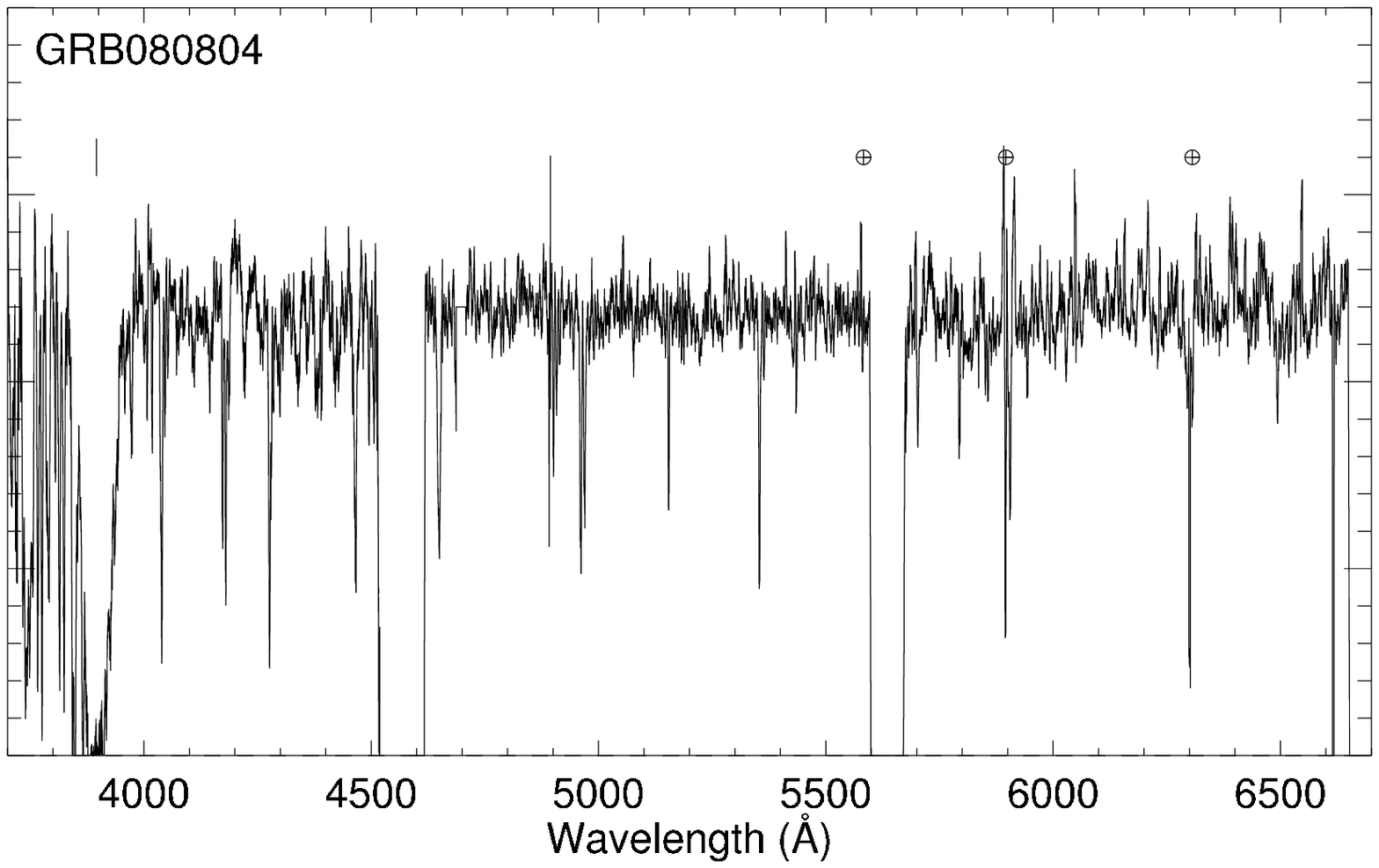}
\end{figure}

\begin{figure}
\epsscale{1.00}
\plotone{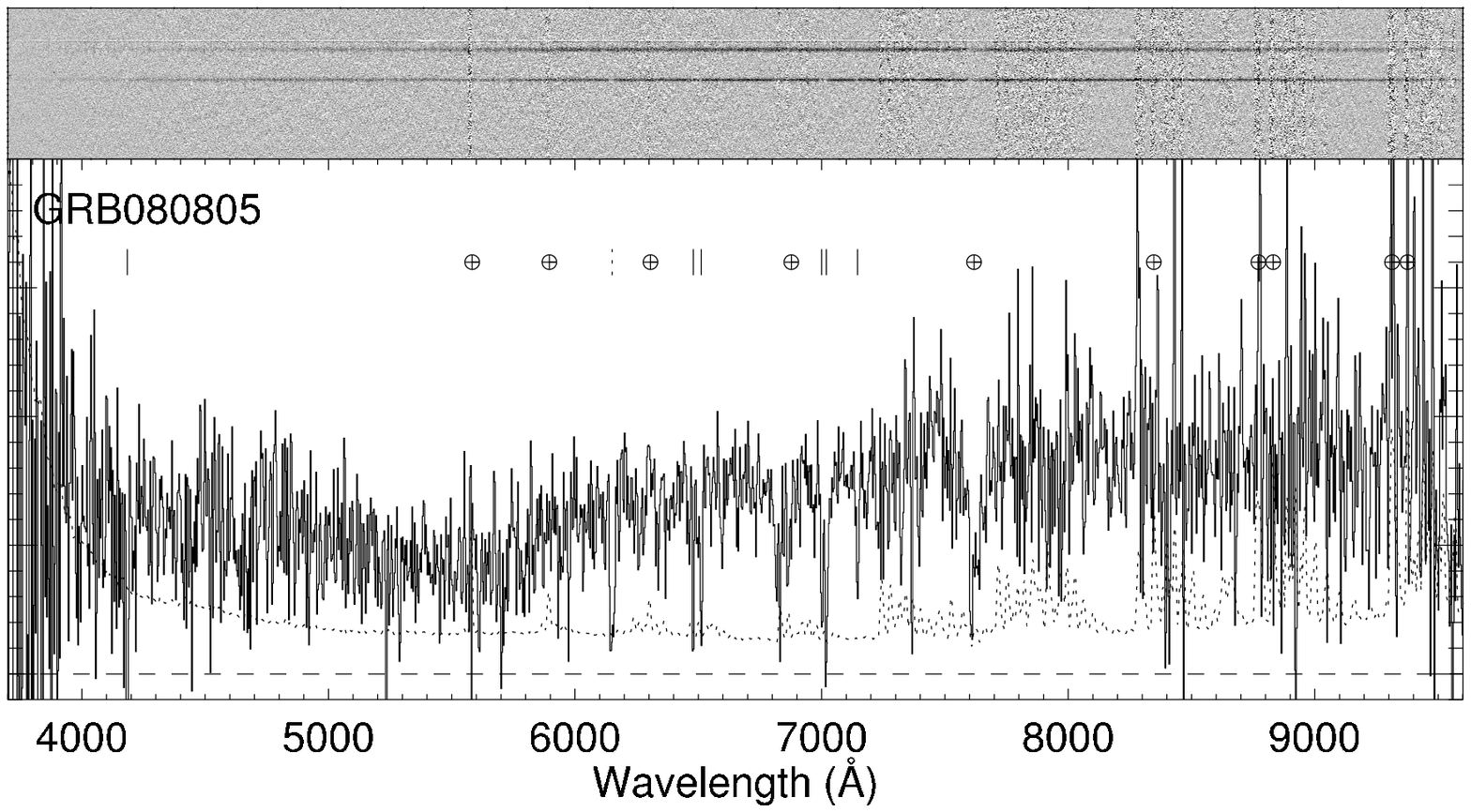}
\end{figure}

\begin{figure}
\epsscale{1.00}
\plotone{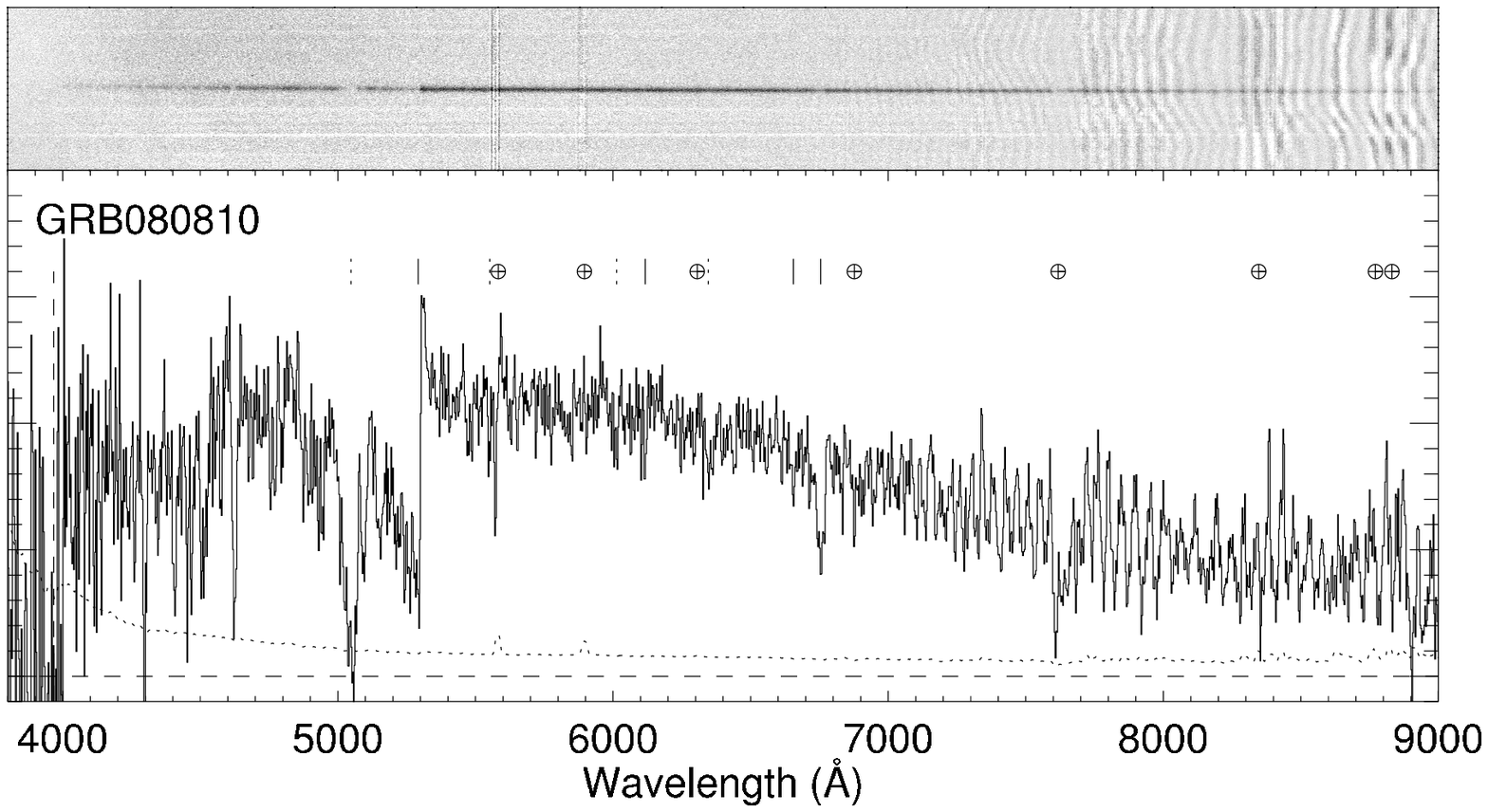}
\end{figure}

\begin{figure}
\epsscale{1.00}
\plotone{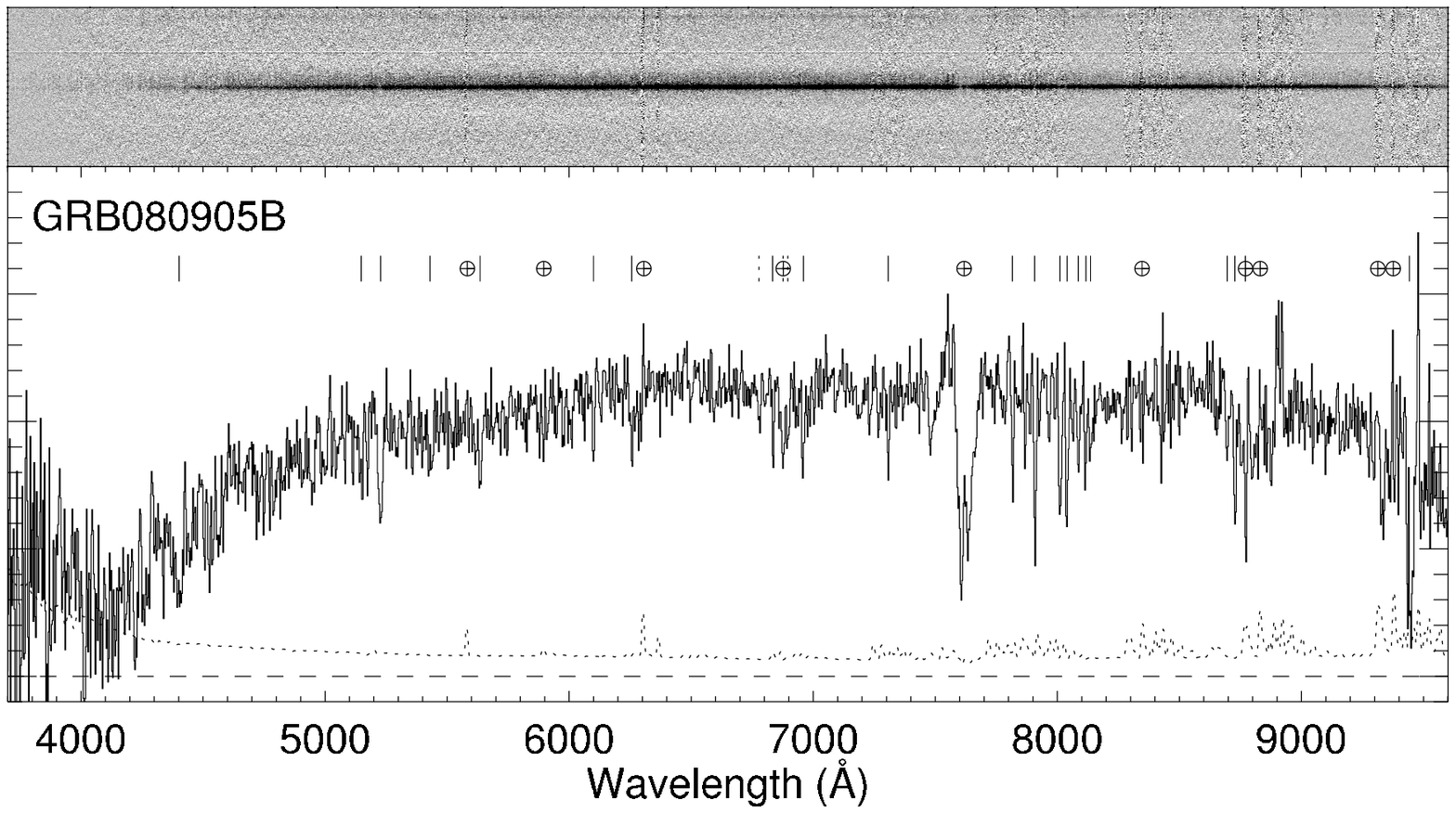}
\end{figure}

\begin{figure}
\epsscale{1.00}
\plotone{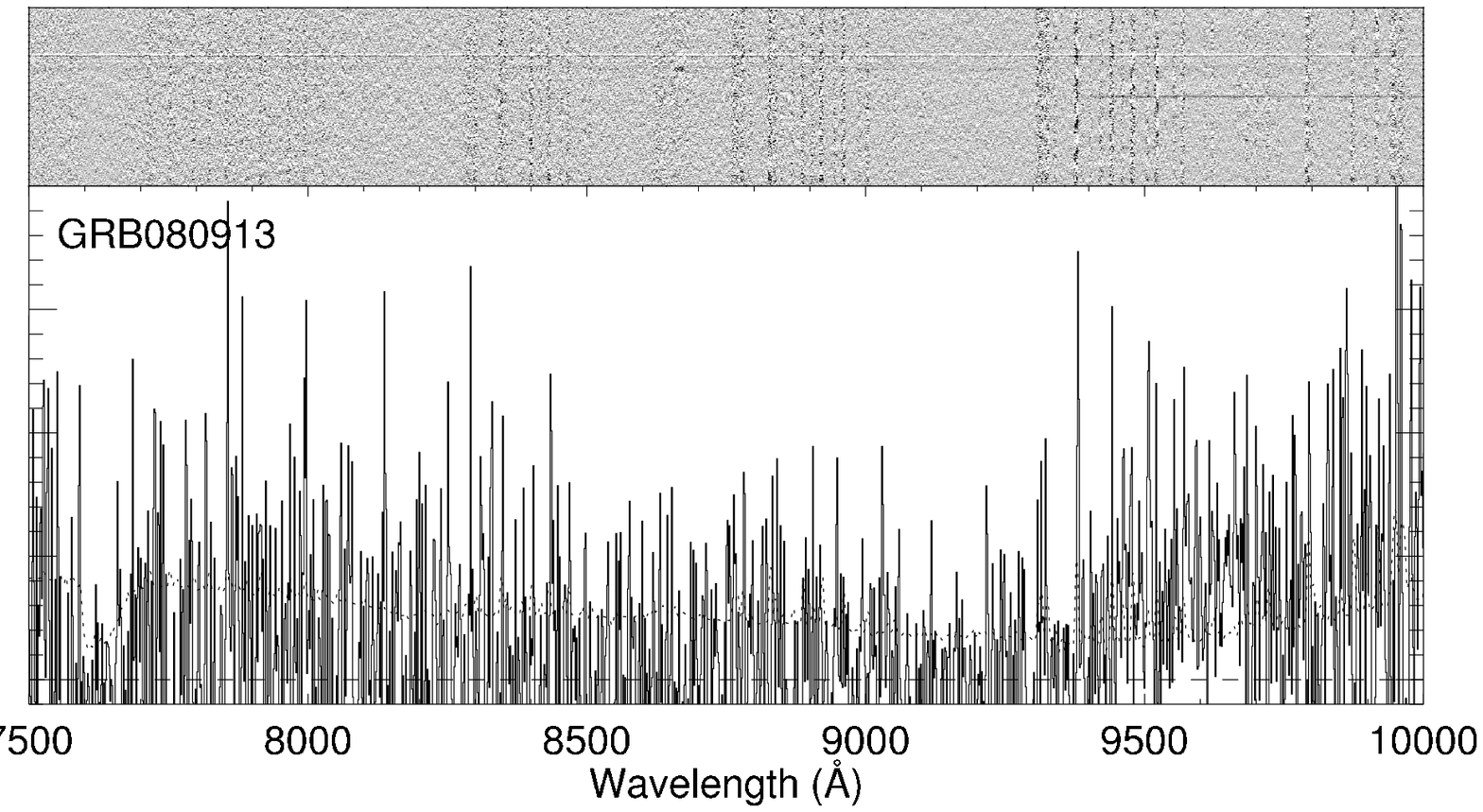}
\end{figure}

\begin{figure}
\epsscale{1.00}
\plotone{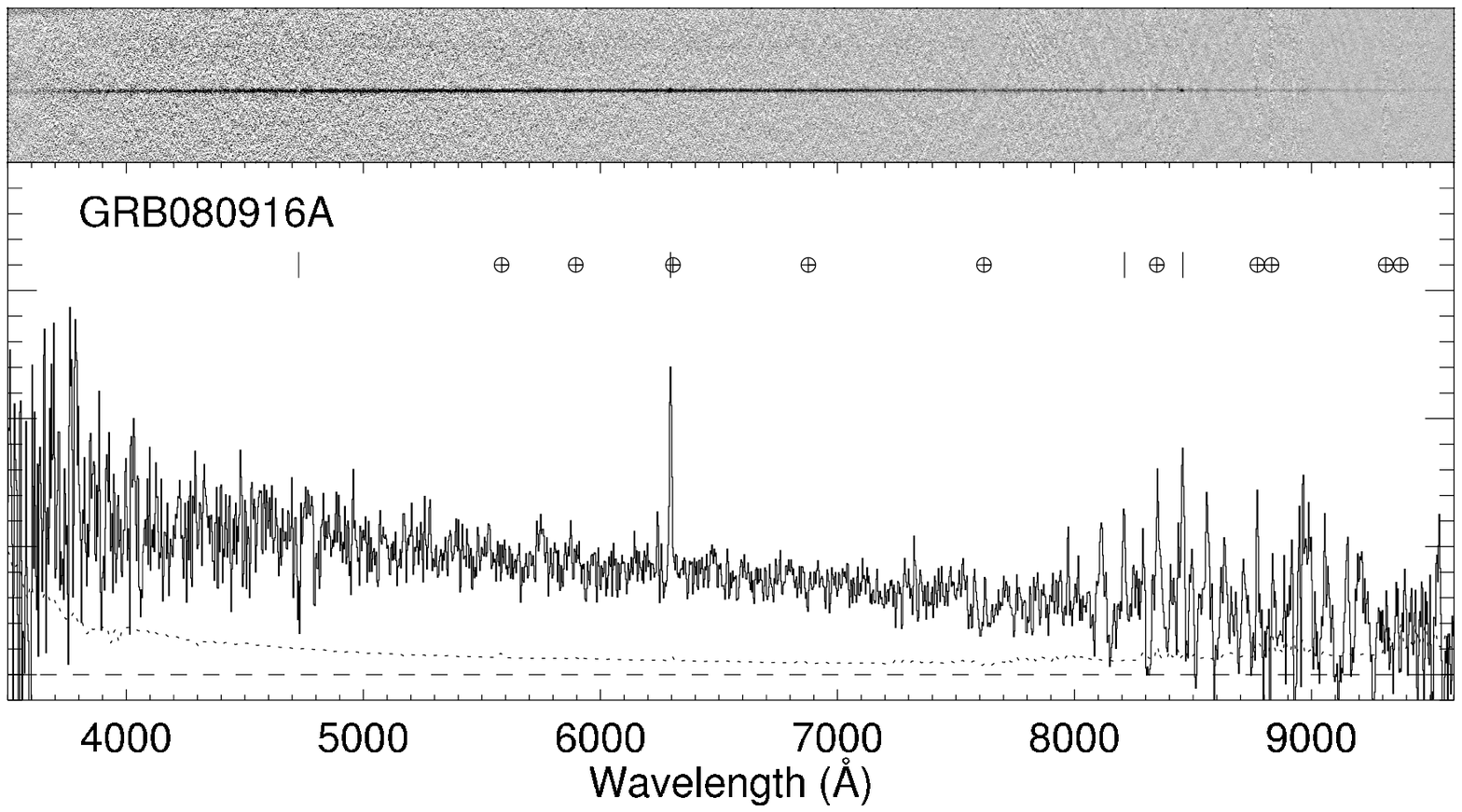}
\end{figure}

\begin{figure}
\epsscale{1.00}
\plotone{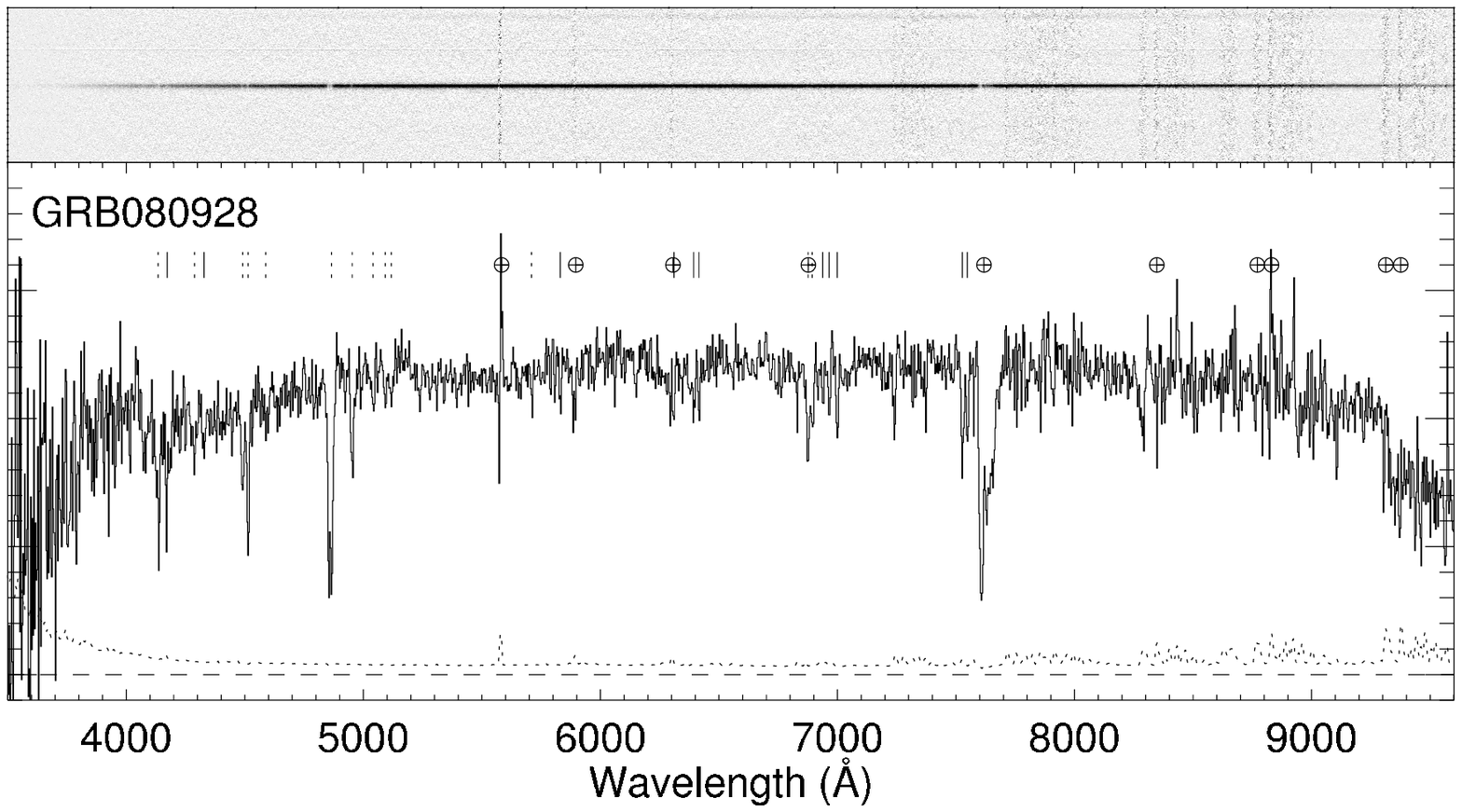}
\end{figure}

\clearpage
\begin{deluxetable}{lrcrrll}
\tablecaption{The statistical sample. The  $z$ classes are a: afterglow 
spectrum; h: host galaxy emission, p: photometric redshift from afterglow 
imaging. For bursts where the optical afterglow is not detected we provide 
the excess X-ray absorption and the corresponding redshift limit when
approriate (see maintext). We also provide references for redshift
measurements.
\label{statsample}}
\tablewidth{0pt}
\tablehead{\colhead{GRB} &
\colhead{\hspace*{2mm} $z$} &
\colhead{$z$ class} &
\colhead{N$_\mathrm{X}$ [10$^{21}$ cm$^{-2}$]} &
\colhead{$\beta_\mathrm{OX}$} &
Ref}
\startdata
080928  & 1.69         & a    & ---  & 1.00     & \citet{vreeswijk:080928} \\
080916B & ---          & ---  &  0   & $<0.90$  & --- \\
080916A & 0.69         & ah   & ---  & 0.69     & \citet{fynbo:080916A} \\
080913  & 6.70         & a    & ---  & $<0.48$  & \citet{greiner:080913} \\
080905B & 2.37         & a    & ---  & 0.64     & \citet{vreeswijk:080905B} \\
080810  & 3.35         & a    & ---  & 0.96     & \citet{antonio:080810} \\
080805  & 1.51         & a    & ---  & 0.40     & \citet{palli:080805} \\
080804  & 2.20         & a    & ---  & 0.78     & \citet{thoene:080804} \\
080727A & ---          & ---  &  0   & $<0.76$  & --- \\
080721  & 2.59         & a    & ---  & 0.60     & \citet{palli:080721} \\
080710  & 0.85         & a    & ---  & 1.04     & \citet{perley:080710} \\
080707  & 1.23         & a    & ---  & 0.83     & \citet{fynbo:080707} \\
080703  & $<5.5$       & a    & ---  & 0.42     & \citet{ward:080703} \\
080613B & ---          & ---  &  0   & $<0.79$  & --- \\
080607  & 3.04         & a    & ---  & 0.24     & \citet{X:080607} \\
080605  & 1.64         & a    & ---  & 0.38     & \citet{palli:080605} \\
080604  & 1.42         & a    & ---  & 0.90     & \citet{wiersema:080604} \\
080603B & 2.69         & a    & ---  & 0.92     & \citet{fynbo:080603B} \\
080602  & ---          & ---  & 1.0  & $<0.28$  & --- \\
080523  & $<3.0$       & a    & ---  & 0.71     & \citet{fynbo:080523} \\
080520  & 1.55         & ah   & ---  & 0.77     & \citet{palli:080520} \\
080430  & 0.77         & a    & ---  & 0.77     & \citet{cucchiara:080430} \\
080413B & 1.10         & a    & ---  & 0.89     & \citet{vreeswijk:080413B} \\
080330  & 1.51         & a    & ---  & 0.99     & \citet{guidorzi} \\
080325  & ---          & ---  & 1.5  & $<0.79$  & --- \\
080320  & $<6.4$       & a    & ---  & $<0.15$  & \citet{nial:080320} \\
080319C & 1.95         & a    & ---  & 0.31     & \citet{wiersema:080319C} \\
080319B & 0.94         & a    & ---  & 0.67     & \citet{vreeswijk:080319B} \\
080319A & $<4.2$       & h    & ---  & 0.49     & \citet{perley09} \\
080310  & 2.43         & a    & ---  & 0.88     & \citet{fox08} \\
080307  & $<6.1$       & a    & ---  & 0.40     & \citet{xin:080307} \\
080212  & $<3.5$       & a    & ---  & 0.66     & \citet{kue:080212} \\
080210  & 2.64         & a    & ---  & 0.74     & \citet{palli:080210} \\
080207  & $<3.5$ (XRT) & ---  & 6.1  & $<0.43$  & --- \\
080205  & $<5.5$       & a    & ---  & 0.79     & \citet{oates:080205} \\
071122  & 1.14         & a    & ---  & 0.83     & \citet{cucchiara:071122} \\
071117  & 1.33         & h    & ---  & 0.58     & \citet{palli:071117} \\
071112C & 0.82         & ah   & ---  & 0.72     & \citet{palli:071112C} \\
071031  & 2.69         & a    & ---  & 0.97     & \citet{cedric:071031} \\
071025  & 5.2          & ap   & ---  & 0.50     & This work \\
071021  & $<5.6$       & a    & ---  & ---      & \citet{sylvia:071021} \\
071020  & 2.15         & a    & ---  & 0.56     & \citet{palli:071020} \\
070808  & $<3.5$ (XRT) & ---  & 6.5  & $<0.48$  & --- \\
070802  & 2.45         & a    & ---  & 0.49     & \citet{ardis:070802} \\
070721B & 3.63         & a    & ---  & 0.72     & \citet{daniele:070721B} \\
070621  & $<3.5$ (XRT) & ---  & 3.2  & $<0.53$  & --- \\
070611  & 2.04         & a    & ---  & 0.73     & \citet{ct:070611} \\
070521  & 1.35         & h    & ---  & $<-0.06$ & \citet{perley09} \\
070520B & ---          & ---  & 0.6  & $<0.88$  & --- \\
070518  & $<2.0$       & a    & ---  & 0.81     & \citet{cucchiara:070518} \\
070506  & 2.31         & a    & ---  & 0.93     & \citet{ct:070506} \\
070419B & $<5.9$       & a    & ---  & 0.25     & \citet{antonio:070419B} \\
070419A & 0.97         & a    & ---  & 0.94     & \citet{cenko:070419A} \\
070412  & ---          & ---  & 0.7  & $<0.17$  & --- \\
070330  & $<5.5$       & a    & ---  & 0.68     & \citet{kuin:070330} \\
070328  & ---          & ---  & 1.8  & $<0.31$  & --- \\
070318  & 0.84         & a    & ---  & 0.78     & \citet{andreas:070318} \\
070306  & 1.50         & h    & ---  & $<0.23$  & \citet{andreas:070306} \\
070224  & $<6.1$       & a    & ---  & 0.92     & \citet{ct:070224} \\
070223  & $<6.1$       & a    & ---  & ---      & \citet{rol:070223} \\
070219  & $<3.5$ (XRT) & ---  & 2.2  & $<0.38$  & --- \\
070208  & 1.17         & ah   & ---  & 0.68     & \citet{cucchiara:070208} \\
070129  & $<3.4$       & a    & ---  & 0.62     & This work \\
070110  & 2.35         & ah   & ---  & 0.77     & \citet{andreas:070110} \\
070103  & $<3.5$ (XRT) & ---  & 2.1  & $<0.48$  & --- \\
061222A & 2.09         & h    & ---  & $<0.22$  & \citet{perley09} \\
061121  & 1.31         & a    & ---  & 0.64     & \citet{bloom:061121} \\
061110B & 3.44         & a    & ---  & 0.55     & \citet{johan:061110B} \\
061110A & 0.76         & ah   & ---  & 0.99     & \citet{johan:061110A} \\
061102  & ---          & ---  &  0   & $<0.91$  & --- \\
061021  & 0.35         & a    & ---  & 0.75     & This work \\
061007  & 1.26         & ah   & ---  & 0.79     & \citet{palli:061007} \\
061004  & ---          & ---  &  0   & $<0.47$  & --- \\
061002  & ---          & ---  & 0.6  & $<0.93$  & --- \\
060929  & ---          & ---  & 0.4  & $<0.67$  & --- \\
060927  & 5.47         & a    & ---  & 0.55     & \citet{alma:060927} \\
060923C & $<11.0$      & a    & ---  & $<0.31$  & \citet{fox:060923C} \\
060923B & ---          & ---  & 1.9   & $<0.68$ & --- \\
060923A & $<2.8$       & h    & ---  & $<0.11$  & \citet{tanvir:08} \\
060919  & $<3.5$ (XRT) & ---  & 5.5  & $<1.02$  & --- \\
060912A & 0.94         & h    & ---  & 0.62     & \citet{levan07} \\
060908  & 1.88         & a    & ---  & 0.38     & This work \\
060904A & ---          & ---  &  0   & $<0.52$  & --- \\
060814  & 0.84         & h    & ---  & $<-0.06$ & \citet{ct:060814} \\
060807  & $<3.4$       & a    & ---  & 0.54     & This work \\
060805A & ---          & ---  & 0.1  & $<1.07$  & --- \\
060729  & 0.54         & a    & ---  & 0.80     & \citet{ct:060729} \\
060719  & $<4.6$       & a    & ---  & $<-0.13$ & This work \\
060714  & 2.71         & ah   & ---  & 0.77     & \citet{palli:NH} \\
060712  & ---          & ---  & 1.1  & $<0.96$  & --- \\
060708  & 1.92         & p    & ---  & 1.04     & \citet{oates09} \\
060707  & 3.43         & a    & ---  & 0.73     & \citet{palli:NH} \\
060614  & 0.13         & h    & ---  & 0.79     & \citet{dellavalle06} \\
060607A & 3.08         & a    & ---  & 0.53     & \citet{fox08} \\
060605  & 3.78         & a    & ---  & 1.00     & \citet{savaglio:060605} \\
060604  & $<2.8$       & a    & ---  & 0.75     & \citet{blustin}\\
060526  & 3.21         & a    & ---  & 1.03     & \citet{palli:NH} \\
060522  & 5.11         & a    & ---  & 0.74     & \citet{cenko:060522} \\
060512  & 2.1          & a    & ---  & 0.98     & This work \\
060502A & 1.51         & a    & ---  & 0.65     & \citet{cucchiara:060502A} \\
060428B & $<5.5$       & a    & ---  & 1.00     & \citet{de:060428B} \\
060427  & ---          & ---  & 0.7  & $<0.81$  & --- \\
060323  & $<4.4$       & a    & ---  & 0.76     & \citet{mar:060323} \\
060319  & 1.15         & h    & ---  & $<0.41$  & D. Perley (private comm.) \\
060306  & $<3.5$ (XRT) & ---  & 3.2  & $<0.54$  & --- \\
060219  & $<3.5$ (XRT) & ---  & 2.6  & $<0.54$  & --- \\
060218  & 0.03         & h    & ---  & ---      & \citet{jesper06} \\
060210  & 3.91         & a    & ---  & 0.37     & \citet{cucchiara:060210} \\
060206  & 4.05         & a    & ---  & 0.95     & \citet{johan:060206} \\
060204B & $<4.8$       & a    & ---  & 0.47     & \citet{johan:060204B} \\
060202  & 0.78         & h    & ---  & $<0.20$  & \citet{butler07} \\
060124  & 2.30         & a    & ---  & 0.80     & \citet{mirabal:060124} \\
060115  & 3.53         & a    & ---  & 0.78     & \citet{piranomonte:060115} \\
060111A & $<5.5$       & a    & ---  & 0.70     & \citet{blustin:060111A} \\
060108  & $<3.2$       & a    & ---  & 0.51     & \citet{oates06} \\
051117B & ---          & ---  & 0.6  & $<1.06$  & --- \\
051016B & 0.94         & h    & ---  & 0.63     & \citet{alicia:051016B} \\
051006  & $<3.5$ (XRT) & ---  & 5.1  & $<1.30$  & --- \\
051001  & ---          & ---  & 0.9  & $<0.56$  & --- \\
050922C & 2.20         & a    & ---  & 0.99     & \citet{palli:NH} \\
050922B & ---          & ---  & 1.1  & $<0.58$  & --- \\
050915A & $<15.0$      & a    & ---  & $<0.44$  & \citet{bloom:050915A} \\
050908  & 3.34         & a    & ---  & 1.14     & \citet{fugazza:050908} \\
050904  & 6.30         & a    & ---  & $<0.41$  & \citet{kawai06} \\
050824  & 0.83         & ah   & ---  & 0.91     & \citet{jesper:050824} \\
050822  & ---          & ---  & 0.6  & ---      & --- \\
050820A & 2.61         & a    & ---  & 0.77     & \citet{fox08} \\
050819  & ---          & ---  &  0   & $<0.90$  & --- \\
050814  & 5.3          & p    & ---  & 0.51     & \citet{jakobsson06a} \\
050803  & ---          & ---  & 1.2  & $<-0.15$ & --- \\
050802  & 1.71         & a    & ---  & 0.51     & \citet{johan:050802} \\
050801  & 1.38         & p    & ---  & 0.95     & \citet{oates09} \\
050730  & 3.97         & a    & ---  & 0.79     & \citet{delia:050730} \\
050726  & $<5.5$       & a    & ---  & $<0.89$  & \citet{poole:050726} \\
050716  & $<11.0$      & a    & ---  & $<0.28$  & \citet{rol07} \\
050714B & $<3.5$ (XRT) & ---  & 2.2  & $<0.48$  & --- \\
050525A & 0.61         & ah   & ---  & 0.92     & \citet{foley:050525A} \\
050505  & 4.28         & a    & ---  & 0.53     & \citet{berger06} \\
050502B & $<6.4$       & a    & ---  & $<0.58$  & \citet{cenko:050502B} \\
050416A & 0.65         & h    & ---  & 0.70     & \citet{alicia07} \\
050412  & ---          & ---  &  0   & $<0.60$  & --- \\
050406  & 2.7          & p    & ---  & 1.02     & \citet{schady06} \\
050401  & 2.90         & a    & ---  & 0.36     & \citet{darach:050401} \\
050319  & 3.24         & a    & ---  & 0.90     & \citet{palli:NH} \\
050318  & 1.44         & a    & ---  & 0.75     & \citet{berger05} \\
050315  & 1.95         & a    & ---  & 0.63     & \citet{berger05} \\
\enddata
\end{deluxetable}

\end{document}